%% file: Hilla_De_Leon_phd_thesis3.tex
\documentclass[twoside,12pt]{book}
\usepackage[inner=2.76cm, outer=2.76cm, bottom=1in]{geometry}
\usepackage{fancyhdr}
\usepackage{graphicx}
\usepackage{amsthm}
\usepackage{amsmath}
\usepackage{longtable}
\usepackage{notoccite}
\usepackage{soul}
\usepackage{graphics}
\usepackage{amssymb}
\usepackage{cite}
\usepackage{epsfig}
\usepackage{verbatim}
\usepackage{epstopdf}
\usepackage{amsfonts}
\usepackage{transparent}
\usepackage{hhline} 
\usepackage{array}
\usepackage{multirow}
\usepackage{soul}
\usepackage{setspace}
\usepackage{pdfpages}
\usepackage{caption}
\usepackage{subcaption}
\usepackage{cancel}
\usepackage{amsmath}
\usepackage{wrapfig}
\usepackage{float}
\usepackage{titlesec}
\usepackage{cleveref}
\usepackage{cite}
\usepackage{enumerate}
\usepackage{array}
\usepackage{colortbl}
\usepackage{framed}
\usepackage{slashbox}
\usepackage[toc, page]{appendix}
\usepackage{pdfpages}

\newcommand{\ket}[1]{\left| #1 \right>} % for Dirac bras
\newcommand{\bra}[1]{\left< #1 \right|} % for Dirac kets

\newcommand{ \pilesseft}{\mbox{$\pi\text{\hspace{-6 pt}/}$}EFT$\, $}

\def\authorname{}
\def\ttitle{}

\newcommand*{\thesistitle}[1]{\def\@title{#1}\def\ttitle{#1}}

\renewcommand*{\author}[1]{\def\authorname{#1}}

\newcommand{\be}{\begin{epuation}}
\newcommand{\ee}{\end{equation}}

\newcommand{\mev}{\, \text{MeV}}
\newcommand{\kev}{\, \text{keV}}

\newcommand{\fm}{\, \text{fm}}

\newcommand{\ceft}{\mbox{$\chi$}EFT$\, $}
\newcommand{\cblack}{\color{black} }

\usepackage[Lenny]{fncychap}
\usepackage{fancyhdr}
\usepackage{fix-cm} 
\usepackage{afterpage}

\makeatletter
\newcommand\HUGE{\@setfontsize\Huge{40}{50}}
\makeatother 

\begin{document}
	\onehalfspace
	%\begin{comment}
	\pagenumbering{gobble}	
	\setlength{\voffset}{0cm}
	\setlength{\hoffset}{0cm}
\includepdf[pages=-,fitpaper=true]{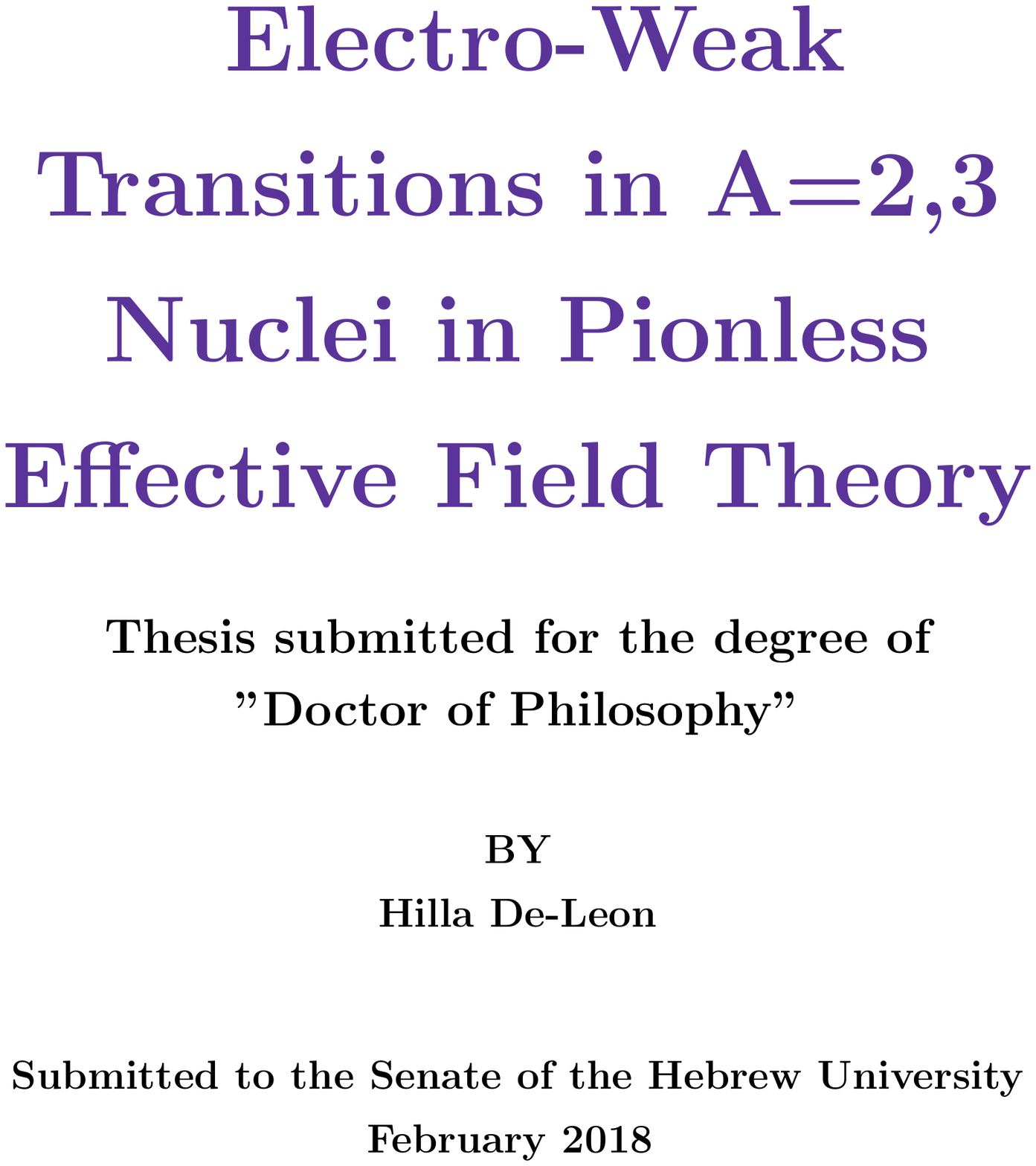}
% \includepdf[pages=1-2]{George_Harrison.pdf}
	%\chapter*{Acknowledgments}
			\thispagestyle{empty} 
	\thispagestyle{empty} 
%		\thispagestyle{empty} 
%		\end{comment}
\pagenumbering{roman}
	\setcounter{page}{0}
%\addcontentsline{toc}{chapter}{Abstract}

\chapter*{Abstract} 
\addcontentsline{toc}{chapter}{Abstract}
\vspace{-1 cm}	
In this thesis, which is in the field of few-body systems, I am studying low-energy electro-weak interactions in light nuclear systems whose number of nucleons, $A$, is smaller than 4 ({\it i.e.,} $d$, $^3$H, $^3$He), where the main purpose of this work is to calculate the proton-proton ($pp$) fusion rate. 

The main source of energy generated in the Sun is a set of exothermic reactions (named the $pp$ chain) in which 4 protons are fused into $^4$He. The first and the slowest reaction of this set is the $pp$ fusion, which is a two-nucleon weak reaction that fuses two protons into deuteron ($d$) and determines the Sun's lifetime ($\tau\sim 10^9$ years). However, due to its long lifetime it cannot be measured, so its rate, which is proportional to the matrix element squared of the weak interaction between initial and final nuclear states, has to be estimated from theory only. 

The relevant theory of physics at low energies is Quantum Chromodynamics (QCD) but unfortunately, a direct calculation of the $pp$ fusion matrix element is not trivial due to the non-perturbative character of QCD at low energies. In recent decades, many nuclear physicists faced the challenge of calculating the $pp$ fusion rate, but these calculations, and in particular those conducted in the last years, are inconsistent with each other. This inconsistency emphasizes the necessity of deriving a new, simple calculation that should benchmark versus other past predictions and would provide a verified and validated prediction.%and reproduce different experimentally measured reactions, 
%, which is the \textbf{goal of this work}. 

In the past two decades, a novel theoretical method, named Effective Field Theory (EFT), revolutionized nuclear physics. EFT is a simple, renormalizable and model independent theoretical method for describing low-energy reactions. The condition for describing a physical process using EFT is that its transfer momentum, $Q$, is small compared to a physical cutoff, $\Lambda_{cut}$ ({\it i.e.,} $Q/\Lambda_{cut}\ll 1$). In this thesis, I focus on electro-weak interactions that are characterized by momentum transfer, $Q$, which is much smaller than $\Lambda_{cut}=m_\pi$, the pion mass. In such cases, the corresponding EFT is the pionless EFT (\pilesseft), where the pions are integrated out and only nucleons are left as effective degrees of freedom. 

 \pilesseft weak Lagrangian includes a two-body low-energy constant (LEC), named $L_{1, A}$, which has to be calibrated from different well-measured, low-energy weak reaction that contains at least two nucleons. An appropriate well-measured, low-energy reaction for that calibration is the $^3$H $\beta$-decay that involves $L_{1, A}$, so using \pilesseft for its calculation, enables the extraction of $L_{1, A}$. 

Nevertheless, in contrast to the nuclear electro-weak interactions containing \textbf{two} nucleons that have been calculated over the last 20 years using \pilesseft, electro-weak interactions which involve \textbf{three} nucleons (such as $^3$H $\beta$-decay) were not introduced until recently due to the complexity in describing these systems using \pilesseft formalism. Consequently, the consistency of \pilesseft for the transition between $A=3$ and $A=2$ nuclear systems, which is essential for the $L_{1, A}$ extraction, could have never been examined. Since the $^3$H $\beta$-decay is the only relevant low-energy %\cblack 
$1<A<4$ \cblack well-measured weak reaction, weak reactions cannot be used for examining the \pilesseft consistency, and 
therefore, another set of well-measured low-energy $A<4$ interactions with similar characteristics to the weak reactions, is needed for validation and verification of \pilesseft. The strong analogy between the \textbf{electromagnetic} to weak observables indicates that the four $A<4$ well-measured electromagnetic observables are the ones that serve as the required candidates. 

Hence, in this thesis I develop a general perturbative diagrammatic approach for calculating the transition between $A=3$ bound-states of three-nucleon. This framework has been examined and found to be accurate not only for calculating electro-weak interactions, but also for other observables such as the binding energy difference between $^3$H and $^3$He due to the Coulomb interaction. 
\cblack
I show that the electromagnetic calculations up to next-to-leading order (NLO) in $Q/\Lambda_{cut}$ reproduce the experimental data very well, where the NLO contributions amount to a few percent only. The order-by-order analysis of those observables along with the high consistency between the $A=2$ and $A=3$ calculations lead to high theoretical uncertainty of less than 1\%. This theoretical uncertainty estimate, accompanied by the strong analogy between the electromagnetic and weak observables, enables the extraction of $L_{1, A}$ from the calculated $^3$H $\beta$-decay and lead to the high accuracy calculation of the $pp$ fusion rate. 

This calculation, which is indeed simple and benchmarks versus past predictions, 
shows that the present prediction for the $pp$ fusion cross-section is up to 5\% bigger than previous estimations and contains 1.2\% total uncertainty with 70\% degree of belief. This might affect current models of solar evolution.
\cblack
%\addcontentsline{toc}{chapter}{Abstract}

	%\pagenumbering{roman}
	\tableofcontents
	\listoffigures
	\newpage
	\listoftables
	\newpage
	\pagenumbering{arabic}

	\chapter{Introduction}
	The evolution of the Sun, as well as other main sequence stars, remains one of the main theoretical questions in astrophysics. Our comprehensive understanding of this evolution has been used as a tool to reach many discoveries in other branches of physics, heralded by the discovery of the mass of the neutrino. The energy generated in the Sun comes from an exothermic set of reactions, 
	the proton-proton ($pp$) chain, by fusing four Hydrogen ions into $^4$He. %$4p\rightarrow ^4\text{He}+2e^++2\nu_e.$ 
	The chain is initiated by the $pp$ fusion reaction: 
	$p+p\rightarrow d+e^++\nu_e$, which rules 99.76\% of the proton reactions. The $pp$ fusion is governed by the weak interaction, which makes it the slowest reaction in the chain ($\tau \sim 10^9$ years), and therefore it determines the Sun's lifetime. However, its slow rate makes the measurement of its cross-section impossible, so it must be calculated theoretically.
	
The $pp$ fusion cross-section, 
$\sigma_{pp}=\frac{S^{11}(E)}{E}\exp[-2\pi\eta(E)]$, consists of both the long-range Coulomb factor ($\eta(E)$), and the short-range astrophysical $S$-factor ($S^{11}(E)$), where $E$ is the kinetic energy of the center of mass of the interacting protons, dictated by the temperature. In solar conditions, the magnitude of $E$ is only a few $\kev$ and therefore $S^{11}(E)$ can be expanded in a power series in $E$ such that:
	\newline
	 $S^{11}(E)=S^{11}(0)+\frac{dS^{11}(E)}{dE}|_{E=0}E+...$. At these solar energies this series expansion is dominated by the $E=0$ threshold value, {\it i.e.,} $S^{11}(0)$ which is proportional to $\Lambda_{pp}^2(0)$, the square of the $pp$ fusion matrix element.

	The relevant theory of physics at these energies is Quantum ChromoDynamics (QCD). The challenge in theoretical calculation of the $pp$ fusion matrix element stems from the non-perturbative character of QCD at the nuclear regime which makes a direct calculation non-trivial. This has led to many theoretical efforts to calculate the reaction rate. 
	
	In 1969, Bahcall and May have found that $\Lambda_{pp}^2(0)=7.08$ \cite{1969ApJ...155..501B}, using standard
	nuclear physics effective range expansion (ERE) \cite{PhysRev.76.38}. A review of all recent calculations of $S^{11}(0)$ appeared in 2011, by Adelberger {\it et al.} \cite{pp_review}, and recommended to use the value
	\begin{equation}
	S^{11}(0)=(4.011\pm0.04)\cdot 10^{-23}\mev\cdot\fm^2 \cdot \left(\frac{(ft)_{0^+\rightarrow0^+}}{3071.4\, \text{sec}}\right)^{-1}\left(\frac{g_A}{1.2695}\right)^2\left(\frac{f^R_{pp}}{0.144}\right)\left(\frac{\Lambda_{pp}^2(0)}{7.035}\right), 	
	\end{equation}
	representing an agreement between three different theoretical approaches that existed at that time, 
	%\color{red} 
	where $(ft)_{0^+\rightarrow0^+}= (3071.4\pm0.8)s$, is the value for
	superallowed $0^+\rightarrow0^+$ transitions that has been determined from a comprehensive analysis of experimental
	rates corrected for radiative and Coulomb effects \cite{PhysRevC.79.055502}. This value determines the weak
	mixing matrix element, $|V_{ud}| = 0.97418(27)$ \cite{gA_1.2695}. Ref.~\cite{solar2} used the then recommended PDG value for the weak axial constant: $g_A = 1.2695\pm
	0.0029$~\cite{gA_1.2695}.
	The phase-space factor $f^R_
	{pp}$ takes into account 1.62\% increase due to radiative corrections to the cross section \cite{PhysRevC.67.035502}.
	\color{black}
	
	However, in 2013, Marcucci {\it et al.} calculations for proton energies up to $100 \kev$ ~\cite{PhysRevLett.110.192503}, have yielded $S^{11}(0)=(4.03\pm0.006)\cdot 10^{-23}\mev\cdot\fm^2$, using the consistency of three-nucleon forces and two-body axial currents \cite{PhysRevLett.103.102502}, while Acharya {\it et al.} recently yielded %using $\chi$EFT up to next-to-next-leading-order (NNLO): 
	$S^{11}(0) = \left(4.081^{+0.024}_{-0.032}
	\right) \cdot 10
	^{-23}
	\text{MeV}\cdot \text{fm}^2$ \cite{Acharya:2016kfl}.
	
 These two predictions \cite{PhysRevLett.110.192503, Acharya:2016kfl} are inconsistent with each other. This inconsistency indicates that the uncertainties might be underestimated, and emphasizes the need for a new calculation which has a comprehensive error estimation. The main purpose of this thesis is to construct such a new calculation that will identify the theoretical and empirical uncertainties, preferably would be simple, should benchmark versus other past predictions, and must reproduce different experimentally measured reactions. This might enable us a verification of the theoretical method that is needed for prediction of the elusive
	$pp$ fusion cross-section.

\begin{figure}
		\vspace{-0.3 cm}
		\centering
			\includegraphics[width=1\linewidth]{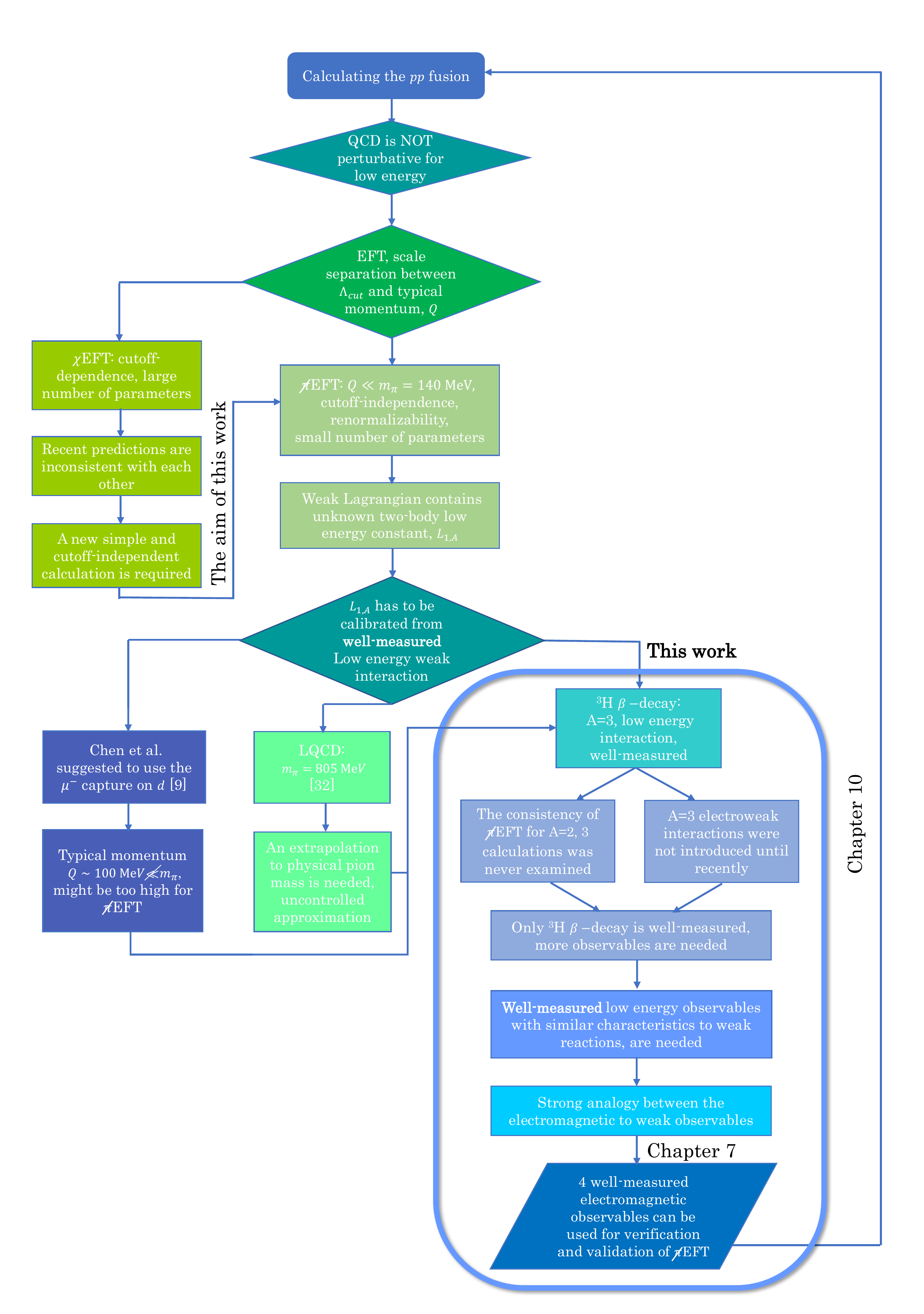}\\
		\caption{\footnotesize{Flow chart that demonstrates the main steps leading to the present \pilesseft calculation of the $pp$ fusion.}}\label{fig_flow_chart}
	\end{figure}

We focus on a novel theoretical method, which revolutionized nuclear physics in the past two decades, named Effective Field Theory (EFT). We are using the pionless version of this approach (\pilesseft) to calculate the $pp$ fusion matrix element.
 In contrast to past works attempted to calculate the $pp$ fusion within \pilesseft, e.g. J.W. Chen {\it et al.} \cite{L1A}, we use $^3$H $\beta$-decay to calibrate the model. This is done by developing a perturbative diagrammatic approach to calculate the transition between $A=3$ bound-states (where $A$ is the number of nucleons). 
 A significant part of this work is dedicated to estimate the theoretical and empirical uncertainties relevant for calculating low-energy observables such as the $pp$ fusion rate.
 We find a successful validation of this approach by studying the magnetic structure of $A=2, 3$ nuclei. 
 The main steps leading to the \pilesseft calculation of the $pp$ fusion matrix element in this thesis are demonstrated in Fig.~\ref{fig_flow_chart}. 

 In the next sections we briefly review the basic principles of effective field theory, with emphasis on its chiral ($\chi$EFT) and pionless (\pilesseft) versions and the challenges of using \pilesseft for calculating the $pp$ fusion matrix element. We examine the analogy between electromagnetic and weak observables in \pilesseft, which along with the comparison between the \pilesseft predictions of the electromagnetic interactions to their known experimental counterparts, are used for verification and validation of \pilesseft in calculating both electromagnetic and weak observables of $A=2, 3$ nuclei.

	\section{Effective field theory: the modern approach to predictions in nuclear physics}
	In recent years, an effort has begun to calculate
	low-energy observables such as the $pp$ fusion matrix element, using EFT approaches. In
	EFT, the ratio of two separated scales is used as an expansion
	parameter in a systematic low-energy expansion in the calculation of
	observables. The first, which represents the low-energy scale is usually the typical momentum scale, 
	$Q$, while the second and higher energy scale, $\Lambda_{\rm cut}$, is frequently related
	to the lightest exchanged particle. The EFT has to preserve all symmetries of
	the fundamental theory, and the resulting Lagrangian includes only the
	relevant degrees of freedom while heavier excitations are integrated
	out of the theory. Thus, one can obtain an expression of
	non-renormalizable interactions which can be organized as a power
	series in $Q/\Lambda_{\rm cut}$ \cite{few, Kaplan1996629, KSW1998_a, 
		Kaplan:1998tg, KSW_c}. This method becomes useful when there is a significant scale separation between $Q$ and $\Lambda_{\rm cut}$, so only a small number of the effective operators corresponding to the leading powers in $\frac{Q}{\Lambda_{\rm cut}}$ needs to
	be retained to reproduce long wavelength observables with the desired accuracy.
	
	Regarding the $pp$ fusion matrix element calculations, one finds two main EFT approaches in the literature, which are relevant for this calculation: the chiral EFT ($\chi$EFT) and the pionless EFT (\pilesseft). In the last 15 years, most of the $pp$ predictions used $\chi$EFT \cite{PhysRevLett.110.192503, Acharya:2016kfl}.

	\section{$\chi$EFT predictions of the $pp$ fusion matrix element}
	Chiral effective field theory ($\chi$EFT) identifies the pion as a Goldstone boson of the spontaneously broken chiral symmetry of QCD, stemming from the small masses of the up and down quarks. The finite mass of the pion
	originates from the explicit chiral symmetry breaking, {\it i.e.,} the non-zero up and down quark masses. This framework has been used to construct a nuclear interaction that describes nuclear spectra and scattering observables to very high accuracy
	 \cite{Epelbaum:2008ga}. It has also been used to construct the nuclear electro-weak current operator and for subsequent calculations \cite{Bacca:2014tla, doi:10.11}. These calculations have highlighted the benefit of an approach in which current operators and the interaction between nucleons are treated consistently.
	
	As mentioned before, $\chi$EFT has been used by Marcucci {\it et al.} \cite{PhysRevLett.110.192503} to calculate the S-factor up to next-to-next-leading-order (NNLO), $S^{11}(0)=(4.03\pm0.006)\cdot 10^{-23}\mev\cdot\fm^2$. The uncertainty of their calculation, which is almost an order of magnitude smaller than a previous recommendation \cite{pp_review}, is based only on the difference between two variants of \ceft potentials. Recently, Acharya {\it et al.} \cite{Acharya:2016kfl} have recalculated the S-factor of $pp$ fusion reaction using $\chi$EFT pu to NNLO: 
	$S^{11}(0) = \left(4.081^{+0.024}_{-0.032}
	\right) \cdot 10
	^{-23}
	\text{MeV}\cdot \text{fm}^2$. Similarly to Marcucci {\it et al.} \cite{Acharya:2016kfl}, Acharya {\it et al.} \cite{Acharya:2016kfl} also used the consistency of three-nucleon forces and two-body axial currents, where the relevant low-energy constants (LECs) were calibrated using the triton $\beta$-decay and binding energies. Their uncertainty estimate is based mainly on the statistical uncertainties in the low-energy coupling constants of $\chi$EFT, on the systematic uncertainty due to limited $\chi$EFT cutoff dependence and on the variations in the database used to calibrate the nucleon-nucleon interaction.

	In both Refs.~\cite{PhysRevLett.110.192503, Acharya:2016kfl}, the theoretical uncertainty due to the truncation of the EFT was taken into account, only in a very limited regime, $\Lambda_{cut}\approx400-600\mev$, thus problematic for robust error estimate~\cite{Melendez:2017phj}\cblack. Moreover, the cutoff dependence of $\chi$EFT, as well as a large number of parameters sharpen the need to use a different approach for predicting the $pp$ fusion and mainly its uncertainties. 
	
 In this work we use a \textbf{renormalizable} and \textbf{cutoff independent} EFT approach named pionless EFT (\pilesseft) which is an alternative EFT approach, 
that is particularly useful at low energies that
are of interest for astrophysical processes, {\it i.e.,}
$Q\sim10\text{ MeV} \ll m_{\pi}=$140 MeV, where $m_\pi$ is the pion mass. The interactions in \pilesseft are only contact vertices, and the pion mass sets the breakdown scale of this approach (%\color{red} 
see for examples: 
	 \cite{Griesshammer_pionless,Griesshammer_3body, few_platter}).\color{black} Using this theory all particles but nucleons are ``integrated out''. The renormalizability which ensures cutoff independence, the small number of parameters as well as their obvious relation to measured quantities (e.g., nucleon-nucleon scattering length and effective ranges), and the natural perturbative expansion are all making \pilesseft ideal for predictions of low-energy nuclear reactions with reliable uncertainty estimates such as the $pp$ fusion \cite{Kong1}.

	\section{The pionless EFT challenges for the $pp$ fusion}
	\pilesseft is ideal for very light nuclei ($A<4$), since their typical momentum $Q$ is smaller than $m_\pi$. However, the \pilesseft is limited in the number of observables that could be used for calibrating the relevant LECs of the interactions. Nuclear electro-weak interactions containing two nucleons have been calculated using \pilesseft with particular emphasis on the radiative capture ($n+p\rightarrow d+\gamma$) and the $pp$ fusion. This theoretical focus is due to the fact that the radiative capture is relevant to big-bang nucleosynthesis in energy regimes characterized by large experimental uncertainties~\cite{Chen:1999bg, Rupak:1999rk, KSW_c, Chen_N_N, ando_deturon}, while the great interest in $pp$ fusion stems from solar physics.
	 
	 For weak interactions, %\cblack 
	 up to NLO, \cblack \pilesseft leaves a universal coupling LEC, $L_{1, A}$, unknown: $L_{1, A}\approx\frac{l_{1, A}}{M\mu^2}$, with $|{l_{1, A}}|$ as a natural dimensionless number, $\mu=m_\pi$, the cutoff of the theory and $M$ as the nucleon mass. %\subsection{Calibrating the two-body axial weal low-energy constant: $L_{1, A}$}
	Since the $pp$ fusion cannot be measured, $L_{1, A}$ has to be calibrated from other many-body low-energy observables. 
	
 Kong and Ravndal have calculated the $pp$ fusion up to NLO \cite{Kong2}. Butler and Chen accomplished a calculation up to N$^4$LO \cite{Proton_Proton_Fifth_Order}, where $L_{1, A}$ was estimated from previous EFT predictions such as Ref.~\cite{fermi_reference_1998}, with $L_{1, A}=6\pm2.5\fm^3$. Ando {\it et al.} have calculated the $\Lambda_{pp}(0)$ %\cblack 
 while resumming range correction to all orders \cblack \cite{Ando_proton} where $L_{1, A}$ was fixed indirectly from the two-body matrix element of Ref.~\cite{PhysRevC.67.055206}, with $L_{1, A}=1.65\pm0.1\fm^3$, for $\mu=m_\pi$. In 2017, the Nuclear Physics with Lattice Quantum Chromo Dynamics (NPLQCD) collaboration has used direct lattice QCD calculation to find $L_{1, A}=3.9\pm1.4 \fm^3$ %using the %triton $\beta$-decay
 \cite{PhysRevLett.119.062002}, for $\mu=m_\pi$. All previous \pilesseft predictions for the $pp$ fusion matrix element are shown in Tab.~\ref{table_compering}. 
 
 \begin {table}[H]
 \begin{center}
 	\begin{tabular}{c|c |c|c| c|c}
 		\centering
 		&ERE \cite{1969ApJ...155..501B}& NLO \cite{Kong2}&N$^4
 		$LO \cite{Proton_Proton_Fifth_Order}& 
 		Ando {\it et al.} \cite{Ando_proton}&NPLQCD \cite{PhysRevLett.119.062002}\\
 		\hline
 		$\Lambda^2_{pp}(0)$&7.08&7.04$\sim$7.7&6.71$\sim$7.03&7.09$\pm$0.02&7.095$\pm0.07$
 	\end{tabular}
 	\caption{\footnotesize{Estimations of $\Lambda^2_{pp}(0)$ from former calculations. The value in the first column is the ERE calculated by Bahcall and May in 1969~\cite{1969ApJ...155..501B}. The second, third, and fourth columns are estimated from the \pilesseft calculations: up to
 			NLO by Kong and Ravndal~\cite{Kong2}, up to N$^4$LO by Butler and Chen~\cite{Proton_Proton_Fifth_Order} and the full calculation by Ando {\it et al.}~\cite{Ando_proton}, respectively. The fifth column is the NPLQCD calculation~\cite{PhysRevLett.119.062002}.}}
 	\label{table_compering}
 \end{center}
 \end{table}

 Tab.~\ref{table_compering} show that all previous \pilesseft $pp$ fusion predictions are consistent with each other while most of them have significant uncertainties. Also, neither one of the above predictions has used \pilesseft precisely to estimate $L_{1, A}$, which raises the question of the reliability of all these calculations.
 
	Chen {\it et al.} \cite{L1A} 
	suggested to calibrate $L_{1, A}$ using muon capture: 
	\begin{equation}\label{muon}
	\mu^-+d\rightarrow \nu_\mu+n+n.
	\end{equation}
	%./;'whose cross-section has been measured recently up to 1\% level at the MuCap experiment \cite{MuCap}.
	The problem in applying \pilesseft to the $\mu^- d$ capture is that both energy and momentum transfer in this process %\cblack 
	are up to $100 \mev$, which \cblack might be too high for \pilesseft, so it might affect the cutoff dependence and requires additional coupling constants, e.g., the axial radius of the nucleon, whose uncertainties were not taken into account. Their suggestion led to the MuSun experiment, measuring this observable, with the aim to reach accuracy of few percents in the determination of $L_{1,A}$, excluding the above mentioned uncertainty contribution. Results of MuSun are expected in the following months \cite{MuSun}.
	
	The NPLQCD collaboration has calibrated $L_{1, A}$ using the triton $\beta$-decay \cite{PhysRevLett.119.062002}, for the non-physical pion mass, $m_\pi= 805\mev$. This method includes an extrapolation to physical pion mass which leads to uncontrolled uncertainties.
		 
	In this work, we calibrate $L_{1, A}$ using the triton ($^3$H) $\beta$-decay into Helium-3 ($^3$He): 
	\begin{equation}\label{tritiom}
	^3\text{H}\rightarrow ^3\text{He}+e^-+\overline{\nu_e}, 
	\end{equation}
	which is a well-measured, low-energy weak reaction, that also involves $L_{1, A}$ and therefore can be used for calibrating it. Unlike $\mu^- d$ capture, this decay is known to per-mill level accuracy \cite{Chou}, and its momentum and energy transfer lies well below the \pilesseft limit, so it is the perfect tool for \pilesseft calibration and predicting the $pp$ fusion matrix element. Also, the similarity to previous $\chi$EFT~\cite{Acharya:2016kfl} and LQCD~\cite{PhysRevLett.119.062002} calculations, can serve as a benchmark for evaluating the accuracy of our calculation. 
	
	The calibration of LEC from a three-body interaction ($^3$H $\beta$-decay) and the prediction of a two-nucleon interaction ($pp$ fusion) is based on a hidden assumption that \pilesseft is consistent for calculating both $A=2$ and $A=3$ systems. However, this assumption is non-trivial for \pilesseft, since the deuteron ($^2$H) wave function differs significantly from that of the three-nucleon ($^3$H, $^3$He) which is regularized by a
	high momentum, {\it i.e.,} ultraviolet cutoff $\Lambda$ \cite{faddeev,Kong1,3bosons,triton}. \pilesseft is renormalizable, {\it i.e.,} the theory has no dependence on the ultraviolet cutoff $(\Lambda)$. However, numerical and theoretical solutions of the integral equations reveal a strong dependence on this cutoff. 
	%\cblack 
	To overcome this problem, a three-body force, which is allowed by symmetry, was added to the theory at leading order (LO) \cite{3bosons, triton} to remove this cutoff dependence.
	\cblack
	 The question whether a similar behavior exists in a three-nucleon system coupled to an electro-weak field, cannot be answered straightforwardly and has never been examined.
	
	The Coulomb interaction in light nuclei is an additional
	complication: the Coulomb interaction is non-perturbative at low
	momenta $\lesssim 10\, \text{MeV}$, as expressed by
	the strong renormalization of the proton-proton scattering length, but should be perturbative in nuclei where the typical momenta are much higher. $^3$He is the lightest, and therefore simplest, nucleus to test the combination of \pilesseft and Coulomb interaction \cite{quartet, 3He, 2016PhLB..755..253K} and many recent works have discussed the problem of such a combination. In particular, it was shown that while LO $^3$He is described correctly within \pilesseft, at NLO the results are not so clear, and some approaches pointed out the need for additional, isospin dependent, three-body forces \cite{konig2,H_NLO}. Therefore, additional three-nucleon observables are needed to obtain predictive power within \pilesseft at NLO
	such as $^3$He binding energy \cite{konig1, konig2, konig3,konig5}.
\section{Goals of this thesis}
The primary goal of this work is to use \pilesseft to predict and estimate the uncertainties in $pp$ fusion rate, using the LEC, $L_{1, A}$, calibrated from the three-nucleon reaction $^3$H $\beta$-decay. The corresponding subgoals needed for this calculations are: 

\begin{enumerate}\item To establish a general framework for calculating electro-weak interaction for three-nucleon systems up to NLO. By presenting the three-nucleon diagrammatic representation of the scattering amplitude, we construct a general $A=3$ matrix element. Apart from electro-weak observables, this matrix element can be used for calculating the energy difference between $^3$H and $^3$He due to the Coulomb interactions, and the NLO corrections to $^3$H and $^3$He scattering amplitudes due to effective range corrections \cite{Big_paper}.
	
	\item To study the consistency of \pilesseft for the transition between $A=2$ and $A=3$ systems and vice-versa up to NLO using the well-measured $A<4$, $M_1$ \textbf{electromagnetic} observables at zero momentum transfer. By using two theoretical approaches for normalizing the deuteron wave function and comparing the two-body low-energy constants for both $A=2, 3$ we can find the nominal method that retains the consistency of the transition between three-nucleon system to two-nucleon system and vice versa. Validity is then checked against experimental data. 
	\item To evaluate \pilesseft uncertainties due to theoretical and empirical sources. 
	%\item To verify and validate the theory, using a parallel experimentally accessible setting, which we find in the electromagnetic structure of $A=2, 3$ observables.
\end{enumerate}
 
	\section{Verification and validation of pionless EFT using electromagnetic observables}
		 The extraction of the weak LEC $L_{1, A}$, could be done by the \pilesseft calculation of the half-life of the $^3$H $\beta$-decay, which is a well-measured, low-energy reaction. This reaction, as well as other low-energy observables, is related to the matrix element of the interaction Hamiltonian between the initial and final nucleon states.
		 However, the \pilesseft predictions of %\cblack 
		 observables \cblack between \textbf{$A=3$ bound-states systems}, were not introduced until recently \cite{magnetic_moments, Vanasse:2015fph, beta_decay, Vanasse:2017kgh}. The primary challenge in describing these interactions stems from the fact that the three-nucleon wave functions are the numerical solution of coupled integral equations \cite{3bosons, triton}. 
		 
		 The consistency of \pilesseft for calculating $A=2, 3$ electro-weak observables, which is crucial for the extraction of the weak LEC $L_{1, A}$, cannot be examined using the weak observables only, due to the small number of appropriate reactions. Instead, we are using the \textbf{four} electromagnetic well-measured observables applicable to the \pilesseft regime, which are very similar to weak observables. These observables are the three light nuclei magnetic moments: $\langle\hat{\mu}_d\rangle$, $\langle\hat{\mu}_{^3\text{H}}\rangle$, $\langle\hat{\mu}_{^3\text{He}}\rangle$ \cite{3He_3H_data, mu_d_data} and the radiative capture process $ n+p \rightarrow d+\gamma$, whose cross-section is denoted by $\sigma_{np}$ \cite{np_data}. 
	 \begin {table}[H]
	 \begin{center}
	 	\begin{tabular}{l| c| c}
	 		%\hline
	 		& Electromagnetic& weak\\
	 		\hline
	 		One-body LECs& $\kappa_0$, $\kappa_1$& $g_A$\\
	 		Two-body LECs& $L_1, L_2$& $L_{1, A}$\\
	 		\hline
	 		One-body operator& $\sigma$, $\sigma\tau^0$& $\sigma\tau^{+, -}$, $\tau^{+, -}$\\
	 		Two-body operator&$L_{1}t^\dagger s$, $L_{2}t^\dagger t$&$L_{1, A}t^\dagger s$\\
	 		\hline
	 		$A=2$, $Q\approx0$ &$\sigma_{np}: n+p\rightarrow d+\gamma$&$pp$ fusion: 
	 		\\
	 		observables & $d$ magnetic moment $\langle\hat{\mu}_d\rangle$& $p+p\rightarrow d+e^+ + \nu_e$\\
	 		\hline
	 		$A=3$, $Q\approx 0$ &$^3$H, $^3$He magnetic moments: & $^3$H $\beta$-decay into $^3$He: 
	 		\\
	 		observables& $\langle\hat{\mu}_{^3\text{H}}\rangle$, $\langle\hat{\mu}_{^3\text{He}}\rangle$&
	 		$^3\text{H}\rightarrow ^3\text{He}+e^-\bar{\nu_e}$\\
	 		%\hline
	 	\end{tabular}
	 	\caption{\footnotesize{Electro-weak LECs, operators and low-energy observables.}}
	 	\label{table_EW}
	 \end{center}
 \end{table} 
	 
Table~\ref{table_EW} shows the similar \pilesseft characteristics of both electromagnetic and weak reactions of $A=2$ and $A=3$ systems. Up to NLO, the \pilesseft 
electro-weak Lagrangian contains two types of operators: one-body and two-body. The one-body operators are the Pauli matrices ($\sigma,\tau,\sigma\tau$) and the two-body operators are the four-nucleon fields in term of dibaryons fields: $t^\dagger t$ and $t^\dagger s$. The \pilesseft \textbf{electromagnetic} Lagrangian contains (up to NLO), two types of LECs: one-body and two-body. The one-body LECs which are coupled the Pauli matrices ($\sigma,\tau$): $\kappa_0 $ (the isoscalar magnetic moment of the nucleon), and $\kappa_1$ (the
isovector magnetic moment of the nucleon), are both known with high accuracy from the proton and neutron magnetic moments. The two-body LECs, 
$L_1$ and $L_2$, are known from $A=2$ electromagnetic reactions: $L_1$, which couples the spin-singlet channel to the spin-triplet channel, %is known from $\sigma_{np}$, \cite{Chen:1999tn, ando_deturon}
and while $L_2$, which couples the spin-triplet to another spin-triplet channel.%, is known from $\langle\hat{\mu}_d\rangle$ \cite{KSW_c, Chen:1999tn, ando_deturon}.
Beside the unknown two-body LEC, $L_{1, A}$ (which is the weak analogue to $L_{1}$), the \textbf{weak} \pilesseft Lagrangian contains the one-body axial low-energy constant, $g_A$, which is the weak analogue to $\kappa_1$.

 the $A=2,3$ $M_1$ observables depend on the values of the two-body electromagnetic LECs, $L_1$ and $L_2$. In past works, the experimental values of the $A=2$ observables ($\sigma_{np}$ and $\langle\hat{\mu}_d\rangle$) were used to fix these LECs \cite{ando_deturon, ando_magntic_BBN}. Here, we calculate consistently the $A=2,3$ $M_1$ observables ($\langle\hat{\mu}_{^3\text{H}}\rangle$, $\langle\hat{\mu}_{^3\text{He}}\rangle$,$\langle\hat{\mu}_{d}\rangle$ and $\sigma_{np}$), which depend on the same LECs, so we can extract these LECs from two observables and then use them to predict the remaining observables. Therefore, we have six independent ways for calibrating the LECs. These calibrations will be used later in this thesis for the evaluating the stability and consistency of \pilesseft for $A=2$ and $A=3$ systems up to NLO and for estimating its theoretical uncertainty. 

We argue that this similarity between the weak reactions to the electromagnetic ones, along with the comparison between the electromagnetic experimental data and the theoretical predictions of these observables, will enable us to examine the \pilesseft consistency for both sectors (electromagnetic and weak) and to estimate its theoretical uncertainties.

\section{Structure of the thesis}
This thesis is organized as follows: The general formalism of
\pilesseft is presented in Chapter 2. The Faddeev equations describing the
three-nucleon wave functions ($^3$H and $^3$He) and a diagrammatic
representation of the three-nucleon wave function are presented in Chapter 3, while the three-nucleon wave function normalization 
is presented in Chapter 4. The general form of the calculation
of a matrix element between two $A=3$ bound-state amplitudes is detailed in Chapter 5, and the NLO corrections to the Faddeev equations are discussed in Chapter 6. The calculation of the $A=2$ and $A=3$ electromagnetic observables as well as the calibrations of the two-body LECs, $L_1$ and $L_2$ are presented in Chapter 7. The calculation of weak matrix elements, which are used for the extraction of the two-body LEC, $L_{1, A}$ is presented in Chapter 8, while the calculation of the $pp$ fusion matrix element up to NLO is given in Chapter 9. We present an analysis of the results and calculate the theoretical uncertainty in Chapter 10, as well as giving our final results for the $pp$ fusion matrix element. We then summarize and provide an outlook in Chapter 11.
%\section{About this thesis}	

%The accuracy and precision of our $M_1$ calculation, as will be show in this thesis, verify and validate \pilesseft, and open a new path to model independent and simple predictions of low-energy electromagnetic and weak reactions, especially the $pp$ fusion. 

\chapter{The general formalism of pionless EFT}\label{formalism} 
In this chapter, we briefly summarize the general formalism of \pilesseft with strong, weak and electromagnetic interactions up to next-to-leading order (NLO). \pilesseft, as any other EFT, is based on scale separation between a typical momentum, $Q$, and high-energy scale, $\Lambda_{cut}$. \pilesseft is applicable for very low-energy interactions in which $Q\ll m_\pi\approx140\mev$, where $m_\pi$, the pion mass, is used as the high-energy scale. 
The \pilesseft Lagrangian is, therefore, expanded in a power series of $\frac{Q}{\Lambda_{cut}}$ \cite{Kaplan1996629}, which includes only the contact interactions between the nucleons. In the next section, we present the typical \pilesseft scales, that are essential for building the \pilesseft Lagrangian and, as a result, the \pilesseft interactions matrix elements, up to NLO. 
\section{A small parameter expansion of pionless EFT}
 For light nuclei, the external momentum, $p$, and the deuteron binding momentum,
$\gamma_t$, are formally considered as $\mathcal{O}(Q)$, while the renormalization scale, $\mu$ (which will be introduced later in this chapter), is considered to be the cutoff, $\Lambda_{\rm cut}$. In contrast to Beane and Savage \cite{rearrange} and similarly to Rupak and Kong power-counting \cite{quartet}, in this work we assume a scale separation between the scattering length, $a$ (which is of the order of $\frac{1}{Q}$) and the effective range, $\rho$ (scaled as $\frac{1}{\Lambda_{cut}}$). This scale separation enables us to use their ratio $\left(\dfrac{\rho}{a}\right)$ as the small parameter that determines our order of expansion up to $\dfrac{\rho}{a}=\mathcal{O}\left(Q/\Lambda_{\rm cut}\right)$.

The \pilesseft power-counting of the present work is: 
\begin{itemize}
	\item The external nucleon momentum, $p$, scales as $Q$;
	\item The deuteron binding momentum, $\gamma_t$, scales as $Q$ ;
	\item The renormalization scale, $\mu$, scales as $\Lambda_{cut}$;
	\item The scattering length, $a$, scales as $\dfrac{1}{Q}$;
	\item The effective range, $\rho$, scales as $\dfrac{1}{\Lambda_{\rm cut}}$;
	\item The kinetic energy scales as $Q^2/M$;
	\item The loop integration: $\int d^3{\bf p}$
	is scaled by $Q^3$ and
	\item The nucleon propagator $S(p_0,p)=\left[p_0-\frac{{\bf
			p}^2}{2M}+i\varepsilon\right]^{-1}\sim \mathcal{O}(M/Q^2)$,
\end{itemize}
where $p$ and $p_0$ refer to the inner momentum and its associated energy, respectively, and $M$ is the nucleon mass. %\cblack 
Similar to Ref.~\cite{quartet}, the integration measure
$ \int dq_0d^3q$ scales as $\mathcal{O}\left(Q^5/4\pi M\right)$, as we include a factor of $1/(4\pi)$ with every loop.
\cblack
\section{Pionless EFT at next-to-leading order}
We consider the most general \pilesseft Lagrangian which involves only two nucleons (external electro-weak currents will be included later). A system with two nucleons with zero angular momentum $L = 0$ can exist in two different channels: a spin singlet, $s, ^1S_0$, (with spin 0 and isospin 1) or spin triplet, $t, ^3S_1$ (deuteron with spin 1 and isospin 0). The \pilesseft Lagrangian up to NLO in $\frac{Q}{\Lambda_{\rm cut}}$ can be written as \cite{Chen_N_N}: 
\begin{multline}\label{lag2}
\mathcal{L}= N^\dagger\left(i\partial_0+\frac{\nabla^2}{2M}\right)N-C_{0}^t\left(N^TP_tN\right)^\dagger\left( N^TP_tN\right)-C_{0}^s\left(N^TP_sN\right)^\dagger\left( N^TP_sN\right)+\\
\frac{C_{2}^t}{8}\left[\left(N^TP_tN\right)^\dagger\left(N^T\left(\overleftarrow{\nabla}-\overrightarrow{\nabla}\right)^2P_tN\right)+h.c\right]+
\frac{C_{2}^s}{8}\left[\left(N^TP_sN\right)^\dagger\left(N^T\left(\overleftarrow{\nabla}-\overrightarrow{\nabla}\right)^2P_sN\right)+h.c\right]~.
\end{multline}
Here $M$ denotes the nucleon mass, $N$ denotes the nucleon field, the subscripts $s,t$ denote the spin-singlet and spin-triplet channels, respectively, and %\cblack 
$C_0^{t,s}, C_2^{t,s}$ are coupling constants defined using the nucleon-nucleon ($N-N$) scattering amplitude, $\mathcal{A}(p)$ which will be discussed later in the section. 
\begin{comment}
\begin{align}
C_0^{t, s} &= \frac{4\pi }{M}\left(\frac{1}{1/a_{t, s}-\mu }\right)\label{eq_C0} \\
C_2^{t, s} &= \frac{4\pi} {M}\left(\frac{1}{1/a_{t, s}-\mu}\right)^2\frac{\rho_{t, s}}{2}\label{eq_C2},
\end{align}
where $a_{t, s}$, the scattering length, is of $\mathcal{O}(Q)$ and both $\rho_{t, s}$, the effective range, and $\mu$, the renormalization scale, are of $\mathcal{O}(\Lambda_{cut})$, so $C_2=C_0\cdot\mathcal{O}\left(\frac{Q}{\Lambda_{cut}}\right)$ \cite{few}. 
\end{comment}
\cblack
The projection operators are defined as:
\begin{equation}\label{eq:projections}
P_t^i=\frac{1}{\sqrt{8}}\sigma ^2\sigma ^i\tau ^2, P_s^A=\frac{1}{\sqrt{8}}\sigma ^2\tau ^2\tau ^A~,
\end{equation}
where $\sigma,\tau$ are Pauli matrices.

The equations of motion of each channel in \cref{lag2} can be written as:

\begin{equation}\label{centermass}
N^T\left(\overleftarrow{\nabla}-\overrightarrow{\nabla}\right)^2P_{t, s}N=-4M\left(i\partial_0+\frac{\nabla^2}{4M}\right)\left(N^TP_{t, s}N\right)~, 
\end{equation}
where both time and space derivatives refer to the center of the mass of each channel, and the mass difference between the proton and the neutron is neglected.

Substituting \cref{centermass} in \cref{lag2} yields:
\begin{multline}\label{eq_lag3}
\mathcal{L}= N^\dagger\left(i\partial_0+\frac{\nabla^2}{2M}\right)N-C_{0}^t\left(N^TP_tN\right)^\dagger\left( N^TP_tN\right)-C_{0}^s\left(N^TP_sN\right)^\dagger\left( N^TP_sN\right)-\\
-\frac{C_{2}^t}{2}M\left[\left(N^TP_{t}N\right)^\dagger\left(i\partial_0+\frac{\nabla^2}{4M}\right)\left(N^TP_{t}N\right)+h.c\right]
-\frac{C_{2}^s}{2}M\left[\left(N^TP_{s}N\right)^\dagger\left(i\partial_0+\frac{\nabla^2}{4M}\right)\left(N^TP_{s}N\right)+h.c\right]~.
\end{multline}

\subsection{$N-N$ scattering amplitude}\label{N_N_amplitude}
The nucleon-nucleon ($N-N$) scattering amplitude, $\mathcal{A}(p)$ (where $p$ is the momentum in the center of mass frame), presented in this subsection, is the building block of the two-nucleon propagator, which is essential for calculating the \pilesseft interactions. 
\begin{figure}[h!]
	\centering
	\includegraphics[width=0.65\linewidth]{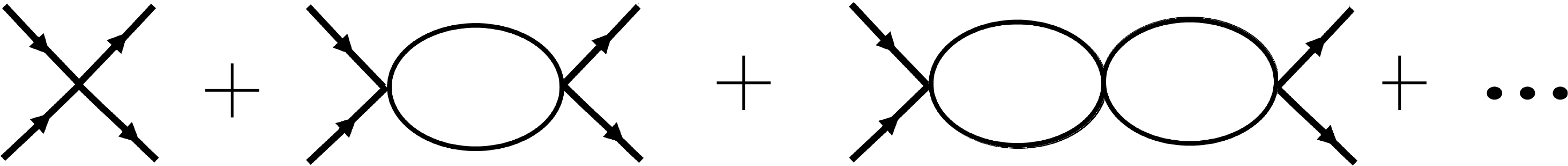}
	\caption{\footnotesize{The $N-N$ scattering amplitude is a sum of the chains of $N-N$ bubbles.}}\label{LO_propagtor}
\end{figure}
 
 $\mathcal{A}(p)$ can be written in terms of the coupling constants, $C_0$ and $C_2$, up to NLO \cite{KSW1998_a, Kaplan:1998tg,Kaplan1996629,Kaplan1997471}. The LO part of this scattering amplitude is given by the sum over all chains of bubbles (shown in Fig.~\ref{LO_propagtor}), which can be written as: 
\begin{equation}\label{eq_A}
\mathcal{A}_{t, s} ^{\text{LO}}(p)=-\frac{C_0^{t, s}}{1 - C_0^{t, s}\mathcal{I}_B(p)}~,
\end{equation}
where $t,s$ denote the nucleon-nucleon channel and $\mathcal{I}_B(p)$ denotes the two-nucleon loop integral evaluated
using the so-called power divergence subtraction (PDS) scheme (see
 Refs.~\cite{KSW1998_a, KSW_c,Kaplan:1998tg}):
\begin{equation}
\mathcal{I}_B(p)=-i\left(\frac{\mu}{2}\right)^{4-D}\int\frac{d^Ddq}{(2\pi)^D}\frac{1}{E/2-p^2/M-p_0+i\epsilon}\frac{1}{E/2-p^2/M+p_0+i\epsilon}, 
\end{equation} where $E=p^2/M$ is the total center-of-mass energy. This integral is linearly divergent for $D>2$, but dimensional regularization in the space of dimensions makes it finite \cite{Kaplan1996629,Kaplan:1998tg}. When $D = 3$, this integral has a simple pole that can be subtracted while using the PDS scheme, which
requires the introduction of the renormalization scale $ \mu$. By analytic continuation of the dimension from $D=3$ to $D = 4$, one finds that: 
\begin{equation}\label{IB}
\mathcal{I}_B(p)=-\frac{M}{4\pi}(\mu+ip)~.
\end{equation}
%\cblack
The (full) scattering amplitude $\mathcal{A}_{t,s}$ is related also to the S-matrix by \cite{Kaplan:1998tg}:
\begin{equation}\label{eq_a1}
S-1=e^{2i\delta_{t,s}}-1=i\frac{pM}{2\pi}\mathcal{A}_{t,s}~,
\end{equation}
where $\delta_{t,s}$ is the phase shift, such that (up to NLO):
\begin{equation}\label{eq_phase_shift}
p\cot\left({\delta}_{t,s}\right)=-\frac{1}{a_{t,s}}+\frac{1}{2}\rho_{{t, s}}p^2~,
\end{equation}
where $a_{t,s}$ is called the scattering length and $\rho_{t,s}$ is the effective range.
Given \cref{eq_a1}, it is convenient to define $p\cot(\delta)$ such that:
\begin{equation}\label{eq_A_delta}
p\cot\left({\delta}_{t,s}\right)=ip + \frac{4 \pi}{\mathcal{A}_{t,s} M}~.
\end{equation} 
By combing \cref{eq_A,eq_phase_shift,eq_A_delta}, one finds that:

\begin{equation}\label{eq_C0_a}
C_0^{t,s}=\dfrac{4\pi}{M}\dfrac{1}{\left(-\mu+\frac{1}{a_{t,s}}\right)}~.
\end{equation}
\cblack

\begin{figure}[h!]
	\centerline{
		\includegraphics[width=0.65\linewidth]{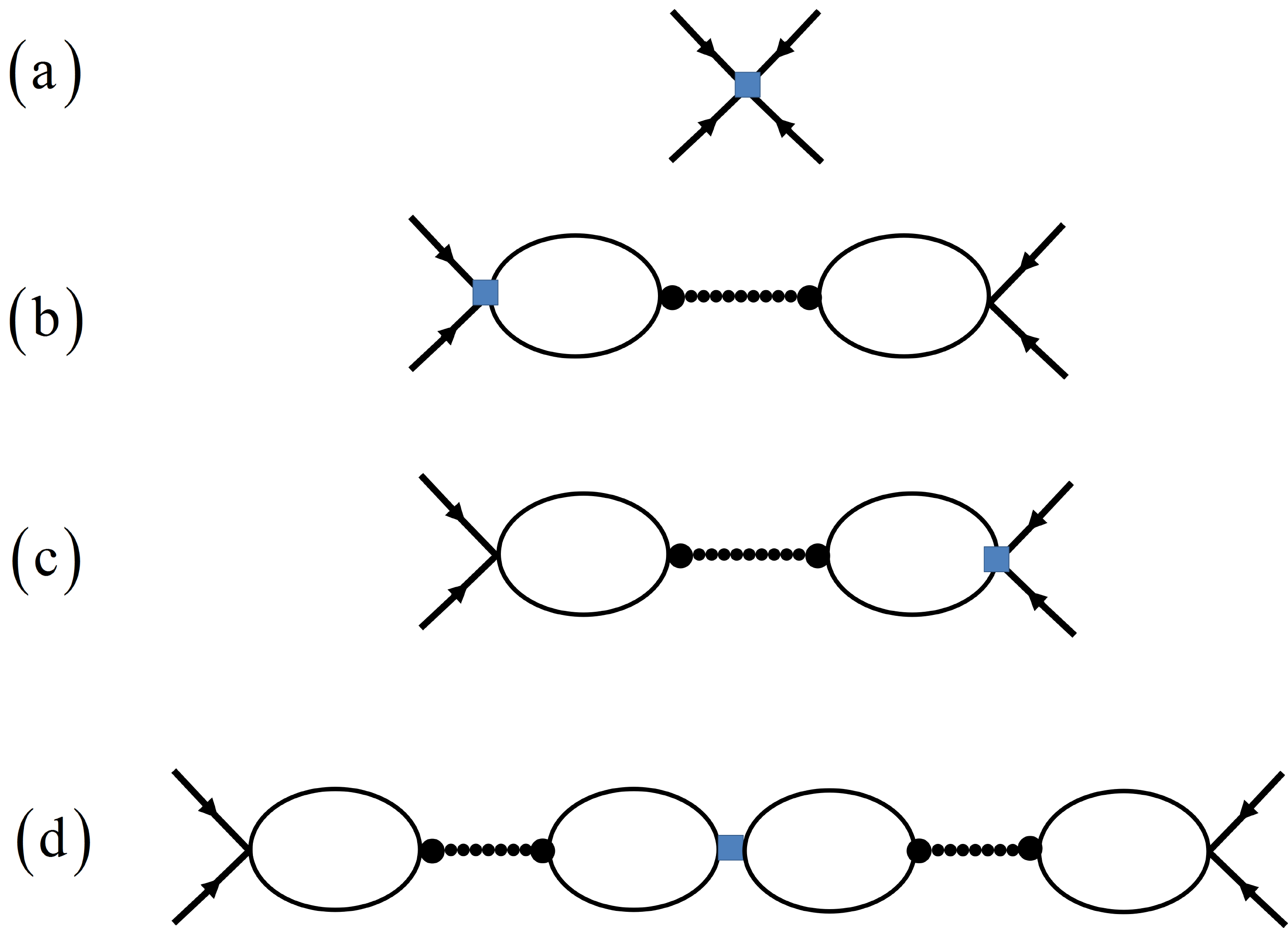}}
	\caption{\footnotesize{The $N-N$ scattering amplitudes up to
			NLO. The blue squares denote the NLO insertion
			of the effective range.}}
	\label{N_N_NLO}
\end{figure}

The NLO correction to the $N-N$ scattering amplitude results in a single NLO insertion $(C_2p^2)$ into the LO scattering amplitude. There are four different options for the locations of these insertions, as shown in Fig.~\ref{N_N_NLO}. 
Summing over all NLO insertions yields \cite{KSW1998_a}:
%\cblack
\begin{equation}
\mathcal{A}^{\text{NLO}}_{t, s}(p)=\mathcal{A}^{\text{LO}}_{t, s}(p)+\mathcal{A}^{(1)}_{t, s}(p)=\mathcal{A}^{\text{LO}}_{t, s}(p)+\frac{-C_2^{t, s}p^2}{\left(1 - C_0^{t, s}\mathcal{I}_B\right)^2}~.
\end{equation} 
Using \cref{eq_a1,eq_phase_shift,eq_A_delta} up to NLO, one finds that:
\begin{multline}\label{eq_C2}
p\cot\left({\delta}_{t,s}^{\text{NLO}}\right)=p\cot\left({\delta}_{t,s}^{\text{LO}}\right)+\frac{1}{2}\rho_{t,s}p^2=ip+\frac{4 \pi}{M\mathcal{A}^{LO}_{t,s}}+\frac{4 \pi {C_2}^{t,s} p^2}{(C_0^{t,s})^2 M}\rightarrow C_2^{t,s}=\dfrac{4 \pi}{M \left(-\mu +\frac{1}{a_{t,s}}\right)^2}\dfrac{\rho_{t, s}}{2}~.
\end{multline}
\cblack

\subsection{Coulomb corrections to N-N scattering amplitude}\label{pp_amplitudes}
Some of the interactions discussed in this thesis, contain besides the strong interaction, the Coulomb interactions ({\it i.e.,} $pp$ fusion and $^3$He). The derivation of the proton-proton ($pp$) scattering amplitude for these reactions is based on Refs.~\cite{Coulomb_effects, proton_proton_scattring, Kong2}.

At LO, the {\it pp} nucleon bubble has an infinite series of ladder diagrams of a Coulomb photon
exchange. 
\begin{figure}[h!]
	\centering
	\includegraphics[width=0.75\linewidth]{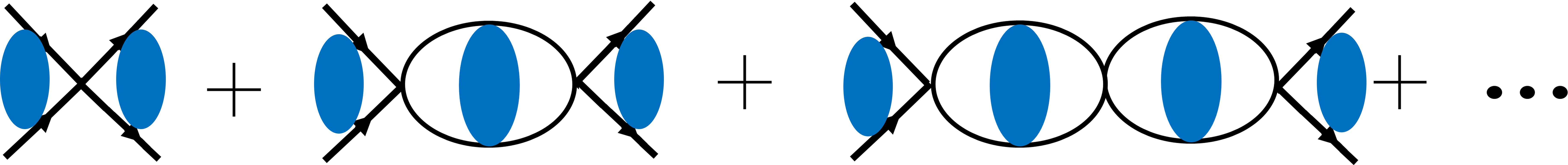}\\
	\caption{\footnotesize{The $pp$ scattering amplitude is a sum of chains of Coulomb-dressed bubbles.}}
	\label{fig_proton_proton}
\end{figure}

The sum of all the bubbles yields the scattering amplitude shown in Fig.~\ref{fig_proton_proton}:
\begin{equation}\label{eq_amplitide}
i\mathcal{A}_{pp}(p)=-iC_\eta^2 C_0^{pp} e^{2i\sigma_0}\frac{1}{1- C_0^{pp} J_0(p)}~, 
\end{equation}
where: \begin{equation}\label{define}
\sigma_l=\arg\left[\Gamma\left(1+i\eta \left(p\right)+l\right)\right],\qquad C_\eta=\sqrt{\frac{2\pi\eta}{-1+e^{2\pi\eta}}},\qquad 
\eta(p)=\frac{\alpha M}{2p}~,
\end{equation}
and $\alpha\sim 1/137$ is the fine-structure constant.
$J_0$(p), the amplitude of an infinite photons exchange (without external energy and momentum transfer) is given by: 
\begin{equation}\label{eq_J0}
J_0(p)=M\int\frac{d^3k}{(2\pi)^3}\frac{\psi_0^*\psi_0}{p^2-k^2}~, 
\end{equation}
where 
\begin{equation}
\psi_0=\int\frac{d^3k}{(2\pi)^3} \langle k\ket{\psi_p}= C_\eta e^{i\sigma_0}~,
\end{equation}
and $\ket{\psi_p}$ is the solution of the Schr\"{o}dinger equation ($\hat{H}-E)\ket{\psi_p} = 0$, where $\hat{H}=\hat{H}+\hat{V_C}$ is the full Hamiltonian with a Coulomb potential.
Hence, \cref{eq_J0} has the form:
\begin{multline}\label{eq_J0_1}
J_0(p)= M\int \frac{d^3k}{(2\pi)^3} \dfrac{{2\pi\eta(k)}}{e^{2\pi\eta(k)}-1}\dfrac{1}{p^2-k^2}=\\
M\int \frac{d^3k}{(2\pi)^3} \dfrac{{2\pi\eta(k)}}{e^{2\pi\eta(k)}-1}\frac{p^2}{k^2}\frac{1}{p^2-k^2}-M\int \frac{d^3k}{(2\pi)^3} \dfrac{{2\pi\eta(k)}}{e^{2\pi\eta(k)}-1}\frac{1}{k^2}~.
\end{multline}
The first integral ($J_0 ^{\text{fin}}(p)$, hereafter) converges and the second integral ($J_0 ^{\text{div}}$, hereafter) has no $p$-dependence, but it diverges with $k\rightarrow\infty$. 
Introducing a new variable $x$, we write:
\begin{equation}\label{eq_J0_4}
J_0 ^{\text{fin}}(p)=M\int \frac{d^3k}{(2\pi)^3} \dfrac{{2\pi\eta(k)}}{e^{2\pi\eta(k)}-1}\frac{p^2}{k^2}\frac{1}{p^2-k^2}=_{k=\frac{\pi \alpha M}{x}}\int dx\frac{\alpha M^2 p^2}{2\pi}\frac{x}{ \left(e^x-1\right) \left(\pi ^2 \alpha ^2 M^2-p^2 x^2\right)},
\end{equation}
and using the known integral
\begin{equation}
\int_0^\infty dx\dfrac{x}{e^x-1}\dfrac{1}{x^2 + a^2} = \frac{1}{2}\left[\log\left(\frac{a}{2\pi}\right)-\frac{\pi}{a}-\psi\left(\frac{a}{2\pi}\right)\right]~,
\end{equation}
we get that: 
\begin{equation}\label{eq_J0_con}
J_0 ^{\text{fin}}(p)=-\frac{\alpha M^2}{4\pi}\Phi(\eta)~,
\end{equation}
where
\begin{equation}
\Phi(x)= \psi(ix) + \frac{1}{2ix}-
- \log(ix),
\end{equation}
and $\psi$ is the derivative of the $\Gamma$ function.

Using $\mu$ as the renormalization scale, the divergent part of $J_0$ for $D=3-\epsilon$ is given by: 
\begin{multline}\label{eq_J0_6}
J_0 ^{\text{div}}=-M\int \frac{d^3k}{(2\pi)^3} \dfrac{{2\pi\eta(k)}}{e^{2\pi\eta(k)}-1}\frac{1}{k^2}=\int{dx}\frac{\alpha M^2}{2 \pi \left(e^x-1\right) x}=\\
-\frac{\alpha M^2}{4\sqrt{\pi}\Gamma\left(\frac{d}{2}\right)}\left(\frac{\mu}{M\alpha\sqrt{\pi}}\right)^\epsilon\int dx\dfrac{x^{\epsilon-1}}{\left(e^x-1\right)}=
-\frac{\alpha M^2}{4\sqrt{\pi}\Gamma\left(\frac{d}{2}\right)}\left(\frac{\mu}{M\alpha\sqrt{\pi}}\right)^\epsilon\zeta(\epsilon)\Gamma(\epsilon)~, 
\end{multline}
where $\zeta(x)$ is the Riemann zeta function. 

For $\epsilon\rightarrow0$, \cref{eq_J0_6} becomes: 
\begin{equation}\label{eq_J0_7}
J_0 ^{\text{div}}=\frac{\alpha M^2}{4\pi}\left[ \log \left(\frac{\mu }{\alpha}\right)+\frac{1}{\epsilon }-\frac{3 C_E }{2}+1+\frac{\log (\pi )}{2}\right]~, 
\end{equation}
which includes a pole for $\epsilon=0$. 
For $D\rightarrow2$, the divergent part of \cref{eq_J0_1} has another pole:
\begin{equation}\label{eq_J0_d_2}
J_0 ^{\text{div}}(D\rightarrow2)=\frac{\mu M }{4 \pi (D-2) }~. 
\end{equation}
According to the PDS regularization scheme, the latter pole has to be subtracted
from \cref{eq_J0_7}, even when $D \rightarrow 3$, the result of which is:
\begin{equation}\label{eq_J0_fineal}
J_0 ^{\text{div}}=\frac{\alpha M^2}{4\pi}\left[ \log \left(\frac{\mu\sqrt{\pi} }{\alpha M}\right)+1- \frac{3}{2}C_E +\frac{1}{\epsilon }\right]-\frac{\mu M}{4\pi}~.
\end{equation}

The Coulomb scattering length is thus contained in the ultraviolet divergent part of the Coulomb bubble: 
\begin{equation}\label{eq_aC}
\frac{1}{a_p}=\frac{4\pi}{M}\left(\frac{1}{ C_0^{pp}}-J_0 ^{\text{div}}\right),
\end{equation}
where $a_p=-7.82\fm$ is the measured $pp$ scattering length. 
Since $J_0 ^{\text{div}}=J_0(p)+\frac{\alpha M^2}{4\pi}\Phi(\eta)$, \cref{eq_aC} can be written as:
\begin{equation}
\frac{1}{a_p}=\frac{4\pi}{M}\left(\frac{1}{ C_0^{pp}}-J_0(p)-\frac{\alpha M^2}{4\pi}\Phi(\eta)\right)~,
\end{equation}
and the $pp$ scattering amplitude (\cref{eq_amplitide}) can be written as in Ref.~\cite{Coulomb_effects}:
\begin{equation}
i\mathcal{A}_{pp}(p)=-iC_\eta^2 C_0^{pp} e^{2i\sigma_0}\frac{1}{1- C_0^{pp} J_0(p)}=-iC_\eta^2 e^{2i\sigma_0}\frac{1}{\frac{1}{C_0^{pp}}- J_0(p)}=-iC_\eta^2 e^{2i\sigma_0}\frac{4 \pi }{M \left(\frac{1}{a_p}+\alpha M \Phi(\eta) \right)}~.
\end{equation}
\begin{comment}
From \cref{eq_J0_fineal,eq_aC} it is natural to introduce the $\mu$-dependent scattering length: 
\begin{equation}
\dfrac{1}{{a_C(\mu)}}=\dfrac{1}{a_p}+\alpha M\left[\log\left(\dfrac{\mu\sqrt{\pi}}{\alpha M}\right)+1-\frac{3}{2}C_E\right],
\end{equation}
and similarly to $C_0^{t, s}$:
\begin{equation}
C_0^{pp}=-\frac{4\pi}{M}\left(\mu-\frac{1}{a_C(\mu)}\right)^{-1}.
\end{equation}
\begin{comment}
By combining \cref{eq_amplitide,eq_J0_con,eq_aC}, we get that: 
\begin{equation}
[ iy_s^2\mathcal{D}_{pp}(p)]^{-1}=\frac{1- C_0^{pp}\left[J_0 ^{\text{fin}}(p)+J_0 ^{\text{div}}\right]}{ C_0^{pp}}=\frac{M}{4\pi}\left[\frac{1}{a_p}+\alpha M \Phi(\eta)\right]
\end{equation}
and the LO Coulomb propagator is given by: 

\begin{equation}\label{eq_pp_LO}
i\mathcal{D}_{pp}(p)=\frac{i4\pi}{My_s^2}\frac{1}{\frac{1}{a_p}+\Phi(\eta)}
\end{equation}
\end{comment}
\subsubsection{The NLO correction to the Coulomb scattering amplitude}
\begin{figure}[h!]
	\centerline{
		\includegraphics[width=0.65\linewidth]{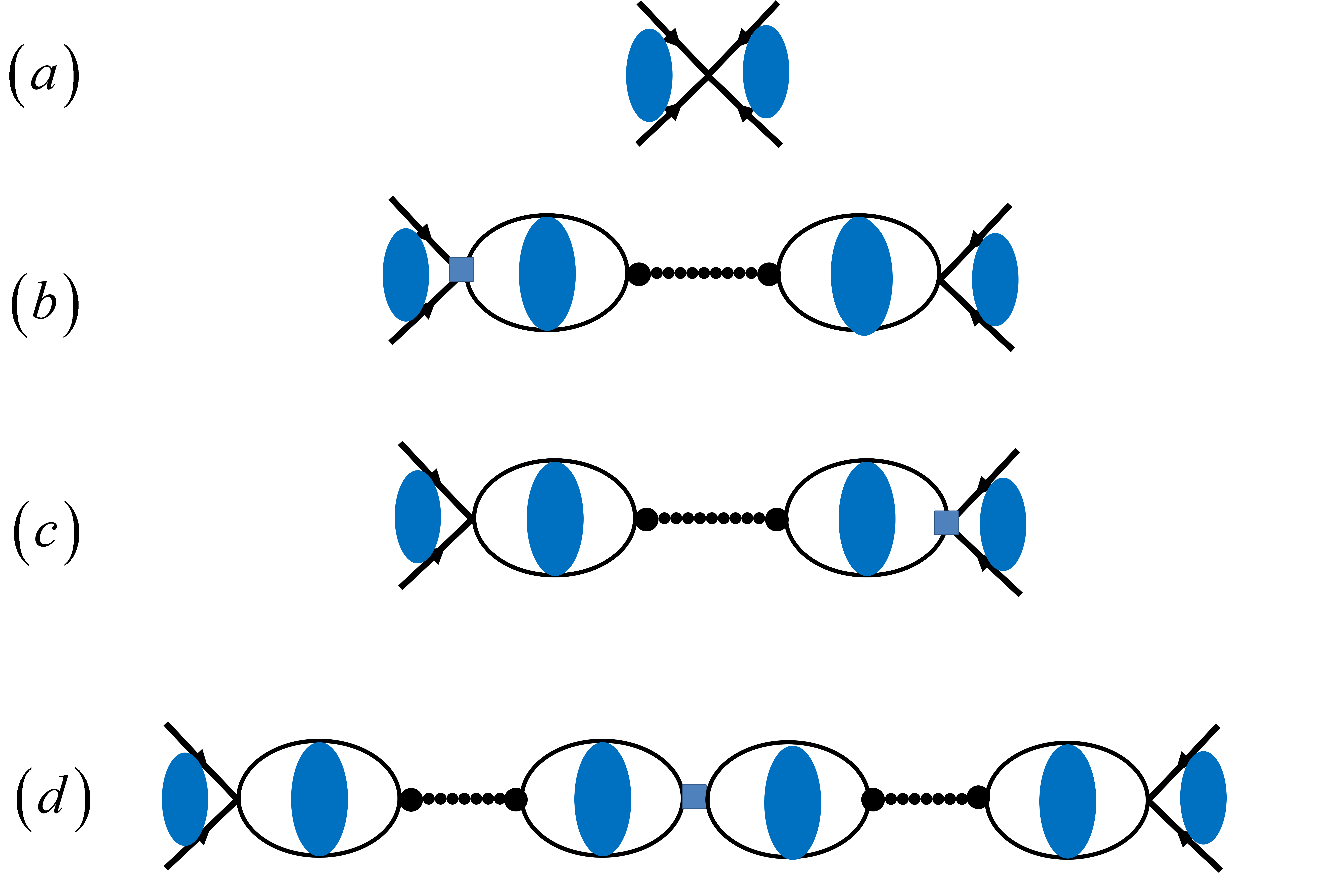}}
	\caption{\footnotesize{The {\it pp} dibaryon propagator up to NLO. The blue squares denote the NLO insertion
			of the effective range.}}
	\label{fig_proton_NLO}
\end{figure}

The NLO correction to the $pp$ scattering amplitude is given by the diagrams shown in Fig.~\ref{fig_proton_NLO}. 
%The contribution to NLO correction are given by: 
%\begin{equation}
%V_2= C_2^{pp}\cdot k^2.
%\end{equation}

The contribution from the first diagram (Fig.~\ref{fig_proton_NLO} (a)) is:
\begin{equation}\label{eq_T_a}
T^a(p)= C_2^{pp}\psi_0(p)\psi_2(p)~,
\end{equation}
where
\begin{equation}
\psi_2(p)=\int\frac{d^3k}{(2\pi)^3} k^2\psi_p(k)= C\eta e^{i\sigma_0}\left[p^2-\mu\alpha M -\frac{1}{2} \left(\alpha M\right)^2\right]~, 
\end{equation}
and \begin{equation}
C_2^{pp}=\dfrac{M\left(C_0^{pp}\right)^2}{8\pi}\rho_C~,
\end{equation}
with $\rho_C$ denoting the Coulomb effective range.

The contribution from the second and third diagrams (Fig.~\ref{fig_proton_NLO} (b+c)) is given by (see \cite{Coulomb_effects}):
\begin{equation}\label{eq_T_b_c}
T^{b+c}(p)= C_2^{pp}\frac{ C_0^{pp}}{1-C_0J_0(p)}\psi_0[\psi_0(p)J_2(p)+\psi_2(p)J_0(p)]~,
\end{equation}
where:
\begin{equation}
\begin{split}
J_2(p)=&M\int \frac{d^3k}{(2\pi)^3} \frac{\psi_2(k)\psi_0^*(k)}{p^2-k^2}=
M\int \frac{d^3k}{(2\pi)^3} \dfrac{e^{2\pi\eta(k)}}{e^{2\pi\eta(k)}-1}\dfrac{k^2-\mu\alpha M -\frac{1}{2} \left(\alpha M\right)^2}{p^2-k^2}=\\
&M\int \frac{d^3k}{(2\pi)^3} \dfrac{e^{2\pi\eta(k)}}{e^{2\pi\eta(k)}-1}\dfrac{k^2-p^2+p^2+\mu\alpha M -\frac{1}{2} \left(\alpha M\right)^2}{p^2-k^2}=\\
&\left[p^2-\mu\alpha M -\frac{1}{2} \left(\alpha M\right)^2\right]J_0(p)-J,
\end{split}
\end{equation}
and
%\color{red}
\begin{equation}
J= M\int \frac{d^3k}{(2\pi)^3} \dfrac{e^{2\pi\eta(k)}}{{2\pi\eta(k)}-1}=M^4\alpha^3\int dx\frac{\pi e^x}{2x^3\left(e^{x}-1\right)}~, 
\end{equation}
which has a $p$-independent ultraviolet divergence. 
 
\color{black}
The contribution from the fourth diagram (Fig.~\ref{fig_proton_NLO} (d)) is: 
\begin{equation}\label{eq_T_d}
T_d= \frac{ C_2^{pp}( C_0^{pp})^2}{ [1- C_0^{pp}J_0(p)]^2}\psi_0^2(p)J_0(p)J_2(p)~.
\end{equation}

The NLO correction to the $pp$ scattering amplitude ($\mathcal{A}^{(1)}_{pp}(p)$) is given by the summation over of the four contributions (\cref{eq_T_a,eq_T_b_c,eq_T_d}): 
%\color{red}
\begin{equation}
\begin{split}\label{eq_all}
\mathcal{A}_{pp}^{(1)}(p)=& \frac{ C_2^{pp}\psi_0(p)}{[1- C_0^{pp}J_0(p)]^2}\left \{\psi_2(p) + C_0^{pp} [\psi_0(p)J_2(p)- \psi_2(p)J_0(p)]\right\}=\\
&\frac{ C_2^{pp}\psi_0^2}{[1- C_0^{pp}J_0(p)]^2} \left[p^2-\mu\alpha M -\frac{1}{2} \left(\alpha M\right)^2- C_0^{pp}J\right]~.
\end{split}
\end{equation}
Since $\mu$ can be chosen such that $\alpha M\ll\mu$; $\dfrac{1}{2} \left(\alpha M\right)^2+ C_0^{pp}J$ is negligible in comparison to $p^2-\mu\alpha M$. Therefore, \cref{eq_all} becomes:
\begin{multline}
\mathcal{A}_{pp}^{(1)}(p)=\frac{ C_2^{pp}\psi_0^2}{[1- C_0^{pp}J_0(p)]^2} \left(p^2-\mu\alpha M\right)= C_\eta^2e^{2i\sigma_0}\frac{M \rho_c}{8 \pi \left(J_0(p)-\frac{1}{C_0^{pp}}\right)^2}\left(p^2-\mu\alpha M\right)=\\
C_\eta^2e^{2i\sigma_0}\frac{4 \pi}{M}\frac{\rho_C}{2}\frac{1}{\left(\frac{1}{a_p}+\alpha M \Phi(\eta) \right)^2}\left(p^2-\mu\alpha M\right)~.
\end{multline}
 Up to NLO, the $pp$ scattering amplitude is given by:
 \begin{multline}\label{eq_A_pp_NLO}
i\mathcal{A}_{pp}^{\text{NLO}}(p)=i\mathcal{A}_{pp}^{\text{LO}}(p)+i\mathcal{A}_{pp}^{(1)}(p)=\\-iC_\eta^2 e^{2i\sigma_0}\frac{4 \pi }{M}\left[ \left(\frac{1}{a_p}+\alpha M \Phi(\eta) \right)^{-1}- \frac{\rho_C}{2}\frac{1}{\left(\frac{1}{a_p}+\alpha M \Phi(\eta) \right)^2}\left(p^2-\mu\alpha M\right)\right]=\\
i\mathcal{A}_{pp}^{\text{LO}}(p)\left[1- \frac{\rho_C}{2}\frac{p^2-\mu\alpha M}{\frac{1}{a_p}+\alpha M \Phi(\eta) }\right]~.
 \end{multline}

\begin{comment}\color{black}
The NLO \cblack correction to the \cblack $pp$ amplitude scattering is then given by:
\begin{equation}
\mathcal{A}_{pp}^{\text{NLO}}=-C_\eta^2e^{2i\sigma_0}\frac{\pi \rho_C}{2}\frac{p^2-\alpha\mu M}{\frac{1}{a_p}+\Phi(\eta)}.
\end{equation}
\end{comment}

\cblack
\section{Pionless EFT Lagrangian coupled to electro-weak fields}
All interactions discussed in this thesis are either weak or electromagnetic, which should be added to \pilesseft Lagrangian, \cref{eq_lag3}.
\subsection{Weak interaction in pionless EFT}\label{weak_section}
For the low-energy process, the weak-interaction Lagrangian is: 
\begin{equation}\label{lweak}
\mathcal{L}_{\text{Weak}}=\frac{G_FV_{ud}}{\sqrt{2}}l_+^{\mu }J_{\mu
}^-~, 
\end{equation}
where $G_F$ is the Fermi constant and $V_{ud}$ is the Cabibbo-Kobayashi-Maskawa (CKM) matrix element. $l^\mu$ is the leptonic
current %$l^\mu=\bar{u}_e\gamma(1-\gamma_5)v_\mu$, 
and 
$J_\mu$ is the hadronic current, which contains two parts, a polar-vector and
axial-vector, $J_\mu=V_\mu-A_\mu$.

Up to NLO, the part of the polar
vector current relevant to $\beta$-decay with a vanishing energy transfer is:

\begin{equation}
\label{eq_weak_V}
V_0^{\pm}=N^{\dagger}\frac{{\tau^\pm}}{2}N~,
\end{equation}
where $\tau^{\pm}=\tau_1\pm i\tau_2$.

Here, we utilized the fact that the Conserved Vector Current (CVC)
hypothesis is accurate at this order of EFT.
The axial-vector part is (see Ref.~\cite{Ando_proton,rearrange}): 
\begin{equation}
\label{eq_weak_A}
\boldsymbol{ A^{\pm}}=\underbrace{\frac{g_A}{2}N^\dagger
	\boldsymbol{\sigma}\tau^{\pm}N}_{\text{LO}}
+\underbrace{L_{1, A} \left[\left (N^T P^i_tN\right)^\dagger \left (N^T P^A_sN\right)+h.c\right]}_{\text{NLO}}~,
\end{equation}
where $g_{A}$ is the axial coupling constant for a single nucleon, known from neutron $\beta$-decay and $L_{1, A}$ is the unknown two-body weak 
LEC and $\boldsymbol{\sigma}$ the Pauli matrix vector.
\subsection{Electromagnetic interaction in pionless EFT}\label{magnetic_section}
The one-body Lagrangian of the electromagnetic interaction is given by: 
\begin{equation}\label{eq_l_magentic_1}
\mathcal{L}^{(1)}_{\text{magnetic}}=\frac{e}{2M}N^\dagger\left[\left(\kappa_0+\kappa_1\tau_3\right)\boldsymbol{\sigma}\cdot B \right]N,
\end{equation}
where $\kappa_0 = 0.439902328(26)$ is the isoscalar magnetic moment of the nucleon and $\kappa_1= 2.352945028(26)$ is the
isovector magnetic moment of the nucleon, both presented in units of nuclear magneton. The magnetic field is conventionally defined as $B = \nabla\cdot A$.

According to the \pilesseft na\"{i}ve power-counting \cite{quartet}, at NLO there are two contributions to the two-body electromagnetic Lagrangian: 
\begin{equation}\label{eq_l_magentic_2}
\mathcal{L}^{(2)}_{\text{magnetic}}=e\left[ L_1\left(N^TP_s^AN\right)^\dagger\left(N^TP_t^iN\right)B_i-%\cblack 
i\epsilon_{ijk}\right.\cblack
\left.L_2\left(N^TP_t^iN\right)^\dagger\left(N^TP_t^jN\right)B_k+h.c\right],
\end{equation}
where $L_1$ and $L_2$ are the two-body LECs of the electromagnetic interaction: $L_1$ couples the spin-singlet channel to the spin-triplet channel and $L_2$ couples %\cblack 
two spin-triplet channels. \cblack
\section{Introducing dibaryon fields}
The Hubbard-Stratonovich (H-S) transformation is an exact mathematical transformation from a nucleon-nucleon field into a field of two nucleons (dibaryon). This transformation is most helpful in writing a three-particle system, as presented first in Refs.~\cite{3bosons, triton, rearrange}.

The transformation is defined using the integral: 

\begin{equation}\label{HS}
\exp\left({ax^2}\right)=\int^{\infty}_{\infty}\dfrac{\exp\left(-\frac{y^2}{a}+2x\cdot y\right)}{\sqrt{\pi a }}dy~.
\end{equation}
A general $\psi^4$ Lagrangian (\cref{lag2}) has the form: 
\begin{equation}\label{lag}
\mathcal{L}= N^T\left(i\partial_0+\frac{\nabla^2}{2M}\right)N-C_0\left(N^\dagger N\right)^2~, 
\end{equation}
and the total path integral of the Lagrangian is defined as: 

%\cblack
\begin{equation}\label{Z1}
Z=\int\mathcal{D}[N^\dagger, N]\exp\left\{-\left[\int dx_0 \int d^dx N^T\left(i\partial_0+\frac{\nabla^2}{2M}\right)N-C_0\left(N^\dagger N\right)^2\right]\right\}~,
\end{equation}
 where: $\mathcal{D}[N]=\prod\limits_{i}dN_i$ over all the partials.
\cblack

Using \cref{HS}, \cref{Z1} can be written as: 
%\cblack
\begin{equation}\label{eq_Z2}
Z=\frac{1}{\sqrt{\pi C_0}}\int\mathcal{D}[N^\dagger, N] da\exp\left\{-\int dx_0 \int d^dx \left[ N^T\left(i\partial_0+\frac{\nabla^2}{2M}\right)N+\frac{1}{C_0}a^2-2\left(N^\dagger N\right)a\right]\right\}, 
\end{equation}
where $a$ is an auxiliary dibaryon field that contains two nucleons.
%from comparing \cref{Z1} to \cref{eq_Z2} we get that (up to some numerical constants) : 

%\begin{equation}\label{eq_HSL}
%N^T\left(i\partial_0+\frac{\nabla^2}{2M}\right)N-C_0\left(N^\dagger N\right)^2= N^T\left(i\partial_0+\frac{\nabla^2}{2M}\right)N+\frac{1}{4C_0}\sigma^2-\left(N^\dagger N\right)\sigma
%\end{equation}
\cblack
\subsection{Hubbard-Stratonovich transformations for pionless EFT}
In this subsection, we apply the H-S transformation to the \pilesseft Lagrangian \cite{few, rearrange,Bedaque:1999vb} that leads to an equivalent Lagrangian in terms of two-body effective degrees of freedom. For simplification, we denote $ \left(i\partial_0+\frac{\nabla^2}{4M}\right)$ of \cref{eq_lag3} by $\mathcal{O}_D$, and by using the following notation: 
\begin{equation}\label{not}
\left(N^TP_{t, s}N\right)=\psi_{t, s}~.
\end{equation}
using this notation,\cref{eq_lag3} has the form: 
\begin{equation}\label{lag4}
\mathcal{L}= N^\dagger\left(i\partial_0+\frac{\nabla^2}{2M}\right)N
-\sum\limits_{a=t, s}\psi_a^\dagger\left(C_{0}^a+{C_2^a\ M}\cdot\mathcal{O}_D\right)\psi_a 
\end{equation}
and the path integral $Z$ is given by: 
%\cblack
\begin{multline}\label{Z2}
Z=\int dNdN^\dagger d\psi_td\psi_s\exp\left\{-\int dx_0 \int d^dx\left[N^\dagger\left(i\partial_0+\frac{\nabla^2}{2M}\right)N\right.
-\left.\sum\limits_{a=t, s}\psi_a^\dagger\left(C_{0}^a+{C_2^a M}\mathcal{O}_D
\right)\psi_a\right]\right\}~. \\
\end{multline}
\cblack
By performing the H-S transformation and introducing
two new dibaryon fields, $t$ (with spin 1 and isospin 0) and $s$ (with spin 0 and isospin 1), one can write path integral $Z$ as: %and assuming that ${\mathcal{O}_D}\ll1$, one gets: 
%\cblack
\begin{multline}
Z=\int dNdN^\dagger d\psi_td\psi_s dt ds\exp\left\{-\int dx_0 \int d^dx\left[N^\dagger\left(i\partial_0+\frac{\nabla^2}{2M}\right)N+\right.\right.\\
\left.\left.\sum\limits_{a=t, s} -a^\dagger(\mathcal{O}_D-\sigma_a)a-y_a(\psi^\dagger a+a^\dagger\psi)\right]\right\}=\\
\int dNdN^\dagger\psi_t, d\psi_s\exp\left\{-\int dx_0 \int d^dx\left[N^\dagger\left(i\partial_0+\frac{\nabla^2}{2M}\right)N\right.
\left.-\sum_{n}\sum\limits_{a=t, s} \psi_a^\dagger\left(\frac{y_a^2}{\sigma_a }\left(\frac{\mathcal{O}_D }{\sigma_a}\right)^n\right)\psi_a^\dagger \right]\right\}.
\end{multline}
Keeping up to first order in $\mathcal{O}_D$, \cref{eq_lag2} becomes:
\begin{equation}\label{eq_lag2}
Z=\int dNdN^\dagger\psi_t d\psi_s\exp\left\{
-\int dx_0 \int d^dx\left[N^\dagger\left(i\partial_0+\frac{\nabla^2}{2M}\right)N-\sum\limits_{a=t, s} \psi_a^\dagger\left(\frac{y_a^2}{\sigma_a }+\frac{\mathcal{O}_D y_a^2}{\sigma_a^2}\right)\psi_a^\dagger\right]
\right\}~.
\end{equation}
\cblack
Comparing \cref{eq_lag2} with \cref{Z2} yields the following relations: 
\begin{align}
C_0^{t, s} &= \frac{y_{t, s}^2}{\sigma_{t, s}}\label{C0}~, \\
C_2^{t, s} &=\frac{y_{t, s}^2}{M\sigma_{t, s}^2}\label{C2}~.
\end{align}
From the comparison of \cref{C0,C2} with \cref{eq_C2,eq_C0_a}, we introduce two new coupling constants for each channel: 
\begin{align}
y_{t, s} &= \frac{ \sqrt{8\pi}}{M \sqrt{\rho_{{t, s}}}}\label{eq_yt}~, \\
\sigma_{t, s} &= \frac{2}{M\rho_{{t, s}}} \left(\frac{1}{a_{{t, s}}}-\mu \right)\label{eq_sigma_H_S}~, 
\end{align}
such that \cref{eq_lag3} has the form: 
%\cblack
\begin{multline}\label{H_S_lag2}
\mathcal{L}=N^{\dagger}\left(i\partial_0+\frac{\nabla^2}{2M}\right)N
-t^{i\dagger}\left[-\sigma_t+\left(i\partial_0+\frac{\nabla^2}{4M}\right)\right]t^i-s^{A\dagger}\left[-\sigma_s+\left(i\partial_0+\frac{\nabla^2}{4M}\right)\right]s^i-
\\ -y_t\left[t^{i\dagger}\left(N^TP_t^iN\right)+h.c\right]-
y_s\left[s^{A\dagger}\left(N^TP_s^iN\right)+h.c\right]~. 
\end{multline}

\cblack

\section{The dibaryon propagator}

\begin{figure}[h!]
	\centering
	\includegraphics[width=0.65\linewidth]{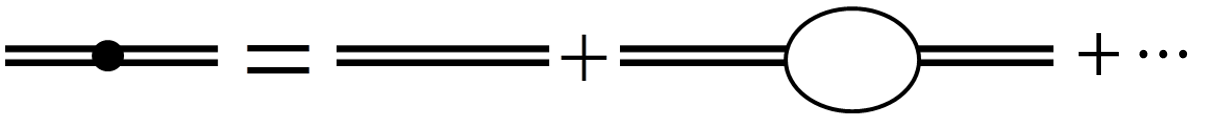}\\
	\caption{\footnotesize{The dressed dibaryon propagator. The bare dibaryon
		propagator is dressed by nucleon bubbles to all
		orders. %**** NOTE: the second diagram should include dressed propagator****
	}}\label{fig_dressed}
\end{figure}

The bare dibaryon propagator arising from
 the \pilesseft dibaryon Lagrangian, \cref{H_S_lag2} is:
\begin{equation}
\label{eq:D_bare}
i\mathcal{D}_{t,s}^{\rm bare} (p_0,{\bf p}) =
-i\left[ p_0 -\frac{{\bf p}^2}{4 M} -\sigma_{t,s}\right]^{-1}~.
\end{equation}
We use a power-counting that is appropriate for systems with a
scattering length $a$ that is large compared to the range of the
interaction $\rho$ \cite{Kaplan:1998tg}. The {\it full} dibaryon
propagator (Fig.~\ref{fig_dressed}) is therefore defined as the geometric sum of nucleon
bubbles connected by bare dibaryon propagators (see
Refs.~\cite{3bosons,triton} for more details):
\begin{equation}
\label{eq_dressed}
i\mathcal{D}_{t, s}^{\rm full} (p_0,{\bf p})=
i\mathcal{D}_{t,s}^{\rm bare} (p_0,{\bf p})
\cdot\sum_{n}\left (\mathcal{D}_{t,s}^{\rm bare} (p_0,{\bf p})\cdot\mathcal{I}_B\cdot\left (-2iy_{t, s}\right)^2\right)^n,
\end{equation}
where $\mathcal{I}_B$ denotes the two-nucleon loop integral evaluated
using the so-called power divergence subtraction (PDS) scheme (see
\cite{KSW1998_a,KSW_c,Kaplan:1998tg}).
The full ({\it unrenormalized}) propagator becomes then
\begin{multline}\label{full}
i\mathcal{D}^{\rm full}_{t,s} (p_0,{\bf p}) =
-i\Bigl[p_0-\frac{{\bf p}^2}{4M}-\sigma_{t,s}+
\frac{My_{t,s}^2}{4\pi} (\sqrt{-Mp_0 + {\bf p}^2/4-i\epsilon}-\mu)\Bigr]^{-1}=\\i\frac{4\pi}{M y_{t,s}^2}\Biggl[\frac{1}{a_{t, s}}-\sqrt{-M p_0+\frac{{\bf p}^2}{4}}+
\frac{\rho_{t, s}}{2}\left ({\bf p}^2/4-Mp_0\right)\Biggr]^{-1}~.
\end{multline}
\begin{comment}
where $\mu$ denotes a renormalization scale introduced through the PDS scheme.

The coupling constants can be obtained by matching to the effective range exasperation:
\begin{gather}
y_{t,s} = \frac{ \sqrt{8\pi}}{M \sqrt{\rho_{{t,s}}}}~,\\
\sigma_{t,s} = \frac{2}{M\rho_{{t,s}}} \left (\frac{1}{a_{{t,s}}}-\mu \right)~,
\end{gather}
where $\rho_{t, s}$ is the effective range and $a_{t,s}$ is the
scattering length. Given that, \cref{eq_D_1} becomes:

\begin{multline}
\label{full}
i\mathcal{D}^{\text{full}}_{t, s} (p_0,{\bf p})=i\frac{4\pi}{M y_{t,s}^2}\Biggl[\frac{1}{a_{t, s}}-\sqrt{-M p_0+\frac{{\bf p}^2}{4}}+
\\\frac{\rho_{t, s}}{2}\left ({\bf p}^2/4-Mp_0\right)\Biggr]^{-1}~. 
\end{multline}
\end{comment}
The dibaryon propagator shown above has two poles. One corresponds to
the {\it physical} bound-state (virtual) pole that results from the
large scattering length in the triplet (singlet) channel. The other
pole is a spurious pole whose energy scale lies beyond the breakdown
scale of the EFT. We expand the propagator in
\cref{full} in powers of the effective range since
the spurious pole causes problems in calculations for few-body
systems. Through this expansion, we can also isolate the pieces that
are dependent and independent of the effective range. Accordingly, we
define the LO dibaryon propagator as:
\begin{equation}
\label{eq:dibaryon_LO}
i\mathcal{D} ^{\text{LO}}_{t,s} (p_0,{\bf p})=
i\frac{4\pi}{M y_{t,s}^2}
\left (\frac{1}{a_{t,s}}-\sqrt{-M p_0+\frac{{\bf p}^2}{4}}\right)^{-1}~,
\end{equation}
which is equal to (see Ref.~\cite{Ando_proton}):
\begin{equation}
i\mathcal{D} ^{\text{LO}}_{t,s} (p_0,{\bf p})=\frac{-i}{y_{t,s}}i\mathcal{A}_{t,s}^{\text{LO}}(p')\frac{-i}{y_{t,s}}~,
\end{equation}
where:
\begin{equation}\label{p_prime}
p'=i\sqrt{{\bf p}^2/4-Mp_0}~.
\end{equation}
%%\cblack 
In the case of a bound-state, we expand the triplet propagator (\cref{full}) about the deuteron pole. Up to NLO, the triplet propagator up is given by \cite{konig2}:
\begin{equation}
\label{NLO_correction_triplet}
\begin{split}
i\mathcal{D} ^{\text{NLO}}_{t} (p_0,{\bf p})=
i\frac{4\pi}{M y_{t}^2}
\left (\gamma_t-\sqrt{-M p_0+\frac{{\bf p}^2}{4}}\right)^{-1}
\cdot\left[1+\frac{\rho_{t}}{2}\left(\frac{{\bf
		p}^2/4-Mp_0-\gamma_t^2}{-\gamma_t+\sqrt{-M p_0+\frac{{\bf p}^2}{4}}}\right)\right].
\end{split}
\end{equation}
For the singlet channel, the singlet propagator up to NLO is given by:
\begin{multline}
\label{NLO_correction_singlet}
i\mathcal{D}^{\text{NLO}}_s (p_0,{\bf p})=
i\frac{4\pi}{M y_s^2}
\left (\frac{1}{a_s}-\sqrt{-M p_0+\frac{{\bf p}^2}{4}}\right)^{-1}
\cdot\left[1+\frac{\rho_{s}}{2}\left(\frac{{\bf
		p}^2/4-Mp_0}{-\frac{1}{a_s}+\sqrt{-M p_0+\frac{{\bf p}^2}{4}}}\right)\right]=\\
	\frac{-i}{y_{s}}i\left[i\mathcal{A}_s^{{LO}}(p')+i\mathcal{A}_s^{(1)}(p')\right]\frac{-i}{y_{s}}~.
\end{multline}

\subsection{The Coulomb propagator}
At LO, the $pp$ propagator contains an infinite series of ladder diagrams of Coulomb photon exchanges. Due to the non-perturbative nature of the Coulomb interaction at low energies, those diagrams have
to be resummed up to infinity like that of the $pp$ scattering amplitude (subsection \ref{pp_amplitudes}). Similarly to \cref{eq:dibaryon_LO,NLO_correction_triplet}, the LO Coulomb propagator is given by \cite{Ando_proton}:

\begin{equation}\label{eq_pp_LO}
i\mathcal{D}_{pp} ^{\text{LO}}(p_0,{\bf p})=-\left(C_\eta^2e^{2i\sigma_0}\right)^{-1}
\frac{1}{y_s^2}i\mathcal{A} ^{\text{LO}}_{pp}(p')=i\frac{4\pi}{M y_s^2}
\left[\frac{1}{a_p}+2\kappa \Phi\left (\kappa /p'\right)\right]^{-1}~,
\end{equation}
where $a_p$ $\sigma_0$, $C_\eta$ and $\Phi$ are all defined in subsection \ref{pp_amplitudes},
\begin{equation}\label{kappa}
\kappa=\frac{\alpha M}{2}~.
\end{equation}
%\begin{comment}
Using \cref{eq_A_pp_NLO}, the Coulomb propagator up to NLO is given by\footnote{\text{Note that for the case of $^3$He, since $p'\geq 70 \mev$, $\mu\alpha M$ can be neglected in comparison to $p'^2.$}}: 
\begin{equation}\label{eq_pp_NLO}
i\mathcal{D}^{\text{NLO}}_{pp}(p_0,{\bf p})=-\left(C_\eta^2e^{2i\sigma_0}\right)^{-1}
\frac{1}{y_s^2}i\mathcal{A}^{\text{NLO}}_{pp}=i\mathcal{D}_{pp}^{\text{LO}}(p_0,{\bf p})\left[1- \frac{\rho_C}{2}\frac{p'^2-\mu\alpha M}{\frac{1}{a_p}+2\kappa\Phi\left(\kappa /p'\right) }\right]~.
\end{equation}
%\end{comment}

\section{The deuteron normalization and $Z$-parameterization}\label{Zd}
%The deuteron normalization is given by \cite{konig1}: 

%Na\"{i}vely, the effective ranges are fixed by scattering experiments. However, since the triplet channel is bound, the deuteron, the triplet effective range can be alternatively be fixed by the long-range properties of the deuteron wave function. These are two alternatives, justified, rearrangements of the perturbative expansion. 

The long range properties of the deuteron wave function are set by its residue \cite{Philips_N_N, Griesshammer_3body}: 
\begin{equation}
Z_d = \left[\frac{i\partial}{\partial_E} \frac{1}{iD_t (E, p) } \Big|_{E=-\frac{\gamma_t^2}{M}, p=0}\right]^{-1}=\frac{1}{1-\gamma_t \rho_t},
\end{equation}
\begin{comment}
\begin{equation}\label{eq_Zd}
Z_d^{-1}=i\frac{\partial}{\partial_{p_0}}\frac{1}{i\mathcal{D}_t(p_0, {\bf p})}\Big|_{p_0=-\frac{\gamma_t^2}{M}, p=0}, 
\end{equation}
\end{comment}
which is a function of both the deuteron binding momentum ($\gamma_t=\sqrt{M E_b(d)}$) and the effective range, $\rho_t$.
\begin {table}[H]
\begin{center}
	\begin{tabular}{c c||c c}
		\centering
		Parameter& Value& Parameter& Value\\
		\hline
		$\gamma_t$& 45.701 MeV \cite{32}& $\rho_t$& 1.765 fm \cite{33}\\
		$a_s$& -23.714 fm \cite{2}& $\rho_s$& 2.73 fm \cite{2}\\
		$a_p$& -7.8063 fm \cite{34}& $\rho_C$& 2.794 fm \cite{34}\\
	\end{tabular}
	\caption{ \footnotesize{Experimental values for parameters used in the numerical calculation}}
	\label{table_1}
\end{center}
\end{table}

Using the effective ranges that are fixed from scattering experiments we get (see Tab.~\ref{table_1}): 
\begin{equation}
Z_d^{\text{exp}} = \frac{1}{1-\gamma_t \rho_t} \approx 1.690(3)~.
\end{equation}
In the effective range expansion (ERE), where $\gamma_t\rho_t$ is the small parameter, the order-by-order expansion of $Z_d$ is: 
\begin{equation}\label{eq_Zd_Q}
\begin{split}
Z_d^{\text{LO}}&=1~, \\
Z_d^{\text{NLO}}&=1+\gamma_t\rho_t \approx 1.408~.
\end{split}
\end{equation}

However, for observables that include deuteron break-up or low-energy scattering, the normalization of the wave function prefactors the cross-section. In these cases, it was suggested that retaining the physical value of $Z_d$ is more essential than fixing $\rho_t$, as this choice converges better and faster in the EFT expansion \cite{Phillips:1999am, Rho:1999bv}. Thus, the alternative NLO parameterization, known as the $Z$-parameterization in which $Z_d-1$ is the small parameter, is given by 
\cite{Griesshammer_3body, Kong2, Philips_N_N, Vanasse, Vanasse:2015fph}: 
\begin{equation}\label{eq_Zd_2}
\begin{split}
Z_d^{\text{LO}}&=1~, \\
Z_d^{\text{NLO}}&=1+\left(Z_d^{\text{exp}}-1\right) = 1.690(3)~.
\end{split}
\end{equation}
The price for retaining $Z_d$ at its experimental value is that the value of the triplet effective range at NLO in this parameterization is: 
\begin{equation}
\rho'_t=\frac{Z_d^{\text{exp}}-1}{\gamma_t}\approx \frac{0.690}{\gamma_t}~,
\end{equation}
rather than $\rho_t$ of Tab.~\ref{table_1}. These two alternatives are valid rearrangements of the perturbative expansion. 
In the following, we use these two NLO parameterizations, and the difference in their predictions is used as one of the measures for the theoretical uncertainty stemming from the EFT truncation.
\newpage

\section{Summary of Feynman Rules}\label{feynman}
The Feynman rules for the electro-weak interactions are given in Tab.~\ref{tbl: feynman_ruls}. These Feynman rules cover all aspects of the three-nucleon electro-weak interactions. Some of these rules were presented in the previous subsections while the others will be derived in the following chapters.
\begin{comment}
where for the $pp$ channel: 
\begin{equation}
a_s=a_p(\mu)=\left\{\frac{1}{a_p}+\alpha M\left[\log{\frac{\mu\sqrt{\pi}}{\alpha M}}+1-\frac{3}{2}\gamma_E\right]\right\}^{-1}
\end{equation}
\end{comment}
\begin{longtable}{|m{0.5 cm}|m{4.4 cm}| m{2.6 cm}|m{8.7 cm}|}
%\centering
\hline
&\begin{center}
	{Field structure}
\end{center} &\begin{center}
	{Diagrammatic structure}
\end{center}&\begin{center}
	{Feynman rule}
\end{center}\\
\hline
\multicolumn{4}{|c|} {\large{Leading order}}\\
\hline
%\rowcolor[rgb]{ 0 0.74902 1}
(a)& single nucleon line
&\includegraphics[width=0.95\linewidth]{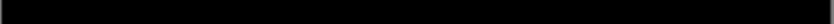}
& \begin{minipage}{\linewidth}
	\begin{equation}
	\nonumber
	iS^{ \alpha a }_{\beta b}=i\delta^{a}_b\delta^\alpha_\beta\left(p_0-\frac{{\bf
			p}^2}{2m}+i\varepsilon\right)^{-1}
	\end{equation}
\end{minipage}\\
\hline
(b)&\begin{tabular}{l}
	LO spin-triplet\\
	dibaryon propagator\\
	\cref{eq:dibaryon_LO}
\end{tabular}
&\includegraphics[width=0.95\linewidth]{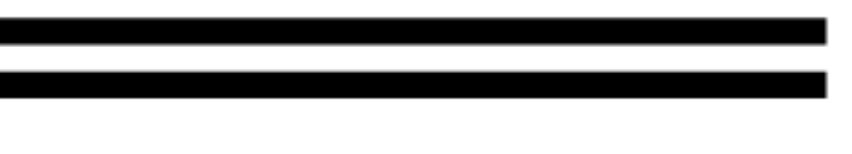}
&\begin{minipage}{\linewidth}
	\begin{equation}
	\nonumber
	i\mathcal{D}^{ij}_t(p_0, {\bf p })=i\delta^{ij}\dfrac{4\pi}{M y_t^2}\left(\dfrac{1}{a_t}-\sqrt{-M p_0+\frac{{\bf p}^2}{4}}\right)^{-1}
	\end{equation}
\end{minipage}\\ 
\hline
%\rowcolor[rgb]{0.529412 0.807843 0.980392}
%\rowcolor[rgb]{0.529412 0.807843 0.980392}
(c)&\begin{tabular}{l}
	LO spin-singlet\\
	dibaryon propagator\\
	\cref{eq:dibaryon_LO}
\end{tabular}
&\includegraphics[width=0.95\linewidth]{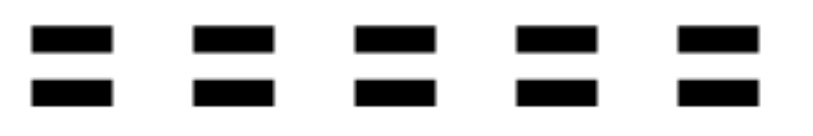}
& \begin{minipage}{\linewidth}
	\begin{equation}
	\nonumber
	i\mathcal{D}^{AB}_s(p_0, {\bf p })=i\delta^{AB}\dfrac{4\pi}{M y_s^2}\left[\dfrac{1}{a_s}-\sqrt{-M p_0+\frac{{\bf p}^2}{4}}\right]^{-1}
	\end{equation}
\end{minipage}\\
\hline
%\rowcolor[rgb]{0.529412 0.807843 0.980392}
(d)&\begin{tabular}{l}
	LO {\it pp} propagator\\
	\cref{eq_pp_LO}
\end{tabular}
&\includegraphics[width=0.95\linewidth]{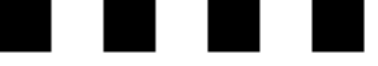}
&\begin{minipage}{\linewidth}
	\begin{equation}
	\nonumber
	i\mathcal{D}^{AB}_{pp}(p_0, {\bf p })=i\delta^{AB}\frac{4\pi}{M y_t^2}\left[\frac{1}{a_p}+2\kappa \Phi\left(\kappa /p'\right)\right]^{-1}
	\end{equation}
\end{minipage}\\ 
\hline
%\rowcolor[rgb]{0.596078 0.984314 0.596078}

%\rowcolor[rgb]{0.596078 0.984314 0.596078}
(e)&\begin{tabular}{l}
	spin-triplet\\ dibaryon vertex\\
	\cref{H_S_lag2}
\end{tabular}
&\includegraphics[width=0.95\linewidth]{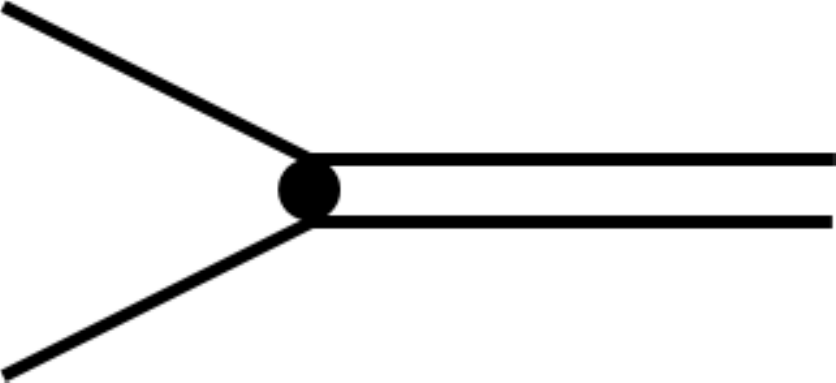}
&\begin{minipage}{\linewidth}
	\begin{equation}
	\nonumber
	- 2 i y_t \frac{1}{\sqrt{8}}\sigma^2\tau^2\sigma^i
	\end{equation}
\end{minipage}\\ 
\hline
(f)&\begin{tabular}{l}
	spin-singlet\\ dibaryon vertex\\
	\cref{H_S_lag2}
\end{tabular}
&\includegraphics[width=0.95\linewidth]{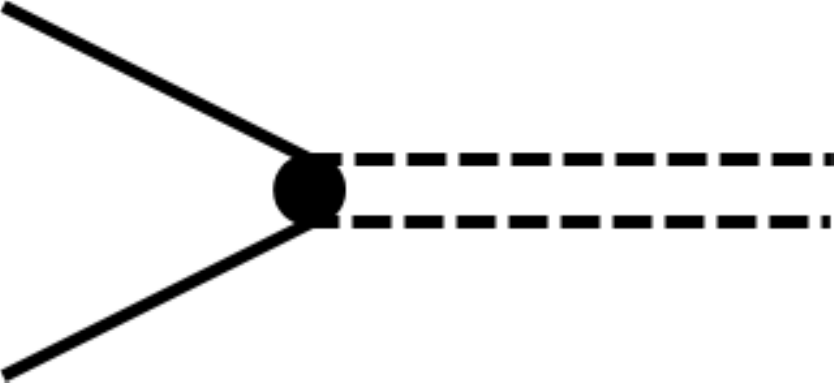}
& \begin{minipage}{\linewidth}
	\begin{equation}
	\nonumber
	- 2 i y_s \frac{1}{\sqrt{8}}\sigma^2\tau^2\tau^A
	\end{equation}
\end{minipage}\\ 
\hline
%\rowcolor[rgb]{0.980392 0.980392 0.823529}
(g)&\begin{tabular}{l}
	three nucleons vertex\\
	\cref{KH}
\end{tabular}
&\raisebox{-\totalheight}{\includegraphics[width=0.95\linewidth]{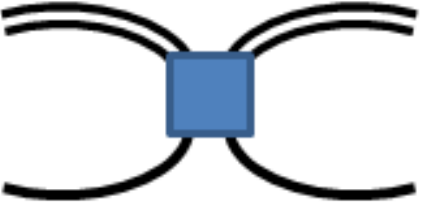}}
&\begin{minipage}{\linewidth}
	\begin{equation}
	\nonumber
	\dfrac{2H(\Lambda)}{\Lambda^2}
	\end{equation}
\end{minipage}\\ 
\hline
\newpage
\hline
\multicolumn{4}{|c|} {\large{Coulomb interaction}}\\
\hline
%\rowcolor[rgb]{1 0.647059 0}
(h)&\begin{tabular}{l}
	Coulomb propagator\\
	\cref{d_photon}
\end{tabular}
&\raisebox{-\totalheight}{\includegraphics[width=0.95\linewidth]{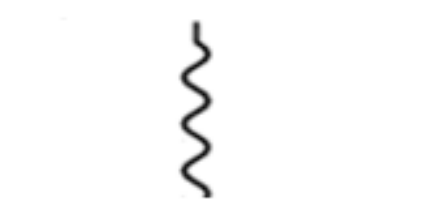}}
&\begin{minipage}{\linewidth}
	\begin{equation}
	\nonumber
	i\Delta_{\text{Coulomb}}(p)=\dfrac{i}{p^2+\lambda^2}
	\end{equation}
\end{minipage}\\ 
\hline
%\rowcolor[rgb]{1 0.647059 0}
(i)&\begin{tabular}{l}
	photon-nucleon vertex
\end{tabular}
&\raisebox{-\totalheight}{\includegraphics[width=0.95\linewidth]{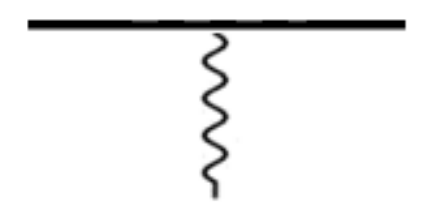}}
& \begin{minipage}{1\linewidth}
	\begin{equation}
	\nonumber
	\pm i e
	\end{equation}
\end{minipage}\\
\hline
%\rowcolor[rgb]{1 0.647059 0}
(j)&\begin{tabular}{l}
	photon-dibaryon\\ vertex
\end{tabular}
&\raisebox{-\totalheight}{\includegraphics[width=0.95\linewidth]{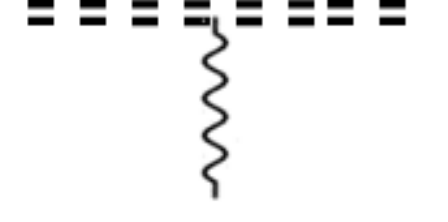}}
& \begin{minipage}{1\linewidth}
	\begin{equation}
	\nonumber
		\pm i e
	\end{equation}
\end{minipage}\\
\hline
\multicolumn{4}{|c|} {\large{Next to leading order}}\\
\hline
%%\rowcolor[rgb]{ 1 0.752941 0.796078}
%%%\rowcolor[rgb]{ 1 0.752941 0.796078}
(k)&\begin{tabular}{l}
	NLO spin-triplet\\
	Effective range\\
	correction\\
	\cref{NLO_correction_triplet}
\end{tabular}
&\includegraphics[width=0.95\linewidth]{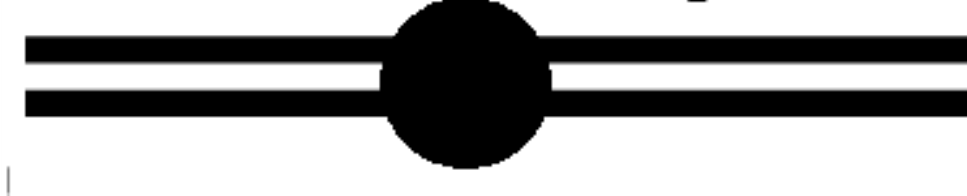}
&\begin{minipage}{\linewidth}
	%\cblack
	\begin{equation}
	\nonumber
	%i\mathcal{D}^{ij}_s(p_0, {\bf p })=
	-i\delta^{ij}\dfrac{4\pi}{M y_t^2}\frac{\rho_{t}}{2}\frac{{\bf
			p}^2/4-Mp_0-\gamma_t^2}{\left(\frac{1}{a_t}-\sqrt{-M p_0+\frac{{\bf p}^2}{4}}\right)^2}
	\end{equation}\end{minipage}\cblack\\
\hline
(l)&\begin{tabular}{l}
	NLO spin-singlet\\
	Effective range\\ correction\\
	\cref{NLO_correction_singlet}
\end{tabular}
&\includegraphics[width=0.95\linewidth]{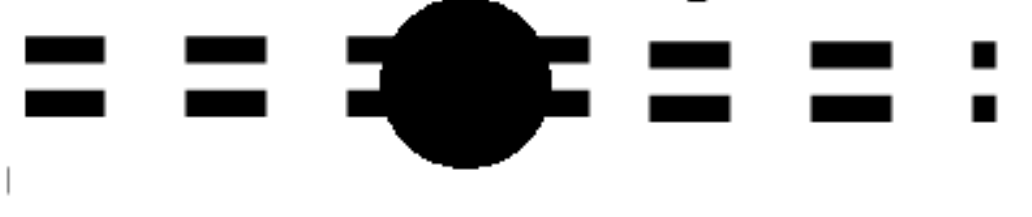}
& \begin{minipage}{\linewidth}
	\begin{equation}
	\nonumber
	%i\mathcal{D}^{AB}_s(p_0, {\bf p })=
	-i\delta^{AB}\dfrac{4\pi}{M y_s^2}\frac{\rho_s}{2}\frac{{-M p_0+\frac{{\bf p}^2}{4}}}{\left(\frac{1}{a_s}-\sqrt{-M p_0+\frac{{\bf p}^2}{4}}\right)^2}
	\end{equation}
\end{minipage}\\
\hline
%%\rowcolor[rgb]{ 1 0.752941 0.796078}
(m)&\begin{tabular}{l}
	NLO Coulomb\\
	Effective range\\ correction\\
	\cref{eq_pp_NLO}
\end{tabular}
&\includegraphics[width=0.95\linewidth]{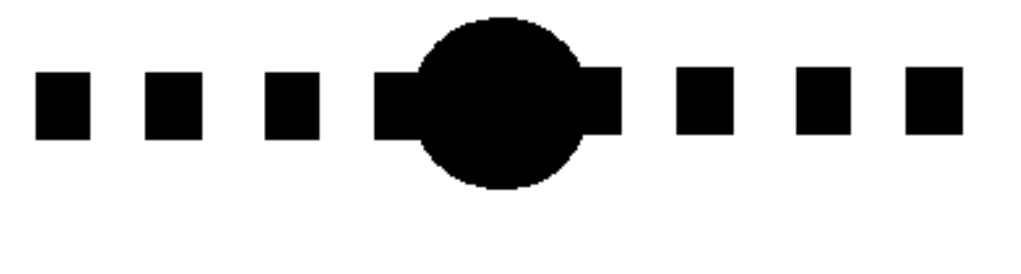}
&\begin{minipage}{\linewidth}
	\begin{equation}
	\nonumber
	%i\mathcal{D}^{AB}_{pp}s(p_0, {\bf p })=
	-i\delta^{AB}\frac{4\pi}{M y_s^2}\frac{\rho_C}{2}\frac{{\bf p}^2/4-M p-\alpha\mu M}{\left[\frac{1}{a_p}+2\kappa \Phi\left(\kappa /p'\right)\right]^2}
	\end{equation}
\end{minipage}\\ 
\hline
\multicolumn{4}{|c|}{\large{Weak Interaction - LO}}\\
\hline
%\rowcolor[rgb]{ 0.901961 0.901961 0.980392}
(n)&\begin{tabular}{l}
	one-body \\
	weak interaction\\
	\cref{eq_weak_A} 
\end{tabular}
&\raisebox{-\totalheight}{\includegraphics[width=0.95\linewidth]{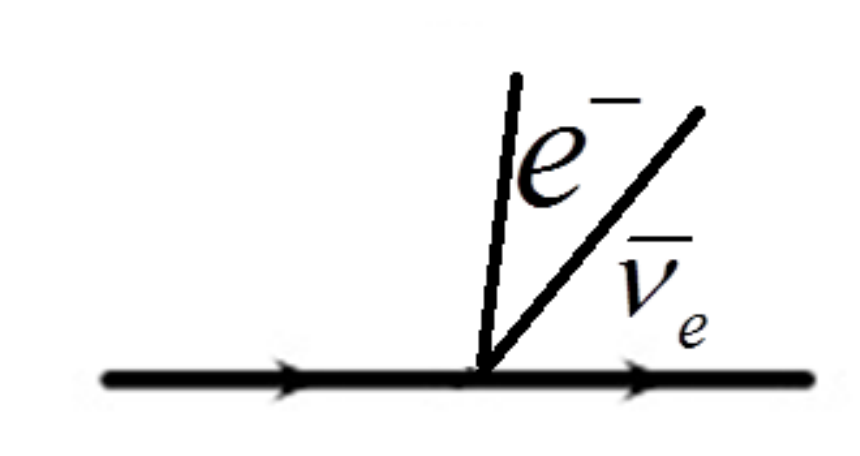}}
& \begin{minipage}{\linewidth}
	\begin{equation}
	\nonumber
	%- i\frac{g_A}{2}\boldsymbol{\sigma}\tau^-
	- i\frac{g_A}{2}\boldsymbol{\sigma}\tau^{\pm}
	\end{equation}
\end{minipage}\\
\hline
\multicolumn{4}{|c|}{\large{Weak Interaction - NLO}}\\
\hline
%\rowcolor[rgb]{ 0.901961 0.901961 0.980392}
(o)&\begin{tabular}{l}
	$t^\dagger (N^TP_sN)$/$(N^TP_sN)^\dagger t$\\ interaction vertex\\ \cref{eq_weak_axial2} 
\end{tabular}
&\raisebox{-\totalheight}{\includegraphics[width=0.95\linewidth]{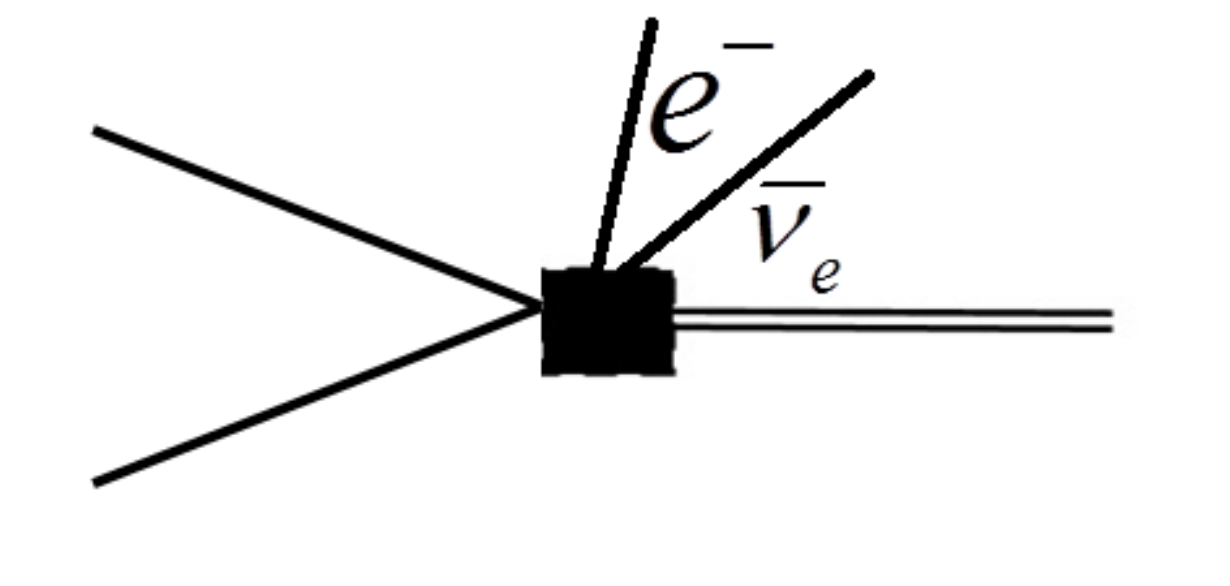}}
& \begin{minipage}{\linewidth}
	%\cblack
	\begin{eqnarray}
	\nonumber
	&i\left[\pi\sqrt{\frac{\rho_t}{2\pi}}\frac{1}{\mu-\frac{1}{a_s}}g_A-L_{1, A}\frac{1}{\sqrt{2\pi\rho_t}}\left(\mu-\frac{1}{a_t}\right)\right]\cdot\\
	\nonumber
	&\left[\frac{1}{\sqrt{8}}\sigma^2\tau^2\tau^A+\frac{1}{\sqrt{8}}\left(\sigma^2\tau^2\tau^A\right)^\dagger \right]
	\end{eqnarray}
\end{minipage}\\ 
\hline
%\rowcolor[rgb]{ 0.901961 0.901961 0.980392}
(p)&\begin{tabular}{l}
	$s^\dagger (N^TP_tN)$/$(N^TP_tN)^\dagger s$\\ interaction vertex\\
	\cref{eq_weak_axial2}
\end{tabular}
&\raisebox{-\totalheight}{\includegraphics[width=0.95\linewidth]{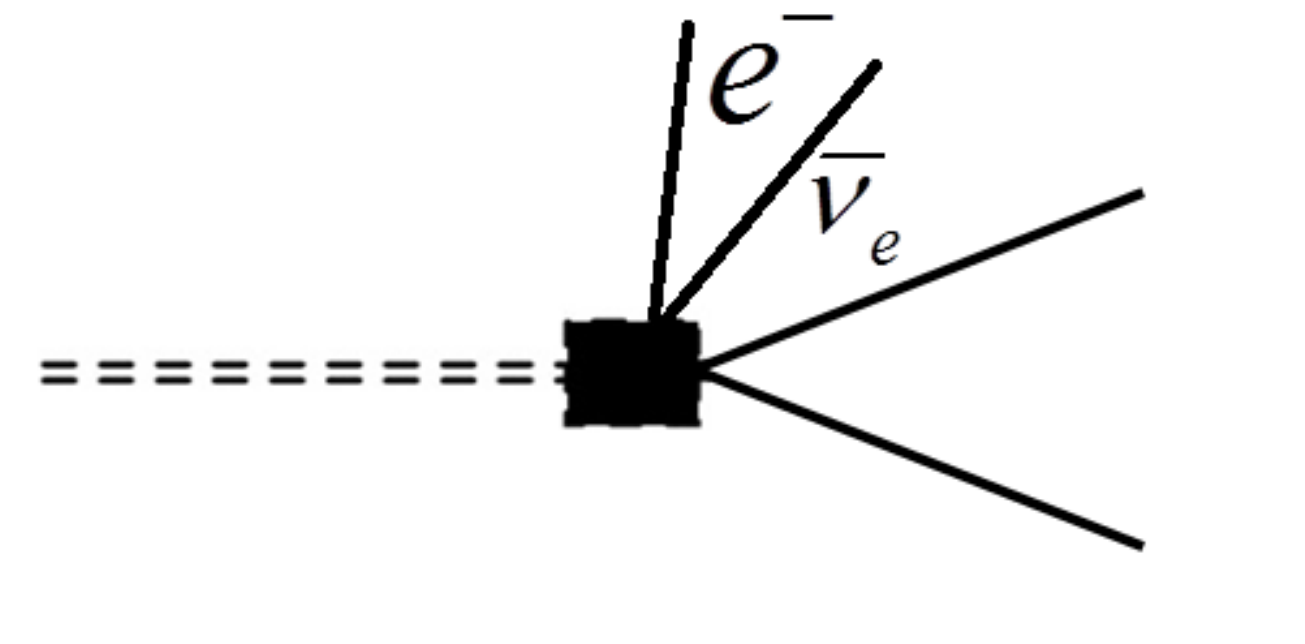}}& \begin{minipage}{\linewidth}
	%\cblack
	\begin{eqnarray}
	\nonumber
	&i\left[\pi\sqrt{\frac{\rho_s}{2\pi}}\frac{1}{\mu-\frac{1}{a_t}}g_A-L_{1, A}\frac{1}{\sqrt{2\pi\rho_s}}\left(\mu-\frac{1}{a_s}\right)\right]\cdot\\
	\nonumber
	&\left[\frac{1}{\sqrt{8}}\sigma^2\tau^2\sigma^i+\frac{1}{\sqrt{8}}\left(\sigma^2\tau^2\sigma^i\right)^\dagger \right].
	\end{eqnarray}
\end{minipage}\\
\hline
%\rowcolor[rgb]{ 0.901961 0.901961 0.980392}
(q)&\begin{tabular}{l}
	$s^\dagger t$/$t^\dagger s$\\ interaction vertex \\
	\cref{eq_weak_axial2}
\end{tabular}
&{\includegraphics[width=0.95\linewidth]{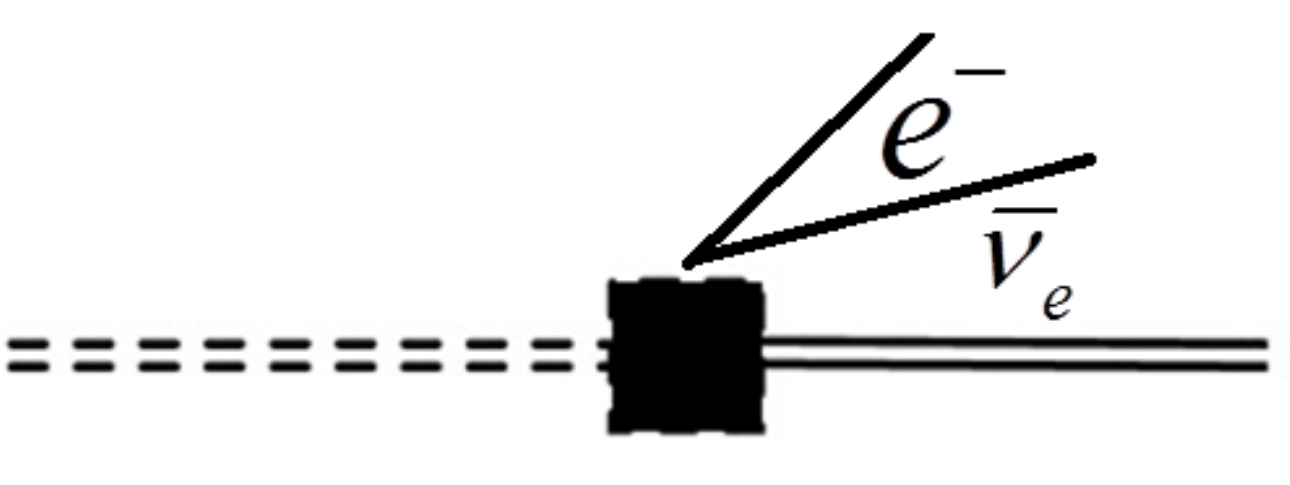}}
& \begin{minipage}{\linewidth}
	%\cblack
	\begin{eqnarray}
	\nonumber
	&i\left[\frac{1}{2}\frac{\rho_t+\rho_s}{\sqrt{\rho_t\rho_s}}g_A-\right.\\
	\nonumber
	&\left.L_{1, A}\frac{1}{2\pi\sqrt{\rho_t\rho_s}}\left(\mu-\frac{1}{a_t}\right)\left(\mu-\frac{1}{a_{s}}\right)\right]
	\end{eqnarray}
\end{minipage}\\
\hline 
\multicolumn{4}{|c|}{\large{Magnetic Interaction - LO}}\\
\hline
%\rowcolor[rgb]{ 0.901961 0.901961 0.980392}
(r)&\begin{tabular}{l}
	one-body \\
	magnetic interactions \\
	\cref{eq_l_magentic_1}
\end{tabular}
&\raisebox{-\totalheight}{\includegraphics[width=0.95\linewidth]{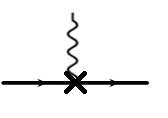}}
& \begin{minipage}{\linewidth}
	%\cblack
	\begin{equation}
	\nonumber
	%- i\frac{g_A}{2}\boldsymbol{\sigma}\tau^-
	- i\frac{e}{2M}\left(\kappa_0+\kappa_1\tau^3\right)\boldsymbol{\sigma}
	\end{equation}
\end{minipage}\\
\hline
\multicolumn{4}{|c|}{\large{Magnetic interactions - NLO, $L_1$ }}\\
\hline
%\rowcolor[rgb]{ 0.901961 0.901961 0.980392}
(s)&\begin{tabular}{l}
	$t^\dagger (N^TP_sN)$\\ interaction vertex\\ 
	\cref{eq_magnetic_dibaryon}
\end{tabular}
&\raisebox{-\totalheight}{\includegraphics[width=0.95\linewidth]{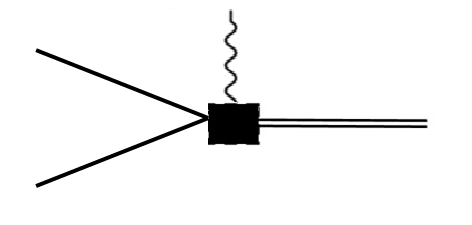}}
& \begin{minipage}{\linewidth}
	%\cblack
	\begin{eqnarray}
	\nonumber 
	&i\left[\pi\sqrt{\frac{\rho_t}{2\pi}}\frac{1}{\mu-\frac{1}{a_s}}\frac{e}{2M}\left(2\kappa_1\right)
	-\right.\\
	\nonumber
	&\left.\qquad
	-\frac{eL_1}{\sqrt{2\pi\rho_t}}\left(\mu-\frac{1}{a_t}\right)\right]
	\left[\frac{1}{\sqrt{8}}\sigma^2\tau^2\tau^A\right]
	\end{eqnarray}
\end{minipage}\\ 
\hline
(t)&\begin{tabular}{l}
	$(N^TP_sN)^\dagger t$\\ interaction vertex\\ 
		\cref{eq_magnetic_dibaryon}
\end{tabular}
&\raisebox{-\totalheight}{\includegraphics[width=0.95\linewidth]{L1_a.png}}
& \begin{minipage}{\linewidth}
		%\cblack
	\begin{eqnarray}
	\nonumber
	&i\left[\pi\sqrt{\frac{\rho_t}{2\pi}}\frac{1}{\mu-\frac{1}{a_s}}\frac{e}{2M}\left(2\kappa_1\right)-\right.
	\\
	\nonumber
	&\left.
	\frac{eL_1}{\sqrt{2\pi\rho_t}}\left(\mu-\frac{1}{a_t}\right)\right]\cdot
	\left[\frac{1}{\sqrt{8}}\left(\sigma^2\tau^2\tau^A\right)^\dagger \right]
	\end{eqnarray}
\end{minipage}\\ 
\hline
%\rowcolor[rgb]{ 0.901961 0.901961 0.980392}
(u)&\begin{tabular}{l}
	$s^\dagger (N^TP_tN)$/$(N^TP_tN)^\dagger s$\\ interaction vertex\\
		\cref{eq_magnetic_dibaryon}
\end{tabular}
&\raisebox{-\totalheight}{\includegraphics[width=0.95\linewidth]{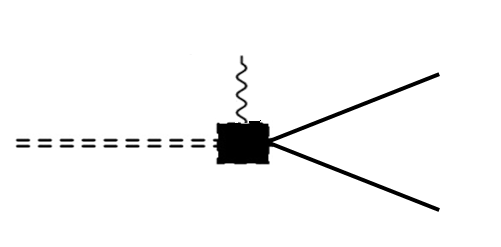}}& \begin{minipage}{\linewidth}
	%\cblack
	\begin{eqnarray}
	\nonumber
	&i\left[\pi\sqrt{\frac{\rho_s}{2\pi}}\frac{1}{\mu-\frac{1}{a_t}}\frac{e}{2M}\left(2\kappa_1\right)-
	eL_{1}\frac{1}{\sqrt{2\pi\rho_s}}\left(\mu-\frac{1}{a_s}\right)\right]\cdot\\
	\nonumber
	&\left[\frac{1}{\sqrt{8}}\sigma^2\sigma^i\tau^A+\frac{1}{\sqrt{8}}\left(\sigma^2\tau^2\sigma^i\right)^\dagger \right]
	\end{eqnarray}
\end{minipage}\\
\hline
%\rowcolor[rgb]{ 0.901961 0.901961 0.980392}
(v)&\begin{tabular}{l}
	$s^\dagger t$/$t^\dagger s$\\ interaction vertex \\
		\cref{eq_magnetic_dibaryon}
\end{tabular}
&{\includegraphics[width=0.95\linewidth]{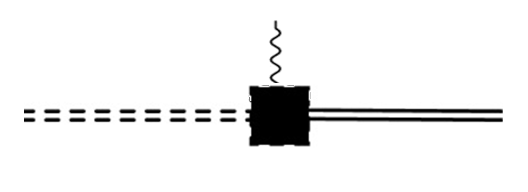}}
& \begin{minipage}{\linewidth}
	%\cblack
	\begin{eqnarray}
	\nonumber
	&i\left[\frac{1}{2}\frac{\rho_t+\rho_s}{\sqrt{\rho_t\rho_s}}\frac{e}{2M}\left(2\kappa_1\right)-\right.\\
	\nonumber
	&-e\left.L_{1}\frac{1}{2\pi\sqrt{\rho_t\rho_s}}\left(\mu-\frac{1}{a_t}\right)\left(\mu-\frac{1}{a_{s}}\right)\right]
	\end{eqnarray}
\end{minipage}\\
\hline
\multicolumn{4}{|c|}{\large{Magnetic interactions - NLO, $L_2$ }}\\
\hline
%\rowcolor[rgb]{ 0.901961 0.901961 0.980392}
(w)&\begin{tabular}{l}
	$t^\dagger (N^TP_tN)$/$(N^TP_tN)^\dagger t$\\ interaction vertex\\ 
		\cref{eq_magnetic_dibaryon}
\end{tabular}
&\raisebox{-\totalheight}{\includegraphics[width=0.95\linewidth]{L1_a.png}}
& \begin{minipage}{\linewidth}
	%\cblack
	\begin{eqnarray}
	\nonumber
	&i\left[\pi\sqrt{\frac{\rho_t}{2\pi}}\frac{1}{\mu-\frac{1}{a_t}}\frac{e}{2M}\left(2\kappa_0\right)-
	eL_{2}\frac{2}{\sqrt{2\pi\rho_t}}\left(\mu-\frac{1}{a_t}\right)\right]\cdot
	\\
	&\nonumber
	\left[\frac{1}{\sqrt{8}}\sigma^2\tau^2\sigma^i+\frac{1}{\sqrt{8}}\left(\sigma^2\tau^2\sigma^i\right)^\dagger \right]
	\end{eqnarray}
\end{minipage}\\ 
\hline
%\rowcolor[rgb]{ 0.901961 0.901961 0.980392}
(x)&\begin{tabular}{l}
	$t^\dagger t$ interaction vertex \\
		\cref{eq_magnetic_dibaryon}
\end{tabular}
&{\includegraphics[width=0.95\linewidth]{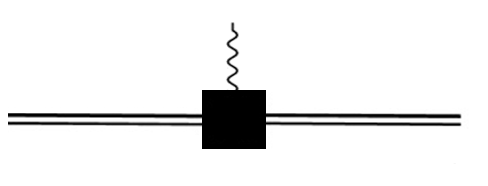}}
& \begin{minipage}{\linewidth}
	%\cblack
	\begin{eqnarray}
	\nonumber
	&i\left[\frac{e}{2M}\left(2\kappa_0\right)-
	eL_{2}\frac{1}{\pi\rho_t}\left(\mu-\frac{1}{a_t}\right)\left(\mu-\frac{1}{a_{s}}\right)\right]
	\end{eqnarray}
\end{minipage}\\ \hline 
% \end{tabular}
\caption{\footnotesize{%(color on line%)
		The Feynman rules for the NLO electro-weak interactions. a: %(blue) : 
		the single nucleon propagator. b-d%(light blue)
		: the dibaryon propagator. e-f: %(green): 
		: the dibaryon- nucleon interaction. g: 
		%(yellow)
		the three-body force. h-j: %(orange)
		the Coulomb interaction. k-m: %(pink)
		the ERE correction to the dibaryon propagator. n-q: %(purple)
		the weak interaction. r-x: the magnetic interaction. For all Feynman rules, a capital letter indicates the isospin index, while a small letter indicates the spin index.}}
\label{tbl: feynman_ruls}
\end{longtable}

\chapter{Three nucleon system at leading order}\label{3_body}
In this chapter, we %\cblack 
review \cblack the derivation of the Faddeev equation for
nucleons and its projections on the quantum numbers relevant for $^3$H ($n-d$) and $^3$He ($p-d$) at LO. %\cblack 
The derivation of the Faddeev equations for the three-nucleon has been widely studied in literature during the last twenty years, e.g., Refs.~\cite{quartet,3He,triton,3bosons,konig1,konig3,Griesshammer:2005ga,Parity-violating,konig5}. In this chapter we redrive the general equations for three-nucleon system. In addition we show a new numerical solution for the unbound $n-d$ scattering. \cblack
%In the following derivation, the three-body force was not taken into account and will be calculated numerically. A detailed analytical calculation of the of the three-body force is given
%in \cite{3bosons, triton, Three_Body_System}.
\section{n-d scattering and $^3$H bound-state}
In this section, we derive the Faddeev equations for nucleons and their projections on the quantum numbers relevant for the doublet and quartet channels of the $n-d$ scattering ($^3$H), based on Refs.~\cite{triton,Parity-violating,konig1,konig5}. 

This n-d scattering can be divided into two separate channels: the quartet channel in which the nucleon and the dibaryon are summed to total angular momentum $J=3/2$ and the doublet channel where the nucleon and the dibaryon are summed to a total angular momentum $J=1/2$, \cblack while the doublet channel is a result of both $n-d$ and $N-s$ scattering. In the latter, one needs to solve coupled integrals equations for the two scattering matrices. 

\subsection{Quartet channel}
\begin{figure}[h!]
	\begin{center}
		\includegraphics[width=0.95\linewidth]{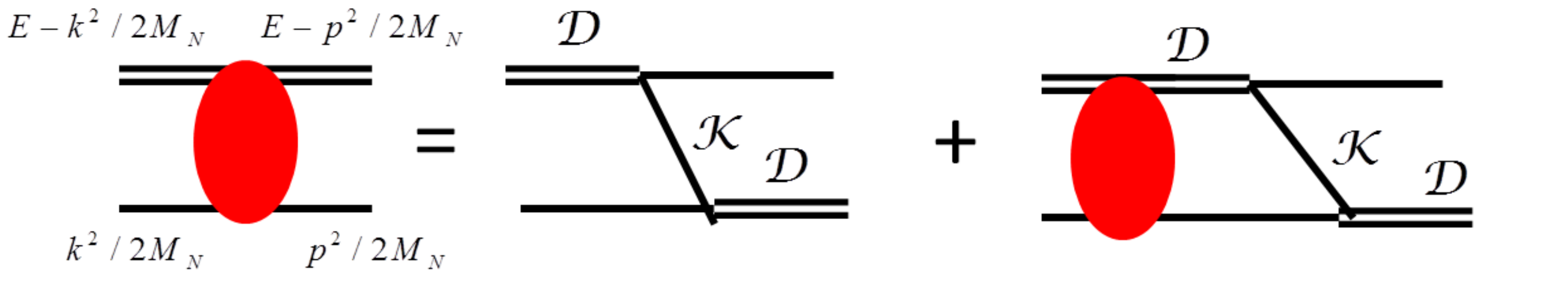}\\
		\caption{\footnotesize{Integral equation for the strong scattering matrix $T$ (red bubble), where $\mathcal{D}$ is the dibaryon propagator and $\mathcal{K}$ is the one-nucleon exchange propagator.}}\label{fig_quartet}
	\end{center}
\end{figure}

In the case of quartet channel, the three nucleons are coupled to a total spin of $J=3/2$ and $T=1/2$. Hence the two-nucleon (dibaryon) channel has to be $^3S_1$ (deuteron) only.
The Feynman diagrams leading $n-d$ scattering integral equation are shown in Fig.~\ref{fig_quartet}.
We use the Feynman rules presented in section~\ref{feynman} to write the three-nucleon scattering matrix, $T$ (see Appendix~\ref{whateverB} %\cblack 
and Ref.~\cite{Bedaque:1999vb,konig1,Parity-violating,triton}):
\cblack
\begin{multline}\label{eq_doublet}
(i T(E;{\bf p}, {\bf k})^{ij})^{\alpha a}_{\beta
	b}=-\frac{iMy_t^2}{2}\left(\sigma^j\sigma^i\right)^\alpha_\beta\delta^a_bi\cdot\frac{1}{{\bf k}^2+{\bf p}^2-{\bf k}\cdot{\bf p}-ME}+\\
\int\frac{\hbox{d}^4q}{(2\pi)^4}\frac{My_t^2}{2} i \mathcal{D}_t(p_0, {\bf p'})i\left(T\left(E;{\bf p}, {\bf p'}\right)^{kj}\right)^{\gamma a}_{\beta b}\cdot\frac{\left(\sigma^i\sigma^k\right)^\alpha_\gamma\delta^a_b}{{\bf p'}^2+{\bf p}^2-{\bf p'}\cdot{\bf p}-ME}, 
\end{multline}
\cblack
where $\alpha,\beta,\gamma$ indicate the elements in the $\sigma$'s matrices, while $a, b$ indicate the elements in the $\tau$'s matrices. $\delta^a_b$ is the Kronecker delta. $\mathcal{D}_t$ and $y_t$ are the dibaryon propagator and coupling constant, respectively and $E$ is the three-nucleon energy. %\cblack $\lambda$ is spin-parameter \cite{Griesshammer:2005ga}, depends on the spins of the three particles and how their combinations. \cblack

Using the projection equation,one finds that: 
\begin{equation}\label{eq_int_field}
\begin{split}
iT(E;{\bf p}, {\bf k})^{\alpha a}_{\beta
	b}=\frac{1}{3}(\sigma^i)^\alpha_{\alpha'}i(T(E;{\bf p}, {\bf k})^{ij})^{\alpha' a}_{\beta'
	b}(\sigma^j)^{\beta'}_\beta
\end{split}
\end{equation}
and for $a=b=1$, $\alpha=\beta$, \cref{eq_int_field} becomes: 
%\cblack
\begin{equation}\label{eq_doublet2}
\begin{split}
T(E,k, p)=&- 2M y_t^2K_0(k, p,E)+2{My_t^2}\int{D_t(E,p') T(E,k, p')K_0(p', p, E)}\frac{p'^2}{2\pi^2}dp',
\end{split}
\end{equation}
\cblack
where the one nucleon exchange propagator is defined as:
\begin{equation}\label{eq_K1}
K_0(k, p, E)=\frac{1}{2pk}Q_0\left(\frac{p^2+k^2-ME}{pk}\right),
\end{equation}
\begin{equation}\label{eq_Q}
Q_0(\text{a})=\frac{1}{2} \int^1_{-1}\frac{1}{x+a}dx
\end{equation}
and
\begin{equation}\label{eq_D}
D_t(E, p)=\mathcal{D}_t\left(E-\frac{p^2}{2M}, {\bf p}\right)~.
\end{equation}
Equation (\ref{eq_doublet2}) can be written in the following form: 
%\cblack
\begin{equation}\label{eq_intergal}
T(E,k, p)=-2 My_t^2K_0(k, p,E)+ 2T(E,k, p')\otimes K_0^S(p', p, E), 
\end{equation}
\cblack
where
\begin{equation}\label{eq_K0_S}
K_0^S(p', p, E)={My_t^2}K_0\left(p', p, E\right)\cdot
D_t(E, p')
\end{equation}
and 
\begin{equation}\label{eq_otimes}
A(..., p)\otimes B(p, ...)=\int A(.., p)B(p, ...)\frac{p^2}{2\pi^2}dp.
\end{equation}

\begin{comment}
In the following derivation, the three-body force was not taken into account and will be calculated numerically. A detailed analytical calculation of the of the three-body force is given in \cite{3bosons, triton, Three_Body_System}
\end{comment}

%The Faddeev equation for the $n-d$ quartet channel are given in \cref{eq_doublet,eq_intergal,eq_K1,eq_int_field,eq_doublet2,eq_Q,eq_delta,eq_D}. Notice that the quartet channel contributes to higher orders calculations; therefore, we will not take it into account in this work.
\subsection{Doublet channel}
\begin{figure}[H]
	\centerline{
		\includegraphics[width=0.65\linewidth]{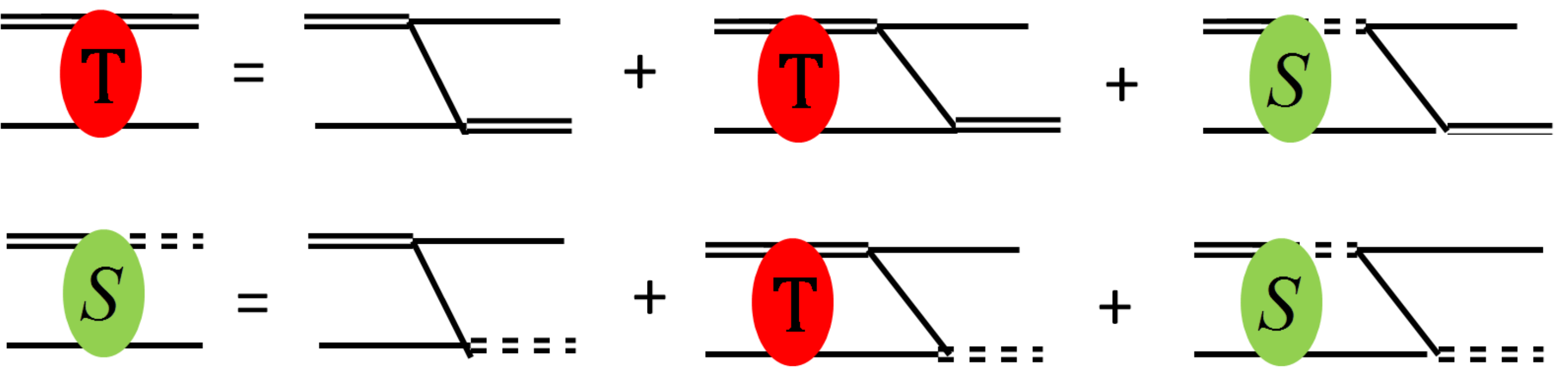}}
	\caption{\label{fig_triton_no_H} \footnotesize{n-d scattering equations. The double lines are the propagators of 
			the dibaryon fields $D_t$ (solid) and $D_s$ (dashed). The red
			bubbles (T) represent the triplet channel T=0, S=1, while the
			green bubbles (S) represent the singlet channel T=1, 
			S=0.}}
\end{figure}

In contrast to the quartet channel, in the doublet channel the spins of the nucleon and the deuteron are coupled to a total spin of $1/2$. The spin-singlet dibaryon can now appear in the intermediate state, which leads to two coupled amplitudes that differ in the type of the outgoing dibaryon as shown in Fig.~\ref{fig_triton_no_H}.
For the $n-d$ scattering we set: $a_{nn}=a_{np}=a_{s}$ and 
$S_{np}=S_{nn}=S$.
The two coupled Faddeev integral equations for $n-d$ scattering in the doublet channel are (see \cite{triton,Parity-violating,konig1,konig5} and Appendix~\ref{whateverB}): 
\begin{multline}
\label{stm_T}
T (E,k,p)=My_t^2 K_0 (k,p,E)
-My_t^2\int{D_t (E,p') T (E,k,p')K_0 (p', p, E)}\frac{p'^2}{2\pi^2}dp'+
\\
3My_ty_s\int{D_s (E,p')S (E,k,p')K_0 (p', p,
	E)}\frac{p'^2}{2\pi^2}dp'~,
\end{multline}
\begin{multline}
\label{stm_S1}
S (E,k,p)=-3My_ty_s K_0
-My_s^2\int{D_s (E,p') S(E,k,p')K_0 (p', p, E)}\frac{p'^2}{2\pi^2}dp'+\\
3My_sy_t\int{D_t (E,p') T (E,k,p')K_0 (p', p,
	E)}\frac{p'^2}{2\pi^2}dp'~.
\end{multline}
Equations (\ref{stm_T}) and (\ref{stm_S1}) can be written in
matrix form:
\begin{equation}\label{stm1}
t ^{nd} (E,k,p)=B_0(E,k,p)+
t ^{nd} (E,k,p')\otimes \hat{K}(p',p,E)~,
\end{equation}
where for n-d scattering:
\begin{equation}
\label{eq:T_LO}
t ^{nd} (E,k,p)=\left (
\begin{array}{c}
T (E,k,p) \\S (E,k,p)
\end{array}\right), 
\end{equation}
the inhomogeneous part of the integral equation is given by:
\begin{equation}\label{B_0}
B_0^{nd}(E,k,p)=K_0(k,p,E)\cdot
\left (
\begin{array}{c}
{My_t^2}\\
{-3y_ty_s}
\end{array}\right)~
\end{equation}
and the kernel is,
\begin{equation}
\label{eq_K_0}
\hat{K}^{nd} (p', p, E)=K_0 (p',p,E)
\cdot\left (\begin{array}{cc}
-My_t^2&3My_ty_s \\
3My_ty_s &-My_s^2\\
\end{array}\right)\cdot\left (\begin{array}{c}
{D}_t (E,p')\\
D_s (E,p')\\
\end{array}\right)\\
~.
\end{equation}
\subsubsection{ The Faddeev equation for the bound-state}
The above sections describe the Faddeev equation for the three-nucleon
system at an arbitrary energy. For energies close to the three-nucleon
binding energy, {\it i.e.,} when $E\sim E_B$, the scattering amplitude
takes the form
\begin{equation}\label{eq_t0}
t (E,k,p)=\frac{\mathcal{B}^\dagger (k)\mathcal{B} (p)}{E-E_B}+\mathcal{R} (E,k,p)~,
\end{equation}
where the $\mathcal{B} (E,k)$ are what we call {\it amputated} wave
functions or vertex factors, whereas the $\mathcal{R} (E,k,p)$ are terms that are regular at $E = E_B$, and thus can be
neglected for $E\rightarrow E_B$~\footnote{Similar to the practice commonly used in the literature (see, for example, Refs.\cite{konig3,H_NLO}), we have neglected the contribution of a regular part for the scattering amplitude normalization for $E \rightarrow E_B$. The question of whether these parts might contribute to the three-nucleon amplitude was discussed in private conversation with Prof. Daniel Phillips of Ohio University and deserves a separate discussion, which is beyond the scope of the current work.}{\label{note1}. By substituting \cref{eq_t0}
	into \cref{stm1}, \cref{stm1} becomes
	\begin{equation}
	\label{eq_gamma1}
	\mathcal{B}^{^3\text{H}} (p)=
	\mathcal{B}^{^3\text{H}} (p')\otimes \hat{K} ^{^3\text{H}}(p', p, E_{^3\text{H}})~,
	\end{equation}
	where $\hat{K} ^{^3\text{H}}(p', p, E_{^3\text{H}})=\hat{K} ^{nd}(p', p, E_{^3\text{H}})$, {\it i.e.,} the homogeneous integral equation has the form of the
	non-relativistic Bethe-Salpeter equation
	\cite{solving_(in)homogeneous_bound_state_equations, bound_state}, with $E_{^3\text{H}}$, the triton binding energy.

	Specifically, for the case of the $^3$H bound-state, we express the amplitude as
	\begin{equation}\label{eq_gamma2}
	\mathcal{B}^{^3\text{H}} (p)=\left (
	\begin{array}{c}
	\Gamma_t^{^3\text{H}} (p)\\\Gamma_s^{^3\text{H}} (p)
	\end{array}\right)~,
	\end{equation}
	where $\Gamma_t$, $\Gamma_s$ denote the two bound-state amplitudes that
	have a spin-triplet or spin-singlet dibaryon,
	respectively.
	
	For the triton, one needs to solve the integral equation:
	\begin{multline}
	\label{eq_triton_H}
	\left (
	\begin{array}{c}
	\Gamma_t^{^3\text{H}} (p) \\
	\Gamma_s^{^3\text{H}} (p) \\
	\end{array}
	\right)=K_0 (p',p,E_{^3\text{H}})\left (
	\begin{array}{cc}
	-M y_t^2 D_t (E_{^3\text{H}},p')&3 M y_ty_s D_s (E_{^3\text{H}},p')\\
	3 M y_ty_s D_t (E_{^3\text{H}},p')&-M y_s^2 D_s (E_{^3\text{H}},p')\\
	\end{array}\right)\\
	\otimes
	\left (
	\begin{array}{c}
	\Gamma^{^3\text{H}}_t (p') \\
	\Gamma^{^3\text{H}}_s (p') \\
	\end{array}
	\right),
	\end{multline}
	which can be written in compact form:
	\begin{equation}\label{eq_triton_no_H_compact}
	\Gamma_\mu^{^3\text{H}}(p)=\sum\limits_{\nu=t,s}My_\mu y_\nu a_{\mu\nu}K_0(p',p,E_{^3\text{H}})\otimes\left[D_\nu(E_{^3\text{H}},p')\Gamma^{^3\text{H}}_\nu(p')\right],
	\end{equation}
	where $\mu =t,s$ are the different triton channels and $y_{\mu,\nu}$ are the nucleon-dibaryon coupling constants for the different channels. The $a_{\mu\nu}$ and $b_{\mu\nu}$ are a result of $n-d$ doublet-channel projection ( \cite{Parity-violating}), for example:
	\begin{subequations}\label{eq_3H_projecation}
		\begin{align}
		a_{tt}&=
		\dfrac{4}{3}\left[(\sigma^i)^{\alpha}_\beta(P_t^i)^\dagger_{\gamma\delta}(P_t^j)^{\delta\beta}(\sigma^j)^{\gamma}_\chi\right]=-1\\
		a_{ts}&=\dfrac{4}{3}\left[(\sigma^i)^{\alpha}_\beta((P_t^i)^\dagger)^{ab}_{\gamma\delta}(P_s^A)_{bc}^{\delta\beta}(\tau^A)_{da}\right]_{c,d=2}=3\\
		a_{st}&=
		\dfrac{4}{3}\left[(\tau^A)^{ab}\left((P_t^A)^\dagger\right)_{\beta\alpha}^{dc}(P_t^i)^{\chi \beta}_{bc}(\sigma^i)_{\alpha}^\delta\right]_{a=d=2}=3\\
		a_{ss}&=\dfrac{4}{3}\left[(\tau^A)^{ab}(P_t^A)^\dagger_{cd}(P_t^B)^{db}(\tau^B)^{ec}\right]_{a=e=2}=-1,
		\end{align}
	\end{subequations}
	where $i,j$ are the different spin projections and $A,B$ are the isospin projections, the same as those in \cref{H_S_lag2}.
\cblack
\begin{comment}
To find the value of the three-body force counterterm $H (\Lambda)$, we solve the homogeneous integral equation (i.e.
\cref{eq_triton_no_H_compact}) numerically and use the known binding
energy of the triton $-E_{^3\text{H}}=-8.48$ MeV to remove the
$\Lambda$ dependence of $-E_{^3\text{H}}$. Our calculations are carried out with the experimental data shown in Tab.~\ref{table_1}. In particular, we will be setting
$a_{nn}=a_{np}=a_{s}$ and $S_{np}=S_{nn}=S$ (where $S$ is the singlet channel of the scattering amplitude).
\end{comment}

All three-body equations in this work can be solved numerically only. We replace each integral by a sum of weighted functions on the interval $[0,\Lambda]$ which is divided into $n$ segments using Gaussian quadrature \cite{numerical}:
\begin{equation}
\int_{0}^\Lambda f(x)\, dx = \sum_{i=1}^n w_i f(x_i),
\end{equation}
where $w_i$ are the weights. The three-body bound-state \cref{eq_triton_no_H_compact} is then the solution of the homogeneous parts of Faddeev equations \cite{solving_(in)homogeneous_bound_state_equations, bound_state}, which can be solved as a coupled eigenvalue problem. %numerically with the following experimental data in Tab.~\ref{table_1}.
\subsection{The three-body force}

\pilesseft is renormalizable, {\it i.e.,} the theory has no dependence on the ultraviolet cutoff $(\Lambda)$. In principle, the loop integration over all possible momenta should not be affected by the upper bound of the integral, $\Lambda$. However, a numerical solution of the integral equations reveals a strong dependence on $\Lambda$. In order to find the energy for each cutoff, we find the value of $E$ for which: 
%\cblack
\begin{equation}
\Gamma_\mu(p)=\sum\limits_{\nu=t,s}My_\mu y_\nu a_{\mu\nu}K_0(p',p,E)
%+b_{\mu\nu}\frac{H(\Lambda)}
%{\Lambda^2}
\otimes\left[D_\nu(E,p')\Gamma_\nu(p')\right].
\end{equation} %
\cblack
\begin{comment}
Then we set $H(\Lambda)$ to be 0 for $\Lambda_0=8.5$ GeV, (i.e $K(E, \Lambda_0)\otimes u=u$). Starting from this cutoff, $\Lambda_0$, we then
determine the values of $H (\Lambda)$ needed for $K(E, \Lambda)\otimes u=u$.
\end{comment}
%\cref{eq_triton_no_H_compact} can be treated as a coupled eigenvectors equation with eigenvalue $c=1$: 
%\begin{equation}\label{eq_cu}
%cu=K\cdot u
%\end{equation}
\begin{figure}
\centering
\includegraphics[width=0.75\linewidth]{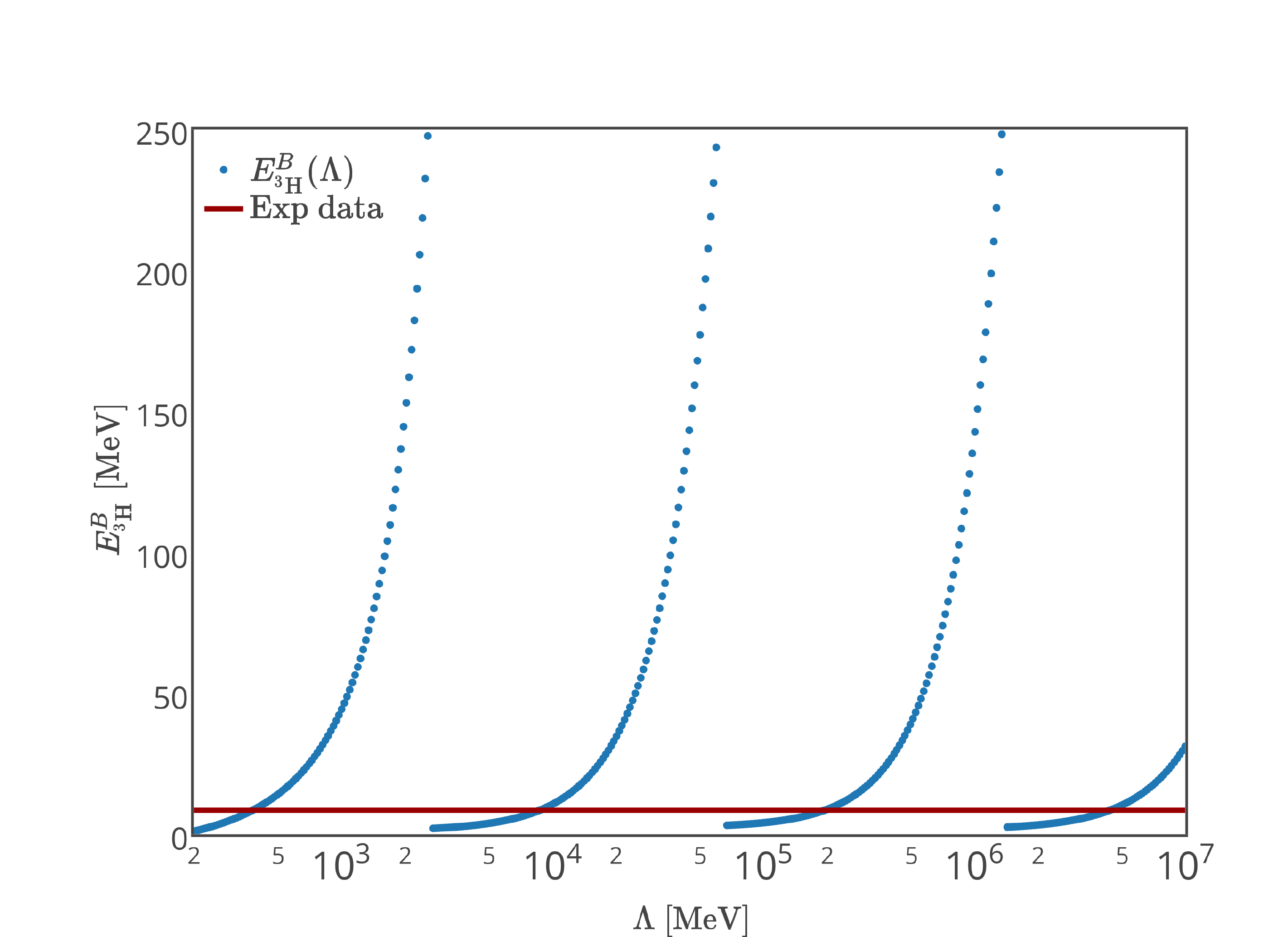}
\caption{\footnotesize{The binding energy which solves \cref{eq_triton_no_H_compact} as a function of cutoff ($\Lambda$) (dots). The solid line is the experimental value of triton binding energy - 8.48 MeV.}}
\label{fig_3H_energy}
\end{figure}
Figure.~\ref{fig_3H_energy} shows that the $^3$H binding energy which solves \cref{eq_triton_no_H_compact} has a strong cutoff dependency. To overcome this problem, one adds a three-body force counterterm to
the exchange nucleon propagator $K_0(k, p,E)$ \cite{3bosons}: 
\begin{equation}\label{KH}
K_0(k, p,E)\rightarrow K_0(k, p,E)+\frac{H(\Lambda)}{\Lambda^2}.
\end{equation}.
\begin{figure}[h!]
	\centering
	\includegraphics[width=0.85\linewidth]{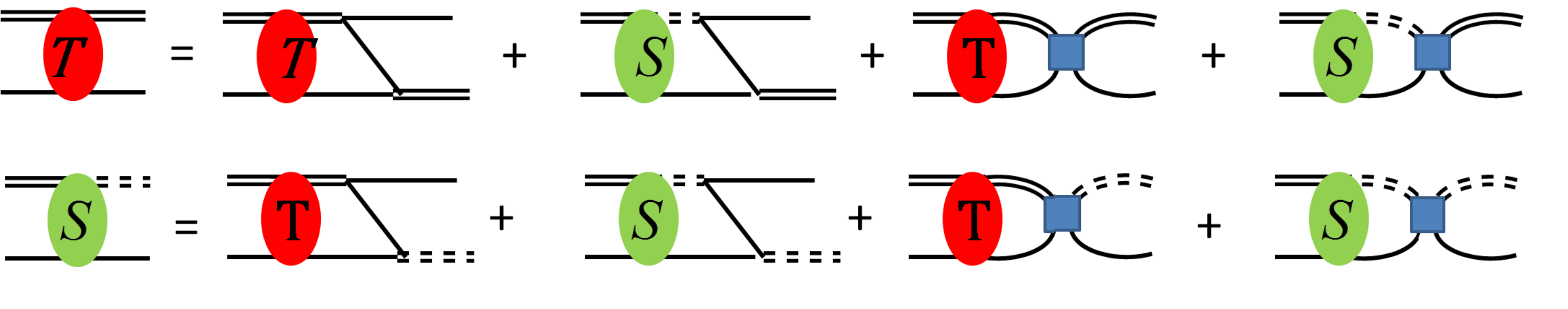}
	\caption{\footnotesize{
			The triton scattering equations with a three-body
			force. The double lines are the propagators of the
			dibaryon fields $D_t$ (solid) and $D_s$ (dashed). The red
			bubbles (T) represent the triplet channel T=0, S=1, while the
			green bubbles (S) represent the singlet channel T=1, 
			S=0. The blue squares are the three-body force}
	}\label{three body_with_H}
\end{figure}

Formally, this three-body force term is obtained by adding \cite{triton}
\begin{equation}\label{three_body_eq}
\begin{split}
 \mathcal{L}_3=M\frac{H(\Lambda )}{3\Lambda ^2}\left\{y_t^2N^{\dagger }\left(\vec{t}\cdot \vec{\sigma }\right)^\dagger\left(\vec{t}\cdot \vec{\sigma }\right)N+y_s^2N^{\dagger }\left(\vec{s}\cdot \vec{\tau
}\right)^{\dagger }\cdot \left(\vec{s}\cdot \vec{\tau }\right)N-y_ty_s\left[N^{\dagger }\left(\vec{t}\cdot \vec{\sigma }\right)^{\dagger }\left(\vec{s}\cdot
\vec{\tau }\right)N+h.c.\right]\right\}~,
\end{split}
\end{equation}to the Lagrangian (\cref{H_S_lag2}).

To find $H(\Lambda)$, we solve the homogeneous equation (\cref{eq_triton_no_H_compact}) numerically and use the known binding energy of triton
$-E_{^3\text{H}}=-8.48$ MeV to remove the $\Lambda$ dependence of $-E_{^3\text{H}}$. For each cutoff we find the appropriate value of $H(\Lambda)$ for which:
%\color{red}
\begin{equation}\label{eq_triton_H_compact}
\Gamma_\mu^{^3\text{{He}}}(p)=\sum\limits_{\nu=t,s}My_\mu y_\nu\left[a_{\mu\nu}K_0(p',p,E_{^3\text{H}})+b_{\mu\nu}\frac{H(\Lambda)}{\Lambda^2}\right]\otimes\left[D_\nu(E_{^3\text{H}},p')\Gamma^{^3\text{H}}_\nu(p')\right]~,
\Gamma^{^3\text{{He}}}_\nu(E_{^3\text{He}},p')\end{equation}
where $b_{tt}=b_{ss}=-1$ and $b_{ts}=b_{st}=1$.

Equation (\ref{eq_triton_H_compact}) can be treated as a coupled eigenvectors equation, where for each $\Lambda$ we find $H(\Lambda)$ that solves \cref{eq_triton_H_compact} numerically. 
The comparison between our numerical calculations of $H(\Lambda)$ and the analytic result %\cblack
 introduced in Refs.~\cite{3bosons,triton}, 
\begin{equation}
 H(\Lambda)=\frac{\sin\left[s_0\log\left(\frac{\Lambda}{{\Lambda^*}}\right)+\arctan\left(\frac{1}{s_0}\right)\right]}{\sin\left[s_0\log\left(\frac{\Lambda}{\Lambda^*}\right)- \arctan\left(\frac{1}{s_0}\right)\right]}~,
 \end{equation}
 with $s_0 =1.00624$ and $\Lambda^*=1.55\mev$
 \cblack multiplied by $c(\Lambda)=0.879$ \cite{Universal_Relations_for_Identical_Bosons}, is shown in Fig.~\ref{fig_H_3H}. Our numerical results are in a good agreement with the analytical value. 
\begin{figure}[h!]
	\begin{center}
		\includegraphics[width=0.75\linewidth]{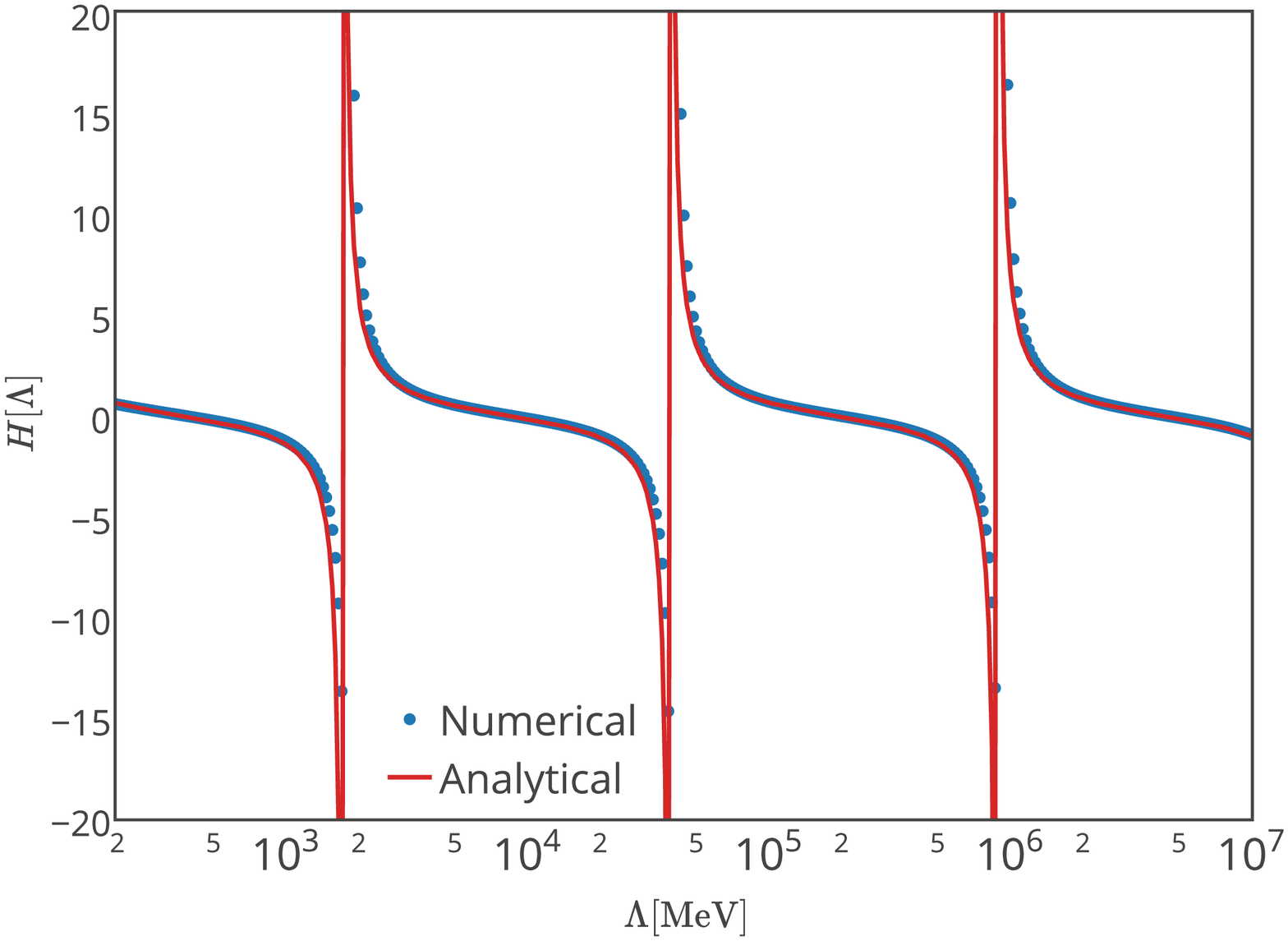}\\
		%\vspace{-0.5 cm}
		\caption{\footnotesize{Values of the three-body force $H(\Lambda)$ as a function of the cutoff $\Lambda$ for $^3$H. The dots show our numerical results for $H(\Lambda)$ and the solid line is the analytical result.}}\label{fig_H_3H}
	\end{center}
\end{figure}

\subsection{The n-d scattering with $E>0$}
In the case of the doublet channel, the spins of the nucleon and the deuteron are coupled to total spin of $J=1/2, T=1/2$. The spin-singlet dibaryon can now appear in the intermediate state, which leads to two coupled amplitudes that differ from each other in the type of the outgoing dibaryon as shown in Fig.~\ref{fig_full_triton}.
In the case of the unbound $n-d$ scattering the three nucleons can be in both $n-d$ and $N-s$ ($n-S_{np}$ states, $p-S_{nn}$), and we set: $a_{nn}=a_{np}=a_{s}$ and $S_{np}=S_{nn}=S$.

Fig.~\ref{fig_full_triton} shows a diagrammatic representation of the coupled-channel integral equations for the scattering amplitudes $ T, S',S$ and $S^\dagger$ (which is the conjugate transpose of $S$) in the doublet channel %\cblack 
for the case that $E>0$ in which \cref{eq_t0} is not valid.

\cblack

The full Faddeev equations for $^3$H can be written as (see Appendix~\ref{whateverB} and Ref.~\cite{Parity-violating}): 
\begin{figure}[h!]
	\begin{center}
		\includegraphics[width=0.95\linewidth]{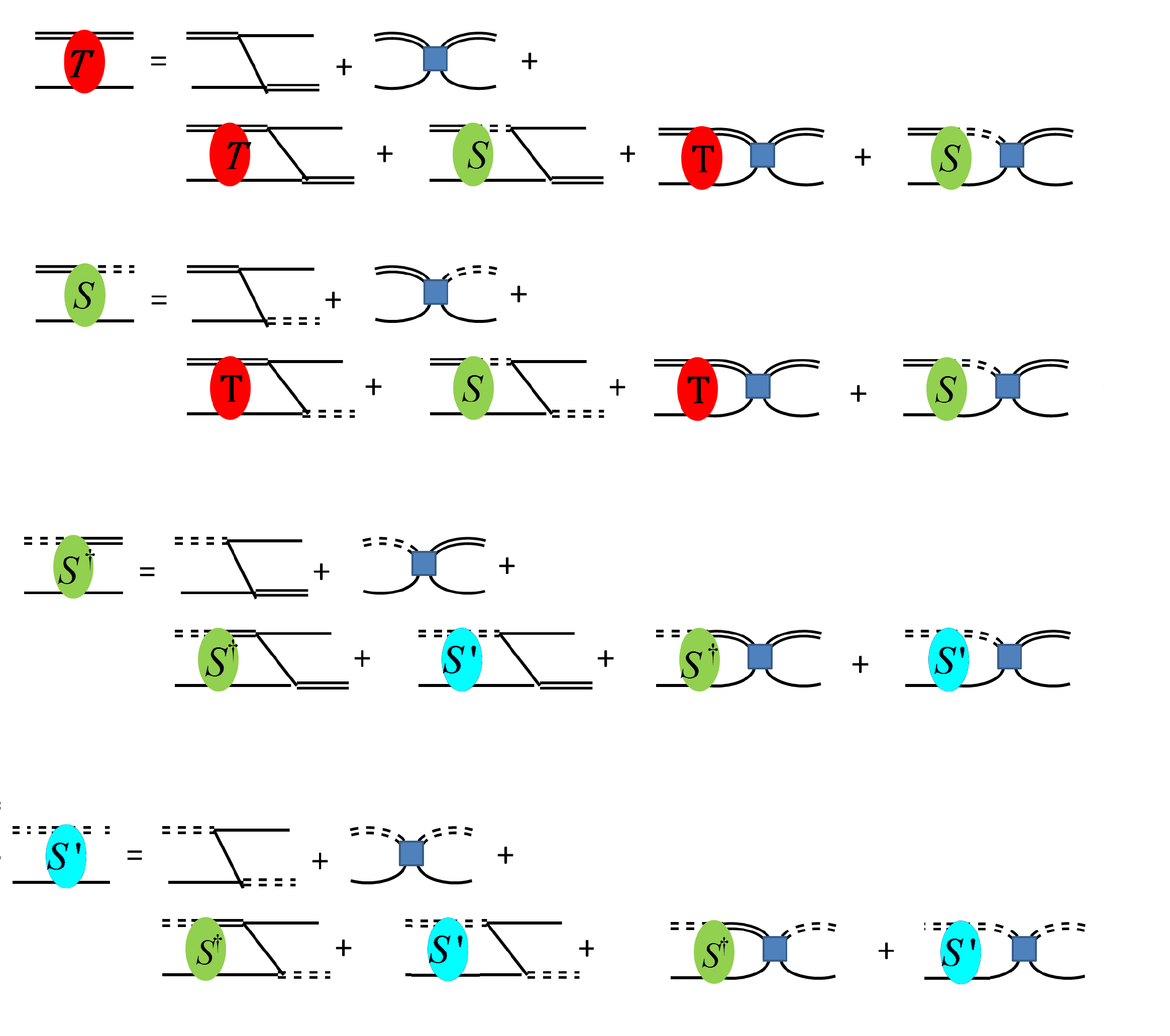}\\
		\caption{\footnotesize{The full scattering equations for triton with a three-body force. The double lines are the propagators of the dibaryon fields $D_t$ (solid) and $D_s$ (dashed). The red bubbles (T) represent the triplet channel T=0, S=1, the green bubbles ($S,S^\dagger$) represent the singlet channel T=1, S=0 while the cyan bubbles (S') represent the singlet channel T=1, S=0. The blue squares are the three-body force.}}\label{fig_full_triton}
	\end{center}
\end{figure}
\begin{equation}\label{stm1_full}
\overleftrightarrow{T}(E,k, p)=B_0+
K_0^S(E,p', p)\otimes \overleftrightarrow{T}(E,k, p),
\end{equation}
where
\begin{equation}\label{T^full}
\overleftrightarrow{T}(E,k, p)=\left(\begin{array}{cc}
T(E,k, p)& S^\dagger(E,k, p)\\S(E,k, p)&S'(E,k, p)
\end{array}\right),
\end{equation}
\begin{equation}\label{B_0_full}
B_0=M\left(\begin{array}{cc}
y_t^2\left[K_0(p,k,E)+\frac{H(\Lambda)}{\Lambda^2}\right]&
-y_ty_s\left[3K_0(p,k,E)+\frac{H(\Lambda)}{\Lambda^2}\right]\\
-y_ty_s\left[3K_0(p,k,E)+\frac{H(\Lambda)}{\Lambda^2}\right]&
y_s^2\left[K_0(p,k,E)+\frac{H(\Lambda)}{\Lambda^2}\right]
\end{array}\right)
\end{equation} and
\begin{equation}\label{K_0_3H_full}
\begin{split}
K_0^{S}(p, k, E)&=MK_0\left(p,k,E\right)\cdot
\left(\begin{array}{cc}
-y_t^2\cdot D_t(E, p)&3y_ty_s\cdot D_s(E, p)\\
3y_ty_s\cdot D_t(E, p)&-y_s^2\cdot D_s(E, p)\\
\end{array}\right)+\\
&M\frac{H(\Lambda)}{\Lambda^2}\left(\begin{array}{cc}
-y_t^2\cdot D_t(E, p)&y_ty_s\cdot D_s(E, p)\\
y_ty_s\cdot D_t(E, p)&-y_s^2\cdot D_s(E, p)\\
\end{array}\right)~.
\end{split}
\end{equation}

 Fig.~\ref{fig_full_triton} shows that both $T(E,k, p)$ and $S'(E,k, p)$ are symmetric while $S(E,k, p)$ and $S^\dagger(E,k, p)$ are not symmetric. However, since the off diagonal terms of $\overleftrightarrow{T}(E,k, p)$ are the transpose of each other, the overall matrix, $\overleftrightarrow{T}(E,k, p)$, is symmetric.

In the case of the inhomogeneous scattering amplitude, \cref{stm1_full} can be solved as a linear equation, where for each $k_i$: 
\begin{equation}\label{eq_sol_full_triton}
\begin{split}
&\left(\begin{array}{c}
T(E,k_i, p)\\
S(E,k_i, p)\\
S^\dagger (E,k_i, p)\\
S'(E,k_i, p)\\
\end{array}\right)=\left(\begin{array}{c}
v_1(E,k_i, p)\\
v_3(E,k_i, p)\\
v_2(E,k_i, p)\\
v_4(E,k_i, p)\\
\end{array}\right)+\\
&\left(\begin{array}{cccc}
M_1(E,p', p)&M_2(E,p', p)&0&0\\
M_3(E,p', p)&M_4(E,p', p)&0&0\\
0&0&M_1(E,p', p)&M_2(E,p', p)\\
0&0&M_3(E,p', p)&M_4(E,p', p)
\end{array}
\right)\otimes
\left(\begin{array}{c}
T(E,k_i, p)\\
S(E,k_i, p)\\
S^\dagger (E,k_i, p)\\
S'(E,k_i, p)\\
\end{array}\right)~,
\end{split}
\end{equation}
%By writing $v_1, v_2, v_3, v_4$ as vectors of $n\cdot1$ and $M_1, M_2, M_3, M_4$ as $n\cdot n$ matrices, \cref{eq_sol_full_triton} can be solved as a liner equation.

%We used the following definitions: 
where
\begin{eqnarray}
M_1(E,p', p)&=&-\frac{My_t^2}{2}\left[\frac{1}{pp'}Q_0\left(\frac{p^2+p'^2-ME}{pp'}\right)+\frac{2H(\Lambda)}{\Lambda^2}\right]D_t(E, p)\frac{p'^2}{2\pi^2}\\
M_2(E,p', p)&=&\frac{My_ty_s}{2}\left[\frac{3}{pp'}Q_0\left(\frac{p^2+p'^2-ME}{pp'}\right)+\frac{2H(\Lambda)}{\Lambda^2}\right]D_s(E, p)\frac{p'^2}{2\pi^2}\\
M_3(E,p', p)&=&\frac{My_ty_s}{2}\left[\frac{3}{pp'}Q_0\left(\frac{p^2+p'^2-ME}{pp'}\right)+\frac{2H(\Lambda)}{\Lambda^2}\right]D_t(E, p)\frac{p'^2}{2\pi^2}\\
M_4(E,p', p)&=&-\frac{My_s^2}{2}\left[\frac{1}{pp'}Q_0\left(\frac{p^2+p'^2-ME}{pp'}\right)+\frac{2H(\Lambda)}{\Lambda^2}\right]D_s(E, p)\frac{p'^2}{2\pi^2} 
\end{eqnarray}
and 
\begin{eqnarray}
v_1(E,k_i, p)&=&\frac{My_t^2}{2}\left[\frac{1}{k_ip}Q_0\left(\frac{p^2+k_i^2-ME}{k_ip}\right)+\frac{2H(\Lambda)}{\Lambda^2}\right]\\
v_2(E,k_i, p)&=&-\frac{My_ty_s}{2}\left[\frac{3}{k_ip}Q_0\left(\frac{p^2+k_i^2-ME}{k_ip}\right)+\frac{2H(\Lambda)}{\Lambda^2}\right]\\
v_3(E,k_i, p)&=&-\frac{My_ty_s}{2}\left[\frac{3}{k_ip}Q_0\left(\frac{p^2+k_i^2-ME}{k_ip}\right)+\frac{2H(\Lambda)}{\Lambda^2}\right]\\
v_4(E,k_i, p)&=&\frac{My_s^2}{2}\left[\frac{1}{k_ip}Q_0\left(\frac{p^2+k_i^2-ME}{k_ip}\right)+\frac{2H(\Lambda)}{\Lambda^2}\right]~,
\end{eqnarray}
and the experimental input parameters used for the numerical calculation, are taken from Tab. \ref{table_1}.

The linear solution of \cref{eq_sol_full_triton} assumes nothing about the symmetry of the matrices $T, S, S^\dagger$ and $S'$ for all $k$ and $p$ a-priori. In order to verify our calculation, we define a new parameter $\alpha$, which enables us to estimate the symmetry of a general matrix $\overleftrightarrow{M}$: 
\begin{equation}
\alpha^{\overleftrightarrow{M}}(E)=\sum_k\sum_p\left|\frac{\overleftrightarrow{M}(E,k, p)-\overleftrightarrow{M}^\dagger(E,k, p)}{\overleftrightarrow{M}(E,k, p)}\right|
\end{equation}
where $\alpha=0$ implies that the matrix is symmetric. Fig.~\ref{fig_sym} shows the symmetry parameter $\alpha$ for all relevant matrices ($T,S,S^\dagger,S'$) as a function of the cutoff $\Lambda$, for two values of energy. For both energies, $\alpha^T$ and $\alpha^{S'}$ are smaller than $10^{-13}$, while
$\alpha^S,\alpha^{S^\dagger}\sim1$. However, the difference between $S$ and $(S{^\dagger})^\dagger$,
$\sum_k\sum_p\left|\frac{
	\overleftrightarrow{S}(E,k, p, )-\left(\overleftrightarrow{S}^\dagger\right)^\dagger(E,k, p)}{\overleftrightarrow{S}(E,k, p)}\right|
$ is smaller than $10^{-13}$, so we conclude that our calculations are accurate also for the unbound case.

\begin{figure}[h!]
	\centering
	\begin{subfigure}[b]{0.60\linewidth}{\label{zero_energy}\includegraphics[width=\linewidth]{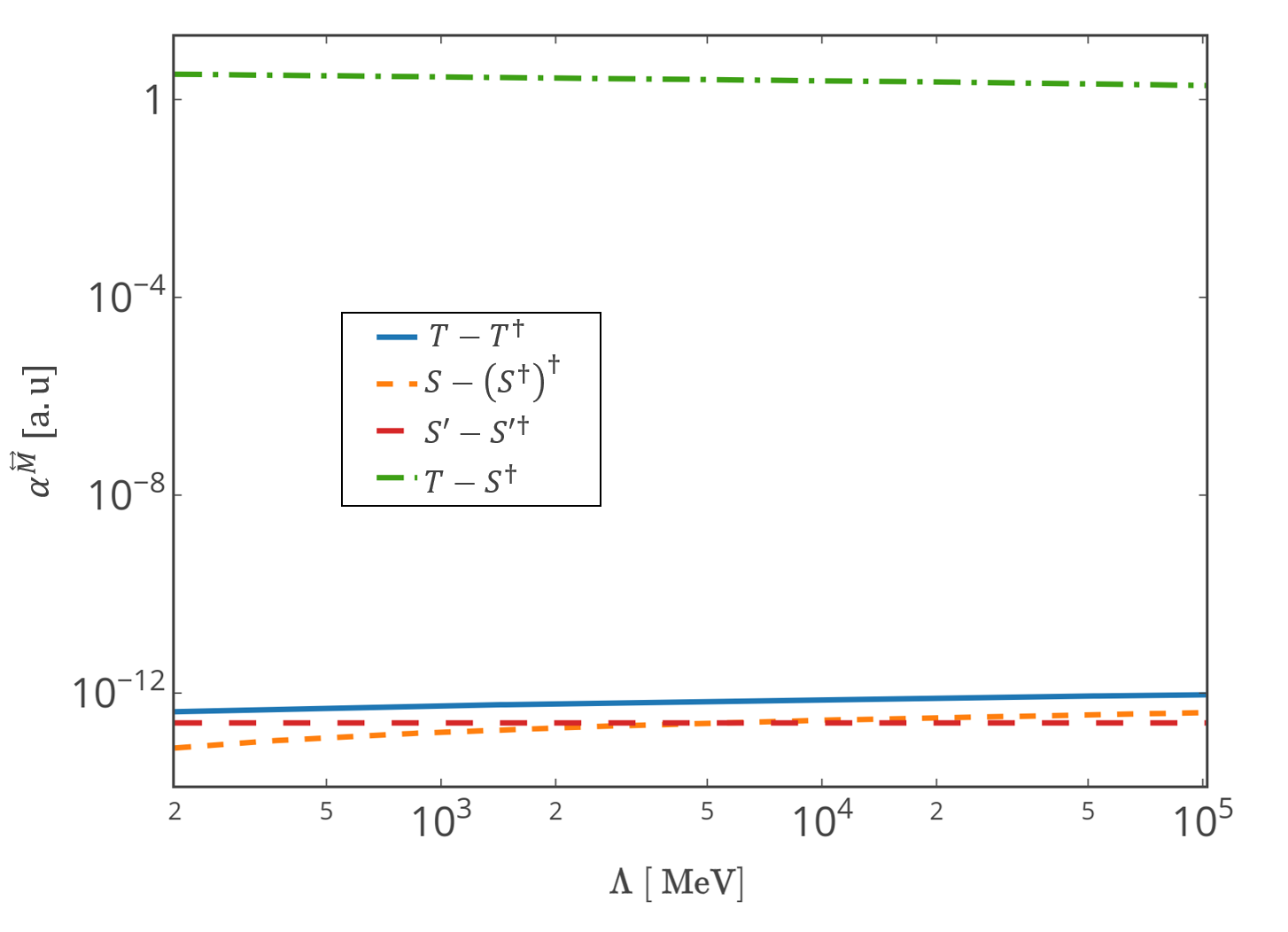}}
		\caption{}
	\end{subfigure}
	\begin{subfigure}[b]{0.6\linewidth}{\label{no_bound}\includegraphics[width=\linewidth]{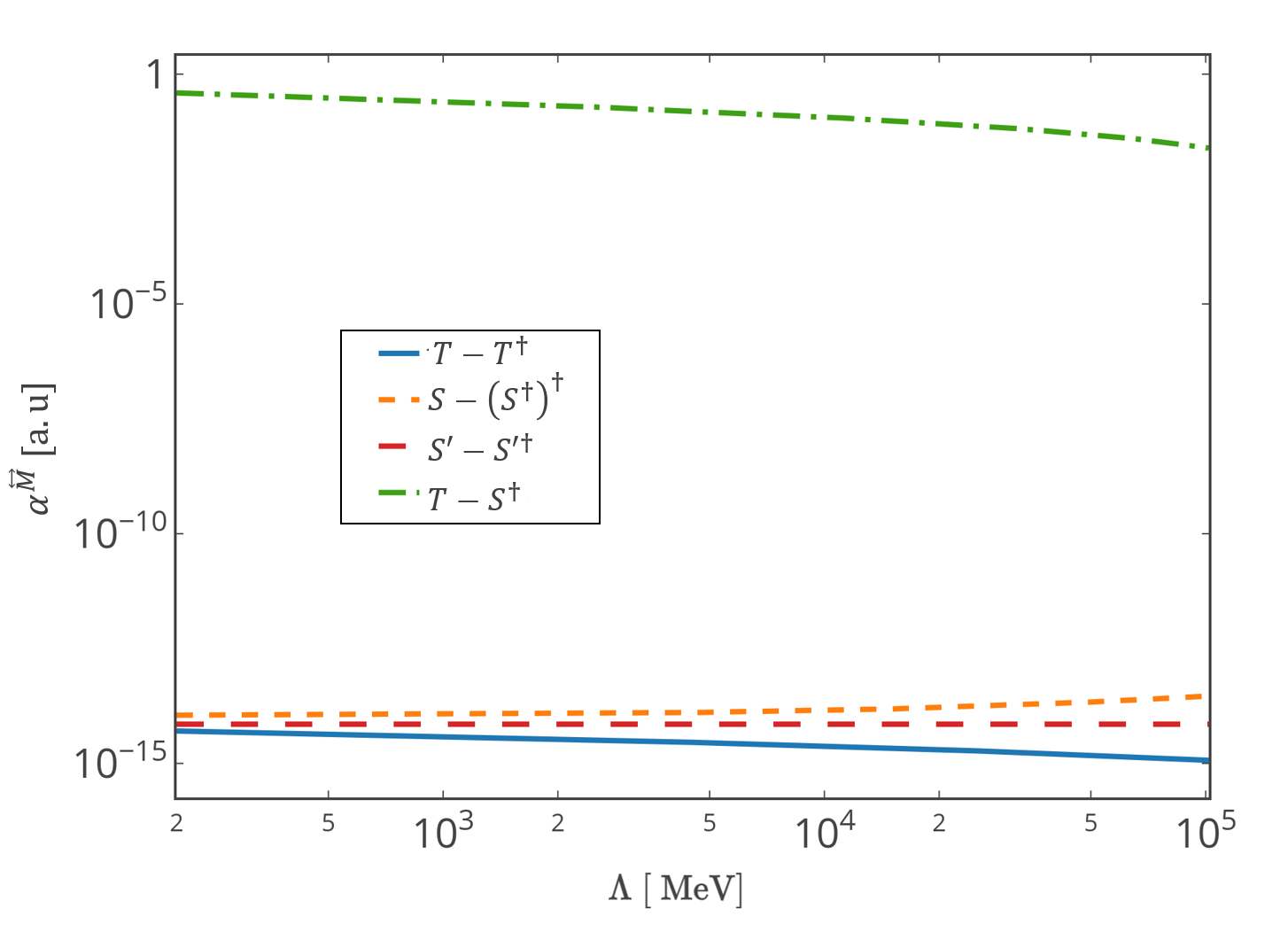}}
		\caption{}
	\end{subfigure}
	\linebreak
	\caption{\footnotesize{The symmetry parameter $\alpha$ as a function of $\Lambda$ for $E=0$ MeV (a) and $E=8.48$ MeV (b). In both panels the solid line is $\alpha^{\protect\overleftrightarrow{T}}$, the long dashed-dotted line is $\alpha^{\protect\overleftrightarrow{S}}$ and the short dashed-dotted line is $\alpha^{\protect\overleftrightarrow{S'}}$. In addition we plotted the difference between $\protect\overleftrightarrow{S}$ and $\left(\protect\overleftrightarrow{S}^\protect\dagger\right)^\protect\dagger$ which is assumed to be zero (dashed line).}}
	\label{fig_sym}
\end{figure}

\section{$p-d$ scattering and $^3$He bound-state}
In this section, we rederive the Faddeev equations for the bound-state $p-d$ scattering ($^3$He) for the doublet channel, based on Ref.\cite{3He,KonigPhd13,konig1,konig2,konig3,konig5}. Similarly to the $n-d$ scattering, the quartet channel, which is of higher orders, is not relevant for this work.
%\subsection{Including the Coulomb interaction}

$^3$He, contains one neutron and two protons, so the Coulomb interaction should be taken into account for accurately describing this system.
The photon Lagrangian of the Coulomb interaction retains only contributions from the Coulomb photon that generate a static Coulomb potential between two charged particles, defined as \nocite{konig1}:
\begin{equation}\label{D_photon}
\mathcal{L}_{photon} =
-\frac{1}{4}F^{\mu\nu}F_{\mu\nu}-\frac{1}{\xi}
\left (\partial_\mu A^\mu-\eta_\mu\eta_\nu\partial^\nu A^\mu\right)^2, 
\end{equation}
where $F^{\mu\nu}$ is the electromagnetic tensor, $A^\mu$ is the electromagnetic four-potential, %\cblack
$\eta^\mu=(1, \boldmath{0})$ is the unit timelike vector \nocite{Pardy:1999ex} \cblack and the parameter $\xi$ determines the choice of gauge. For convenience, we introduce the Feynman rule corresponding to the Coulomb photon propagator:\begin{equation}\label{d_photon}
i\mathcal{D}_{photon} ({\bf k})=\frac{i}{{\bf k}^2+\lambda^2}, 
\end{equation} where $\lambda$ is an artificial small photon mass, added to regulate the singularity of the propagator when the momentum
transfer vanishes \cite{konig1}. %

%\subsubsection{Coulomb diagrams power-counting}
% 
Na\"{i}vely, proton-deuteron $(p-d)$ scattering should contain an infinite sum
of photon exchanges \cite{3He}.
The typical momentum scale for the $^3$He bound-state is $
Q\geq\sqrt{-2ME^B_{^3\text{He}}/A}$ and the Coulomb parameter $\eta$ \cite{Coulomb_effects} is defined as:
\begin{equation}
\eta (Q)=\frac{\alpha M}{2Q}.
\end{equation}
Therefore, for $^3$He, since $E^{B}_{^3\text{H}}=-7.72\mev$, $Q\simeq 70\text{MeV}$ and $\eta (Q)\ll 1$, the Coulomb interaction can be treated as only one-photon exchange
diagrams. The Coulomb diagrams that contribute to $p-d$ scattering
are shown in Fig.~\ref{Coulomb_correction}, while the fine-structure constant $\alpha\sim 1/137$ can be used
as an additional expansion parameter.

%%%%%%%%%%%%%%%%%%%%%%%%%%%%%%%%%%%%%%%%%%%%%%%%%%%	
\begin{figure}[h!]
	\centerline{
		\includegraphics[width=0.9\linewidth]{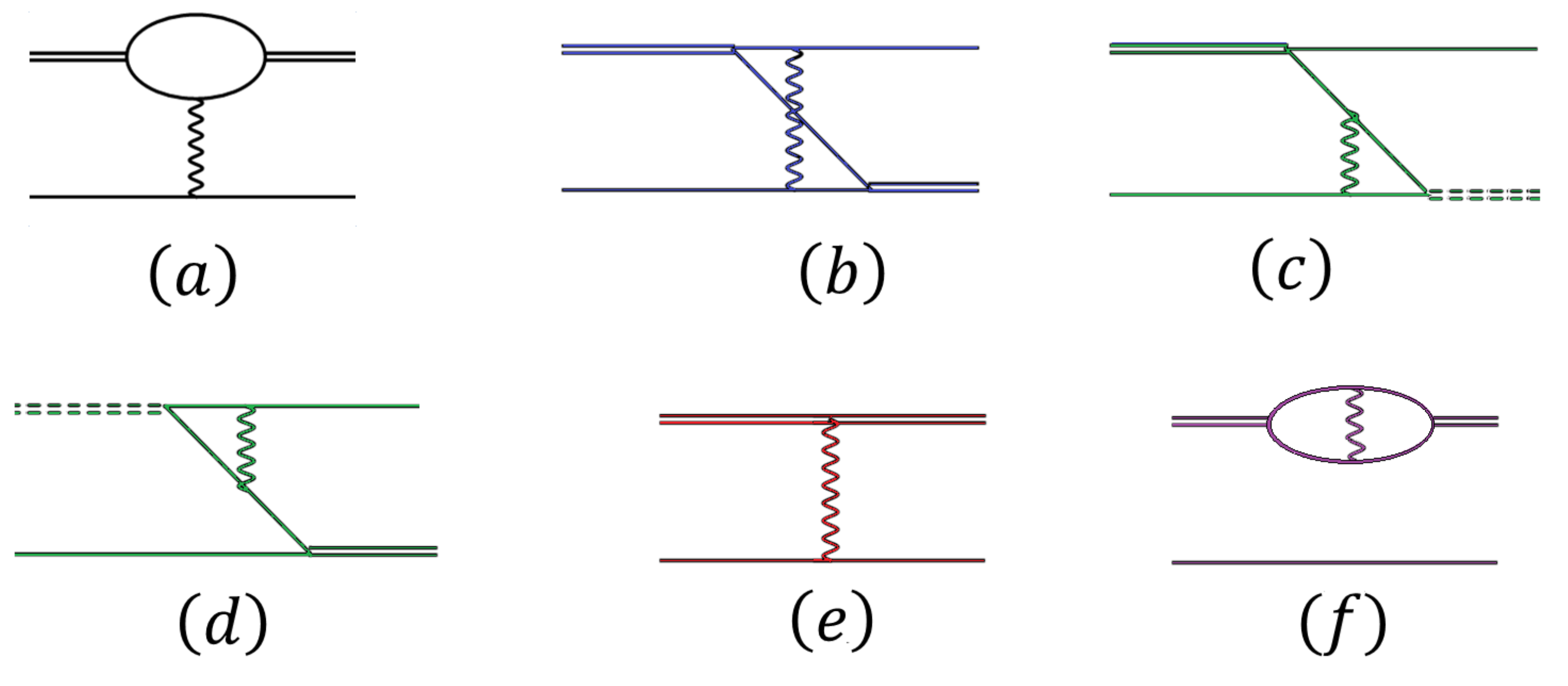}}
	\caption{\label{Coulomb_correction} \footnotesize{The possible one-photon exchange
			diagrams.}}
\end{figure}
%%%%%%%%%%%%%%%%%%%%%%%%%%%%%%%%%%%%%%%%%%%%%%%%%%%%%%%%%

The power-counting for the diagrams shown in
Fig.~\ref{Coulomb_correction} was discussed in Refs.~\cite{konig3,quartet}
whereas in Ref.~\cite{konig5} it was shown that diagram (e) is of a higher order than diagrams (a)-(d), and need not
be taken into account at NLO.

Based on Chapter~\ref{formalism}, we find that:
\begin{itemize}
\item Every nucleon-dibaryon vertex gives a factor of $y_{t,s}=\frac{\Lambda_{\rm cut}^{1/2}}{M}$
\item Every nucleon-photon vertex gives a factor of $\pm ie$
\item Every $q_0$ integral gives a factor of $\frac{Q^2}{4\pi M}$
\item Every $q^2dq$ integral gives a factor of ${Q^3}$
\item Every photon propagator gives a factor of $\frac{1}{Q^2}$
\item Every nucleon propagator gives a factor of $\frac{M}{Q^2}$~,
\end{itemize}
therefore, we find that the power-counting for diagram (a) is:
\begin{equation}
\mathcal{O}(a)=\frac{\Lambda_{\rm cut}}{M^2}\times\frac{\alpha}{Q^2}\times\left(\frac{M}{Q^2}\right)^3\times Q^3\times\frac{Q^2}{4\pi M}=\frac{\alpha}{Q^2}\times\frac{\Lambda_{\rm cut}}{Q}~,
\end{equation}
similarly, we find that the power-counting for diagrams (b-c) is:
\begin{equation}
\mathcal{O}(b-d)=\frac{\Lambda_{\rm cut}}{M^2}\times\frac{\alpha}{Q^2}\times\left(\frac{M}{Q^2}\right)^3\times Q^3\times\frac{Q^2}{4\pi M}=\frac{\alpha}{Q^2}\times\frac{\Lambda_{\rm cut}}{Q}~,
\end{equation}
and the power-counting for diagrams (e) is:
\begin{equation}
\mathcal{O}(e)=\frac{\alpha}{Q^2}=\mathcal{O}(a-d)\times\frac{Q}{\Lambda_{\rm cut}}~.
\end{equation}
Diagram $(f)$ is the contribution from
the non-perturbative proton-proton propagator, which affects the
$^3$H-$^3$He binding energy difference, as discussed in detail in
\cite{konig5}, and will be shown later.

% *** The doublet channel
\subsection{The doublet channel}
The doublet channel in $p-d$ scattering contains three
coupled amplitudes as shown in Fig.~\ref{helium}. In contrast to the triton, for the $p-d$ scattering the spin-singlet
dibaryon has two distinct isospin projections, {\it i.e.,} the $np$ and $pp$ spin-singlet states \cite{konig1}. The Faddeev equations for $p-d$ scattering, at LO, can be written as:
\begin{multline}\label{eq_3He}
t^{pd} (E, k, p)=B_0^{pd} (E, k, p)
+ t^{pd}(E, k, p')\otimes\left[\hat{K}^{pd} (p', p, E)+\hat{K}_0^C (p', p, 
E)\right], 
\end{multline}
where the three individual components of the amplitude $t$ are
\begin{equation}
\label{T_LO_He}
t ^{pd}(E, k, p)=\left (
\begin{array}{c}
T (E, k, p) \\S (E, k, p)\\
P (E, k, p)
\end{array}\right)~,
\end{equation}
%The superscript $\alpha$ indicates that the $p-d$ nuclear amplitudes
%($T, S, P$) are calculated in the presence of a Coulomb interaction (of
%strength $\alpha$).
and:
	\begin{equation}
	\label{B_0_3He}
	B_0 ^{pd}(E, k, p)=\left[K_0(k,p,E)+\frac{H}{\Lambda^2}\right]\cdot
	\left (
	\begin{array}{c}
	{My_t^2}\\
	-y_ty_s\\
	-2y_ty_s
	\end{array}\right)+
	\left (
	\begin{array}{c}
	My_t^2\left[K_C^a (k, p, E)+K_C^b (k, p, E)\right]\\
	-My_ty_sK_C^c (k, p, E)\\
	-2My_ty_sK_C^b (k, p, E)
	\end{array}
	\right)~, 
	\end{equation}
	\begin{multline}\label{eq_K_S_He}
	\hat{K}^{pd} (p', p, E)=MK_0 (p', p, E) \left (
	\begin{array}{ccc}
	-y_t^2\cdot D_t (E, p')&3y_ty_s\cdot D_s (E, p')&3y_ty_s\cdot D_{pp} (E, p')\\
	y_ty_s\cdot D_t (E, p')&y_s^2\cdot D_s (E, p')&-y_s^2\cdot D_{pp} (E, p')\\
	2y_ty_s\cdot D_t (E, p')&-2y_s^2\cdot D_s (E, p')&0\\
	\end{array}
	\right)+\\
	{M}\frac{H (\Lambda)}{\Lambda^2}\left (
	\begin{array}{ccc}
	-y_t^2\cdot {D}_t (E, p')&y_ty_s\cdot D_s (E, p')&y_ty_s\cdot D_{pp} (E, p')\\
	\frac{1}{3}y_ty_s\cdot{D}_t (E, p')&-\frac{1}{3}y_s^2\cdot D_s (E, p')&-\frac{1}{3}y_s^2\cdot D_{pp} (E, p')\\
	\frac{2}{3}y_ty_s\cdot{D}_t (E, p')&-\frac{2}{3}y_s^2\cdot D_s (E, p')&-\frac{2}{3}y_s^2\cdot D_{pp} (E, p')\\
	\end{array}\right)~, 
	\end{multline}
	and
	\begin{equation}\label{eq_K_C_0}
	\hat{K}_0^C (p', p, E)=MK^C(p',p,E)\cdot\left (
	\begin{array}{ccc}
	-y_t^2D_t (E, p') & 3y_ty_sD_s (E, p') & 3y_ty_sD_{pp}(E,p') \\
	y_ty_sD_t (E, p') & y_s^2D_s (E, p') & -y_s^2D_{pp}(E,p') \\
	2y_ty_sD_t (E, p') & -2y_s^2D_s (E, p') & 0 \\
	\end{array}
	\right)~, 
	\end{equation}
	where 
	\begin{equation}\label{eq_K_C}
	K^C (p', p, E)=\left(
	\begin{array}{ccc}
	K_C^a(p', p, E)+K_C^b(p', p, E) & K_C^b(p', p, E)& K_C^c (p', p, E)\\
	K_C^b(p', p, E) & -K_C^a(p', p, E)+K_C^b(p', p, E) & K_C^c (p', p, E)\\
	K_C^d(p', p, E) & K_C^d(p', p, E) & 0 \\
	\end{array}
	\right)
	\end{equation}
	and where:
	\begin{equation}\label{Dpp}
	D_{pp} (E, p)=\mathcal{D}_{pp} \left(E-\frac{p^2}{2M}, {\bf p}\right)~
	\end{equation}
	is the Coulomb propagator
	\cite{Ando_proton, Coulomb_effects}. 
	
	The different one-photon exchange diagrams contributing to the Coulomb interaction (obtained by integrating \cref{D_photon} multiplied by \cref{eq_K1}) are:
	\begin{equation}
	\label{K_C_a}
	K_C^a (p', p, E)=\frac{M\alpha}{2p'p}
	Q_0\left (-\frac{p'^2+p^2+\lambda^2}{2p'p}\right)\cdot\left[\frac{ \arctan\left (\frac{p'+2 p}{\sqrt{3 p'^2-4 ME}}\right)-\arctan\left (\frac{2 p'+p}{\sqrt{3p-4 ME}}\right) }{p'-p}\right]
	\end{equation}
	for Fig.~\ref{Coulomb_correction} (a), 	
	\begin{multline}\label{K_C_b}
	K_C^b (p', p, E)=\\\frac{M^2\alpha}{4\left (p'^2-ME+p'p+p^2\right)}
	Q_0\left (\frac{p'^2+{p}^2-ME}{p'p}\right)\left[\frac{
		\arctan\left (\frac{p'+2 p}{\sqrt{3 p'^2-4
				ME}}\right)-\arctan\left (\frac{2 p'+p}{\sqrt{3p-4
				ME}}\right) }{p'-p}\right]~ 
	\end{multline}
	for Fig.~\ref{Coulomb_correction} (b), and
	\begin{multline}\label{K_C_c}
	K_C^c (p', p, E)=K_C^d (p', p, E)=\alpha K_0 (p', p, E)\cdot\\\frac{1}{4 \pi (p'- p)}\log \Biggl[\frac{2 ME-2 p'^2+2 p' p-2 p^2}{-p' \sqrt{4 ME-3 p'^2+2 p' p-3 p^2}+
		p \sqrt{4 ME-3 p'^2+2 p' p-3 p^2}+2
		ME-p'^2-p^2}\Biggr]~ 
	\end{multline}
	for Fig.~\ref{Coulomb_correction} (c and d), where $K_0$ was defined in \cref{eq_K1}.
	
	%%%%%%%%%%%%%%%%%% Fig.~%%%%%%%%%%%%%%%%%%%%%%%%%%%%%%
	\begin{figure}[h!]
		\begin{center}
			% Requires \usepackage{graq'hicx}
			\includegraphics[width=0.85\linewidth]{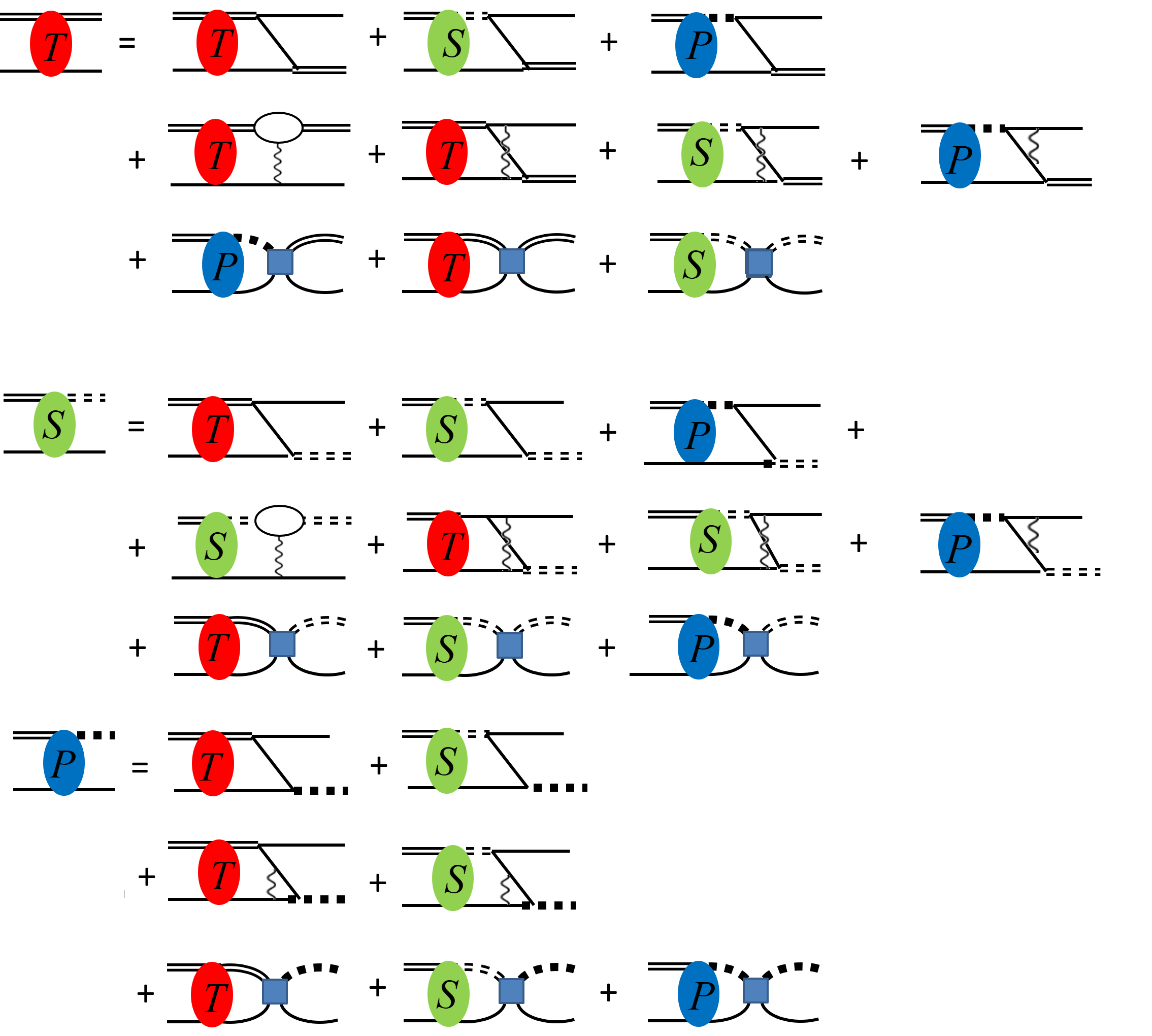}\\
			\caption{ \footnotesize{Diagrammatic form of the homogeneous part of $p-d$ scattering that includes a three-body
					force. The double lines denote the dibaryon propagators $D_t$
					(solid), $D_{np}$ (dashed) and $D_{pp}$ (doted). The red
					bubbles (T) represent the triplet channel (T=0, S=1), the green
					bubbles (S) represent the singlet channel (T=1, S=0) with $np$
					dibaryon while the blue bubbles (P) represent the singlet
					channel T=1, S=0 with a $pp$ dibaryon. The blue squares represent the three-body force.}}\label{helium}
		\end{center}
	\end{figure}
	%%%%%%%%%%%%%%%%%%%%%%%%%%%%%%%%%%%%%%%%%%%%%%%%%%%%%
\subsection{ $^3$He bound-state amplitude and three-body force}\label{limit}
The above subsection provides all the information necessary to solve the
homogeneous Faddeev equations for $^3$He, similarly to those corresponding to $^3$H.
For $^3$He, the homogeneous part of \cref{eq_3He} can be written as:
\begin{multline}\label{eq_3He_compact}
\Gamma^{{^3\text{He}}}_\mu(p)=\sum\limits_{\nu=t,s,pp}My_\mu y_\nu\Bigl[a'_{\mu\nu}K_0(p',p,E_{^3\text{He}})
+b'_{\mu\nu}\frac{H(\Lambda)}{\Lambda^2}+\\
a'_{\mu\nu}K^C_{\mu\nu}(p',p,E_{^3\text{He}})\Bigr]\otimes D_\nu(E_{^3\text{He}},p')\Gamma^{^3\text{{He}}}_\nu(p')~,
\end{multline}
\begin{comment}}
where:
\begin{equation}
\psi_\mu^{^3\text{He}}(E,p)\rangle=
D_\mu(E_{^3\text{He}},p)\Gamma_\mu(E_{^3\text{He}},p)\\
\end{equation}
\begin{comment}
the three-nucleon wave function has the form:

\begin{equation}
|\psi^{^3\text{He}}(E,p)\rangle=\left(\begin{array}{c}
D_t(E_{^3\text{He}},p)\Gamma_t(E_{^3\text{He}},p)\\
D_s(E_{^3\text{He}},p)\Gamma_s(E_{^3\text{He}},p)\\
D_{pp}(E_{^3\text{He}},p)\Gamma_{pp}(E_{^3\text{He}},p)\\
\end{array}\right)=\left(\begin{array}{c}
\psi_t(E_{^3\text{He}},p)\\
\psi_s(E_{^3\text{He}},p)\\
\psi_{pp}(E_{^3\text{He}},p)
\end{array}\right),
\end{equation}

\begin{eqnarray}
\hat{I}&=&\text{diag}\left(I_t, I_s, I_{pp}\right)\\
I_{t, s,pp}(q,q',E)&=&\frac{2\pi^2}{{ q}^2}\delta\left(q-q'\right)D_{t,s,pp}(E,q)^{-1}.
\end{eqnarray}
and
\begin{eqnarray}
\psi(E,p)&=&\left(\begin{array}{c}
\Gamma_t(E,p)D_T(E,p)\\\Gamma_s(E,p)D_S(E,p)\\
\Gamma_{pp}(E,p)D_P(E,p)
\end{array}
\right)
\end{eqnarray}
and
\begin{equation}
\begin{split}
&\mathcal{K}(q,q',E)=K^S_0(q,q',E)+K_0^C(q,q',E),
\end{split}
\end{equation}
\end{comment}
where $\mu=t,s,pp$ are the different channels of $^3$He and $K_0(p,p',E)$ and $K^C_{\nu\mu}(p,p',E)$ is the $\mu,\nu$ index of $K^C(p,p',E)$ (\cref{eq_K_C}).
Notice that for the $p-d$ doublet-channel projection, the electromagnetic interaction does not couple to isospin eigenstates \cite{konig1,konig5}, such that:
\begin{subequations}\label{eq_3He_projecation}
	\begin{align}
	a'_{ts}&=\dfrac{4}{3}\left[(\sigma^i)^{\alpha}_\beta((P_t^i)^\dagger)^{ab}_{\gamma\delta}(P_s^A)_{bc}^{\delta\beta}(\boldmath{1}\cdot i\delta^{A,3})_{da}\right]_{c,d=1}=3,\\
	a'_{tpp}&=\dfrac{4}{3}\left[(\sigma^i)^{\alpha}_\beta((P_t^i)^\dagger)^{ab}_{\gamma\delta}(P_s^A)_{bc}^{\delta\beta}(\boldmath{1}\cdot\delta^{A,1}+\boldmath{1}\cdot i\delta^{A,2})_{da}\right]_{c,d=1}=3~.
	\end{align}
\end{subequations}

The three-body force $H (\Lambda)$ has
no isospin dependence, {\it i.e.,}
$H (\Lambda)_{^3\text{H}}=H (\Lambda)_{^3\text{He}}$. 
Therefore, it is possible to calculate the binding energy of $^3$He using the three-body force $H (\Lambda)$ obtained in the triton system. Similar to Ref.~\cite{konig3} we find the binding energy which solves the Faddeev
equations for $^3$He, \cref{eq_3He_compact}, numerically, using the three-body force known from $^3$H at
LO with for a large range of cutoffs as shown in Fig.~\ref{fig_helium_energy}. From now on, similarly to
Ref.~\cite{konig5}, we will use the numerical binding energies $E_{^3\text{He}}(\Lambda)$ at LO as the binding energy
of $^3$He rather than the experimental $E_{^3\text{He}} =7.72 \mev$. The high similarity between our
results and Ref.~\cite{konig5} has been used as an additional check for our numerical calculations. Note that the non-perturbative calculation is the same as in Ref.~\cite{konig3}.
The perturbative calculation is similar to the one in Ref.~\cite{konig5}, except for the difference of definition of the two-body term, as presented in Chapter.~\ref{general_matrix}.
\cblack

\begin{figure}[h!]
	\centering
	\vspace{-.5 cm}
	\includegraphics[width=.8\linewidth]{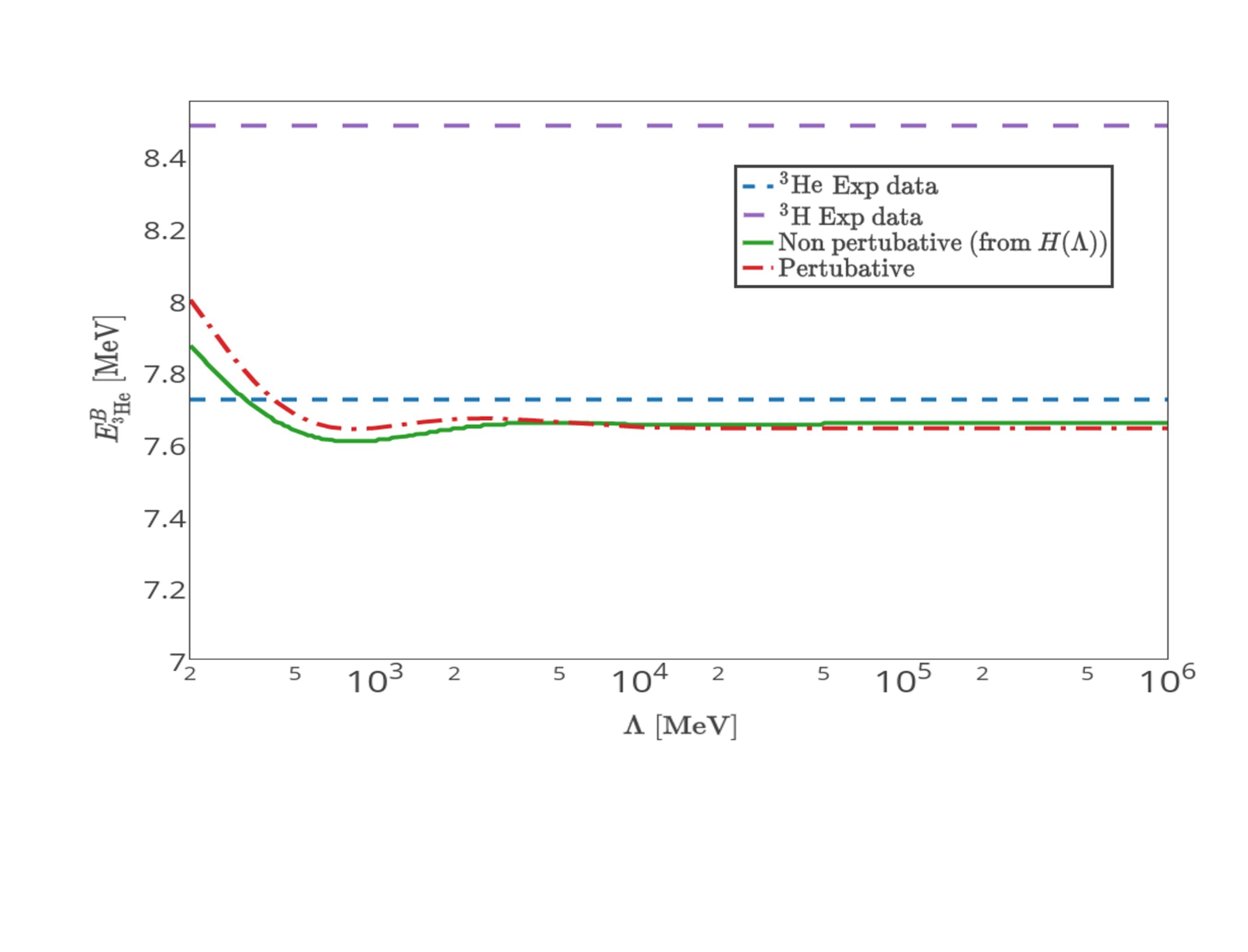}\\
	\vspace{-1.2 cm}
	\caption{\footnotesize{Predictions for
			$E_{^3\text{H}},E_{^3\text{He}}$ binding energy as a function of the
			cutoff $\Lambda$. The solid line is the binding energy calculated using the non perturbative solution (subsection~\ref{limit}), 
			% the sport dashed-doted line is the predication from the
			%perturbation theory \cref{}, 
			the dashed-dotted line is the
			 binding energy predicted using perturbation theory (\cref{eq_delta_E1}). The short-dashed line is the experimental
			value, $E_{^3\text{He}}=7.72$ MeV, and the long-dashed line is the experimental
			value, $E_{^3\text{H}}=8.48$ MeV.}} %where the long dashed-doted
	%line is is binding energy form the three-body force under the
	%assumption that $\alpha=0$ but the $pp$ channel has different
	%scattering length ($a_{np}\neq a_{pp}$).}
	\label{fig_helium_energy}
\end{figure}

\chapter{Normalization of the three-nucleon amplitude}\label{Norm}
In this chapter, we present the normalization of the three-nucleon ({\it i.e.,} $^3$H and $^3$He) bound-state amplitude, in the form of the non-relativistic Bethe-Salpeter (BS) equation. This normalization, as introduced in Refs.~\cite{bound_state, norm1, norm2} is found to have a diagrammatic representation as well. This diagrammatic representation is equivalent to the matrix element of the identity operator, and is found to be the sum over all the possible connections between two identical three-nucleon amplitudes. A similar diagrammatic representation is used later in this work for calculating the $^3$H-$^3$He binding energy shift (Chapter 5), for the NLO contributions to the three-nucleon bound-state amplitude (Chapter 6) and for the three-nucleon electro-weak matrix elements (Chapters 7-9).
\section{The Bethe-Salpeter wave function normalization}\label{B_S_normalization}
The three-nucleon Faddeev equations (\cref{stm1,eq_3He}) have the same form as the non-relativistic BS equation \cite{bound_state, norm1, norm2, KonigPhd13}: 
\begin{equation}\label{eq_norm1}
\mathcal{M}=V-VG_{BS}\mathcal{M}=V-\mathcal{M}G_{BS}V~, 
\end{equation}
where $\mathcal{M}$ is the scattering matrix, $V$ is the two-body interaction kernel and $G_{BS}$ is the free two-body propagator.\begin{figure}[h!]
	\centering
	\includegraphics[width=0.65\linewidth]{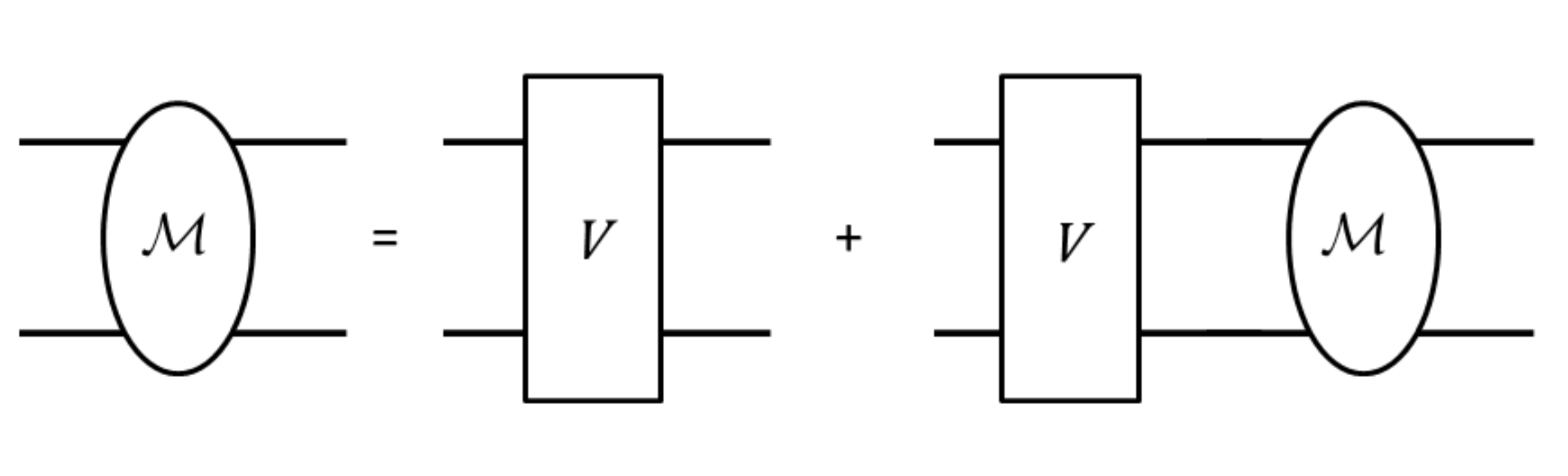}\\
	\caption{\footnotesize{Diagrammatic representation of the two-body BS equation for the scattering matrix $\mathcal{M}$.}}
\end{figure}
From \cref{eq_norm1} we find: 
\begin{equation}\label{eq_norm2}
V=\mathcal{M}+\mathcal{M}G_{BS}V~, 
\end{equation}
and upon substituting $V$ in \cref{eq_norm1}, we get: 
\begin{equation}\label{eq_norm3}
\mathcal{M}=V-\mathcal{M}G_{BS}\mathcal{M}-\mathcal{M}G_{BS}VG_{BS}\mathcal{M}~. 
\end{equation}
For a bound-state, $\mathcal{M}$ has a general form: 
\begin{equation}\label{eq_norm4}
\mathcal{M}=\frac{\ket{\Gamma}\bra{\Gamma}}{E-E_B}+\mathcal{R}~,
\end{equation}
where $\ket{\Gamma}$ is the wave function amplitude, $\bra{\Gamma}$ is
the representation of $\ket{\Gamma}$ in the dual space (including
Dirac conjugate) and $\mathcal{R}$ is a regular part which is finite at $E=E_B$ and therefore can be neglected for $E\rightarrow E_B$ (see the footnote at subsection~\ref{note1}). %The first term of \cref{eq_norm4} originates from the bound
%state propagator and $\mathcal{R}$ is a
Substituting \cref{eq_norm4} into (\cref{eq_norm1}), and
equating residues at $E=E_B$ yields the wave equation for $\ket{\Gamma}$: 
\begin{equation}\label{eq_norm5}
\ket{\Gamma}=-VG_{BS}\ket{\Gamma}~.
\end{equation}
Substituting \cref{eq_norm4} into \cref{eq_norm3}, multiplying the resulting equation by $E-E_B$ and taking the limit $E\rightarrow E_B$, one finds that: 
\begin{equation}\label{eq_norm7}
\begin{split}
&\ket{\Gamma}\bra{\Gamma}=
-\lim\limits_{E\rightarrow E_B}\ket{\Gamma}\frac{\bra{\Gamma}G_{BS}
	\left (1+VG_{BS}\right){\ket{\Gamma}}}{E-E_B}\bra{\Gamma}\\
&\Rightarrow
1=-\lim\limits_{E\rightarrow
	E_B}\frac{\bra{\Gamma}G_{BS}\left (1+VG_{BS}\right){\ket{\Gamma}}}{E-E_B}=-\lim\limits_{E\rightarrow
	E_B}\frac{\bra{\Gamma}G_{BS}\left (G_{BS}^{-1}+V\right)G_{BS}{\ket{\Gamma}}}{E-E_B}~.
\end{split}
\end{equation} 
From \cref{eq_norm5}
$ \lim\limits_{E\rightarrow
	E_B}\bra{\Gamma}G_{BS}\left (1+VG_{BS}\right){\ket{\Gamma}}=0$, so the RHS of \cref{eq_norm7} is of the form $0/0$, so one can use the l'H\^opital's rule to evaluate the limit (which equals -1) explicitly: 
%\begin{equation}\label{eq_diff1}
%\dfrac{\lim\limits_{E\rightarrow
%		E_B}\frac{\partial}{\partial_E}\bra{\Gamma}G_{BS}\left (G_{BS}^{-1}+V\right)G_{BS}{\ket{\Gamma}}}{\lim\limits_{E\rightarrow
%		E_B}\partial_E(E-E_B)}=\lim\limits_{E\rightarrow
%	E_B}\frac{\partial}{\partial_E}\bra{\Gamma}G_{BS}\left (G_{BS}^{-1}+V\right)G_{BS}{\ket{\Gamma}}. 
%\end{equation}
\begin{multline}\label{eq_norm_full}
\dfrac{\lim\limits_{E\rightarrow
		E_B}\frac{\partial}{\partial_E}\bra{\Gamma}G_{BS}\left (G_{BS}^{-1}+V\right)G_{BS}{\ket{\Gamma}}}{\lim\limits_{E\rightarrow
		E_B}\partial_E(E-E_B)}=\lim\limits_{E\rightarrow
	E_B}\frac{\partial}{\partial_E}\bra{\Gamma}G_{BS}\left (G_{BS}^{-1}+V\right)G_{BS}{\ket{\Gamma}}=\\
\bra{\Gamma}G_{BS}'G_{BS}^{-1}G_{BS}+G_{BS}\left(G_{BS}^{-1}\right)'G_{BS}+G_{BS}G_{BS}^{-1}G_{BS}'{\ket{\Gamma}}|_{E=E_B}+\\
\bra{\Gamma}G_{BS}'VG_{BS}+G_{BS}V'G_{BS}+G_{BS}VG_{BS}'{\ket{\Gamma}}|_{E=E_B}=\\
\bra{\Gamma}G_{BS}'+G_{BS}\left(G_{BS}^{-1}\right)'G_{BS}+G_{BS}'{\ket{\Gamma}}|_{E=E_B}+\bra{\Gamma}-G_{BS}'+G_{BS}V'G_{BS}-VG_{BS}'{\ket{\Gamma}}|_{E=E_B}=\\
\bra{\Gamma}G_{BS}\left(G_{BS}^{-1}\right)'G_{BS}+G_{BS}V'G_{BS}{\ket{\Gamma}}|_{E=E_B}=\bra{\Gamma}G_{BS}\frac{\partial}{\partial_E}\left(G^{-1}_{BS}+V\right)G_{BS}{\ket{\Gamma}}|_{E=E_B}=-1~,
\end{multline}
where the terms proportional to $\frac{\partial}{\partial_E}\ket{\Gamma}$ vanish, due to the BS equation \cite{norm2}.

According to our notation, 
$G_{BS}=-{D}(E,p)$, the two-body propagator, and $V={My^2}K_0(p, p',E)$, the one-nucleon exchange matrix, % so \cref{eq_norm_full} becomes:
%\begin{equation}\label{eq_BS_final}
%\begin{split}
%1=&\bra{\mathcal{B} }\mathcal{D}\frac{\partial}{\partial E}\left(D(E,p)^{-1}-My^2K_0(p, p',E)\right)\mathcal{D}{\ket{\mathcal{B} }}|_{E=E_B}, 
%\end{split}
%\end{equation
such that the three-nucleon normalization condition is:
\begin{equation}\label{eq_BS_final}
\begin{split}
1=&\bra{\mathcal{B} }{G_{BS}}\frac{\partial}{\partial E}\left(-G_{BS}-V\right){G_{BS}}{\ket{\mathcal{B} }}|_{E=E_B}, 
\end{split}
\end{equation}

\section{The non-relativistic Bethe-Salpeter wave-function normalization}

The three-nucleon homogeneous integral equation (\cref{eq_gamma1}) was
found to have the same form as the non-relativistic bound-state
Bethe-Salpeter equation (\cref{eq_norm5}):
\begin{equation}\label{eq_Gamma}
\Gamma(p)=My^2K_0(p,p',E)D(E,p')\otimes \Gamma(p')~.
\end{equation}
The normalization condition for the equation is given in Appendix A and
in \cite{KonigPhd13,norm1,norm2}.

This is thus a representation of the normalization operator, $\hat{Z}$, such that: 
\begin{multline}
\label{eq_identity_operator}
\hat{Z}^{-1}=
\int\frac{d^4p}{(2\pi)^4}
\int\frac{d^4p'}{(2\pi)^4}\Gamma(p)S(-p_0,{\bf -p})\mathcal{D}(E+p_0,{\bf p})
\times
\frac{\partial}{\partial E}\left[\hat{I}(E, p, p')-{My^2}K_0(p, p',E)\right]_{E=E_B}\\
\times\mathcal{D}(E+p'_0,{\bf p'})S(-p'_0,{\bf -p'})\Gamma(p')~.
\end{multline}
Carrying out the angular and energy integrations gives
\begin{multline}
\hat{Z}^{-1}=\int\frac{p^2dp}{2\pi^2}\int\frac{p'^2dp'}{2\pi^2}\Gamma(p)D(E,p)
\times M^2y^2\Bigg\{\frac{1}{4 \pi \sqrt{3 p^2-4 E M}}\frac{2\pi^2}{p'^2}\delta(p-p')\\
\frac{-1}{2\left[p'^2 \left(p^2- 2E M\right)+\left(p^2-E M\right)^2+p'^4\right]}\Bigg\} D(E,p')\Gamma(p')~,
\end{multline}
\cblack
with:
\begin{eqnarray}
\hat{I}(E, p, p')&=&\frac{2\pi^2}{p'^2}\delta (p-p')
{D}^{-1}(E, p), 
\end{eqnarray}
and $S(p_0,{\bf p})$ as the one-nucleon propagator:
\begin{equation}
S(E,{\bf p})=\dfrac{1}{p_0-\frac{{\bf p}^2}{2M}}~.
\end{equation}

\section{The normalization of $^3$He,$^3$He wave-functions}
The homogeneous part of the Faddeev equation of both $^3$H and $^3$He
has the form of a non-relativistic BS equation, which couples
different channels.

Using \cref{eq_triton_H_compact}, the normalization condition that
determines the wave-function factor $Z^{^3\text{H}}$ has the form:
\begin{multline}\label{eq_norm_3H}
1={Z^{^3\text{{H}}}}\int\frac{d^3p}{(2\pi)^3}\int\frac{d^3p'}{(2\pi)^3}\sum\limits_{\mu,\nu=t,s} 
{\Gamma^{^3\text{{H}}}_\mu(p)}{D_\mu(E_{^3\text{H}},p)}\\
\times\left\{\frac{\partial}{\partial_E}
\left [\hat{I}_{\mu\nu}(E,p,p')-\hat{\mathcal{K}}^{^3\text{H}}_{\mu\nu}(p,p',E)\right]_{E=E_{^3\text{H}}}\right\}
\times {D_\nu(E_{^3\text{H}},p')} {\Gamma^{^3\text{{H}}}_\nu(p')}~.
\end{multline}
We rewrite the above equation in terms of the {\it wave-functions} $\psi^{^3\text{{H}}}_\mu(p)$ and obtain
\begin{multline}
1=\int\frac{d^3p}{(2\pi)^3}\int\frac{d^3p'}{(2\pi)^3}\sum\limits_{\mu,\nu=t,s} 
{\psi^{^3\text{{H}}}_\mu(p)} \times
\left\{\frac{\partial}{\partial_E}\left [\hat{I}_{\mu\nu}(E,p,p')-\hat{\mathcal{K}}^{^3\text{H}}_{\mu\nu}(p,p',E)\right]_{E=E_{^3\text{H}}}\right\}\times {\psi^{^3\text{{H}}}_\nu(p')}~.
\end{multline}
We recall that $\psi^{^3\text{H}}$ is the \textbf{normalized} three-nucleon wave-function
\begin{equation}\label{eq_psi_3H}
\langle\psi^{^3\text{H}}_\mu|p\rangle=
{\sqrt{Z^{^3\text{H}}}}\int dp_0\mathcal{D}_\mu(E_{^3\text{H}}+p_0,p)\Gamma^{^3\text{{H}}}_\mu(p)S(-p_0,-p)~,
\end{equation}
and
\begin{eqnarray}
\hat{I}_{\mu\nu}(E,p,p')&=&\frac{2\pi^2}{{ p}^2}\delta\left(p-p'\right)D_{\mu}(E,p)^{-1}\delta_{\mu,\nu}~,\\
\hat{\mathcal{K}}^{^3\text{H}}_{\mu\nu}(p,p',E)&=&My_\mu y_\nu a_{\mu\nu}K_0(p',p,E)~,
\end{eqnarray}
where $\delta_{\mu,\nu}$ is the Kronecker delta.
%and $Z^{^3\text{{H}}}$ is $^3$H normalization.

For $^3$He, 
the normalization condition that determines the wave-function factor $Z^{^3\text{He}}$ has the form:
\begin{multline}\label{eq_norm_3He}
1=\int\frac{d^3p}{(2\pi)^3}\int\frac{d^3p'}{(2\pi)^3}\sum\limits_{\mu,\nu=t,s,pp} 
{\psi^{^3\text{{He}}}_\mu(p)}
\times\left\{\frac{\partial}{\partial_E}\left [\hat{I}_{\mu\nu}(E,p,p')
-\hat{\mathcal{K}}^{^3\text{He}}_{\mu\nu}(p,p',E)\right]_{E=E_{^3\text{He}}}\right\}{\psi^{^3\text{{He}}}_\nu(p')}~,
\end{multline}
where:
\begin{align}\label{eq_psi_3He}
\nonumber
\langle\psi^{^3\text{He}}_\mu|p\rangle&=
{\sqrt{Z^{^3\text{He}}}}\int dp_0\mathcal{D}_\mu(E_{^3\text{He}}+p_0,p)
&\times\Gamma^{^3\text{{He}}}_\mu(p)S(-p_0,-p)~,
\end{align}
\begin{align}
\hat{\mathcal{K}}^{^3\text{He}}_{\mu\nu}(p,p',E)&=My_\mu y_\nu a'_{\mu\nu}\left[K_0(p',p,E)+K_{\mu\nu}^C(p',p,E)\right]~,
\end{align}
and $K_{\mu\nu}^C(p',p,E)$ is the $\mu,\nu$ index of the matrix $K^C$ (\cref{eq_K_C}).
\section{The diagrammatic form of the normalization}
The implication of the one-body unit operator is turning a single
nucleon operator into two one-nucleon propagators under the
assumption of energy and momentum conservation in the center-of-mass
system:

\begin{eqnarray}
\nonumber
\sum_i^A {\bf p}^i&=&\sum_i^A{\bf p'}^i=0~,\\
\sum_i^A {p_0}^i&=&\sum_i^A{p_0'}^i=E~,
\end{eqnarray}
where $i,j$ are the different nucleons indexes, ${\bf p^i, ({p'}^i)}$
refers to the one-nucleon incoming (outcoming) momentum and
$p^i_0 ({p'}_0^i)$ refers to the i's nucleon incoming (outcoming)
energy.

The Jacobi momentum $\bf{p}$ is defined as the relative momentum
between the dimer and the one-nucleon of the incoming (outcoming)
three-nucleon wave-function,
$\bf{p}(\bf{p'})=\frac{1}{2}\left[\bf{p}(\bf{p'})-\left(\bf{-p}\left(\bf{-p'}\right)\right)\right]$
and $E$ is the total three-nucleon energy.

Let us note that an energy derivative acting on a single nucleon
propagator that contains the energy $E$ can be written as two
propagators:
\begin{equation}\label{eq_S}
\frac{\partial}{\partial E} S (E, {\bf p})= -\int \frac{\hbox{d}^3
	p'}{(2\pi)^3}S(E, {\bf p})\times S(E, {\bf p'})(2\pi)^3 \delta({\bf p-p'})~.
\end{equation}
Therefore, the normalization operator for \cref{eq_Gamma} can be
written as a multiplication of the one-nucleon propagators and the corresponding delta functions, under the
assumption of energy and momentum conservation: \cblack
%\begin{widetext}
	%\cred
	\begin{multline}\label{z_eq}
	Z^{-1}=-\int\frac{dp_0}{2\pi}\int\frac{d^3p}{(2\pi)^3}
	\int\frac{dp'_0}{2\pi}\int\frac{d^3p'}{(2\pi)^3}\Gamma(p)iS(-p_0,{\bf -p})i\mathcal{D}(E+p_0,{\bf p})\\
	\times\Bigg\{y^2\int \frac{dk_0}{2\pi}\int\frac{d^3k}{(2\pi)^3}
	\int\frac{dk'_0}{2\pi}\int\frac{d^3k'}{(2\pi)^3}
	iS(E+p_0+k_0,{\bf p+k})iS(E+p'_0+k'_0,{\bf p'+k'})iS(-k_0,{\bf -k})iS(-k'_0,{\bf -k'})\\
	\times\delta^3\left[{\bf p+k-(p'+k')}\right]\left[\delta^3({\bf p'-p})\delta^3({\bf k-k'})+\frac{1}{2}\delta^3({\bf k'-p})\delta^3({\bf k-p'})\right]\Bigg\}
	i\mathcal{D}(E+p'_0,{\bf p'})iS(-p'_0,{\bf -p'})\Gamma(p')~.\\
	\end{multline}
	By performing the energy integration, \cref{z_eq} becomes:
	\begin{multline}
	\int\frac{p^2dp}{2\pi^2}\int\frac{p'^2dp'}{2\pi^2}\Gamma(p)D(E,p)
	\times M^2y^2\Bigg[\frac{-i}{{4 \pi (p-p')}}\log\left(\frac{ i \sqrt{3 p'^2-4 E M}-2 p-p'}{ i \sqrt{3 p^2-4 E M}-p-2 p'}\right)\frac{2\pi^2\delta{(p-p')}}{p'^2}\\-\frac{1}{2p'^2 \left(p^2-2 E M\right)+2\left(p^2-E M\right)^2+2p'^4}\Bigg]
	\times D(E,p')\Gamma(p'),
	\end{multline}
%\end{widetext}
which is identical to \cref{eq_identity_operator}. %with $i=\sqrt{1}$.\cblack
%Given \cref{eq_S}, the normalization operator \cref{eq_identity_operator}, represents all possible insertions of one-nucleon propagators, coupled to the corresponding delta function, in order to ensure energy and momentum are conservation.

Figure~\ref{fig_B_S_premutation} shows in detail the two topologies of
the normalization diagrams. For the case in which the normalization
insertion connects the two dimers in the three-nucleon systems, it is
proportional to $\frac{\partial}{\partial E}\hat{{I}}$
(Fig.~\ref{fig_B_S_premutation} (a)). For the case in which the
one-nucleon exchange propagator connects both one of the dimer
nucleons and the single nucleon, the diagram is proportional to
$\frac{\partial}{\partial E}\hat{\mathcal{K}}$
(Fig.~\ref{fig_B_S_premutation} (b)).
\begin{figure}[h!]
	\centering
	\begin{tabular}{@{}c@{}}
		\includegraphics[width=.65\linewidth]{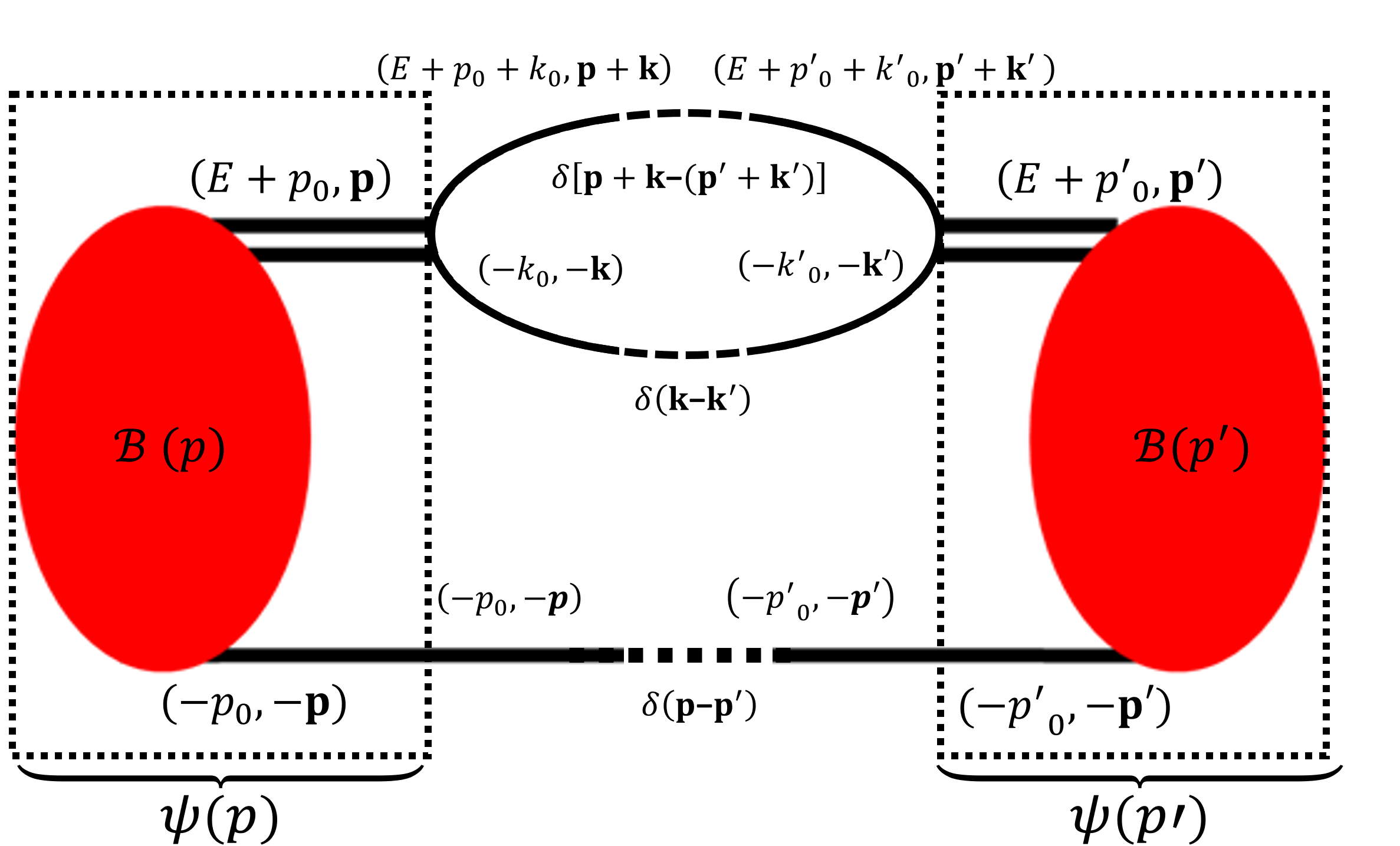} \\[\abovecaptionskip]
		\small (a) 
	\end{tabular}
	\begin{tabular}{@{}c@{}}
		\includegraphics[width=.65\linewidth]{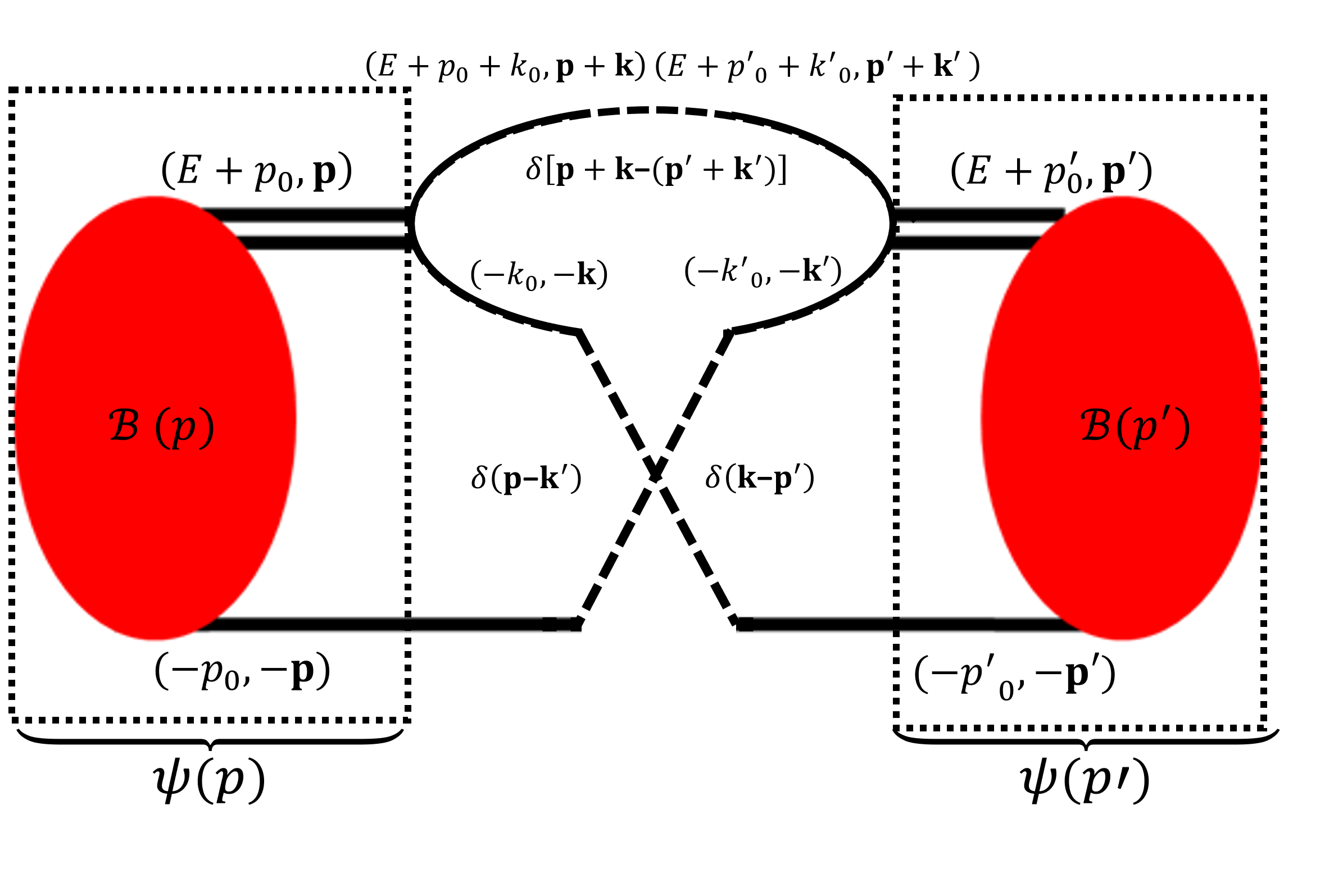} \\[\abovecaptionskip]
		\small (b) 
	\end{tabular}
	\caption{\footnotesize{Diagrammatic representation of two possible connections between two identical three-nucleon wave-functions, $\psi$. The
			double lines are the propagators of the two dibaryon fields,
			$\mathcal{D}$. 
			%\cblue
			The solid lines represent one-nucleon propagators, while the %vertical 
			dashed lines denote the delta functions. Diagram (a) is
			proportional to $\dfrac{\partial}{\partial E}{\hat{I}}$, while
			diagram (b) is proportional to
			$\dfrac{\partial}{\partial
				E}\hat{\mathcal{K}}$.}}\label{fig_B_S_premutation}
\end{figure}

Note that since for both $^3$H and $^3$He, $\hat{\mathcal{K}}_{\mu\nu}$ is
not diagonal, \cref{eq_norm_3H,eq_norm_3He} involve different
channels.\begin{comment} The diagrammatic form of $^3$H and $^3$He normalizations without a one-photon diagram is shown in Fig.~\ref{fig_norm1}, while the Coulomb diagrams which contribute to $^3$He normalization only, are shown in Fig.~\ref{fig_norm1_photon}.

\begin{figure}[h!]
	\centering
	\includegraphics[width=1\linewidth]{norm3.pdf}\\
	\caption{\footnotesize{The diagrammatic representation of the normalization of the homogeneous three body Bethe-Salpeter Wave Functions without one photon exchange. The double lines are propagators of the dibaryons fields $D_t$ (solid),
			$D_s$ (dashed) and $D^{pp}$ (dotted). The red bubbles represent the deuterium channel T=0, S=1, the green bubbles represent the triplet channel T=1, S=0 with $np$ dibaryon while the purple bubbles represent the triplet
			channel T=1, S=0 with pp dibaryon.}}\label{fig_norm1}
\end{figure}
\begin{figure}[h!]
	\centering
	\includegraphics[width=0.85\linewidth]{norm_photon.pdf}\\
	\caption{\footnotesize{The diagrammatic representation of the normalization of the homogeneous three body Bethe-Salpeter Wave Functions with one photon exchange. The double lines are propagators of the dibaryons fields $D_t$ (solid),
			$D_s$ (dashed) and $D^{pp}$ (dotted). The red bubbles represent the deuterium channel T=0, S=1, the green bubbles represent the triplet channel T=1, S=0 with $np$ dibaryon while the purple bubbles represent the triplet
			channel T=1, S=0 with pp dibaryon. }}\label{fig_norm1_photon}
\end{figure}

\end{comment}

\cblack
\chapter {$A=3$ matrix element in pionless EFT}\label{general_matrix} 
In this chapter, we present the general method for calculating an
$A=3$ matrix element in \pilesseft. This method is used in this work for 
calculating the $A=3$ electro-weak observables, as well as the $^3$He energy shift perturbatively and the NLO
contribution to the three-nucleon wave functions.

\section{The general form of an $A=3$ matrix element}
In Chapter~\ref{Norm}, we showed that the three-nucleon normalization can be written as:
\begin{multline}
1=\sum\limits_{\mu,\nu} 
{\psi^{i}_\mu(p)}\otimes
\biggl\{\frac{\partial}{\partial_E}\bigl [\hat{I}_{\mu\nu}(E,p,p')
-\hat{\mathcal{K}}^{i}_{\mu\nu}(p,p',E)\bigl]_{E=E_{i}}\biggr\}\otimes {\psi^{i}_\nu(p')}~,
\end{multline}
which can be written in terms of a matrix element:
\begin{equation}
1= \sum\limits_{\mu,\nu}
\bra{\psi^{i}_\mu}
\mathcal{O}_{\mu\nu}^{\text{norm}}(E_i)\ket{\psi_\nu^{i}}~,
\end{equation}
where $\mathcal{O}_{\mu\nu}^{\text{norm}}(E_i)$ is the normalization operator such that: 
\begin{equation}\label{eq:1b:Onorm}
\mathcal{O}_{\mu\nu}^{\text{norm}}(E_i)=
\frac{\partial}{\partial_E}\left [\hat{I}_{\mu\nu}(E,p,p')-My_\mu y_\nu a^i_{\mu\nu}\hat{K}_{\mu\nu}^i(p',p,E)\right]\bigg|_{E=E_i}~,
\end{equation}
where:
\begin{equation}
\hat{K}_{\mu\nu}^i=\begin{cases}
K_0(p',p,E)&i=^3\text{H}\\
K_0(p',p,E)+K_{\mu\nu}^C(p',p,E)&i=^3\text{He}
\end{cases}
\end{equation}
and
\begin{equation}
a_{\mu\nu}^i=\begin{cases}
a_{\mu\nu}&i=^3\text{H}\\
a'_{\mu\nu}&i=^3\text{He}~,
\end{cases}~
\end{equation}
which are a result of $N-d$ doublet-channel projection
(\cref{eq_3H_projecation,eq_3He_projecation}). Note that we are considering
here one-body operators that do not have additional momentum
dependence. However, the formulas given here could easily be
extended also to this case.

Equation (\ref{eq:1b:Onorm}) can be generalized to any
operator, $\mathcal{O}_{j,i}$, between the initial (i) and final
(j) $A=3$ bound-state wave-functions ($\psi_{i,j})$, whose matrix element is evaluated as
%\cred
\begin{equation}\label{eq_general_operator}
\langle\mathcal{O}_{j,i}(q_0,q)\rangle=\bra{S, S'_{z}, I, I'_{z}, E'}\mathcal{O}_{j,i}(q_0,q)\ket{S, S_{z}, I, I_{z}, E}~,
\end{equation}
where:
\begin{itemize}
	\item $S$ denotes the total spin $\left(\frac{1}{2}\right)$ of the three-nucleon
	system.
	\item $S_{z},S'_{z}$ denote the initial and final spin projections,
	respectively.
	\item $I$ denotes the total isospin $\left(\frac{1}{2}\right)$ of the
	three-nucleon
	\item $I_{z},I'_{z}$ denote the initial and final isospin projections,
	respectively.
	%\item $E$ and $E'$ are the initial and final energies, respectively.
	\item $q$ is the momentum transfer of such an operator (assuming that
	for the initial state, the three-nucleon total momentum is zero).
	\item The energy transfer is defined as: $q_0=E'-E$.
\end{itemize} 
Therefore, a general operator that connects two three-nucleon bound-states
with $I=\frac{1}{2}$, $S=\frac{1}{2}$, factorizes into the
\cblack following parts:
\begin{equation}
\mathcal{O}_{j,i}=\mathcal{O}^{J}\mathcal{O}^{T}\mathcal{O}_{j,i}(q_0,q), 
\end{equation}
where $\mathcal{O}^{J}$, the spin part of the operator whose total
spin is $J$, and $\mathcal{O}^{T}$, the isospin part of the operator, depend on the initial and final quantum numbers. The spatial part
of the operator, $\mathcal{O}_{j,i}(q_0,q)$, is a function of the
three-nucleon wave-function's binding energies ($E_i$, $E_j$) and the energy and
momentum transfer ($q_0,q$, respectively).

%The matrix element of a general operator between initial (i) and final
%(j) $A=3$ bound-state wave-functions ($\psi_{i,j})$ has the form:
%\begin{multline}\label{eq_general_operator}
% \langle \mathcal{O}_{j,i}(q_0,q)\rangle=\\
%\bra{{S, S'_{z}, I, I'_{z}, E'} }\mathcal{O}^{J}\mathcal{O}^{I}\mathcal{O}_{j,i}(q_0,q)
% \ket{S, S_{z}, I, I_{z}, E}~.
%\end{multline}
The observable associated with the above matrix element is also
related to a reduced matrix element between $A=3$ bound-state wave
functions:
\begin{equation}
\nonumber
\langle\| \mathcal{O}_{j,i}(q_0,q)\|\rangle=
\langle{{S,I, E',q}\| }\mathcal{O}^{J}\mathcal{O}^{T}\mathcal{O}_{j,i}(q_0,q)\|{S, I, E}\rangle~.
\end{equation}
In the next subsection, we write explicitly the reduce matrix
element term for a general one-body operator. Note that the amplitude
$\Gamma^i(p)$ (and as a result, $\psi^i(p)$) still carries implicit spin
and isospin indices, $S_z$ and $I_z$, respectively. We calculate the reduced matrix
element shown above by performing the spin algebra with the
afore-mentioned spin and isospin projectors and the spin- and isospin
part of the operator under consideration for one particular choice of
external spin projections. Then we use the Wigner-Eckhart theorem to combine this matrix element with a Clebsch-Gordan
coefficient to obtain the reduced matrix element.

\section{Matrix elements of one-body operators}
The one-body normalization operator,
$\mathcal{O}_{\mu\nu}^{\text{norm}}(E_i)$ (\cref{eq:1b:Onorm}), is a result of $N-d$ doublet-channel projection
(\cref{eq_3H_projecation,eq_3He_projecation}). For the case that the one-body spin and isospin operators are
combinations of Pauli matrices, the general matrix element will be a
result of the different $N-d$ doublet-channel projections coupled to a
spin-isospin operator. To evaluate the reduced matrix
element of a general one-body operator, one needs to calculate
explicitly one component of the spin operator, $\mathcal{O}^J$. For an
operator whose spin part is proportional to $\pmb{\sigma}$, for example, the zero-component of $\langle\mathcal{O}^J\rangle$, $\langle\mathcal{O}^J_0\rangle$ is given by:

\begin{multline}\label{eq_operator}
\langle \mathcal{O}_{j,i}^{\text{1B}}(q_0,q)\rangle_0=
\sum\limits_{\mu,\nu}y_\mu y_\nu
\bra{\psi^j_\mu}\Bigl\{d^{ij}_{\mu\nu} \hat{\mathcal{I}}(q_0,q)
+a^{ij}_{\mu\nu}\left[\hat{\mathcal{K}}(q_0,q)+{\hat{\mathcal{K}}^C_{\mu\nu}}(E, q_0,q)\right]\Bigr\} \ket{\psi^i_\nu}~,
\end{multline}
where $\hat{\mathcal{I}}(E, q_0,q)$ and $\hat{\mathcal{K}}(E, q_0,q)$ represent
all the possible connections between two three-nucleon wave-functions
($\psi^i, \psi^j$) that contain a \textbf{one-body} insertion of 
momentum and energy transfer without a Coulomb interaction. The
spatial parts that do not contain a one-nucleon exchange are denoted by
$\hat{\mathcal{I}}(q_0,q)$, and the spatial parts that do contain a
one-nucleon exchange are denoted by $\hat{\mathcal{K}}(q_0,q)$; the full
expressions for $\hat{\mathcal{I}}(E, q_0,q)$ and $\hat{\mathcal{K}}(q_0,q)$
are given \cref{eq_hat_I,eq_hat_K}. $a^{ij}_{\mu\nu}$ and $d^{ij}_{\mu\nu}$ are
a result of the $N-d$ doublet-channel projection coupled to
$\mathcal{O}^J_0\mathcal{O}^T$. ${\hat{\mathcal{K}}^C_{\mu\nu}}(E, q_0,q)$
are the diagrams that contain a one-photon interaction in addition to
the energy and momentum transfer. A derivation of an analytical
expression for these diagrams is too complex, so they were calculated
numerically only.

\begin{multline}\label{eq_hat_I}
\hat{\mathcal{I}}(q_0,q)=\frac{M^2i}{4 \pi (q-p+p')} \cdot\\
\Biggl\{\log \left[q \left(\sqrt{4 E M-3 p'^2}-2 p+p'\right)+(p'-p) \left(\sqrt{4 E M-3 p'^2}-2 p-p'\right)+q^2+2 q_0 M\right]\\-\sqrt{4 E M-q^2+2 q p-4 q_0 M-3 p^2} \sqrt{\frac{-1}{-4 E M+q^2-2 q p+4 q_0 M+3 p^2}} \cdot\\\log \left[-(p-p') \left(\sqrt{4 E M-q^2+2 q p-4 q_0 M-3 p^2}-p-2 p'\right)+q \left(\sqrt{4 E M-q^2+2 q p-4q_0 M-3 p^2}-p'\right)+2q_0 M\right]\Biggr\}\cdot\\
\delta(q_0-E+E')\delta(p'-p)\frac{2\pi^2}{p'^2}
\end{multline}
and:
\begin{multline}\label{eq_hat_K}
\hat{\mathcal{K}}(q_0,q)=\frac{M^2}{2p p' \left[q \left(2 (p+p')-q\right)-2 M q_0\right]}\times\\
\left\{\log \left[\left(-E M+p^2+p p'+p'^2\right) \left(2 E M-2 M q_0-2 \left(p^2-p p'+p'^2\right)+2 q (p+p')-q^2\right)\right]\right.\\-
\left.\log \left[ \left(-E M+p^2-p p'+p'^2\right) \left(2 E M-2 M q_0-2 \left(p^2+p p'+p'^2\right)+2 q (p+p')-q^2\right)\right]\right\}\cdot\\
\delta(q_0-E+E')\delta(q-p+p')
\end{multline}
\begin{figure}[h!]
	\centering
	% \begin{comment}
	\begin{tabular}{@{}c@{}}
		\includegraphics[width=.65\linewidth]{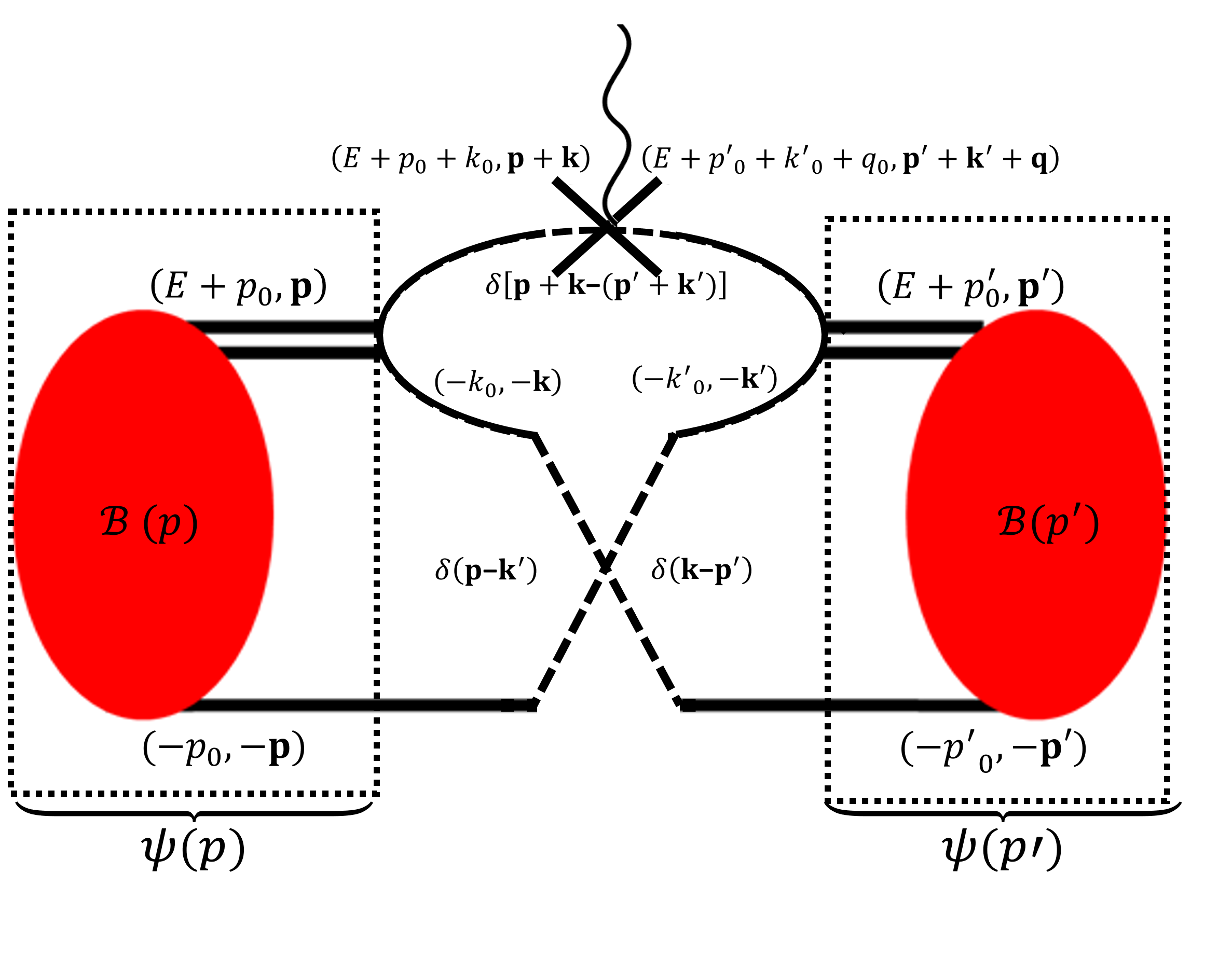}
		\vspace{-0.5 cm}
		\\[\abovecaptionskip]
		\small (a) 
	\end{tabular}
	\vspace{0.7 cm}
	\begin{tabular}{@{}c@{}}
		\includegraphics[width=.65\linewidth]{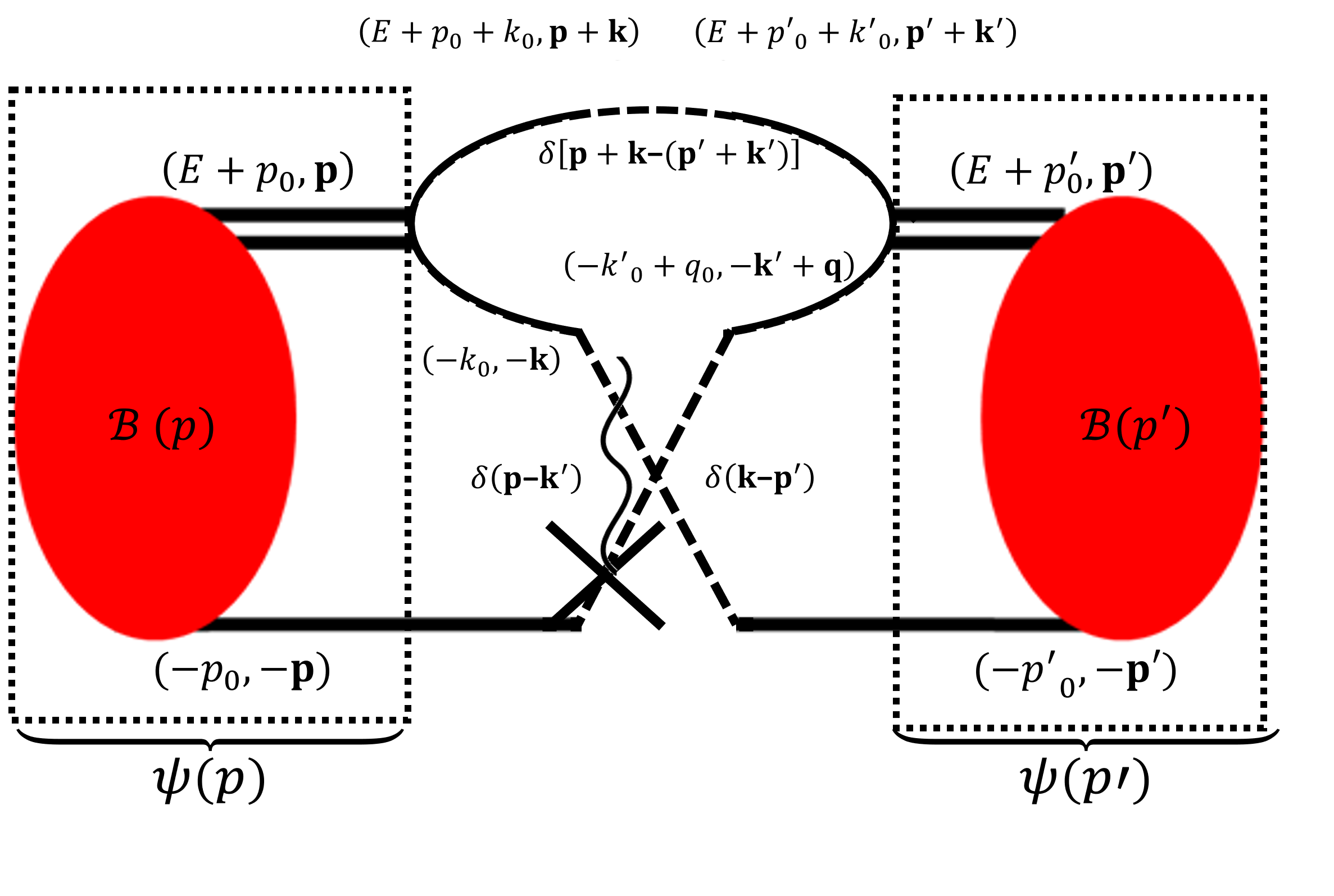}
		\vspace{-0.5 cm}
		\\[\abovecaptionskip]
		\small (b) 
	\end{tabular}
	\vspace{0.5 cm}
	\begin{tabular}{@{}c@{}}
		\includegraphics[width=.65\linewidth]{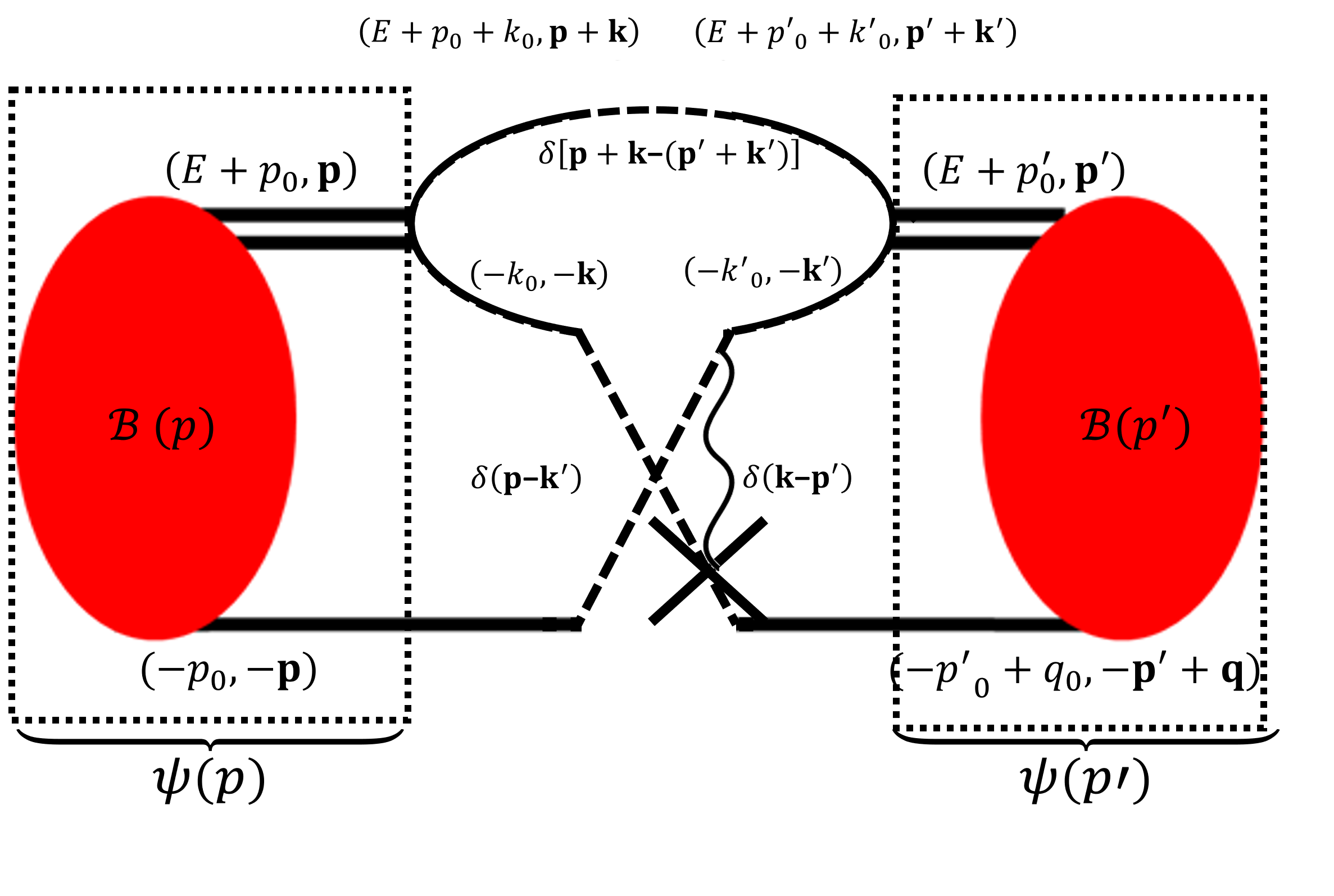}
		\vspace{-0.5 cm}
		\\[\abovecaptionskip]
		\small (c) 
	\end{tabular} 
	%\end{comment}
	\caption{\footnotesize{Diagrammatic representation of all the possible
			variations of $\mathcal{O}^q$ between two three-nucleon wave-functions that involve a one-nucleon exchange. The RHS of each diagram is the final state, $\psi^j$, while the LHS is the initial state, $\psi^i$. The double lines are the propagators of the two dibaryon fields, $\mathcal{D}$. The
			probe represents the momentum and energy transfers due to
			the interaction.}}
	\label{fig_general_matrix element} 
\end{figure}

Figure \ref{fig_general_matrix element} shows all possible diagrams of a {one-body} insertion of momentum and energy transfer between two three-nucleon wave-functions that contain a one-nucleon exchange. 

The one-body reduced matrix
element, $\langle\| \mathcal{O}_{j,i}^{\text{1B}}(E, q_0,q)\|\rangle$, can be easily calculated as a function of the three-nucleon
quantum total spin and isospin numbers, using the Wigner-Eckart theorem. Since these calculations are
not dependent on the spatial structure of the three-nucleon wave
function, one can isolate the spin and isospin matrix elements in
terms of the three-nucleon quantum numbers such that (again for the zero component):}
%\cred
\begin{multline}\label{eq_general_operator_reduced}
\langle\| \mathcal{O}_{j,i}^{\text{1B}}(q_0,q)\|\rangle=\dfrac{\sqrt{2}}{\bra{\frac{1}{2}S_zJ0}\frac{1}{2}S'_z\rangle}
\sum\limits_{\mu,\nu}
\bra{\psi^j_\mu} y_\mu y_\nu\left[{d^{ij}_{\mu\nu}} \hat{\mathcal{I}}(q_0,q)\right. +\left.{a^{ij}_{\mu\nu}}\hat{\mathcal{K}}(q_0,q)\right] \ket{\psi^i_\nu}~,
\end{multline}
that can be written as:
\begin{multline}
\langle\| \mathcal{O}_{j,i}^{\text{1B}}(q_0,q)\|\rangle=\left\langle\frac{1}{2}\left\|\mathcal{O}^J\right\|\frac{1}{2}\right\rangle\left\langle\frac{1}{2},I'_z\left|\mathcal{O}^T\right|I_z,\frac{1}{2}\right\rangle\\ \times
\sum\limits_{\mu,\nu}
\bra{\psi^j_\mu}y_\mu y_\nu\Bigl\{d'^{ij}_{\mu\nu} \hat{\mathcal{I}}(q_0,q)+
a'^{ij}_{\mu\nu}\left [\hat{\mathcal{K}}(p,p',E,q_0)+{\hat{\mathcal{K}}^C_{\mu\nu}}(q_0,q)\right]\Bigr\} \ket{\psi^i_\nu} 
~,
\end{multline}
such that for $i=j$:
\begin{align}\label{eq_cases1}
d'^{ii}_{\mu\nu}&=\delta_{\mu,\nu}\\\label{eq_cases2}
a'^{ii}_{\mu\nu}&=\begin{cases}
a_{\mu\nu}&i=^3\text{H}\\
a'_{\mu\nu}&i=^3\text{He}\\
\end{cases}~,
\end{align}
where ${\hat{\mathcal{K}}^C_{\mu\nu}}(q_0,q)=0$ for $^3$H.

The reduced matrix element of the spin part of the operator,
$\left\langle\frac{1}{2}\left\|\mathcal{O}^J\right\|\frac{1}{2}\right\rangle$,
is a function of the initial and final total spin of the $A=3$ nucleon
wave-function. For the case that $\mathcal{O}^J=\boldsymbol{\sigma}$,
the reduced matrix element, $\left\langle\frac{1}{2}\left\|\mathcal{O}^J\right\|\frac{1}{2}\right\rangle$,
is calculated using the Wigner-Eckart theorem such that:
\begin{equation}
\left\langle\frac{1}{2}\left\|\boldsymbol{\sigma}\right\|\frac{1}{2}\right\rangle=2\left\langle\frac{1}{2}\left\|\boldsymbol{s}\right\|\frac{1}{2}\right\rangle=
%\\2\sqrt{2}\left\langle\frac{1}{2},\frac{1}{2}\left|{s}_z\right|\frac{1}{2},\frac{1}{2}\r%ight\rangle\times %\frac{1}{\langle\frac{1}{2},\frac{1}{2},1,0|\frac{1}{2},\frac{1}{2}\rangle}=\\
%2\sqrt{2}\frac{1}{2}\left\langle\frac{1}{2},\frac{1}{2}\bigg|\frac{1}{2},\frac{1}{2%}\right\rangle\times \frac{1}{1/\sqrt{3}}={\sqrt3}{\sqrt{2}}
%=
\sqrt{6}~.
\end{equation}

\subsection{Two-body matrix element}
In contrast to the normalization operator given in \cref{eq:1b:Onorm},
which contains only one-body interactions, a typical \pilesseft
electroweak interaction contains also the following two-body
interactions up to NLO:
\begin{equation}
t^\dagger t,\,s^\dagger s,\,(s^\dagger t+h.c)~,
\end{equation}
under the assumption of energy and momentum conservation. %\cblue
The
diagrammatic form of the different two-body interactions, given in
Tab.~\ref{tbl: feynman_ruls} ($(o)-(q)$ are the two-body weak interactions while $(s)-(x)$ are the two-body magnetic interactions), is a result of the
Hubbard-Stratonovich transformation of a four-nucleon interaction vertex
(see, for example, Refs.~\cite{ando_deturon,Ando_proton} and
Appendix~\ref{whatever}). \cblack

\section{Deuteron normalization and the matrix element in pionless EFT}
The calculation of matrix elements is significantly harder in the
three-body sector than in the two-body sector due to the more
complicated structure of three-body diagrams \cite{3bosons,
	triton}. However, a closer look at the deuteron wave-function
normalization reveals that the deuteron wave-function normalization
can also be written in the same manner as discussed above, since
\begin{equation}
Z_d^{-1}=i\frac{\partial}{\partial E} \frac{1}{iD_t(E,
	p)}\bigg|_{E=\frac{\gamma_t^2}{M}, p=0}, \, Z_d
=\frac{1}{1-\gamma_t\rho_t}, 
\end{equation}
where the energy derivative of $i\frac{1}{iD_t(E, p)}$ is equivalent
to the addition of a one-nucleon propagator, as discussed in
Chapter~\ref{Norm}. Hence, a general deuteron matrix element that
contains energy and momentum transfer (such as the deuteron magnetic
moment) can be written as the sum over all possible connections
\cite{KSW_c}:
\begin{equation}\label{eq_deutron_matrix}
\langle S,p'\|\mathcal{O}_{j,i}(q_0,q)\|S,p\rangle=\langle 1\|\mathcal{O}^J\|1\rangle \left(\frac{M^2}{8 \pi\gamma _t}\right)^{-1}\langle p'|\hat{\mathcal{I}}(q_0,q)|p\rangle~.
\end{equation}
For the case that ${O}^J=1 $ and
$q_0,q=0$, \cref{eq_deutron_matrix} gives the deuteron form factor, $F_C(0)$, which is
equal to 1:
\begin{equation}
\langle\|\mathcal{O}_{j,i}(q_0,q)\|\rangle=\left(\frac{M^2}{8 \pi\gamma _t}\right)^{-1} \hat{\mathcal{I}}(E_d,0,0)
=\left(\frac{M^2}{8 \pi\gamma _t}\right)^{-1}\frac{M^2}{8\pi\gamma_t}=1~.
\end{equation}
This matrix element form, which is very similar to the general
three-body matrix element (\cref{eq_general_operator_reduced}), implies that
in the case of bound-state matrix elements, our {\it wave-function}
approach can be applied in the two- and the three-nucleon systems,
consistently.

\section{Example: $^3$He-$^3$H binding energy difference with a perturbative Coulomb}\label{energy_shift}
In this subsection, we apply the formalism introduced above to
the so-called Coulomb energy shift in the three-nucleon system. We
define the Coulomb-induced energy shift, $\Delta E$, as
\cite{konig1,konig5,konig3}:
\begin{equation}\label{Eb}
-E_{^3\text{He}}=-E_{^3H}+\Delta E.
\end{equation}
The energy difference between $^3$H and $^3$He due to the Coulomb
interaction can be calculated perturbatively at LO as a matrix element
of one-photon exchange diagrams (Fig.~\ref{Coulomb_correction} a-d) and the $pp$ propagator (diagram f)
between two triton bubbles, as described in detail in \cite{konig3,
	konig5}. In our notation, these Coulomb interactions can be treated
as a special case of a general matrix element, despite the fact that
the Coulomb interaction does not conserve the three-nucleon
isospin. This representation is possible since we
divided the contribution to the energy shift into a one-body (1B) term
and a two-body (2B) term. The one-body term originates from the
one-photon exchange diagrams being calculated as a one-body interaction
between two $^3$H bound-state wave-functions and does not affect the three-nucleon isospin. The two-body term originates from the
difference between the proton-proton propagator and the spin singlet
propagator (which is a two-body operator).

In terms of \cref{eq_general_operator}, $\Delta E$ has the form: 

	\begin{align}\label{eq_delta_E1}
	\nonumber
	&\Delta E(\Lambda)=
	{Z^{^3\text{H}}}\sum_{\mu,\nu=t,s}y_\mu y_\nu\left[\Gamma^{^3\text{H}}_{\mu}(p)D_\mu(E_{^3\text{H}},p)\right]
	\otimes c_{\mu\nu}K^C_{\mu\nu}(p,p',E_{^3\text{H}})\otimes\left[D_\nu(E_{^3\text{H}},p')\Gamma^{^3\text{H}}_{\nu}(p')\right]\\+
	&{Z^{^3\text{H}}}\sum_{\mu=t,s}\left[\Gamma^{^3\text{{H}}}_\mu(p)D_\mu(E_{^3\text{H}},p)\right]\otimes\left[a_{\mu s}K_0(p,p',E_{^3\text{H}})+b_{\mu s}\frac{H(\Lambda)}{\Lambda^2}\right]\otimes
	\left\{\left[D_{pp}(E_{^3\text{H}},p')-D_s(E_{^3\text{H}},p')\right]\Gamma^{^3\text{H}}_{s}(p')\right\}~.
	\end{align}
	Using the fact that:
	\begin{equation}
	\Gamma^{^3\text{{H}}}_s(p')=\sum_{\mu=t,s}\left[\Gamma^{^3\text{{H}}}_\mu(p)D_\mu(E_{^3\text{H}},p)\right]\otimes\left[a_{\mu s}K_0(p,p',E_{^3\text{H}})+b_{\mu s}\frac{H(\Lambda)}{\Lambda^2}\right]~,
	\end{equation}
	\cref{eq_delta_E1} becomes:
	\begin{multline}
	\Delta E(\Lambda)=
	\sum_{\mu,\nu=t,s}{\psi^{^3\text{H}}_{\mu}(p)}\otimes \underbrace{c_{\mu\nu}K^C_{\mu\nu}(p,p',E_{^3\text{H}})}_{\text{one body}}\otimes{\psi^{^3\text{H}}_{\nu}(p')}\\
	+\sum_{\mu=t,s}{\psi^{^3\text{H}}_{\mu}(p)}\otimes\underbrace{\left[\frac{D_{pp}(E_{^3\text{H}},p)-D_s(E_{^3\text{H}},p)}{D_s(E_{^3\text{H}},p)^{2}}\right]\times
		\frac{\delta(p-p')}{p'^2}2\pi^2\delta_{\mu,s}}_{\text{two-body}}\otimes{\psi^{^3\text{H}}_{s}(p')}\\
	=\sum_{\mu=t,s}{\psi_{\mu}^{^3\text{H}}(p)}\otimes\left[\mathcal{O}_{\mu\nu}^{q(\text{1B})}(E_{^3\text{H}},p,p')+\mathcal{O}_{\mu\nu}^{q(\text{2B})}(E_{^3\text{H}},p,p')\right]\otimes{\psi_{\mu}^{^3\text{H}}(p')}~,
	\end{multline}

where $Z^{^3\text{H}}$ is the $^3$H normalization, and
\begin{align}\label{eq_q_C}
\mathcal{O}^{q(\text{1B})}_{\mu\nu}(E, p, p')&=c_{\mu\nu}K^C_{\mu\nu}(p,p',E)\delta(q-p+p'),\\
{\mathcal{O}}^{q(\text{2B})}_{\mu\nu}(E, p, p')&=\left[\frac{D_{pp}(E,p)-D_s(E,p)}{D_s(E,p)^{2}}\right]\\
\nonumber
&\qquad\times\frac{\delta(p-p')}{p'^2}2\pi^2\delta_{\mu,s}\delta_{\nu,s}~, 
\end{align}
where $K_{\mu\nu}^C(p,p',E)$ is given in % (see Appendix E and
% (see \cite{konig1, KonigPhd13} ).
\cref{eq_K_C} and $c_{\mu\nu}=a_{\mu\nu}$ under the assumption that $\Gamma_s=\Gamma_{np},\Gamma_{pp}$; $D_{pp}(E,p),D_s(E,p)$ were defined in Chapter~\ref{formalism}.

Figure.~\ref{fig_helium_energy} shows that summing over all possible one- and two-body 
Coulomb diagrams (\cref{eq_delta_E1}) is consistent with the non-perturbative calculation presented in
section~\ref{limit}. Both calculations reproduce the predictions presented
in Refs.~\cite{konig3,konig5}, and this fact serves as a validation of the
numerical calculation presented here.

\chapter{Perturbative correction to the three-body nuclear wave functions}\label{NLO_corrections}
The components needed for a consistent calculation of an $A=3$ matrix
element up to NLO ({\it i.e.}, retaining terms of order
$\frac{Q}{\Lambda_{cut}}$) are the interaction operator,
$\mathcal{O}_{j,i}(q_0,q)$, and the bound-state amplitudes up to this
order. In this section, we present how the NLO contributions to the
three-nucleon bound-state amplitude can be calculated using the method
presented in Chapter~\ref{general_matrix}. Specifically, we follow the
NLO bound-state calculation of Vanasse et al. \cite{konig2,
	Vanasse:2015fph}, except that we consider $y_t\neq y_s$.

In our notation we distinguish between the NLO correction to
the scattering %\st{amplitude} %%\cblack 
matrix \cblack $t (E, k, p)$ and the NLO correction to the bound-state
scattering amplitude ($\mathcal{B} (k)$) which is the homogeneous solution of the Faddeev equations.

\section{The NLO correction to the full scattering amplitude}
In this section, we use the formalism introduced in
Chapter~\ref{general_matrix} to calculate the NLO correction to the full
scattering amplitude.

For simplicity, in a similar manner to that presented in Chapter~\ref{Norm}, we first write the NLO correction for
the case that the t-matrix contains only one channel, {\it i.e.},
$t(E, k, p) = T (E, k, p)$, and then extend this formalism
for $^3$H and $^3$He.

\begin{figure}[h!]
	\centering
	\includegraphics[width=0.65\linewidth]{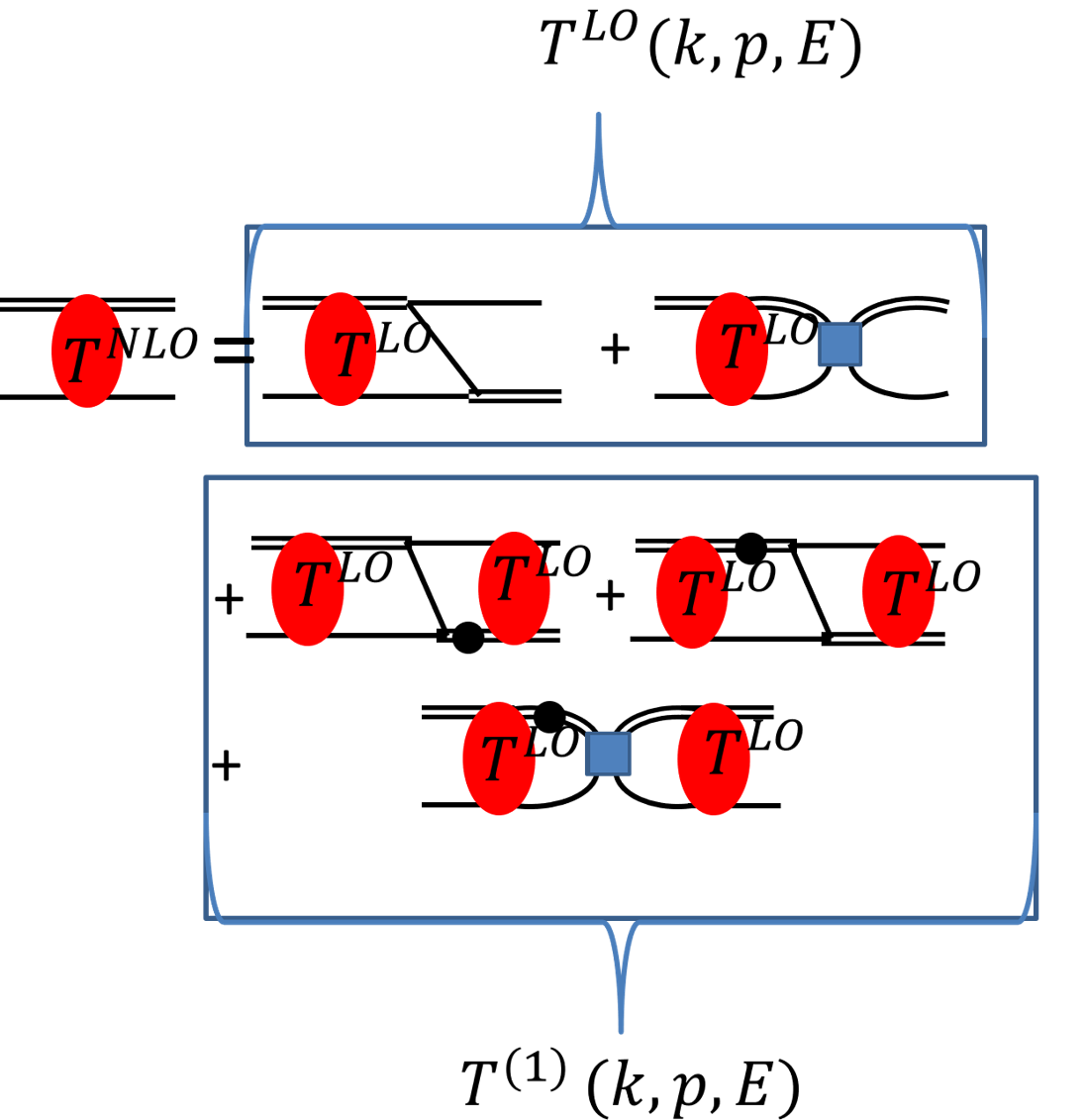}
	\caption{\footnotesize{The $t$-matrix describing a bound-state up
			to NLO (red bubbles). The LO $t$-matrix is the result of the LO homogeneous
			Faddeev equation - \cref{stm1}. The NLO correction to the
			$t$-matrix includes %both three-body force $H^{\text{NLO}} (\Lambda)$ and
			the effective range $\rho$ (black dot). The double lines are the propagators of the two dibaryon fields and the blue square is the three-body force.}}
	\label{fig_T_NLO}
\end{figure}

The full $t$-matrix can be expanded order-by-order:
\begin{equation}\label{eq_full_t}
T (E,k,p)=T ^{\text{LO}} (E,k,p) + T ^{(1)} (E,k,p)+\ldots~,
\end{equation}
where $T^{\text{LO}} (E, k, p)$ is given by \cref{stm1} and
$T^{(1)}$, which contains the effective range corrections up to NLO,
is derived next. Based on Chapter.~\ref{general_matrix} and
Ref.~\cite{konig2}, \cref{eq_full_t} for a bound-state (\cref{eq_t0})
can be written as:
\begin{multline}\label{eq_stm_t_NLO}
T (E,k,p)=T ^{\text{LO}} (E,k,p)+T ^{\text{LO}} (E,k,p')D ^{\text{LO}} (E,p')\otimes
\mathcal{O} ^{(1)} (E,p',p'')\otimes D ^{\text{LO}} (E,p'') T ^{\text{LO}} (E,p'',p),
\end{multline} where the operator $\mathcal{O}^{(1)} (E, p', p'')$ contains all NLO corrections to the
$T$-matrix (see Fig.~\ref{fig_T_NLO}).
Using \cref{eq_stm_t_NLO}, the NLO correction to the T-matrix is given by:
%\cred
\begin{multline}\label{eq_stm_T_NLO_no_H}
T^{(1)} (E, k, p)=
-T^{\text{LO}} (E, k, p')D^{\text{LO}}(E,p')
\otimes
\Bigl\{
\frac{My^2}{2}\left[K_0 (p', p'', E)+\frac{H(\Lambda)}{\Lambda^2}\right]
\\\ \times 
\left[\Delta(E,p'')+\Delta(E,p')\right]\Bigr\}
\otimes
D^{\text{LO}}(E,p'')T^{\text{LO}} (E, p'', p)~.
\end{multline}
By using the STM equation (\cref{stm_T}), \cref{eq_stm_T_NLO_no_H} becomes:
\begin{multline}
T^{(1)} (E, k, p)=\int\frac{p'^2dp'}{2\pi^2}{T^{\text{LO}} (E, k, p')}
\Delta(E,p')
D^{\text{LO}}(E,p')T^{\text{LO}} (E, p', p)\\=
-\frac{2\pi}{M y^2}\rho\int\frac{p'^2dp'}{2\pi^2}T^{\text{LO}} (E, k, p')\frac{3 p^2/4-E M-1/a_2 ^2}{\left(\sqrt{{3 p'^2/4}-E M}-1/a_2 \right)^2}
\times T^{\text{LO}} (E, p', p)~,
\end{multline}
%\
%\frac{4}{My_t^2}\frac{H ^{\text{NLO}}}{\pi^2\Lambda^2}\left\{\left[\int_0^\Lambda
%T ^{\text{LO}} (E,k,p)\delta_t (E,p)q^2dq\right]\cdot
%\right.\\
%\
%\qquad\qquad\qquad\left.\left[\int_0^\Lambda
%T ^{\text{LO}} (E,q,p)\delta_t (E,q)q^2dq\right]\right\}.

\cblack
where $\rho$ is the effective range, $a_2$ is the dibaryon scattering length, $K_0$ is defined in \cref{eq_K1},
\begin{eqnarray}
\Delta(E, p)=
\frac{D^{\text{NLO}} (E, p)-D^{\text{LO}} (E, p)}{D^{\text{LO}}(E, p)},
\end{eqnarray}
and $D^{\text{NLO}}(E,p)$ is defined in \cref{NLO_correction_triplet}.

\subsection{The NLO corrections to the three-nucleon bound-state pole position}\label{pole_position}
Following Vanasse et al. \cite{konig2}, we use
\cref{eq_stm_T_NLO_no_H} to predict the NLO correction to the
three-nucleon binding energy. We extend the method developed by Ji,
Phillips and Platter \cite{H_NLO} to include complications due to the
isospin. The scattering amplitude possesses a pole at the binding
energy and can be written as:
\begin{align}\label{eq_ful_T}
&T (E,k,p)=T ^{\text{LO}} (E,k, p) + T ^{(1)} (E,k,p)\\
\nonumber
&=\frac{\mathcal{Z}^{\text{LO}} (k, p) + \mathcal{Z} ^{(1)} (k,p)}{E - (E_B + \Delta E_B)}+\mathcal{R}_0 (E,k,p)+\mathcal{R}_1 (E,k,p),
\end{align}
where $\mathcal{Z} ^{\text{LO}},\mathcal{Z} ^{(1)}$ are the residue
vector functions and $\Delta E_B$ is the NLO correction to the binding
energy. Both $\mathcal{R}_0 (E, k, p)$ and $\mathcal{R}_1 (E, k, p)$
are regular at $ E = E_B$, so they can be neglected. At the first order
in ERE (NLO) of \cref{eq_ful_T}, one finds that \cite{phillips}:
\begin{equation}\label{eq_T_NLO}
T^{(1)} (E, k, p)=\\
\frac{\mathcal{Z}^{(1)} (k, p)}{E - E_B }+\Delta E_B\frac{\mathcal{Z}^{\text{LO}} (k, p)}{(E - E_B)^2 },
\end{equation}

where $\mathcal{Z} ^{\text{LO}}$ is
defined around the pole ($E\rightarrow E_B$) from \cref{eq_ful_T} as: 
\begin{equation}
\label{eq_Z_0}
\mathcal{Z}^{\text{LO}} (k, p)=\lim_{E\rightarrow E_B} (E-E_B)T^{\text{LO}} (E, k, p).
\end{equation} 

For $E\rightarrow E_B$, $\Delta E_B$ is given by: 
\begin{equation}\label{eq_B1}
\Delta E_B = \lim_{E\rightarrow E_B}\dfrac{ T^{(1)} (E, k, p)(E -E_B)^2}{Z^{\text{LO}} (k, p)}.
\end{equation}
It might seem that the binding energy correction ($\Delta E_B$)
depends on the incoming and outgoing momenta ($k, p$). However, we
would expect the NLO binding energy,
$E_B^{\text{NLO}}=E_B+\Delta E_B$, to depend on the cutoff $\Lambda$
only, similarly to LO (as shown in Fig.~\ref{fig_helium_energy}), so
it is essential to examine its momentum dependence. Since for a
bound-state (\cref{eq_t0}):

\begin{equation}
T(E,k,p)=\frac{\Gamma(k)\Gamma(p)}{E-E_B}~, 
\end{equation} \cref{eq_stm_T_NLO_no_H} becomes: 
\begin{multline}\label{eq_T_NLO_E}
T^{(1)} (E, k, p) (E-E_B)^2=
\Gamma^{\text{LO}} (k)\Gamma^{\text{LO}} (p')D^{\text{LO}} (E, p')
\otimes\mathcal{O}^{(1)} (E, p', p'')\otimes
D^{\text{LO}} (E, p'')\Gamma^{\text{LO}} (p'')\Gamma^{\text{LO}} (p)=\\
\Gamma^{\text{LO}} (k)\psi^{\text{LO}} (p')
\otimes\mathcal{O}^{(1)} (E, p', p'')\otimes \psi^{\text{LO}} (p'')\Gamma^{\text{LO}} ( p)~,
\end{multline}
where $\psi^{\text{LO}} (p)$ is the three-nucleon wave-function (\cref{eq_psi_3H} for $^3$H and \cref{eq_psi_3He} for $^3$He) and $D_t^{\text{LO}}(E,p)$ is the dibaryon propagator at LO (\cref{eq:dibaryon_LO}). Since $\mathcal{Z}^{\text{LO}} (k, p)=\lim_{E\rightarrow E_B}\Gamma_t^{\text{LO}} (k)\Gamma_t^{\text{LO}}(p)$, substituting \cref{eq_T_NLO_E} into \cref{eq_B1} yields: 
\begin{equation}\label{eq_B1_a}
\Delta E_B =\psi^{\text{LO}} (p')
\otimes\mathcal{O}^{(1)} (E, p', p'')\otimes \psi^{\text{LO}}=f(\Lambda),
\end{equation} %Combining \cref{eq_B1_a,eq_T_NLO_E}, one gets: 
which is a function of the cutoff $\Lambda$ only, {\it i.e.}, it has no dependence on the momenta $k$ and 
$p$.
%\begin{equation}\label{eq_T_NLO}
%T^{\text{NLO}} (E, k, p)=a (\Lambda)\frac{T^{\text{LO}} (E, k, p)}{E-E_B},
%\end{equation}
%where: 
%\begin{equation}\label{eq_a_Lambda}
%a (\Lambda)=\mathcal{B} (E, p)\otimes \mathcal{O}^{\text{NLO}} (p, p', E)\otimes\mathcal{B} (p', E)
%\end{equation}
%For $^3$H, $\Delta E_B$ is set to zero, where for $^3$He we set $\Delta E_B$ such: 
%\begin{equation}
%E=E_B+\Delta E_B(\Lambda)=E_B^{\text{LO}} (\Lambda)+\Delta E_B(\Lambda)=7.72, \text{ MeV}.
%\end{equation}

\subsection{NLO three-body force}\label{H_NLO_scation}
From \cref{eq_B1_a}, we find that the NLO correction to $^3$H has a
cutoff dependence that needs to be removed
\cite{HAMMER2001353}. Similarly to the LO case, this
$\Lambda $-dependence is removed by adding a term that includes an NLO
correction to the LO \cblack three-body force, $H^{(1)}(\Lambda)$,
such that $T^{(1)}$ becomes:
\begin{multline}\label{eq_stm_T_NLO}
T^{(1)} (E, k, p)=
-T^{\text{LO}} (E, k, p')D^{\text{LO}}(E,p')
\otimes
\Bigl\{
\frac{My^2}{2}\left[K_0 (p', p'', E)+\frac{H(\Lambda)}{\Lambda^2}\right]
\\\otimes
\left[\Delta(E,p'')+\Delta(E,p')\right]\Bigr\}
\otimes
D^{\text{LO}}(E,p'')T^{\text{LO}} (E, p'', p)
\\-T^{\text{LO}} (E, k, p')D^{\text{LO}}(E,p')
\otimes\frac{H^{(1)}(\Lambda)}{\Lambda^2}{My^2}\otimes
D^{\text{LO}}(E,p'')T^{\text{LO}} (E, p'', p)~.
\end{multline}
Using the STM equation (\cref{stm_T}) yields:
\begin{multline}
T^{(1)} (E, k, p) =
-\frac{\rho}{M y^2}\int\frac{p'^2dp'}{\pi}T^{\text{LO}} (E, k, p') \times\frac{3 p^2/4-E M-1/a_2^2}{\left(\sqrt{{3 p'^2/4}-E M}-1/a_2 \right)^2}T^{\text{LO}} (E, p', p)\\-
T^{\text{LO}} (E, k, p')D^{\text{LO}}(E,p')
\otimes\frac{H^{(1)}(\Lambda)}{\Lambda^2}{My^2}\otimes
D^{\text{LO}}(E,p'')\\\ \times T^{\text{LO}} (E, p'', p)~.
\end{multline} 
Using \cref{eq_stm_T_NLO}, $\Delta E_B$ is now given
by: %(see Fig.~\ref{fig_NLO_matrix_element}):
%\cred
\begin{multline}\label{eq_delta_E_full}
\Delta E_B(\Lambda)=
-\frac{1}{2}My^2\psi^{\text{LO}} (p')\otimes\bigg\{\left[{K}_0 (p', p'', E)+\frac{H(\Lambda)}{\Lambda^2}\right]\\ \times
\left[\Delta(E,p'')+\Delta(E,p')\right]\bigg\}
\otimes\psi^{\text{LO}} (p'')-
M y^2 \psi^{\text{LO}} (p')
\otimes\frac{H^{(1)}(\Lambda)}{\Lambda^2}\otimes
\psi^{\text{LO}} (p'')~.
\end{multline} 

\begin{figure}[h!]
	\begin{center}
		\includegraphics[width=0.7\linewidth]{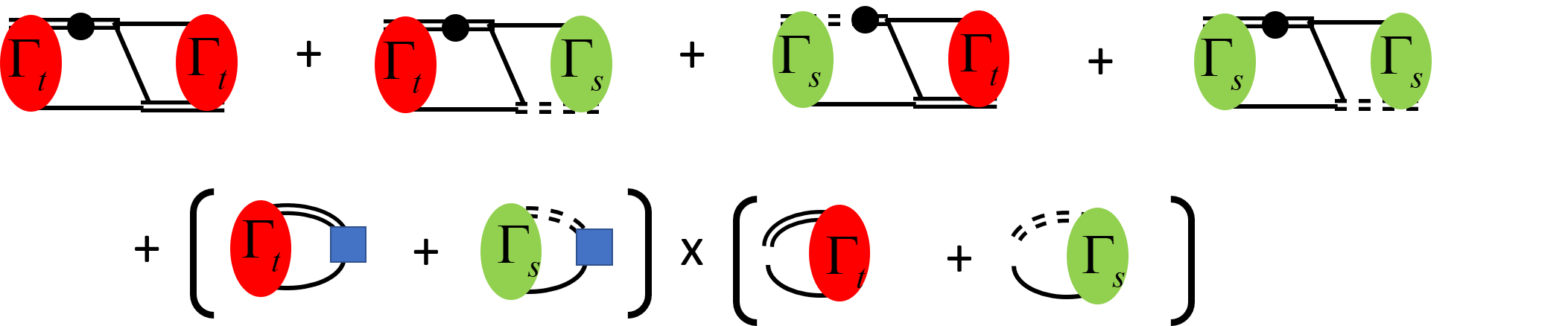}\\
		\caption{\footnotesize{The NLO correction for $^3$H binding energy. The double lines
				are propagators of the two intermediate auxiliary fields,
				$D_t$ (solid) and $D_t^{np}$ (dashed). The red bubbles ($\Gamma_t$) represent the triplet channel T=0,
				S=1, the green bubbles represent ($\Gamma_s$) the singlet channel
				T=1, S=0. The black circles denote the NLO correction to the
				dibaryon propagator, while the blue squares denote
				the NLO correction to the three-body force
				($H^{(1)}(\Lambda)$).}}\label{3H_NLO}
	\end{center}
\end{figure}

Let us now consider the three-nucleon case and set
\cref{eq_delta_E_full} to zero for $^3$H \cite{konig2}, with
\begin{equation}
\mathcal{B}(p)= \mathcal{B}^{^3\text{H}}(p)
\end{equation}
and
\begin{equation}
{\psi}(p)={\psi^{_{^3\text{H}}}}(p)=\left(
\begin{array}{c}
\psi_t^{{^3\text{H}}}( p)\\
\psi_s^{{^3\text{H}}}(p)
\end{array}\right)~.
\end{equation} 
The NLO correction to the $^3$H binding energy is given by:
\begin{equation}
\Delta E_B(\Lambda)=
\sum_{\mu,\nu}\psi^{\text{LO}} (p)\otimes
\mathcal{O}^{(1)}_{\mu\nu}(E_{^3\text{H}},p,p')\otimes
\psi^{\text{LO}} (p')~,
\end{equation} 
with
\begin{multline}\label{eq:O:NLO}
\mathcal{O}^{(1)}_{\mu\nu}(E_{^3\text{H}},p,p')=My_{\mu}y_{\nu}
\Biggl\{\frac{1}{2}\left[a_{\mu\nu}{K}_0 (p, p', E_{^3\text{H}})+b_{\mu\nu}\frac{H(\Lambda)}{\Lambda^2}\right]\\\ \times \Biggl[
\Delta_\mu(E_{^3\text{H}}, p)+\Delta_\nu(E_{^3\text{H}}, p')\Biggr]
+b_{\mu\nu}\frac{H^{(1)}(\Lambda)}{\Lambda^2} \Biggr\}~.
\end{multline}
Where the diagrammatic representation of $\Delta E_B(\Lambda)$ for the case of $^3$H, is shown in Fig. \ref{3H_NLO}. 

Therefore, we find that the NLO three-body force has the form: 
\begin{multline}\label{H_NLO}
-\frac{H^{(1)}(\Lambda)}{\Lambda^2}=
M\sum_{\mu,\nu=t,s}
{\psi_\mu^{^3\text{H}}(p)}\otimes
\Biggl\{\frac{1}{2}y_{\mu}y_{\nu}\left[a_{\mu\nu}{K}_0 (p, p', E_{^3\text{H}})+b_{\mu\nu}\frac{H(\Lambda)}{\Lambda^2}\right] \\\times \Biggl[
\Delta_\mu(E_{^3\text{H}}, p)+\Delta_\nu(E_{^3\text{H}}, p')\Biggr]
\Biggr\}
\otimes{\psi_\nu^{^3\text{H}}( p')} \times 
\left[M \sum_{\mu,\nu=t,s}{y_\mu y_\nu}
{\psi_\mu^{^3\text{H}}(p)}\otimes
b_{\mu\nu}
\otimes{\psi_\nu^{^3\text{H}}( p')}\right]^{-1}~.
\end{multline}
Using the fact that:
\begin{equation}
\Gamma_{\nu}^{^3\text{H}}(p')=M\sum\limits_{\mu=t,s}
y_{\mu}y_{\nu}{\psi_\mu^{^3\text{H}}(p)}\otimes
\left[a_{\mu\nu}{K}_0 (p, p', E_{^3\text{H}})+b_{\mu\nu}\frac{H(\Lambda)}{\Lambda^2}\right]
\end{equation}
and
\begin{equation}
\Gamma_{\mu}^{^3\text{H}}(p)=M\sum\limits_{\nu=t,s}
y_{\mu}y_{\nu}\left[a_{\mu\nu}{K}_0 (p, p', E_{^3\text{H}})+b_{\mu\nu}\frac{H(\Lambda)}{\Lambda^2}\right]\otimes{\psi_\nu^{^3\text{H}}(p')}~,
\end{equation} 
Equation~(\ref{H_NLO}) becomes:
\begin{multline}
-\frac{H^{(1)}[\Lambda]}{\Lambda^2}=\frac{1}{2}M\sum_{\mu=t,s}
{\psi_\mu^{^3\text{H}}(p)}
\otimes\Bigg\{
\Biggl[
\frac{\Delta_\mu(E_{^3\text{H}}, p)}{D_\mu(E_{^3\text{H}}, p)}+\frac{\Delta_\mu(E_{^3\text{H}}, p')}{D_\mu(E_{^3\text{H}}, p')}\Biggr]
2\pi^2\frac{\delta(p-p')}{p^2}
\Bigg\}\otimes{\psi_\mu^{^3\text{H}}( p')}\\\ \times 
\left[ M \sum_{\mu,\nu=t,s}{y_\mu y_\nu}
{\psi_\mu^{^3\text{H}}(p)}\otimes
b_{\mu\nu}
\otimes{\psi_\nu^{^3\text{H}}(E_{^3\text{H}}, p')}\right]^{-1}~.
\end{multline}
A comparison of the analytical \cite{H_NLO} and the numerical results
of the NLO three-body force, $H^{(1)}(\Lambda)$ of \cref{H_NLO},
reveals that they are in good agreement, as shown in
Fig.~\ref{fig_H_NLO}. The diagrammatic representation of
$\Delta E_B(\Lambda)$ is given in Appendix~D.

\begin{figure}[ht]
	\begin{center}
		\includegraphics[width=.75\linewidth]{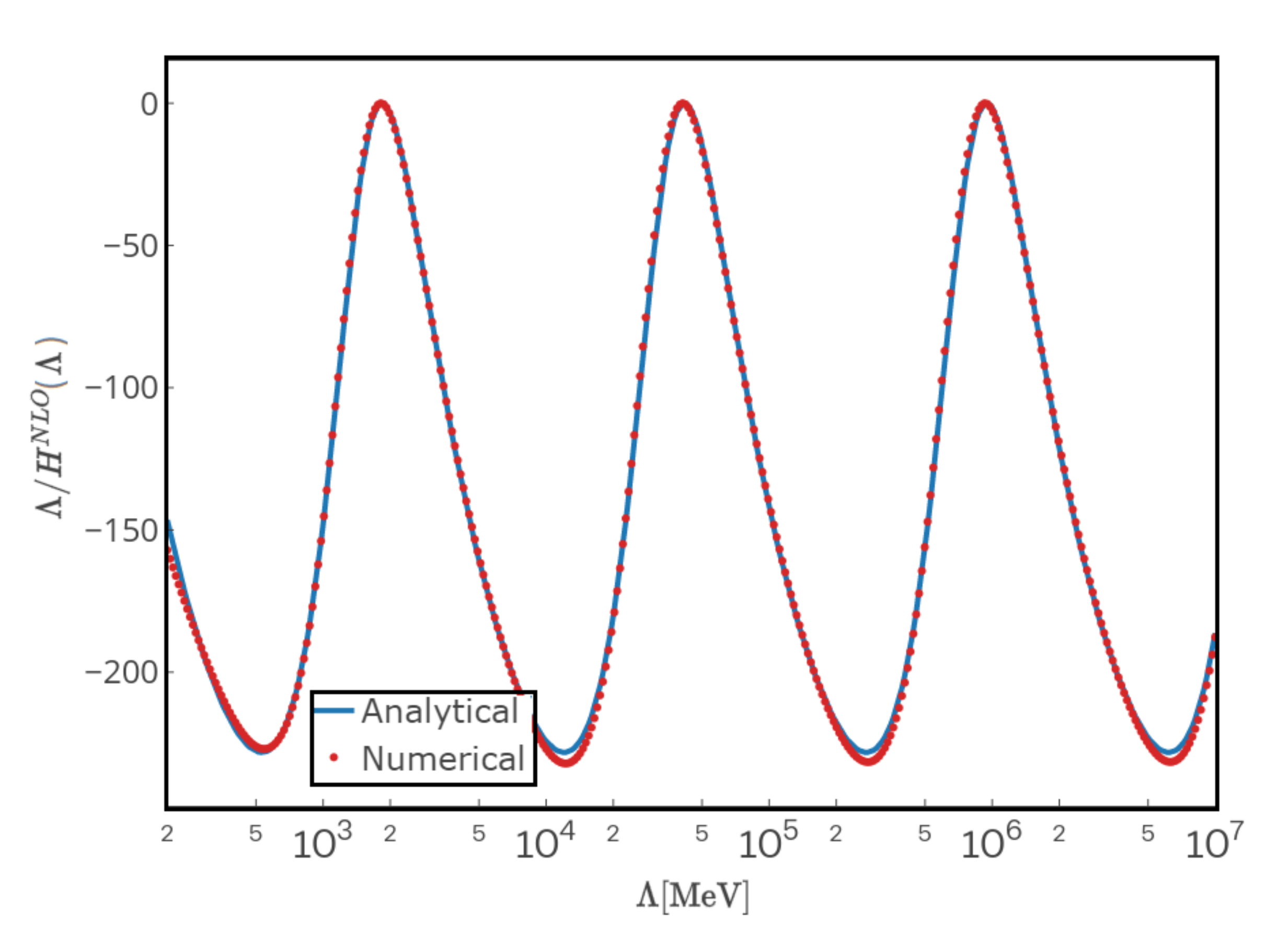}
		\caption{\footnotesize{The three-body force,
				$H(\Lambda)$, at NLO as a function of the cutoff $\Lambda$ in
				MeV for $^3$H. The solid curve is the analytical expression for
				$H(\Lambda)$ taken from \cite{H_NLO}, while the dots are the
				numerical results based on \cref{H_NLO}.}}\label{fig_H_NLO}
	\end{center}
\end{figure}

\cblack
\begin{comment}
%	\begin{figure}[h!]
\centering \includegraphics[width=0.6\linewidth]{B1_a.png}
\caption{\footnotesize{The diagrammatic form of the NLO correction to the binding energy - $\Delta E_B$. The red bubble is the amplitude $\Gamma$ at LO and the double lines are the propagators of the two dibaryon fields. The black circles are the effective range
insertion and the blue square is the NLO three-body force.}}
\label{fig_NLO_matrix_element}
\end{figure}
\end{document}
\end{comment}

\subsection{$^3$He - correction to the three-body force and binding energy}
\begin{figure}[h!]
	\begin{center}
		\includegraphics[width=0.9\linewidth]{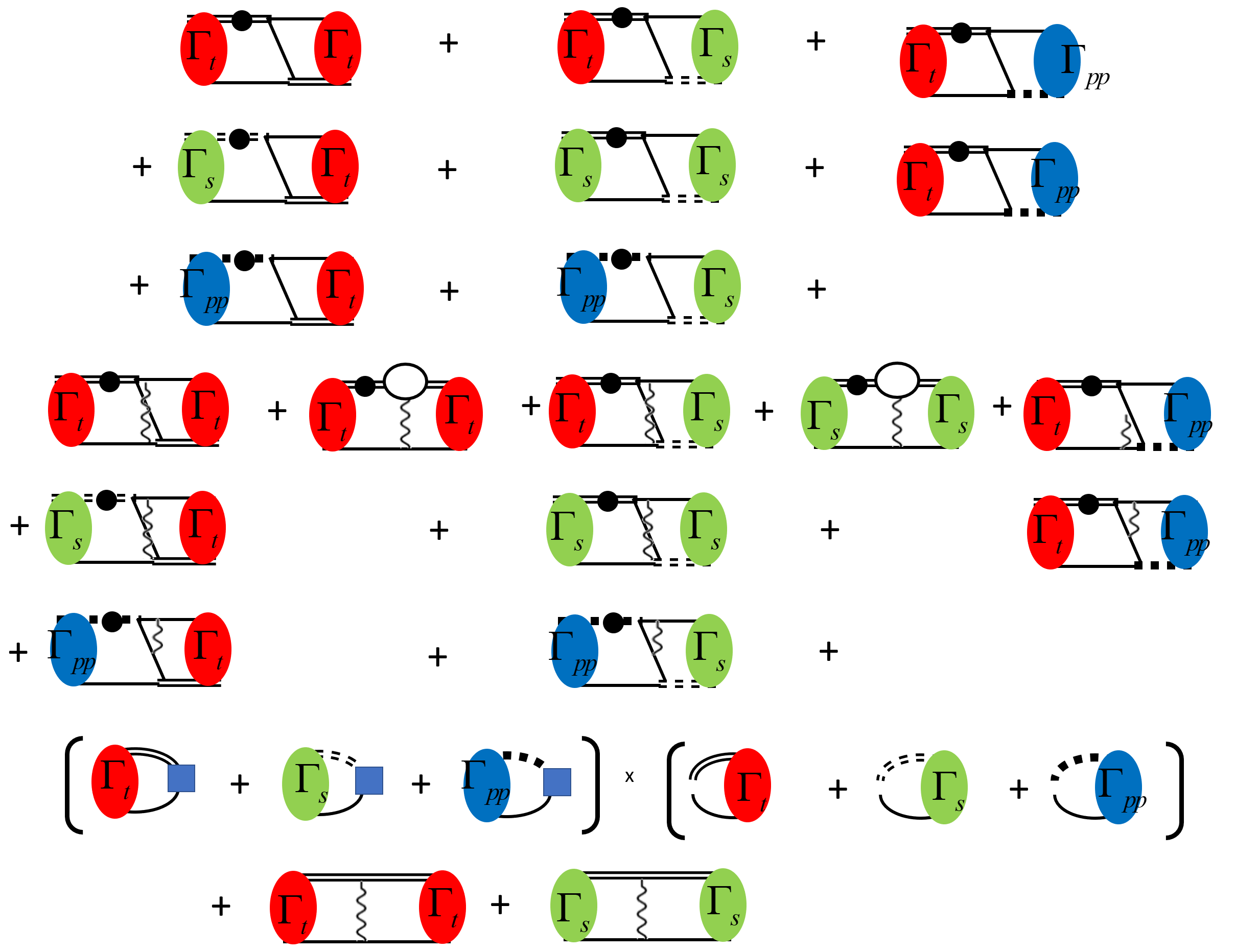}\\
		\caption{\footnotesize{ The NLO correction for $^3$He binding energy. The double lines
				are propagators of the two intermediate auxiliary fields,
				$D_t$ (solid) and $D_t^{np}$ (dashed) and $D^{pp}$ (dotted). The red bubbles ($\Gamma_t$) represent the triplet channel (T=0,
				S=1), the green bubbles represent ($\Gamma_s$) the singlet channel
				(T=1, S=0) with an $np$ dibaryon, while the blue bubbles ($\Gamma_{pp}$)
				represent the singlet channel (T=1, S=0) with $pp$
				dibaryon. The black circles denote the NLO correction to the
				dibaryon propagator, while the blue squares denote
				the NLO correction to the three-body force
				($H^{(1)}(\Lambda)+H^{(\alpha)}(\Lambda)$).}}\label{helium_NLO}
	\end{center}
\end{figure}

The prediction of $H^{(1)}(\Lambda)$ for $^3$H (see subsection~\ref{H_NLO_scation}) enables us to calculate the NLO corrections to $^3$He as well. Similarly to the LO calculation, we are using the three-body force to determine the NLO correction to $^3$He binding energy, by assuming that $H^{(1)}(\Lambda)$ has no isospin dependence \cite{konig2}.
The diagrammatic representation of $\Delta E_B(\Lambda)$ for the case of $^3$He, is shown in Fig. \ref{helium_NLO}. 
%\subsubsection{NLO binding energy and NLO three-body force}
\begin{comment}
For $^3$He, the NLO corrections to the one-nucleon exchange matrix are:
\begin{equation}
\label{eq_K_1_S_3He}
\begin{split}
&K_1^S(p', 
p, E)=-\frac{M}{2}K_0(p', p, E)
\left(
\begin{array}{ccc}
-y_t^2&3y_ty_s&3y_ty_s\\
y_ty_s&y_s^2&-y_s^2\\
2y_ty_s&-2y_s^2&0\\
\end{array}\right)
\left(
\begin{array}{ccc}
\Delta_t(E, p)&0&0\\0& \Delta_s(E, p)&0\\0&0&\Delta_{pp}(E, p)\\
\end{array}\right), \\
\end{split}
\end{equation}
and
\begin{equation}\label{eq_K_1_C}
\begin{split}
K_1^C(p', p, E)=&\left(
\begin{array}{ccc}
-y_t^2\left(2K_C^a+K_C^b\right) & 3y_ty_s\left(K_C^b\right)& 3y_ty_s \left(K_C^c\right) \\
y_ty_s \left(K_C^b\right) & y_s^2 \left(-K_C^a+K_C^b\right) & -y_s^2 \left(K_C^c\right) \\
2y_ty_s \left(K_C^d\right) & -2y_s^2 \left(K_C^d\right) & 0 \\
\end{array}
\right)\cdot\\
&\left(
\begin{array}{ccc}
\Delta_t(E, p)&0&0\\
0&\Delta_s(E, p)&0\\
0&0&\Delta_{pp}(E, p)\\
\end{array}\right), 
\end{split}
\end{equation}
where $K_C^a$, $K_C^b$, $K_C^c$ and $K_C^d$ are defined in \cref{K_C_a,K_C_b,K_C_c}.
\end{comment}
Similarly to $^3$H, the correction to the binding energy of
$^3$He is a function of $\Lambda$ only \cite{konig2}: 
\begin{equation}
\Delta E_B(\Lambda)=\sum\limits_{\mu,\nu=t,s,pp}{\psi^{^3\text{He}}_\mu(p)}\otimes \mathcal{O}^{(1)}_{\mu\nu}(E_{^3\text{He}},p,p')\otimes{\psi^{^3\text{He}}_\nu(p')}~,
\end{equation}
\begin{equation}\label{B1_NLO}
\begin{split}
\mathcal{O}^{(1)}_{\mu\nu}(E_{^3\text{He}},p,p')=&My_\mu y_\nu\Biggl\{ \frac{1}{2}\left[a'_{\mu\nu}K_0(p,p',E_{^3\text{He}})+a'_{\mu\nu}K_{\mu\nu}^C(p,p',E_{^3\text{He}})+b'_{\mu\nu} \frac{2H(\Lambda)}{\Lambda^2}\right]\\
&\times
\left[\Delta_{\nu}(E_{^3\text{He}},p)+\Delta_{\nu}(E_{^3\text{He}},p')\right]+ b'_{\mu\nu}\frac{H^{(1)}(\Lambda)}{\Lambda^2}\Biggr\}\\+
&\alpha Q_0
\left(\frac{p^2+p'^2+\lambda^2}{2pp'}\right)\times(\delta_{\mu,t}\delta_{\nu,t}+3\delta_{\mu,s}\delta_{\nu,s})~,
\end{split}
\end{equation}
where $\alpha Q_0
\left(\frac{p^2+p'^2+\lambda^2}{2pp'}\right)$ originates from diagram (f) in Fig.~\ref{Coulomb_correction}.

\begin{comment}
\begin{multline}\label{B1_NLO}
\Delta E_B=
\left(
\begin{array}{c}
\psi_T(p, E)\\
\psi_S(p, E)\\
\psi_P(p, E)\\
\end{array}
\right)^T\otimes \left[K_1^S(p, p', E)+K_1^C(p, p', E)\right]\otimes
\left(
\begin{array}{c}
\psi_T(p', E)\\
\psi_S(p', E)\\
\psi_P(p', E)\\
\end{array}\right)+\\
\frac{4}{M}\frac{H^{\text{NLO}}(\Lambda)}{\pi^2\Lambda^2}\left[\frac{1}{y_t}\int_0^\Lambda\Gamma_T(p, E)\delta_t(E, p)
p^2dq+\right.\\
\left.\frac{1}{y_s}\int_0^\Lambda\Gamma_S(p, E)\delta_s(E, p)p^2dq+\frac{1}{y_s}\int_0^\Lambda\Gamma_P(p, E)\delta_P(E, p)p^2dq\right]^2+\\
\frac{4}{M}\left\{-\frac{\alpha\rho_tM}{4\pi y_t^2}\int_0^\Lambda qdq\int_0^\Lambda \frac{p'^2dp'2\pi^2}{p'2\pi^2}\Gamma_T(p, E)\delta_t(E, p)Q_0
\left(\frac{p^2+p'^2+\lambda^2}{-2pp'}\right)\Gamma_T(p', E)\delta_t(E, p')+\right.\\
\left.-\frac{3\alpha\rho_sM}{4\pi y_s^2}\int_0^\Lambda qdq\int_0^\Lambda
p'dp'\Gamma_S(p, E)\delta_s(E, p)Q_0
\left(\frac{p^2+p'^2+\lambda^2}{-2pp'}\right)\Gamma_S(p', E)\delta_s(E, p')\right\}.\end{multline}
\cblack
\end{comment}
In contrast to $^3$H and to $^3$He at LO, the numerical result of \cref{B1_NLO} reveals that $\Delta E_B$ for $^3$He diverges with the cutoff $\Lambda$ (see Ref.~\cite{konig2}) and does not vanish. This contradicts the assumption that the addition of an isospin independent $H^{\text{NLO}}(\Lambda)$ to $T^{\text{NLO}}$ removes the cutoff dependence of $\Delta E_B$ for both $^3$H and $^3$He. The solution to this issue is obtained by defining a different three-body force for $^3$He, such that: 
\begin{equation}
E_{^3\text{He}}^{\text{NLO}}=E_{^3\text{He}}^{\text{LO}}(\Lambda)+\Delta E_B(\Lambda)=7.72\mev,
\end{equation}
which equals to the experimental binding energy of $^3$He, where $E_{^3\text{He}}^{\text{LO}}(\Lambda)$ is shown in Fig.~\ref{fig_helium_energy}.

Accordingly, the new three-body force, $H^\alpha(\Lambda)$, is defined and can be calculated numerically as:
%\cred
\begin{multline}\label{H_NLO_3He}
\frac{H^\alpha(\Lambda)}{\Lambda^2}=\Biggl [\frac{7.72\mev-E_{^3\text{He}}^{\text{LO}}(\Lambda)}{\Lambda^2}-\sum\limits_{\mu,\nu=t,s,pp}{\psi^{^3\text{He}}_\mu(p)}\otimes \mathcal{O}^{(1)}_{\mu\nu}(E_{^3\text{He}},p,p')\otimes{\psi^{^3\text{He}}_\nu(p')}\Biggr]\
\\
\times\Biggl[\sum\limits_{\mu,\nu=t,s,pp}{\psi^{^3\text{He}}_\mu(p)}\otimes b'_{\mu\nu}(E_{^3\text{He}},p,p')\otimes{\psi^{^3\text{He}}_\nu(p')}\Biggr]^{-1}
\end{multline}
\cblack
while its analytical form is given in Refs.~\cite{H_NLO, konig2}.

The diagrammatic representation of $\Delta E_B(\Lambda)$ for the case of $^3$He, is shown in Fig. \ref{helium_NLO}.   

\section{NLO corrections to the three-body wave function} %and 
The \textbf{full} (non-perturbative in $\alpha$) Faddeev equations for
$^3$He (at LO in ERE) consist of two parts - the strong part and the
Coulomb interaction part:
\begin{multline}\label{eq_compact2}
\Gamma^{{^3\text{He}}}_\mu(p)=\sum\limits_{\nu=t,s}My_\mu y_\nu\ \times \left[{a'_{\mu\nu}K_0(p',p,E_{^3\text{He}})+b'_{\mu\nu}\frac{H(\Lambda)}{\Lambda^2}}+{c'_{\mu\nu}K^C_{\mu\nu}(p',p,E_{^3\text{He}})}\right]\\\otimes D_\nu(E_{^3\text{He}},p')\Gamma^{^3\text{{He}}}_\nu(E_{^3\text{He}},p')\\+My_\mu y_s\left[a'_{\mu pp}{K_0(p',p,E_{^3\text{He}})+b'_{\mu pp}\frac{H(\Lambda)}{\Lambda^2}}\right]\otimes D_{pp}(E_{^3\text{He}},p')\Gamma^{^3\text{{He}}}_{pp}(p').
\end{multline}

Faddeev equations for $^3$H at LO are:
\begin{multline}
\Gamma^{^3\text{{H}}}_\mu(p)=\sum\limits_{\nu=t,s}My_\mu y_\nu
\bigl[a_{\mu\nu}K_0(p',p,E_{^3\text{H}})
+b_{\mu\nu}\frac{H(\Lambda)}{\Lambda^2}\bigr]
\otimes D_\nu(E_{^3\text{H}},p')\Gamma^{^3\text{{H}}}_\nu(p')~.
\end{multline}

Using \cref{eq_compact2}, the $^3$He-$^3$H binding energy difference,
defined in subsection~\ref{energy_shift}, is a function of the Coulomb
part of \cref{eq_compact2}, using $^3$H wave-functions and assuming
that $\psi_s(E,p)=\psi_{nn}(E,p)=\psi_{np}(E,p)=\psi(E,p)_{pp}$.

This implies that the $^3$He-$^3$H binding energy difference can be
written as a first-order perturbation in $\alpha$:
\begin{equation}\label{eq_delta_EC}
\Delta E(\Lambda)=\sum_{\mu,\nu}\psi_\mu^0(p)\otimes\mathcal{O}^C_{\mu\nu}(E,p,p')\otimes{\psi_\nu^0}(p'),
\end{equation}
where $\psi_{\mu,\nu}^0=\psi^{^3{H}}_{\mu,\nu}$ is the three-nucleon wave-function without
the Coulomb interaction, and $\mathcal{O}^C_{\mu,\nu}(E,p,p')$ are the
Coulomb parts of \cref{eq_compact2}:

\begin{equation}
\mathcal{O}^C_{\mu\nu}(E,p,p')=c_{\mu\nu}K^C_{\mu\nu}(p,p,E)
+\left[a_{\mu\nu}K_0(p,p',E)+b_{\mu\nu}\frac{H(\Lambda)}{\Lambda^2}\right]
\times\left[\frac{D_{pp}(E,p)-D_s(E,p)}{D_s(E,p)^{2}}\right]\delta_{\nu,s}.
\end{equation}

The NLO correction to the binding energy can also be written as a first-order perturbation in $Q/\Lambda_{\rm cut}$:
\begin{multline}\label{eq_delta_E}
\Delta E_B(\Lambda)=\sum_{\mu,\nu}{\psi_\mu^{\text{LO}}}(p)\otimes\mathcal{O}^{\text{(1)}}_{\mu\nu}(E,p,p')\otimes\psi^{\text{LO}}_\nu(p')\\
=Z^{\text{LO}}\sum_{\mu,\nu}\left[\Gamma_{\mu}^{\text{LO}}(p)D_{\mu}^{\text{LO}}(E,p)\right]\otimes\mathcal{O}^{\text{(1)}}_{\mu\nu}(E,p,p')\otimes
\left[D_{\nu}^{\text{LO}}(E,p')\Gamma_{\nu}^{\text{LO}}(p')\right]~,
\end{multline}
where $\mu,\nu$ are the different dibaryon channels and
$\mathcal{O}^{(1)}_{\mu\nu}(E,p,p')$ (defined in \cref{eq:O:NLO}) is the NLO correction to the
binding energy in terms of the different dibaryon channels. Since
\cref{eq_delta_E,eq_delta_EC} have the same form, we can define the
homogeneous scattering amplitude up to NLO such that for $^3$H:

\begin{multline}\label{eq_full_Gamma_3H}
\Gamma_{\mu}^{\text{NLO}}(p)={\Gamma_{\mu}^{\text{LO}}}(p)+{\Gamma_{\mu}^{(1)}}(p)=\\\sum_{\nu=t,s}
\left[a_{\mu\nu}K_0(p,p',E_{^3\text{H}})+b_{\mu\nu}\frac{H(\Lambda)}{\Lambda^2}+\mathcal{O}^{(1)}_{\mu\nu}(E_{^3\text{H}},p,p')\right]
\otimes D^{\text{LO}}_\nu(E_{^3\text{H}},p')\Gamma^{\text{LO}}_\nu(p')~,
\end{multline}
and for $^3$He,
\begin{multline}\label{eq_full_Gamma_3He}
\Gamma_{\mu}^{\text{NLO}}(p)={\Gamma_{\mu}^{\text{LO}}}(p)+{\Gamma_{\mu}^{(1)}}(E_{^3\text{He}},p)\\=\sum_{\nu=t,s,pp}
\Bigl\{a'_{\mu\nu}\left [K_0(p,p',E_{^3\text{He}})+K^C_{\mu\nu}(p,p',E_{^3\text{He}})\right]+b'_{\mu\nu}\frac{H(\Lambda)}{\Lambda^2}+\mathcal{O}^{(1)}_{\mu\nu}(E_{^3\text{He}},p,p')\Bigr\}
\otimes D^{\text{LO}}_\nu(E_{^3\text{He}},p')\Gamma^{\text{LO}}_\nu(p')~,
\end{multline}
which are no longer Bethe-Salpeter equations, therefore, the
Bethe-Salpeter normalization condition is not valid.

Having defined the NLO correction for the bound-state scattering
amplitude, $\Gamma_{\mu}$, it is now possible to define the general
form of a three-nucleon matrix element (such as an electroweak (EW)
interaction) up to NLO: %(see Fig.~\ref{fig_NLO_matrix_element_full}):
\begin{equation}\label{eq:EW_element}
\begin{split}
&\langle\mathcal{O}^{\text{LO}}_{\text{EW}}\rangle +\langle\mathcal{O}^{(1)}_{\text{EW}}
\rangle=\sum_{\mu,\nu}\underbrace{\langle{\psi} ^{\text{LO}}_\mu |\mathcal{O}^{\text{LO}}_{\mu\nu}|
	\psi ^{\text{LO}}_{\nu}\rangle }_{\mathcal{O}^{\text{LO}}_{\text{EW}}}+
\underbrace{\langle\psi ^{\text{LO}}_\mu|\mathcal{O}^{(1)}_{\mu\nu} |
	\psi ^{\text{LO}}_{\nu}\rangle+ \langle\psi ^{(1)}_{\mu} |\mathcal{O}
	^{\text{LO}}_{\mu\nu} | \psi ^{\text{LO}}_\nu\rangle+\langle\psi ^{\text{LO}}_\mu
	|\mathcal{O} ^{\text{LO}}_{\mu\nu}|\psi ^{(1)}_{\nu}\rangle}_{\mathcal{O}
	^{(1)}_{\text{EW}}},
\end{split}
\end{equation}
where: 
\begin{multline}
\psi^{(1)}_\mu (p)=\sqrt{Z_{1}}\Bigl\{\left[D^{\text{NLO}}_\mu (E, p)-D^{\text{LO}}_\mu (E, p)\right]{\Gamma}_\mu^{\text{LO}} (p)+
D^{\text{LO}}_\mu (E, p){\Gamma}_\mu^{(1)} (p)\Bigr\},\label{eq_psi_NLO}
\end{multline} 
where $Z_{1}$ is the NLO correction to the three-nucleon
normalization, which is determined by the A=3 form factor, as will be
discussed next.

\begin{comment}
as shown in Fig.~\ref{fig_NLO_matrix_element_full}.
\begin{figure}[h!]
\centerline{
\includegraphics[width=0.85\linewidth]{full_NLO_operator.png}}
\caption{\footnotesize{The diagrammatic form of the full NLO matrix
element.}}
\label{fig_NLO_matrix_element_full}
\end{figure}
\end{comment}
\section{The NLO normalization}
%\cblack
Charge conservation puts strong constraints on the zero-momentum limit
of the electric form factor. In this subsection, we, therefore, want to
relate the three-nucleon charge form factor to the three-nucleon normalization procedure discussed here. Following Ref.~\cite{KSW_c},
we expand the charge form factor of the deuteron up to NLO
\begin{equation}
F_C=F_C^{\text{LO}}+F_C^{(1)},
\end{equation}
where for the deuteron:
\begin{equation}
F_C^{\text{LO}}(0)=Z_d^{\text{LO}}\lim_{q\rightarrow 0}\frac{4\gamma_t}{q}\arctan\left(\frac{q}{4\gamma_t}\right)=1. 
\end{equation}
Up to NLO, one finds that:
\begin{equation}\label{eq_FC_NLO}
F_C(0)=Z_d^{\text{NLO}}\lim_{q\rightarrow 0}\frac{4\gamma_t}{q}\arctan\left(\frac{q}{4\gamma_t}\right)-Z_d^{\text{LO}}\lim_{q\rightarrow 0}\gamma_t\rho_t\frac{4\gamma_t}{q}\arctan\left(\frac{q}{4\gamma_t}\right)~.
\end{equation}
We can rewrite this as:
\begin{equation}
\left(Z_d^{\text{LO}}+Z_d^{(1)}\right)F_C^{(0)}-Z_d^{\text{LO}}\gamma_t\rho_tF_C^{(0)}=1
\rightarrow Z_d^{(1)}-\gamma_t\rho_t=0. 
\end{equation}
From \cref{eq_FC_NLO}, it is easy to show that up to NLO:
\begin{equation}\label{eq_Zd_NLO}
Z_d^{\text{NLO}}-\gamma_t\rho_t=1\rightarrow Z_d^{\text{NLO}}=1+\gamma_t\rho_t,
\end{equation}
which equals 1.408, %(1.69) for the ERE (Z)- parameterization, 
as discussed in Chapter~\ref{formalism}.

Similarly, the $A=3$ NLO normalization is obtained from the $^3$H and
$^3$He form factor up to NLO
\cite{Vanasse:2015fph,Vanasse:2017kgh,KSW_c}. Based on
\cref{eq_general_operator_reduced}, it is easy to show that at LO, the
$A=3$ form factor is given by:
%\cred
\begin{equation}\label{eq_form_factor}
F_C^{(0)}(q)=
\sum\limits_{\mu,\nu}
{\psi^{\text{LO}}_\mu(E',p')}\otimes \mathcal{O}_{\mu\nu}^{FC(\text{1B})}(q) \otimes{\psi^{\text{LO}}_\nu(E,p)}~,
\end{equation}
where:
\begin{equation}
\mathcal{O}_{\mu\nu}^{FC(\text{1B})}(q)=
y_\mu y_\nu\Bigl\{{d'^{ii}_{\mu\nu}} \hat{\mathcal{I}}(q_0,q)+
{a'^{ii}_{\mu\nu}}\left[\hat{\mathcal{K}}(q_0,q)+{\hat{\mathcal{K}}^C_{\mu\nu}}(q_0,q)\right]\Bigr\}~,
\end{equation}
and $d'^{ii}_{\mu\nu},a'^{ii}_{\mu\nu}$ were defined in 
\cref{eq_cases1,eq_cases2}.

Based on \cref{eq:EW_element}, up to NLO, $F_C(0)$ is given by:
\begin{multline}\label{fc_NLO}
F_C^{\text{NLO}}(0)=F_C^{(0)}(0)+F_C^{(1)}(0)=
\sum\limits_{\mu,\nu}
{\psi^{\text{LO}}_\mu(E,p')}\otimes \mathcal{O}_{\mu\nu}^{FC(\text{1B})}(0) \otimes{\psi^{\text{LO}}_\nu(E,p)}\\+
\frac{1}{2}\Bigl[ {\psi^{(1)}_\mu(E,p')}\otimes \mathcal{O}_{\mu\nu}^{FC(\text{1B})}(0)\otimes{\psi^{\text{LO}}_\nu(E,p)}+
{\psi^{\text{LO}}_\mu(E,p')}\otimes \mathcal{O}_{\mu\nu}^{FC(\text{1B})}{(0)}\otimes{\psi^{(1)}_\nu(E,p)}\Bigr]\\+
{\psi^{\text{LO}}_\mu(E,p')}\otimes \mathcal{O}_{\mu\nu}^{FC(\text{2B})}(0) \otimes{\psi^{\text{LO}}_\nu(E,p)}=F_C^{(0)}(0)=1~,
\end{multline}
where: 
\begin{equation}\label{eq:1b:fc}
\mathcal{O}_{\mu\nu}^{FC(\text{1B})}(0)=\mathcal{O}_{\mu\nu}^{\text{norm}}(E_i)~,
\end{equation}
and $i=^3\text{H},\,^3\text{He}$.

Since the two-body term is a result of the $A_0$ photons, which couple
only the triplet channel, the two-body term can be written as
\cite{Vanasse:2017kgh}:
\begin{equation}\label{eq:2b:fc}
\mathcal{O}_{\mu\nu}^{FC(\text{2B})}(0)=\frac{2\pi^2}{p'^2}\delta(p-p')\delta_{\mu,t}\delta_{\nu,t}~.
\end{equation} 
By substituting \cref{eq:2b:fc,eq:1b:fc} in \cref{fc_NLO}, one finds
that the NLO correction to the triton form factor, $F_C^{(1)}(0)$, is
given by: \cblack
\begin{multline}\label{eq_norm_3H_NLO}
F_C^{(1)}(0)=\frac{1}{2}\sum\limits_{\nu=t,s} \Bigg\{
{\psi_{\mu}^{(1)}(p)}\otimes
\mathcal{O}_{\mu\nu}^{\text{norm}}(E_{^3\text{H}})
\otimes {\psi^{\text{LO}}_{\nu}(p')}+
{\psi^{\text{LO}}_\mu(p)}\otimes
\mathcal{O}_{\mu\nu}^{\text{norm}}(E_{^3\text{H}})\otimes {\psi^{(1)}_{\nu}(p')}\Bigg\}
\\
-\frac{2}{3}
{\psi^{\text{LO}}_t(p)}\otimes
\frac{2\pi^2}{p'^2}\delta(p-p')\otimes {\psi^{\text{LO}}_{t}(p')}=0~,
\end{multline}
and similarly for $^3$He:
\begin{multline}\label{eq_norm_3He_NLO}
F_C^{(1)}(0)=\frac{1}{2}\sum\limits_{\nu=t,s} \Bigg\{
{\psi_{\mu}^{(1)}(p)}\otimes
\mathcal{O}_{\mu\nu}^{\text{norm}}(E_{^3\text{He}})
\otimes {\psi^{\text{LO}}_{\nu}(p')}+
{\psi^{\text{LO}}_\mu(p)}\otimes
\mathcal{O}_{\mu\nu}^{\text{norm}}(E_{^3\text{He}})\otimes {\psi^{(1)}_{\nu}(p')}\Bigg\}
\\
-\frac{2}{3}
{\psi^{\text{LO}}_t(p)}\otimes
\frac{2\pi^2}{p'^2}\delta(p-p')\otimes {\psi^{\text{LO}}_{t}(p')}=0~,
\end{multline}
where for both $^3$H and $^3$He, $\psi_{\mu}^{(1)}(p)$ is the
\textbf{normalized} NLO correction to the three-nucleon wave-function
\footnote{Note that by defining \cref{eq_norm_3H_NLO,eq_norm_3He_NLO}
	to be equal to 0, we are consistent with Refs.~
	\cite{Vanasse:2015fph,Vanasse:2017kgh} in which
	$F_1(0)=F^{\text{LO}}(0)+F^{\text{NLO}}(0)=1$, where $F_1(0)$ is the
	three-nucleon triton charge form factor up to NLO.}.
\chapter{$A=2,3$ Electromagnetic reactions}\label{Magnetic}
In this chapter, we present our calculations for low-energy electromagnetic reactions, which will be used later (chapter 10) to study the consistency of \pilesseft in the transition from $A=2$ to $A=3$ nuclei and vice-versa. These reactions are essential for understanding the nuclear structure and the dynamics of light nuclei, and therefore a comparison between \pilesseft prediction and available experimental data can confirm the validity of \pilesseft for both physical and non-physical pion mass.

In this thesis, we focus on four well-measured low-energy electromagnetic ``$M_1$'' reactions, {\it i.e.,} the magnetic moments of the bound nuclei $\langle\hat{\mu}_{d}\rangle$, $\langle\hat{\mu}_{^3\text{H}}\rangle$, $\langle\hat{\mu}_{^3\text{He}}\rangle$ \cite{3He_3H_data,mu_d_data}, and the cross-section $\sigma_{np}$ for the radiative capture $n+p \rightarrow d+\gamma$ for thermal neutrons \cite{np_data}. While the matrix elements of the $A=2$ reactions ( $\langle\hat{\mu}_d\rangle$ and $\sigma_{np}$) were predicted using \pilesseft up to NLO \cite{KSW_c, Chen:1999tn, ando_deturon}, the matrix elements of the $A=3$ reactions ($^3$H and $^3$He magnetic moments) haven't been predicted using \pilesseft until recently. However, recent studies have shown much progress in this front. First, a recent calculation employing configuration space Schr\"{o}dinger equation representation of \pilesseft hinted a regular behavior of the $A=3$ magnetic properties, at least at small cutoffs of the theory \cite{Kirscher:2017fqc}. Second, a new method to calculate $A=3$ form-factors has been used to calculate $A=3$ magnetic moments, using an SU(4) symmetry approximation for the nucleon, without including the Coulomb interactions for $^3$He \cite{Vanasse:2017kgh,Vanasse2017}. 

In this chapter we use the formalism developed in Chapter~\ref{general_matrix} to calculate the $A=3$ matrix elements of the magnetic moments within \pilesseft.

\section{$M_1$ observables in the $A=2, \, 3$ systems}

$M_1$ observables at vanishing momentum transfer are related to the electromagnetic nuclear current density $\mathcal{\hat{J}} (\vec{q})$ at vanishing momentum transfer $\vec{q}$. Explicitly, the magnetic moment of a state is just the expectation value of the operator:
\begin{equation} \label{Eq:mu_def}
\hat{\mu}=-\frac{i}{2} \vec{\nabla}_q \times \mathcal{\hat{J}} (\vec{q})\big{|}_{q=0},
\end{equation}
while the cross-section of the thermal neutron radiative capture 
$n+p \rightarrow d+\gamma $ is proportional to the transition matrix element of the same operator between the neutron and proton, $S=0$ state, to the deuteron, $S=1$ state \cite{rearrange,KSW_c,ando_deturon}.

Thus, to calculate these observables, one needs the nuclear amplitudes of the $A=2,\,3$ nuclei, as well as the nuclear current in response to a magnetic photon. The resulting $M_1$ strengths are just matrix elements of the nuclear current between the nuclear amplitudes. 

\section{Magnetic photon currents in pionless EFT} 

A magnetic photon interaction with a nucleus can be modeled effectively as an interaction with ever growing clusters of nucleons. In \pilesseft, LO includes a single nucleon interaction with a photon, while the interaction of a magnetic photon with two-body clusters appearing first at NLO. The electromagnetic Lagrangian is given in Chapter \ref{formalism} by \cref{eq_l_magentic_2,eq_l_magentic_1}.

Applying the H-S transformation on \cref{eq_l_magentic_2} leads to the interaction in terms of the dibaryon fields (see Ref.~\cite{ando_deturon} and Appendix \ref{whatever}): 
\begin{equation}\label{eq_magnetic_dibaryon}
\begin{split}
&\mathcal{L}^{\mu }_{\text{magnetic}}=\frac{e}{2M}\left[N^\dagger\left(\kappa_0+\kappa_1\tau_3\right)\vec{\boldsymbol{\sigma}}\cdot \vec{B} N-L_1'(t^\dagger s+s^\dagger t)\cdot \vec{B}+L_2'(t^\dagger t)\cdot \vec{B}\right],
\end{split}
\end{equation}
where the two-dibaryons-one-photon coefficients are given by: 
\begin{align}
&L_1'(\mu)=-\frac{\rho_t+\rho_s}{\sqrt{\rho_t\rho_s}}\kappa_1+l_{1}(\mu), \\
&L_2'(\mu)=l_2(\mu)-2\kappa_0,
\end{align}
where $l_1$ and $l_2$ are RG invariant combinations:
\begin{eqnarray}
l_1(\mu) &=& \frac{M}{\pi\sqrt{\rho_t\rho_s}}L_{1}\left(\mu-\frac{1}{a_t}\right)\left(\mu-\frac{1}{a_{s}}\right)~, \\
l_2(\mu) &=& \frac{2M}{\pi\rho_t}L_2\left(\mu-\frac{1}{a_t}\right)^2
\end{eqnarray}
and $a_{t, s}$
are the spin-triplet and -singlet scattering lengths, $\rho_{t, s}$ are the respective effective ranges and $\mu$ is the renormalization scale. Note that since $[L_1,L2]=\fm^{-4}$, $l_1(\mu)$, $l_2(\mu)$ are dimensionless. 
Here $\mu$ is taken consistently as the three-body regularization scale $\Lambda=\mu$, contrary to past works, where it was taken at the \pilesseft breakdown scale $\mu=m_\pi$. We note that since the photon field $\mathcal {\vec{A}}$ fulfills $\vec{B}=\vec{\nabla}\cdot {\mathcal{\vec{A}}}(\vec{x})$, then the scattering operator $\hat{\mu}$ is given by the prefactor of $\vec{B}$ in \cref{eq_l_magentic_1,eq_magnetic_dibaryon}. Feynman rules are extracted trivially, using this fact.
\subsection{Diagrammatic representation of the electromagnetic matrix elements}
%By calibrating $L_1, L_2$ from $A=3$ magnetic moments, it is possible to examine the consistency of \pilesseft for the transitions between $A=3$ to $A=2$. This can be done by calculating $\sigma_{np}$ and $\langle\hat{\mu}_d\rangle$ using $L_1$ and $L_2$, respectively.

\subsubsection{Two-nucleon electromagnetic matrix elements}
\begin{figure}[h!]
	\vspace{-0 cm}
	\centering
	\includegraphics[width=0.99\linewidth]{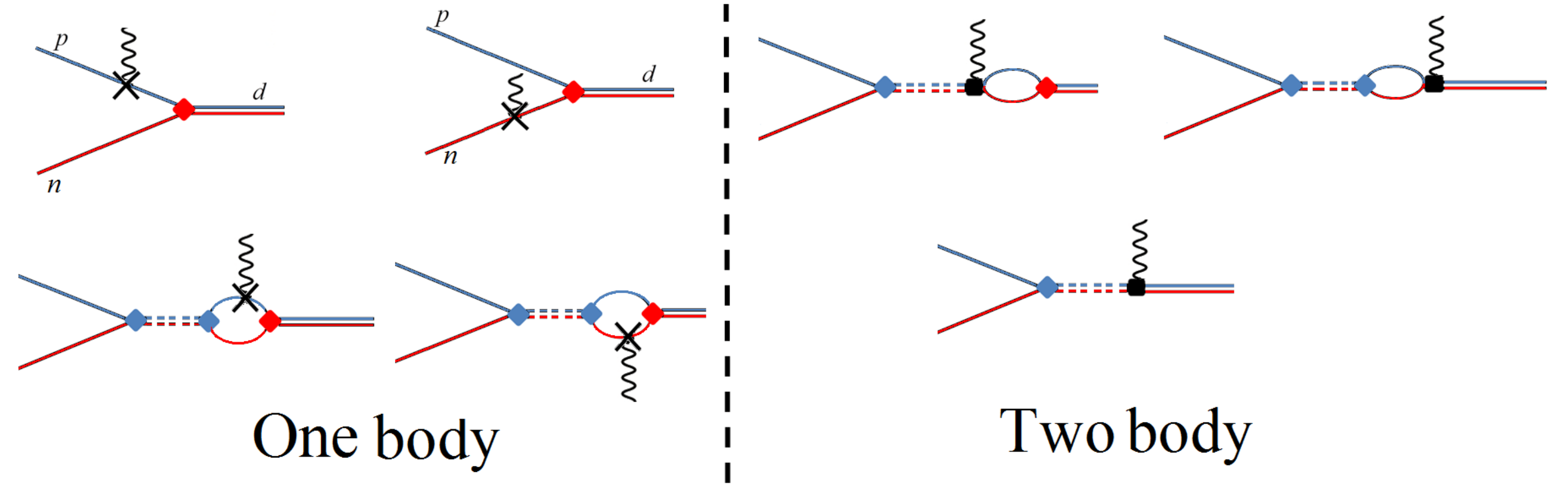}
	\captionsetup{justification=raggedright, singlelinecheck=false}
	\caption{\footnotesize{Diagrammatic representation of $\hat{\mu}$ between two-body states, up to NLO. $\hat{\mu}$ insertion is represented by the photon vertex. The double lines are the NLO propagator of the two dibaryon fields. $D_t$ (solid) and $D_s$ (dashed). The red lines represent the neutron propagator while the blue lines represent the proton propagator. A spin-singlet dibaryon-nucleon-nucleon vertex is proportional to $y_s$ (blue diamond), while a spin-triplet dibaryon-nucleon-nucleon vertex is proportional to $y_t$ (red diamond).}}
	\label{fig_np_capture}
	
\end{figure}

The matrix element of $\hat{\mu}$ (Eq.~\ref{Eq:mu_def}) between two-nucleon states is represented diagrammatically in Fig.~\ref{fig_np_capture}. This matrix element represents $\sigma_{np}$ ($\langle\hat{\mu}_{d}\rangle$) if the initial state is in relative $^1\text{S}_0$ ($^3\text{S}_1$) state, respectively.

From Fig.~\ref{fig_np_capture}, one concludes that up to NLO, the magnetic moment of the deuteron is given by: 
\begin{equation} \label{eq_mu_fineal}
\begin{split}
\langle\hat{\mu}_d\rangle =&Z_d^{\text{NLO}}(2\kappa_0)+Z_d^{\text{LO}}\left[\gamma_t\rho_t L_2'(\mu)\right]=2\kappa_0+\gamma_t\rho_tl_2(\mu)=2\kappa_0\left(1+l'_2(\mu)\right),
\end{split}
\end{equation}
where we implicitly define the net NLO two-body contribution as: $l'_2(\mu)\equiv\gamma_t\rho_t\frac{l_2(\mu)}{2\kappa_0}$. We note that $l'_2$ and $l_2$ are not of the same order of magnitude, since $\frac{\gamma_t\rho_t}{2\kappa_0}=\mathcal{O}\left(\frac{Q}{\Lambda_{\rm cut}}\right)$.

The cross-section of thermal neutron capture on a proton is related to the matrix element $Y$ by: 
\begin{equation}\label{eq_sigma}
%\sigma =\frac{e^2\left(\gamma_t^2+q^2/4\right)^3}{2\pi M^2q}Y^2, 
\sigma_{np} =2\alpha\pi \dfrac{\left(\gamma_t^2+q^2/4\right)^3a_s^2}{ M^4q\gamma_t}Y_{np}^2 \approx 2\alpha\pi \dfrac{\gamma_t^5a_s^2}{ M^4q}(2\kappa_1)^2(Y'_{np})^2, 
\end{equation}
where $Y_{np}$ is the sum over all the diagrams of Fig.~\ref{fig_np_capture} and $q=0.0069 \mev$ is the momentum transfer \cite{Vanasse:2017kgh,PhysRevLett.115.132001} \footnote{In this work, similar to Refs.~\cite{Vanasse:2017kgh,PhysRevLett.115.132001}, we use $q=2Mv_{\text{lab}}=2\cdot M 2200 \text{M/s}=0.0069\mev$. This value is higher than the value used in Refs.~\cite{Chen:1999tn,ando_deturon}, $q=0.068\mev.$ We found that $l_1' (q=0.0069\mev)$ is higher than $l_1' (q=0.0068\mev)$ by 20\%(10\%) for the Z- (ERE-) parameterization.}.

Since $\langle\hat{\mu}_{d}\rangle$ is approximately 1, we define the normalized matrix element, $Y_{np}'=Y_{np}/ (2\kappa_1)$, so that both $A=2$ matrix elements and their associated LECs, $l_1' (\mu)$ and $l_2' (\mu)$ can be compared with each other. This normalized matrix element is also obtained by Fig.~\ref{fig_np_capture} to yield: 
\begin{equation}
\begin{split}
Y'_{np}=
&\sqrt{Z_d}\left[\left(1-\frac{1}{\gamma_ta_s}\right)+
\frac{\gamma_t\sqrt{\rho_t\rho_s}}{4\kappa_1} L_1'(\mu)\right]
\end{split},
\end{equation}
and up to NLO is given by: 
\begin{equation} \label{eq_Y_fineal}
\begin{split}
Y'_{np}=&\sqrt{Z_d^{\text{NLO}}}\left(1-\frac{1}{\gamma_ta_s}\right)+\sqrt{Z_d^{\text{LO}}} \frac{\gamma_t\sqrt{\rho_t\rho_s}}{4\kappa_1} L_1'(\mu)=
\sqrt{Z_d^{\text{NLO}}}\left(1-\frac{1}{\gamma_ta_s}\right)-\gamma_t\frac{\rho_t+\rho_s}{4}+l'_1(\mu),
\end{split}
\end{equation}
where here the net NLO two-body contribution is defined by $l'_1(\mu)\equiv\gamma_t\sqrt{\rho_t\rho_s}\frac{l_1(\mu)}{4\kappa_1}$. We note that $l'_1$ is about an order of magnitude smaller than $l_1$, due to a (numerical) suppression originating in $4\kappa_1\approx 10$.

The above expressions up to higher order corrections can be found in the literature (see, e.g., Refs.~\cite{Chen:1999tn, ando_deturon}). 
\subsubsection{Three-nucleon electromagnetic matrix elements}
\begin{figure}[h!]
	\begin{center}
		\includegraphics[width=1\linewidth]{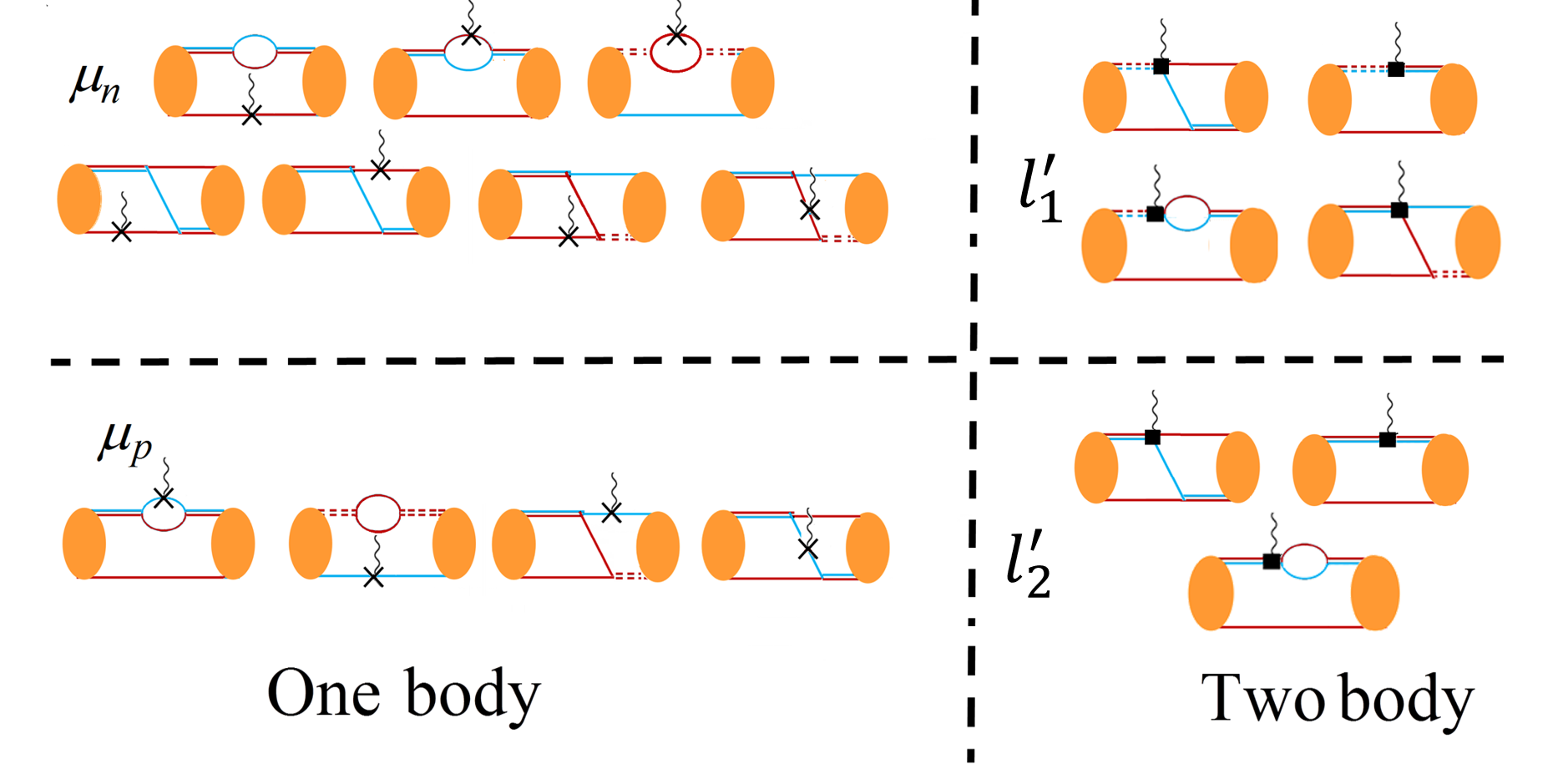}\\
		\caption{\footnotesize{Different topologies of the diagrams contributing to the triton magnetic moment. The double lines are the propagators of the two dibaryon fields. $D_t$ (solid) and $D_s$ (dashed for $nn$ and $np$), where the red line is the neutron and the blue line is the proton. Most of the diagrams couple both the triplet and the singlet channels. The diagrams with one-body interactions are coupled to $\left (\kappa_0-\kappa_1\right)=\mu_n$ (upper) and 
				$\left (\kappa_0+\kappa_1\right)=\mu_p$
				(lower). The two-body interactions are coupled to both $l_1'$ (upper) and $l_2'$ (lower).}}\label{magnetic_topo}
	\end{center}
\end{figure}

The three-body diagrams representing the matrix elements of $\hat{\mu}$ appear in Fig.~\ref{magnetic_topo}. At LO, only the diagrams that include a one-body electromagnetic interaction coupled to $\kappa_0-\kappa_1=\mu_n$ and $\kappa_1+\kappa_0=\mu_p$ contribute to the magnetic moment (left panels of Fig.~\ref{magnetic_topo}). At NLO, these interactions are augmented with a single NLO insertion, in the nuclear amplitude or in the propagators, as well as with the diagrams that include the two-body interactions coupled to the two-body LECs: $l_1'$ and $l_{2}'$ (right panels of Fig.~\ref{magnetic_topo}). As the latter interactions are at NLO, the bubbles and propagators are taken at LO. 
%\cblack

The $A=3$ magnetic moment expectation value, $\langle\hat{\mu}\rangle=\underbrace{\langle\hat{\mu}^{ (\text{1B})}\rangle}_{\text{one-body}}+\underbrace{\langle\hat{\mu}^{ (\text{2B})}\rangle}_{\text{two-body}}$, can be calculated numerically (with the experimental input parameters shown in Tab. \ref{table_1}) by summing over all diagrams up to NLO, as shown in Fig.~\ref{magnetic_topo}.

\begin{multline}\label{eq_one_body}
\langle \hat{\mu}^{(\text{1B})}\rangle=\dfrac{\left\langle\frac{1}{2}\left\|\boldsymbol{\sigma}\right\|\frac{1}{2}\right\rangle}{\sqrt{6}}\times\\
\sum\limits_{\mu,\nu}
{\psi^j_\mu(p')}\otimes y_\mu y_\nu\left(\kappa_0+\kappa_1\tau_3\right)\left[d'^{ij}_{\mu\nu} \hat{\mathcal{I}}(0,0)\right.\\
+\left. a'^{ij}_{\mu\nu}\hat{\mathcal{K}}(0,0)\right] \otimes{\psi^i_\nu(p)} ~,
\end{multline}
where for $^3$H:
\begin{equation}\label{eq_d_mu_3H}
d'^{ii}_{\mu\nu}=\begin{array}{c|ccc}
\mbox{\backslashbox{$\nu$\kern-1em}{\kern-1em$\mu$}}&t&np&nn \\
\hline
t&\frac{2}{3}\mu_n+\frac{1}{3}\mu_p&\mu_n-\mu_p&0\\
np&\frac{\mu_n-\mu_p}{3}&\mu_p&0\\
nn&0&0&\mu_n
\end{array}~,
\end{equation}
and \begin{equation}\label{eq_a_3H}
\mbox{ $a'^{ii}_{\mu\nu}$}=\begin{array}{c|ccc}
\mbox{\backslashbox{$\mu$\kern-1em}{\kern-1em$\nu$}}&t&np&nn \\ \hline
t&-\left(\frac{5}{3}\mu_p-\frac{2}{3}\mu_n\right)&\mu_p+2\mu_n&3\mu_p\\
np&\frac{2}{3}\mu_n+\frac{1}{3}\mu_p&2\mu_n-\mu_p&-\mu_p\\
nn&2\mu_p&-2\mu_p&0
\end{array}~,
\end{equation}
and for $^3$He:
\begin{equation}\label{eq_d_3He}
d'^{ii}_{\mu\nu}=\begin{array}{c|ccc}
\mbox{\backslashbox{$\nu$\kern-1em}{\kern-1em$\mu$}}&t&np&nn \\
\hline
t&\frac{2}{3}\mu_p+\frac{1}{3}\mu_n&\mu_p-\mu_n&0\\
np&\frac{\mu_p-\mu_n}{3}&\mu_n&0\\
nn&0&0&\mu_p
\end{array}
\end{equation}
and \begin{equation}\label{eq_a_mu_3He}
\mbox{ $a'^{ii}_{\mu\nu}$}=\begin{array}{c|ccc}
\mbox{\backslashbox{$\mu$\kern-1em}{\kern-1em$\nu$}}&t&np&nn \\ \hline
t&-\left(\frac{5}{3}\mu_n-\frac{2}{3}\mu_p\right)&\mu_n+2\mu_p&3\mu_n\\
np&\frac{2}{3}\mu_p+\frac{1}{3}\mu_n&2\mu_p-\mu_n&-\mu_n\\
nn&2\mu_n&-2\mu_n&0
\end{array}~.
\end{equation}

Note that for $^3$H and $^3$He, \cref{eq_one_body} is defined such that:
\begin{multline}\label{eq_one_body_1}
\langle \hat{\mu}^{(\text{1B})}\rangle=\kappa_0
\sum\limits_{\mu,\nu}
{\psi^j_\mu(p')}\otimes y_\mu y_\nu\left[d'^{ii}_{\mu\nu} \hat{\mathcal{I}}(0,0)\right.\\
+\left. a'^{ii}_{\mu\nu}\hat{\mathcal{K}}(0,0)\right] \otimes{\psi^i_\nu(p)} =\kappa_0~,
\end{multline}
for the case that $\mu_n=\mu_p=\kappa_0$, since:
\begin{multline}
\sum\limits_{\mu,\nu}
{\psi^i_\mu(p')}\otimes y_\mu y_\nu\left[d'^{ii}_{\mu\nu} \hat{\mathcal{I}}(0,0)+\right.\\
\left. a'^{ii}_{\mu\nu}\hat{\mathcal{K}}(0,0)\right] \otimes{\psi^i_\nu(p)} =1.
\end{multline}

\begin{figure}[h!]
	\begin{center}
		\begin{subfigure}[b]{0.32\linewidth}
			\includegraphics[width=1\linewidth]{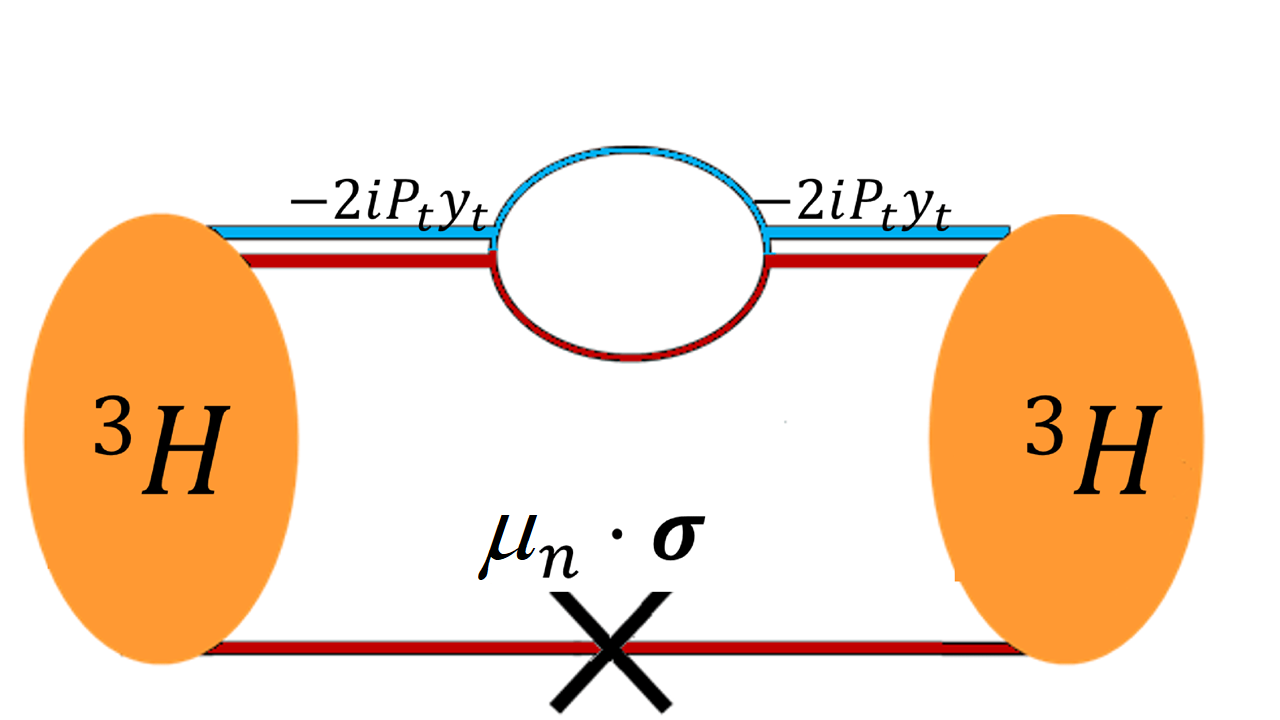}\\
			\caption{}
		\end{subfigure}
		\vspace{1 cm}
		\begin{subfigure}[b]{0.32\linewidth}
			\includegraphics[width=1\linewidth]{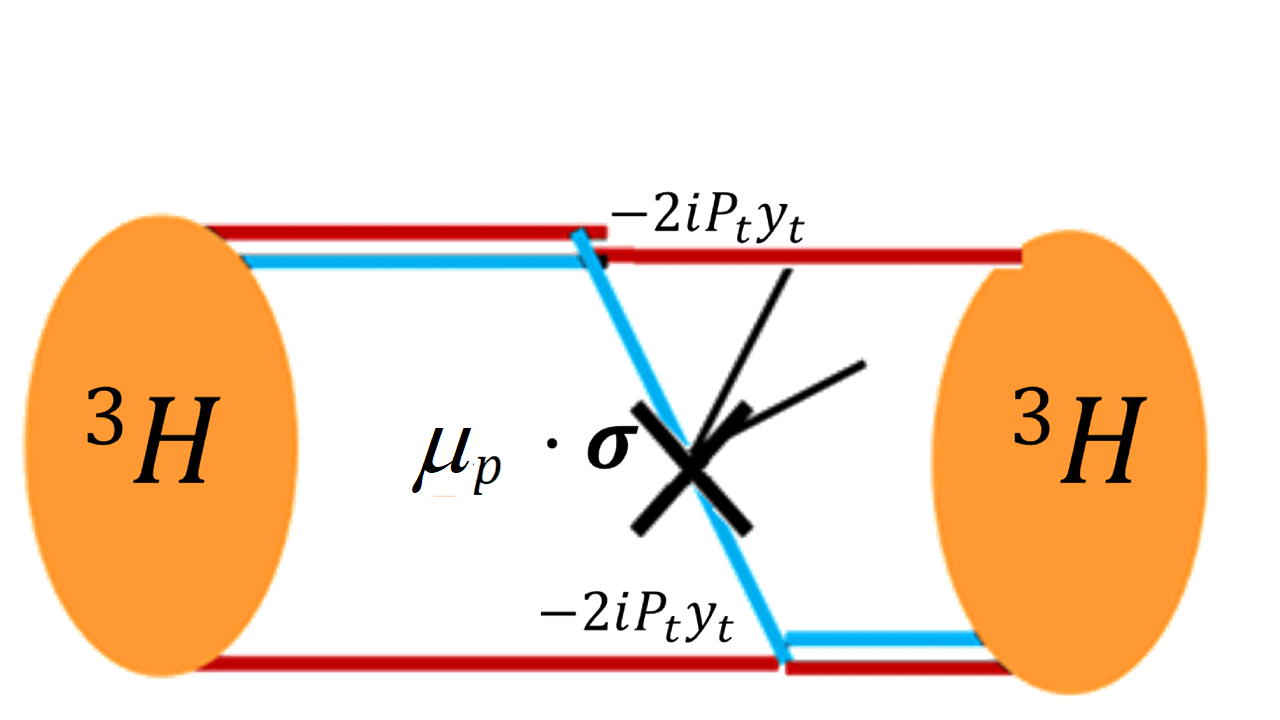}\\
			\caption{}
		\end{subfigure}
		\begin{subfigure}[b]{0.32\linewidth}
			\includegraphics[width=1\linewidth]{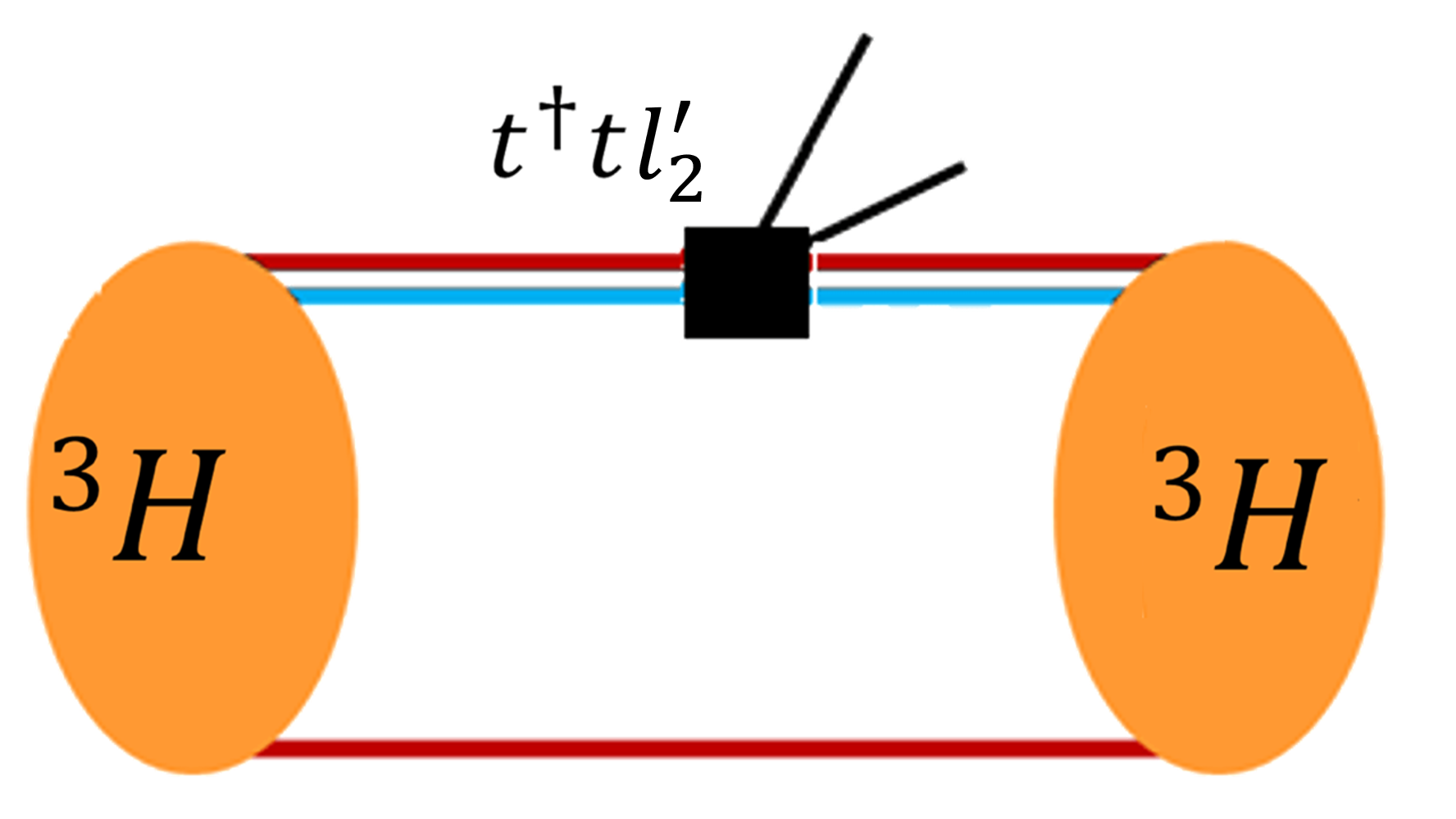}\\
			\caption{}
		\end{subfigure}	
		\caption{\footnotesize{Three of the diagrams contributing to the triton magnetic moment. 
				The one-body diagrams which are coupled to $\mu_n\cdot \boldsymbol{\sigma}$ (a) and $\mu_p\cdot \boldsymbol{\sigma}$ (b), diagram (c) is coupled to the two-body LEC $l'_2$. The double lines are the propagators of the two dibaryon field $D_t$ (solid). The red lines represent the neutron propagator while the blue lines represent the proton propagator.}}\label{fig_magnetic_intercation}
	\end{center}
\end{figure}

The three-nucleon magnetic moment matrix element that contains \textbf{two-body} interactions, $\hat{\mu}^{(\text{2B})}$, can be written as the sum of all two-body interactions (see Ref.~\cite{Big_paper}): 
\begin{equation}
\begin{split}
\langle \mu^{(\text{2B})} \rangle =&+\frac{2}{3}L'_2\langle\psi_t(p)|\psi_t(p)\rangle
\\&+L_1'a^{(2)}_{t,s}\left[\langle\psi_t(p)|\psi_{np}(p)\rangle+h.c\right],
\end{split}
\end{equation}
where
\begin{equation}
a^{(2)}_{ts}=
\begin{cases}-\frac{2}{3} &^3\text{H}\\
1 &^3\text{He}.
\end{cases}
\end{equation}
\cblack
Figure~\ref{fig_magnetic_intercation} (a)+(b) present a one-body $A=3$ matrix element coupled to $i\frac{e}{2M}\left(\kappa_0+\kappa_1\tau_3\right)\boldsymbol{\sigma},
$
$\mu_n$ in the case of neutron and 
$\mu_p$ in the case of proton.

At LO, Fig.~\ref{fig_magnetic_intercation} (a) is given by: 
\begin{multline}
\mu_ny_t^2\int\dfrac{d^3 p}{(2\pi)^3}\dfrac{d^3p'}{(2\pi)^3}\dfrac{2\pi^2}{p^2}\psi_t(E_{^3\text{H}}, p)
\dfrac{M^2}{4 \pi \sqrt{3 p^2-4ME_{^3\text{H}}}}\delta(p-p')\cdot
\psi_t(E_{^3\text{H}}, p')=\\\mu_nMy_t^2
\bra{\psi^i_\mu(E_{^3\text{H}},p')}\mathcal{I}^{q=0}(E_{^3\text{H}},0,p,p') \ket{\psi^i_\nu(E_{^3\text{H}},p)}~~,
\end{multline}
Fig.~\ref{fig_magnetic_intercation} (b) is given by: 
\begin{multline}
\mu_py_t^2\int\dfrac{d^3 p}{(2\pi)^3}\dfrac{d^3p'}{(2\pi)^3}\psi_t(E_{^3\text{H}}, p)\cdot
\frac{M^2}{p^2 \left(p^2-2 ME_{^3\text{H}}\right)+\left(p^2-ME_{^3\text{H}}\right)^2+p'^4}
\psi_t(E_{^3\text{H}}, p')=\\
\\\mu_pMy_t^2
\bra{\psi^i_\mu(E_{^3\text{H}},p')}\mathcal{K}^{q=0}(p,p',E_{^3\text{H}},0) \ket{\psi^i_\nu(E_{^3\text{H}},p)}~,
\end{multline}
where $i=^3\text{H}$.

Figure \ref{fig_magnetic_intercation} (c) (two-body operator) is given by: 
\begin{align}
&l'_2\int\dfrac{d^3 p}{(2\pi)^3}\psi_t(E_{^3\text{H}}, p)^2.%\cdot
%\dfrac{\left\langle\frac{1}{2}\Big\|\boldsymbol{\sigma}\Bi%g\|\frac{1}{2}\right\rangle}{\sqrt{3}}, 
\end{align}
The $A=3$ magnetic moment expectation value, $\langle\hat{\mu}\rangle=\langle\hat{\mu}^{(\text{1B})}\rangle+\langle\hat{\mu}^{(\text{2B})}\rangle$, can be calculated numerically (with experimental input parameters shown in Tab. \ref{table_1})) by summing over all diagrams up to NLO as shown in Fig.~\ref{magnetic_topo}. %The diagrams which contain a one-body magnetic interaction that are coupled to $\mu_n$ and $\mu_p$ and will contain one ERE insertion up to NLO. The two-body diagrams are coupled to the two-body LECs: $L_1'$ and $L_{2}'$. 
%By comparing this sum for either $^3$H or $^3$He to the experimentally known magnetic moments from Ref.~\cite{3He_3H_data}, $l_1$ and $l_2$ can be found simultaneously. %Similarity to Gamow-Teller transition, the magnetic moments also contains one-body LO matrix element and one-body and two-body NLO matrix elements. 

\begin{comment}
$\langle\hat{\mu}\rangle_{LO}$ originates from \cref{eq_l_magntic_1}: \[\frac{e}{2M}N^\dagger\left(\kappa_0+\kappa_1\tau_3\right)N\cdot\sigma \]
its relevant diagrams are shown in left side of Fig.~\ref{magnetic_topo}.

$\langle\hat{\mu}\rangle_{NLO}$ originates from \cref{eq_l_magntic_1}, where in that case the three-body amplitude and the two-body propagator are in NLO. 
Its relevant diagrams are shown in left side of Fig.~\ref{magnetic_topo}

The two-body NLO contribution to $\langle\hat{\mu}\rangle$ originates from both effective range correction which are coupled to $\kappa_1, \kappa_0$ and the two-body LEC $L1, L_2$. 

$\langle\hat{\mu}\rangle_{\text{ERE}}$ originates from
\begin{equation} 
-\frac{e}{2M}\left[\frac{\rho_t+\rho_s}{\sqrt{\rho_t\rho_s}}\kappa_1\left(t^\dagger s+s^\dagger t\right)+2\kappa_0\left(t^\dagger t\right)\right]
\end{equation}
RHS of Fig.~\ref{magnetic_topo}\\
$\langle\hat{\mu}\rangle_{L_{1, 2}}$ originates from
\begin{equation}
\frac{e}{2M}\left[l_{1}\left(t^\dagger s+s^\dagger t\right)+l_2\left(t^\dagger t\right)\right]\cdot 
\end{equation}
RHS of Fig.~\ref{magnetic_topo}

\end{comment}
\cblack
\section{Results}
Evidently, the $A=2,3$ $M_1$ observables depend on the physical (RG invariant) values of the LECs, {\it i.e.,} ${l'}^\infty_{1}\equiv l'_1 (\mu=\Lambda\rightarrow \infty)$ and ${l'}^\infty_{2}\equiv l'_2 (\mu=\Lambda\rightarrow \infty)$. In past works, the experimental values of the $A=2$ observables ($\sigma_{np}$ and $\langle\hat{\mu}_d\rangle$) were used to fix these LECs \cite{ando_deturon, ando_magntic_BBN}. Here, we calculate consistently the $A=2,3$ $M_1$ observables ($\langle\hat{\mu}_{^3\text{H}}\rangle$, $\langle\hat{\mu}_{^3\text{He}}\rangle$,$\langle\hat{\mu}_{d}\rangle$ and $Y'_{np}$), which depend on the same LECs, so we can extract these LECs from two observables and then use them to predict the remaining observables. Therefore, we have six independent ways for calibrating the LECs. These calibrations will be used later in this thesis for the evaluating the stability and consistency of \pilesseft for the simultaneous description of $A=2$ and $A=3$ systems up to NLO and for estimating theoretical uncertainty (Chapter \ref{discussion}).

Table~\ref{table_all_M1} summarizes our predictions for ${l'}^\infty_{1}$, ${l'}^\infty_{2}$ and $M_1$ observables up to NLO {in both $Z$- and ERE- parameterization}. For each row, the $'\star'$ denotes the $M_1$ observables used for ${l'}^\infty_{1}$, ${l'}^\infty_{2}$ calibration, by comparing our calculation to the experimental data. For example, the first row of Tab.~\ref{table_all_M1} shows the LECs fixed from $A=3$ observables and our prediction of $A=2$ magnetic observables, while the second row of Tab.~\ref{table_all_M1} shows the LECs fixed from $A=2$ observables and the prediction of $A=3$ magnetic observables. Note that for each $M_1$ observable we have three predictions.

\begin {table}[H]
\centering
\begin{tabular}{c|| c| c|c|c|c|c}
	\centering
	& ${l'_{1}}^{\infty}/ 10^{-2}$&${l'_{2}}^{\infty}/ 10^{-2}$&$\langle\hat{\mu}_{^3\text{H}}\rangle$[nNM]& $\langle\hat{\mu}_{^3\text{He}}\rangle$[nNM]&
	$\langle\hat{\mu}_{d}\rangle$[nNM]&$Y'_{np}$\\ 
	\hhline{=|=|=|=|=|=|=}
	&4.72 (14.2)&	-1.55 (	4.1	)&	$\star$		&	$\star$				&	0.87 (0.92)	&1.253 (1.31)	\\
	&4.66 (9.0)	&-2.55 (-2.55)	 &	2.978 (2.76)	&	$-$2.145 ($-$1.89)	&	$\star$		&$\star$	\\
	&4.66 (9.0)	&	-2.4 (29)	 &	$\star$		&	$-$2.144 ($-$1.66)	&	0.86 (1.17)	&$\star$	\\
	&4.66 (9.0)	&	-0.13 (-31) &	2.996 (2.59)	&	$\star$				&	0.88 (0.61)	&$\star$	\\
	&4.92 (15.2)&	-2.55 (-2.55)&$\star$		&	$-$2.143 ($-$2.23)	&	$\star$		&1.255 (1.32)	\\
	&4.60 (13.4)&	-2.55 (-2.55)&	2.967 (2.91)	&	$\star$				&	$\star$		&1.253 (1.30)	\\
	\hline								
	\multicolumn{1}{c||}{Mean}	 & 4.73 (13.0)	&	-1.66 (	-0.04)&	2.98 (2.75)	&-2.144 (-1.93)&	0.87 (	0.89)&	1.253 (1.31)	\\
	\hline								
	\multicolumn{1}{c||}{$\Delta$}&0.2 (2.8)	&	1.11 (25)	&0.015 (	0.16)&	0.001 (	0.28)	&0.01 (0.26)&	0.001 (0.01)	\\
	\hline
	%\multicolumn{1}{c||}{$\%\frac{\text{NLO}}{\text{LO}}$}&&	&7\% (0.1\%)	&14\% (4\%)&	1\% (1\%)	&5\% (9\%)\\
	\hline
	\multicolumn{1}{c||}{Exp data}&&&	2.979&	-2.128&		0.857&	1.253
\end{tabular}
\vspace{0.2 cm}
\caption{ \footnotesize{Numerical results for our prediction for ${l'_{1}}^{\infty},{l'_{2}}^{\infty}$ and $A=2,3$ $M_1$ observables. The nominal value is calculated using $Z$-parameterization, while the number in brackets is calculated using the ERE-parameterization. Mean denotes the mean value of the $M_1$ observable based on its three predictions while $\Delta$ denotes the standard divination of these predictions.}}
\label{table_all_M1}
\end{table}

\begin{figure}[h!]
\begin{center}
	\begin{subfigure}[b]{0.55\linewidth}
		\includegraphics[width=\linewidth]{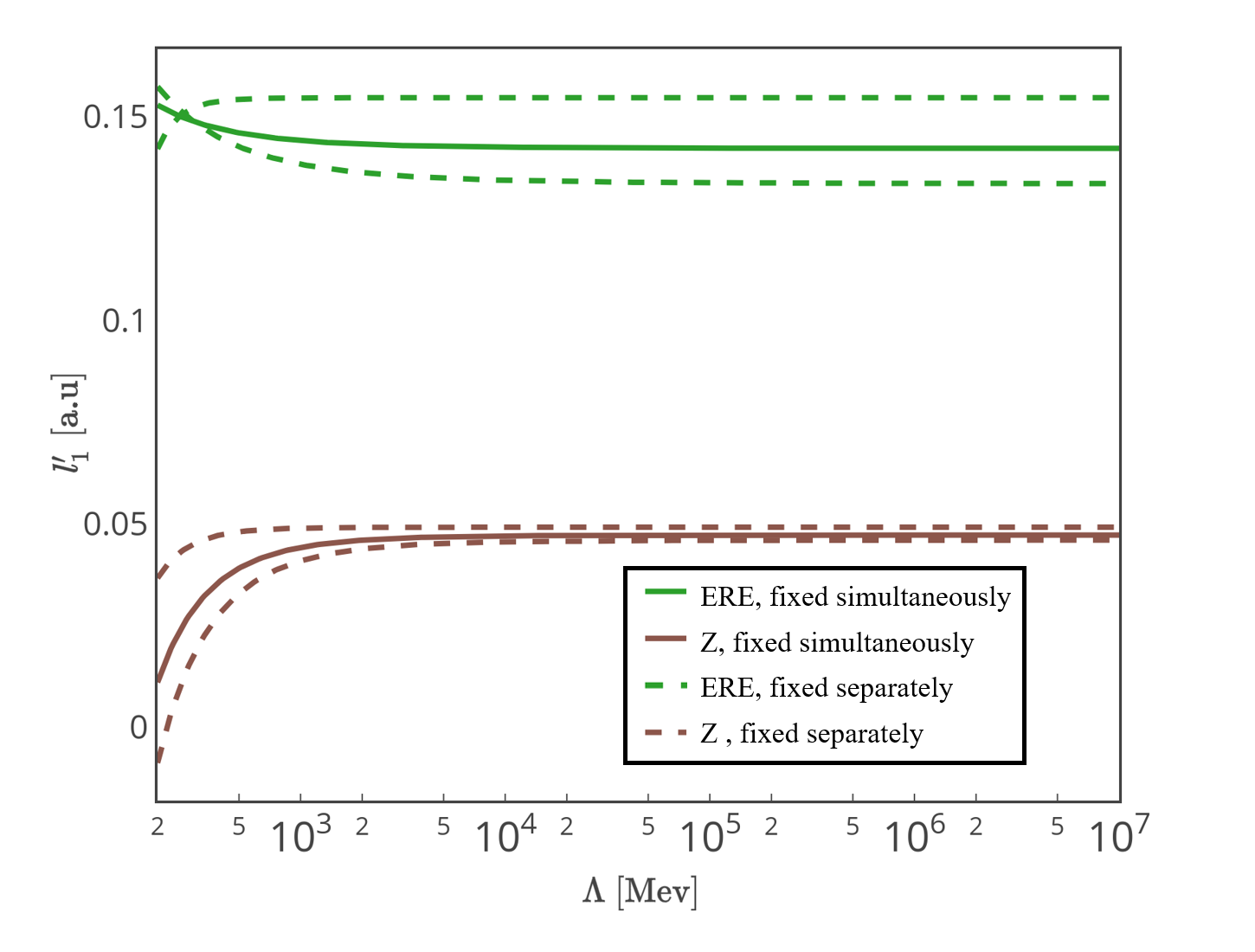}\\
		\caption{}
	\end{subfigure}
	\begin{subfigure}[b]{0.55\linewidth}
		\includegraphics[width=\linewidth]{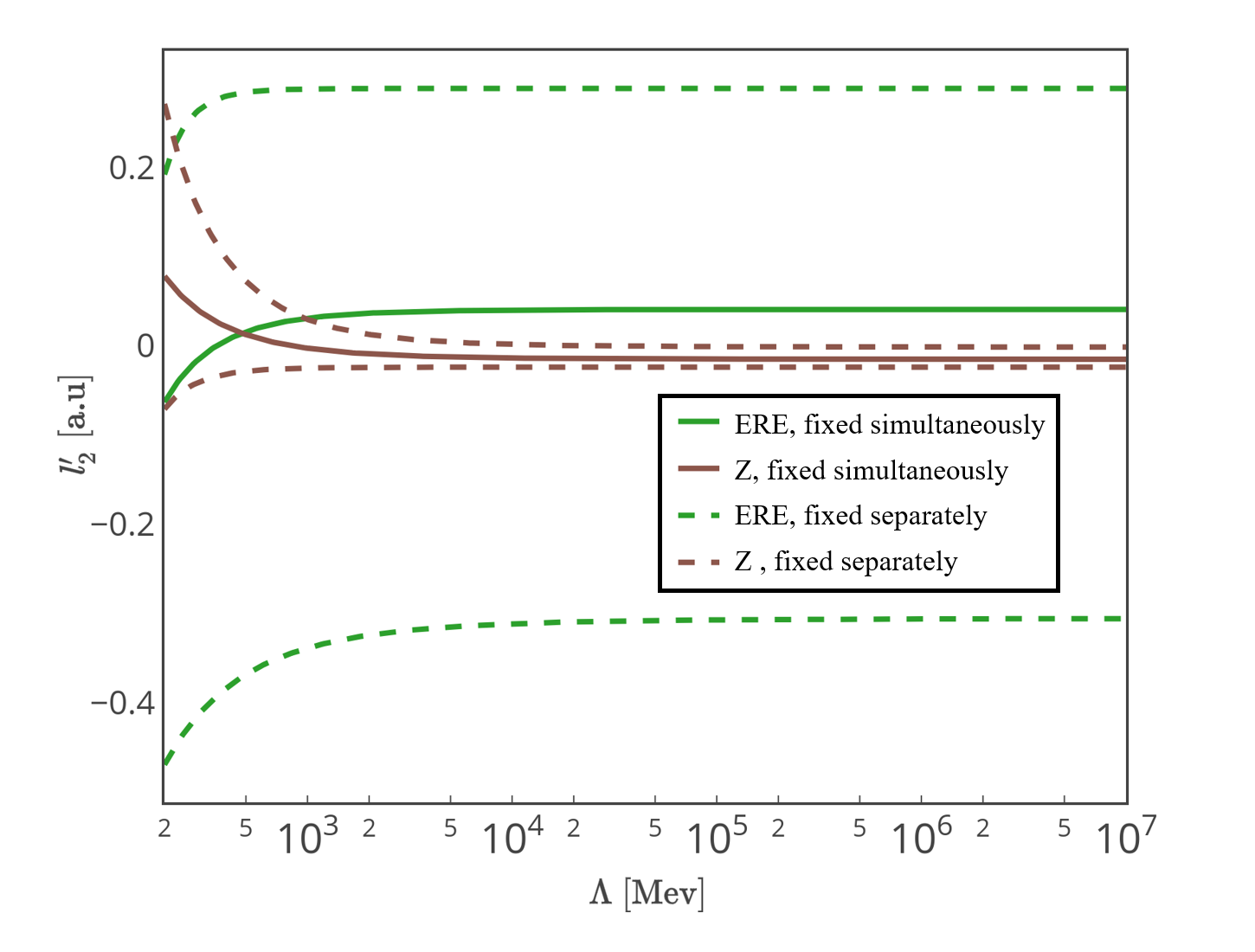}\\
	\end{subfigure}
\vspace{1 cm}
	\caption{\footnotesize{Numerical results for LECs $l'_1$ (up) and $l'_2$ (button), calibrated from $\langle\hat{\mu}_{^3\text{H}}\rangle$ and $\langle\hat{\mu}_{^3\text{He}}\rangle$ as a function of the cutoff $\Lambda$. The solid lines are the numerical results obtained from $A=3$ observables simultaneously, while the dashed lines are the numerical results obtained from $A=3$ observables separately.}}\label{fig_L1_L2}
\end{center}
\end{figure} 
%\end{document}רקג
A closer examination of the LECs mean and standard deviation revels big differences between the LECs. {For the case of the Z-parameterization} we find that while ${l'_{1}}^{\infty}$ has minor dependence on the $M_1$ observables used for its calibration, \textit{i.e.,} $ \Delta{l'_{1}}^{\infty}/{l'_1}\approx3\%$, the stranded deviation of ${l'_{2}}^{\infty}$ is of the same order of magnitude as ${l'_{2}}^{\infty}$, \textit{i.e.,} $ \Delta{l'_{2}}^{\infty}/{l'_{2}}^{\infty}\approx70\%$. For the case of the ERE-parameterization, we find that the differences between ${l'_{1}}^{\infty}$ and ${l'_{2}}^{\infty}$ are more significant; while $ \Delta{l'_{1}}^{\infty}_{ERE}/{l'_{1}}^{\infty}_{ERE}\approx21\%$, $\Delta{l'_{2}}^{\infty}_{ERE}$ is two orders of magnitude larger than ${l'_{2}}^{\infty}_{ERE}$. This behavior of ${l'_{1}}^{\infty}$,${l'_{2}}^{\infty}$ will be discussed later in this thesis (Chapter~\ref{discussion}), and will use for estimating the theoretical uncertainty of \pilesseft. 
%\end{document}
\cblack
\section{Magnetic interactions in Lattice QCD}
\cblack
In previous sections, we used experimental data to calibrate the value of the electromagnetic LECs, $l'_1$ and $l'_2$. An alternative way was accomplished recently by the Nuclear Physics with Lattice Quantum Chromo Dynamics (NPLQCD) collaboration, as reported in Ref.~\cite{PhysRevLett.115.132001,Beane:2014ora}. In this section we will calculate $Y'_{np}$ using $l'^\infty_1$ extracted from lattice QCD data (subsection~\ref{lattic_l1}) and will extract $l'^\infty_2$ using the lattice magnetic moments of both $^3$H and $^3$He (subsection~\ref{lattic_l2}). 
\cblack
\subsection{Calculation of $\sigma_{np}\, (Y'_{np})$ using $l'^\infty_1$ from lattice QCD data}\label{lattic_l1}

%ZZZZZ - change to $l'_1$ and not $l_1$. ZZZZZZZ

In previous sections, we used experimental data to calibrate the value of the electromagnetic LECs, $l'_1$ and $l'_2$. An alternative way was accomplished recently by the Nuclear Physics with Lattice Quantum Chromo Dynamics (NPLQCD) collaboration, as reported in Ref.~\cite{PhysRevLett.115.132001}. They have performed large scale lattice QCD calculations of a system of neutron and a proton coupled to a strong electromagnetic field $\vec{B}$. The energy difference between the $^1S_0$ and $^3S_1$ eigenstates of this two-nucleon system is then given by $\Delta E=2(\kappa_1+\bar{L}_1)\frac{e}{M}|\vec{B}|$ \cite{Detmold:2003rq, Athanassopoulos:2003qe, Chang:2015qxa}, with
%\begin{eqnarray}\label{eq_l1}
%l'_1&=&\frac{\gamma_t\tilde{l_1} \sqrt{\rho_s\rho_t}}{2 \left(2\kappa_1\right)}-\frac{1}{4} \text{$\gamma $t} \text{$\rho $s} \text{$\rho $t}\\\label{eq_L1_bar}
$\bar{L}_1=Z_d\kappa_1\left[2l'_1(\mu)+\gamma_t \sqrt{\rho_s\rho_t}-\frac{\gamma_t}{2} \left(\rho_s+\rho_t\right)\right]$.
%\end{eqnarray}

The results of the simulations allow them to extract $\bar{L}_1=0.285^{+63}_{-60}\text{ nNM}$, or: 
\begin{equation}
{l'_1}^{\text{NPLQCD}} = {6.06^{+14}_{-13}}_{Z} ({6.66^{+14}_{-13}}_{\text{ERE}})/10^{-2}.
\end{equation} 
The uncertainties here are related to the lattice calculation and to the uncertainty introduced in the extrapolation of the simulations (done at large non-physical quark masses) to the natural values of the quark masses. Comparing to Tab.~\ref{table_all_M1}, a $2 \sigma$ deviation is found. This translates to about $2\%$ deviation from experimental matrix element, $Y'_{np}$: 
\begin{equation}
{Y'_{np}}=1.267^{+0.013}_{-0.013}\left(1.230^{+0.013}_{-0.013}\right),
\end{equation}
%\hilight{ZZZZ - mistake?? is this Y' or Y? -ZZZZZ - $Y'_{np}$} 
where nominal value is the one produced using the $Z$-parameterization, and the ERE-parameterization value is given in brackets. This result is in some contrast with the conclusions of Ref.~\cite{PhysRevLett.115.132001}, where a perfect fit with experiment was found, since the power-counting used in that paper is different than the one we use here. e.g., the deuteron normalization is taken at its nature value ($Z_d\approx 1.69$) at all orders, and not perturbatively, as in \cref{eq_Y_fineal}. In addition, no theoretical uncertainty due to the EFT truncation in the calculation of $\sigma_{np}$ is given in that paper. Interestingly, when we use recent lattice QCD calculation of $d$, $^3$H and $^3$He magnetic moments \cite{Beane:2014ora}, we find that the result is inconsistent with the experiment by more than two standard deviations of the numerical calculations. The LQCD ``experimental" value for $\langle \hat{\mu}_{^3\text{H}}\rangle$ is $3.56 \pm 0.23$ nNM so it contains $\sim 7\%$ uncertainty, in addition to the inconsistency with the measured magnetic moments of $2.9789$ nNM \cite{3He_3H_data}. The difference given here is not sufficient to draw any definite conclusions. Therefore, in the next section, we will try to reproduce the lattice QCD magnetic moments using the \pilesseft diagrammatic method introduced in this work, with the non-physical pion mass: $m_\pi=805$ MeV, up to NLO. 

\subsection{Magnetic moments in Lattice QCD}\label{lattic_l2}
 In recent years, physical observables of low-energy electromagnetic interactions have been calculated (with large uncertainties) using lattice QCD for a non-physical pion mass (and then have been extrapolated to the physical pion mass) \cite{Beane:2014ora, Beane:2013br}. They found that for the non-physical pion mass, $m_\pi=805$ MeV, the magnetic moments are: $\kappa_p$=3.119(33)(64) nNM and $\kappa_n$ =
 -1.981(05)(18) nNM. Their calculations yielded the following results: 
 $\langle\hat{\mu}_d\rangle$
 = 1.218(38)(87)
 nNM for the deuteron, $\langle\hat{\mu}_{^3\text{He}}\rangle$ = 2.29(03)(12) nNM for
 $^3$He and $\langle\hat{\mu}_{^3\text{H}}\rangle$ 3.56(05)(18) nNM for the triton. Since for non-physical pion mass, the Coulomb interaction is ignored, the radiative capture, $\sigma_{np}$, cannot be accounted for consistently {\it i.e.,} the Coulomb interaction is ignored.

 In this section, we use our numerical and analytical methods for the extraction of $l'_1$ and $l_2'$ from our prediction to the $A=3$ magnetic moment in \pilesseft by repeating all numerical calculations with the non-physical pion mass, $m_\pi=805$ MeV. In contrast to the physical world, there are different approaches for calculating physical inputs. The values of the lattice nuclear parameters used in the calculation are detailed in Tab.~7.3.
 %\cblack
 \begin{table}[H]\label{table_1_QCD}
 	\centering
 	\begin{tabular}{l|c|c}
 		Parameter& Nature $ m_{\pi}$ = 139 MeV (Tab. \ref{table_1})& LQCD $
 		m_{\pi}$ = 805$\mev$ \cite{Beane:2012vq, Chang:2015qxa,Kirscher:2017fqc}\\
 		\hline
 		$M$ [MeV]& 938.9& 1634(18)\\
 		$B_{np}$ [MeV]&& 15.9(40)\\
 		$a^{s}_
 		{np} $ [fm]&-23.75&\\
 		$a_{pp }$ [fm]&-7.806&\\
 		$\gamma_t$ [MeV]& 45.701 &180(44)\\
 		$E^B_{^3\text{H}}$ [MeV]& 8.482& 54(11)\\
 		$E^B_{^3\text{He}}$ [MeV]& 7.718& \\
 		$\rho_t$&1.765 fm&3.96(20)$
 		/m_{\pi}$\\
 		$\rho_s$&2.73 fm& 5.89(45)$
 		/m_{\pi}$ \\
 		$\rho_C$&2.794 fm& 5.89(45)$
 		/m_{\pi}$ \\
 	\end{tabular}
 	\caption{ \footnotesize{Lattice nuclear parameters used in the numerical calculations. The values for the singlet and triplet effective ranges were taken from private communication with CalLAT.}}
 \end{table}
 
 \begin{comment}
 \begin {table}[H]
 \begin{center}
 \begin{tabular}{c c||c c}
 \centering
 Parameter& Value& Parameter& Value\\
 \hline
 $M$&$1634(18)$ MeV \cite{Beane:2013br}&$\alpha$&0\\
 $\gamma_t$& 87.9(38) 
 MeV & $\rho_t$& 0.93(04) fm \\
 $a_s$& 2.61(02) fm & $\rho_s$& 0.82(05) fm \\
 $a_p$& 2.61(02) fm & $\rho_C$& 0.82(05) fm \\
 $E^B_{^3\text{H}}$& 55.8(62) MeV& $E^B_{^3\text{He}}$& 55.8(62) MeV
 \end{tabular}
 \caption{ \footnotesize{Lattice nuclear parameters used in the numerical calculations as discussed in private conversations with J. Kirscher (HUJI) and CalLAT.}}
 \label{table_lattice}
 \end{center}
 \end{table}
 \end{comment}
 \cblack
 These new non-physical inputs require a new calculation for both three-body force and three-body scattering amplitudes, up to NLO. Repeating the same calculations for both $^3$H and $^3$He magnetic moments with the physical input taken from Tab.~7.3, along with the non-physical magnetic moments: $\mu_p$=3.119(33)(64) nNM and $\mu_n$=-1.981(05)(18) nNM, we find that the $A=3$ magnetic moments are (Tab.~\ref{table_magntic_lattice}): 
 %\cblack
 \begin {table}[H]
 \begin{center}
 	\begin{tabular}{l| c| c}
 		\centering
 		&$^3$H[nNM]& $^3$He[nNM]\\ 
 		\hline
 		One-body, LO& 3.1195$\pm$0.002& -1.981$\pm$0.001\\ 
 		NLO range corrections &2.38$\pm$0.02&-0.92$\pm$0.02\\ 
 		Experimental ``data" \cite{Beane:2014ora} &3.56$\pm0.23$&-2.29$\pm0.15$
 	\end{tabular}
 	\caption{ \footnotesize{Numerical results for $\langle\hat{\mu}_{ ^3\text{H}}\rangle$ and $\langle\hat{\mu}_{^3\text{He}}\rangle$ with the non-physical pion mass.}}
 	\label{table_magntic_lattice}
 \end{center}
\end{table}
%\cblack

In this section, we use the difference between our numerical results of NLO with range corrections to the ''experimental'' values of Beane {\it et al.} \cite{Beane:2014ora}, to extract $l'_1$ and $l'_2$ and to recalculate the deuteron magnetic moment, $\langle\hat{\mu}_d\rangle$:
\begin{equation}
\langle\hat{\mu}_d\rangle=1.5\pm0.3\pm0.13\text{[nNM]}
\end{equation}
with 
\begin{equation}\label{eq_l2_LQCD}
l_2'=3.5\pm3\pm1.3\cdot10^{-2}, 
\end{equation}
where the first uncertainty originates from the $^3$H magnetic moment uncertainty, while the second uncertainty originates from the $^3$He magnetic moment uncertainty (\cite{Beane:2014ora}). We note that $^3$H and $^3$He magnetic moment uncertainties are translate into large uncertainties of the LEC, $l'_2$. 

The LEC, $l'_2$ as being calibrated from $^3$H and $^3$H magnetic simultaneously, holds many uncertainties originating from the different LQCD parameters given in Tab.~7.3 in addition to the ones presented in \cref{eq_l2_LQCD}. \cblack In the next subsection, we will analyze the effects of these uncertainties on the $A=3$ magnetic moments and as a result, on $l'_2$ and $\langle\hat{\mu}_d\rangle$.
\subsection{LQCD empirical uncertainties}
In contrast to the physical pion mass, for the non-physical pion mass the physical parameters ($\gamma_t, a_s, E_3^B, \rho_t, \rho_s$) hold large uncertainties, originating from the character of the LQCD calculation as shown in Tab.~7.3.

In Tab.~7.5, we show the contribution of each such uncertainty to the uncertainty propagated to the electromagnetic observables.
%\cblack
\begin{table}[H]\label{lattice_error_bar}
\begin{tabular}{ccccccccc}
	&Physical observable&$E^B_{^3\text{H}}$&$a_s$&$\gamma_t$&$\rho_t$&$\rho_s$&$\mu_p$&$\mu_n$\\
	\hline 
	&Quoted uncertainty &18.5\%& 25.16\%& 22.22\%& 5.05\%& 7.64\% &3.11\%& 1.16\%\\
	\hline
	\multirow{4}{*}{\rotatebox[origin=c]{90}{\parbox[c]{2cm}{\centering Effect on magnetic observable}}} &$\Delta\langle\hat{\mu}_{^3\text{H}}\rangle$
	&0.01\%& 0.05\%& 0.2\%& 0.01\%& 0.02\%& 3.11\%& 0\%\\
	&$\Delta\langle\hat{\mu}_{^3\text{He}}\rangle$
	&0.02\%& 0.08\%& 0.31\%& 0.02\%& 0.03\%& 0\% & 1.16\%\\
	&$\Delta\langle\hat{\mu}_{d}\rangle$&2.76\%& 0.83\%& 6.06\%& 1.5\%& 0.75\%& 18.42\%& 9.54\%\\
	&$\Delta l'_2$&11.34\%& 3.44\%& 24.94\%& 6.48\%& 3.15\%& 91.33\%& 22.77\%\\
\end{tabular} \caption{\footnotesize{The contribution of the calculated uncertainty in the physical observables to the uncertainty propagated to the electromagnetic observables.}}
\end{table}

%\cblack 
The physical inputs of Tab~.7.5 are divided into two categories: on-shell parameters ($\mu_n$ and $\mu_p$) and off-shell parameters ($E^B_{^3\text{H}}$\,$a_s$\,$\gamma_t$\,$\rho_s$\,$\rho_t$), which effect the three-body force and three-nucleon wave functions. It clear that for the on-shell parameters, $\langle\hat{\mu_{^3\text{H}}}\rangle$ and $\langle\hat{\mu}_{^3\text{He}}\rangle$ uncertainties are proportional to the relevant parameter has a minor effect on $\langle\hat{\mu_{^3\text{H}}}\rangle$ and $\langle\hat{\mu}_{^3\text{He}}\rangle$ uncertainty, {\it i.e.,} $\Delta\langle\hat{\mu}_{^3\text{H}}\rangle\propto\Delta\kappa_p$ and $\Delta\langle\hat{\mu}_{^3\text{He}}\rangle\propto \Delta\kappa_n$. Since $\langle\hat{\mu_{^3\text{H}}}\rangle$ depends mostly on $\kappa_p$ and $\mu_{ ^3\text{He}}$ depends mostly on $\kappa_n$, the total effect of $\kappa_p,\kappa_n$ uncertainties on $l_2'$ (and as a result, on $\langle\hat{\mu}_d\rangle$) is much larger than the effect of $E_B, \rho_{s},a_s$. 
Since $l'_2=\gamma_t\rho_t\frac{l_2}{2\kappa_0}$ $\rho_{t}$; $\gamma_t$ uncertainty more affects $\Delta\langle\hat{\mu}_{d}\rangle$ and $\Delta l'_2$ than $E_B, \rho_{s},a_s$ as shown in Tab.~7.3.
Also, we note that for the off-shell parameters, the uncertainty stems mainly from the $A=3$ wave function, and therefore, we expect that off-shell parameters uncertainty will effect $\langle\hat{\mu_{^3\text{H}}}\rangle$ and $\langle\hat{\mu}_{^3\text{He}}\rangle$ in a similar way. The fact that $\mu_p$ and $\mu_n$ uncertainty is not symmetric to $\langle\hat{\mu_{^3\text{H}}}\rangle$ and $\langle\hat{\mu}_{^3\text{He}}\rangle$, increases their effect on $l'_2$ and $\langle\hat{\mu}_d\rangle$ comparing to the off-shell parmters. 
\begin{comment}
A closer examination of Tab.~\ref{table_lattice} and Tab.~7.5 reveals that the values of $\gamma_t$ differ by more than 200\% from each other according to the different approaches. A comparison of effect of the different $\gamma_t$ values ($87$ MeV and $190$ MeV) on $\langle\hat{\mu}_{^3\text{H}}\rangle, \langle\hat{\mu}_{^3\text{He}}\rangle$ shows that this uncertainty in $\gamma_t$ leads to $\sim3\%$ uncertainty in $\langle\hat{\mu}_{^3\text{H}}\rangle, \langle\hat{\mu}_{^3\text{He}}\rangle$ and to $10\%$ uncertainty in $\langle\hat{\mu}_d\rangle$.
\end{comment}
\cblack

Summarizing, the \pilesseft procedure of calculating the electromagnetic observables creates a significant sensitivity of these observables to uncertainties in the calculations of the binding energies, and, mainly in the deuteron case, in calculated lattice values of the single nucleon magnetic moments.

\chapter{ Weak interactions and $^3$H $\beta$-decay in pionless EFT}\label{beta_decay}
In this chapter we outline the calculation of the matrix
element of the weak reaction: 
	\begin{equation}\label{eq_beta}
^3\text{H}\rightarrow ^3\text{He}+e^-+\overline{\nu_e}, 
\end{equation}
which is well-measured weak reaction that contains more than one nucleon and its energy transfer is applicable for the \pilesseft regime. As mention before, we are using this calculation of $^3$H $\beta$-decay for extracting the unknown LEC, $L_{1, A}$, which is essential for predicting the $pp$ fusion rate.
\section{The weak interactions in pionless EFT }
The weak interaction Lagrangian is given in Chapter \ref{formalism} by \cref{lweak,eq_weak_A,eq_weak_A}.
\begin{comment} 
\begin{equation}\label{Hweak}
\mathcal{L}_{\text{Weak}}=\frac{G_FV_{ud}}{\sqrt{2}}l_+^{\mu }J_{\mu
}^-, 
\end{equation}
where $G_F$ is the Fermi constant and $V_{ud}$ is the Cabibbo–Kobayashi–Maskawa (CKM) matrix element. $l^\mu$ is the leptonic
current %$l^\mu=\bar{u}_e\gamma(1-\gamma_5)v_\mu$, 
and 
$J_\mu$ is the hadronic current. We calculate the two-body hadronic current $J_\mu$ from the \pilesseft effective Lagrangian \cref{H_S_lag2}) with dibaryon fields up
to NLO using the Hubbard-Stratonovich transformation.

The hadronic current contains two parts, with polar-vector and
axial-vector symmetries: $J_\mu=V_\mu-A_\mu$. The part of the polar
vector current relevant to $\beta$-decay with vanishing energy transfer is: 
\begin{equation}
\label{V}
V_0^{-}=N^{\dagger}\frac{{\tau^-}}{2}N~.
\end{equation}
Here, we utilized the fact that the Conserved Vector Current (CVC)
hypothesis is accurate at this order of EFT.
\end{comment}
The axial-vector part (after Hubbard-Stratonovich transformation) is (see Appendix \ref{whatever}): 
\begin{equation}
\label{eq_weak_axial2}
\boldsymbol{ A^{\pm}}=\frac{g_A}{2}N^\dagger
\boldsymbol{\sigma}\tau^{\pm}N 
+g_AL'_{1, A} \left (t^\dagger s+s^\dagger t\right)~,
\end{equation}
where $g_{A}$ is the axial coupling constant for a single nucleon, known from neutron $\beta$-decay.
In two-nucleon calculations, the renormalization scale $\mu$ is
frequently assumed to be comparable to the breakdown scale of theory, 
usually at $m_\pi$. In the current case, %similarly to that of the electromagnetic interactions, 
it is set to $\mu=\Lambda$. One expects that by taking
$\Lambda\rightarrow \infty$ numerically, regularization effects from
the treatment of the $A=3$ system vanish. This causes an asymptotic
behavior of $L_{1, A}$, which cancels the cutoff dependence in the NLO
term, leaving it RG invariant. Testing this is part of the validation
process of our numerical calculation. We therefore define:

\begin{equation}
{L_{1, A}'}(\mu)=\frac{\rho_t+\rho_s}{2\sqrt{\rho_s\rho_t}}-l_{1,A}(\mu)~,
\end{equation} where $l_{1,A}$ is an RG invariant combination (in arbitrary units). 
\begin{equation}
l_{1,A}(\mu)=\frac{L_{1, A}}{2\pi g_A}\frac{1}{\sqrt{\rho_s\rho_t}}\left(\mu-\frac{1}{a_t}\right)\left(\mu-\frac{1}{a_s}\right)~,
\end{equation}
and $\mu$ is the renormalization scale.
Similar to the electromagnetic LEC (Chapter.~\ref{Magnetic}), we take the limit $\mu=\Lambda\rightarrow \infty$, such that:
\begin{equation}
l_{1,A}(\Lambda)=\frac{L_{1, A}}{2\pi g_A}\frac{1}{\sqrt{\rho_s\rho_t}}\left(\Lambda-\frac{1}{a_t}\right)\left(\Lambda-\frac{1}{a_s}\right)~.
\end{equation}
The operator $\mathbf{A}^\pm$ (\cref{eq_weak_axial2}) is written up to NLO. To maintain consistency, the LO interaction is coupled to the NLO three-body wave functions, while the NLO interactions are coupled to the LO wave functions as discussed in Chapter \ref{NLO_corrections}.
% ** triton beta-decay matrix element
\section{$^3$H $\beta$-decay matrix elements}

The triton $\beta$-decay to $^3$He matrix element \cref{eq_beta}, can be calculated using the LO and NLO $A=3$ bound-state wave functions as introduced in Chapter \ref{general_matrix}.

\subsection{$^3$H $\beta$-decay observables}
The half-life of $^3$H $\beta$-decay can be expressed as
 \cite{fermi_reference_1998}: 
\begin{equation}\label{eq_GT}
fT_{1/2}=\frac{K/G_V^2}{\langle F\rangle^2+\frac{f_A}{f_V}\langle GT\rangle^2}~;
\end{equation}
where $ (fT_{1/2})_t=1129.6\pm 3$~s \cite{t_reference} is the triton comparative half-life, $K=2\pi^3\log2/m_e^5$ (with $m_e$ denoting the electron
mass), $G_V$ is the weak interaction vector coupling constant (such
that $K/G_V^2=6146.66 \pm 0.6$ \cite{K_reference}), 
$f_V=2.8355 \cdot 10^{-6}$ and $f_A=2.8506 \cdot 10^{-6}$ are the
Fermi functions calculated by Towner, as reported by Simpson in Ref.~\cite{fv_reference}. %\cblack 
The weak transitions $\langle GT \rangle, \langle F \rangle$ are a result of weak interaction matrix, which can be calculated using the method introduced in Chapter \ref{general_matrix}. 

\subsection{Fermi and Gamow-Teller matrix element:}
For the case of triton $\beta$ decay, it is better to separate the triton into 3 different channels -$d,np,nn$, under the the assumption that for the scattering lengths $a_{nn}=a_{np}=a_s$). Given that, one finds that:
\begin{align}
1&= \sum\limits_{\mu,\nu=t,s,nn}
\bra{\psi^{^3\text{H}}_\mu(E_{^3\text{H}},p')}
\frac{d}{dE}\left [\hat{I}_{\mu\nu}-\right.\left.
My_\mu y_\nu a_{\mu\nu}K_0(p'',p',E)\right]_{E=E_{^3\text{H}}} \ket{\psi^{^3\text{H}}_\nu(E_{^3\text{H}},p'')}~,
\end{align}
%\cblack
where:
\begin{equation}\label{eq_a_mu_nu}
a_{\mu\nu}=\begin{array}{c|ccc}
\mbox{\backslashbox{$\mu$\kern-1em}{\kern-1em$\nu$}}&t&np&nn \\ \hline
t&-1&3&3\\
np&1&1&-1\\
nn&2&-2&0
\end{array}.
\end{equation} is a result of the different projection operators \cite{konig3}. 

\cblack
The Gamow-Teller operator of the triton beta decay, ($\langle GT \rangle$) matrix element is given by:
\begin{multline}\label{Eq_GT}
\langle GT\rangle=\frac{\langle\psi^{^3\text{He}}\|\boldsymbol{A}^{+}\|\psi^{^3\text{He}}\rangle}{\sqrt{2}}=\\{\Bigl\langle\frac{1}{2}\Bigl\|\tau^{+}\Bigr\|\frac{1}{2}\Bigr\rangle}\dfrac{\langle\frac{1}{2}\|\boldsymbol{\sigma}\|\frac{1}{2}\rangle}{\sqrt{2}}g_A\sum\limits_{\mu,\nu}
\bra{\psi^{^3\text{He}}_\mu(E_{^3\text{He}},p')}y_\mu y_\nu\left[d'^{ij}_{\mu\nu} \mathcal{I}^q(E_{^3\text{H}},q_0,p,p')+\right.\\ \left.a'^{ij}_{\mu\nu}\mathcal{K}^q(p,p',E_{^3\text{H}},q_0)\right] \ket{\psi^{^3\text{H}}_\nu(E_{^3\text{H}},p)}
-{g_A}L_{1, A}'\left(\langle\psi^{^3\text{H}}_{nn}|\psi^{^3\text{He}}_t\rangle+\langle\psi^{^3\text{H}}_t|\psi^{^3\text{He}}_{pp}\rangle\right),
\end{multline}
%\cblack
where: \begin{equation}
d'_{\mu\nu}=\begin{array}{c|ccc}
\mbox{\backslashbox{$\nu$\kern-1em}{\kern-1em$\mu$}}&t&np&pp \\
\hline
t&1/3&0&-1\\
np&0&1/3&0\\
nn&-2/3&0&0
\end{array}
\end{equation}
and \begin{equation}
\mbox {$a'^{ij}_{\mu\nu}$}=\begin{array}{c|ccc}
\mbox{\backslashbox{$\mu$\kern-1em}{\kern-1em$\nu$}}&t&np&pp \\ \hline
t&-7/3&1&3\\
np&1&1&-1\\
nn&2/3&-2&-2
\end{array}
\end{equation}
denotes the Gamow-Teller transition.

\begin{figure}[h!]
\begin{center}
\includegraphics[width=1\linewidth]{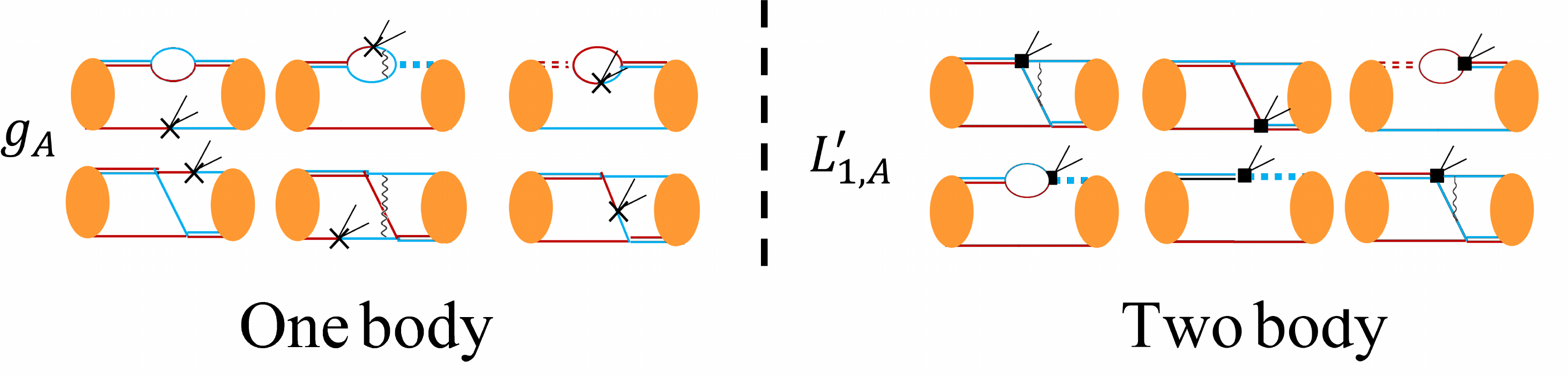}\\
\caption{\footnotesize{Different topologies of the diagrams
contributing to the triton $\beta$-decay amplitude. The LHS of
each diagram is $^3$H while the RHS is $^3$He. The double line
are the propagators of the two dibaryon fields $D_t$
(solid), $D_s$ (dashed for $nn$ and $np$, dotted for $pp$), 
where the red lines are the neutron and the blue lines are the
proton. Most of the diagrams couple both the triplet and the
singlet channels. The diagrams with one-body interactions are
coupled to $g_A$ while the two-body interaction are coupled to
the effective ranges $\rho_t$ and $\rho_s$ and to $l_{1, A}$.}}\label{fig_weak_topo}
\end{center}
\end{figure}

The Fermi matrix element ($\langle F \rangle$) is given by:
\begin{multline}\label{Eq_F}
\langle F\rangle=\frac{\langle\psi^{^3\text{He}}\|V^+\|\psi^{^3\text{He}}\rangle}{\sqrt{2}}=\\{\Bigl\langle\frac{1}{2}\Bigl\|\tau^+\Bigr\|\frac{1}{2}\Bigr\rangle}\sum\limits_{\mu,\nu}
\bra{\psi^{^3\text{He}}_\mu(E_{^3\text{He}},p')}y_\mu y_\nu\left[d'^{ij}_{\mu\nu} \mathcal{I}^q(E_{^3\text{H}},q_0,p,p')+a'^{ij}_{\mu\nu}\mathcal{K}^q(p,p',E_{^3\text{H}},q_0)\right] \ket{\psi^{^3\text{H}}_\nu(E_{^3\text{H}},p)}
\end{multline}
where: \begin{equation}\label{eq_d_fermi}
d'^{ij}_{\mu\nu}=\begin{array}{c|ccc}
\mbox{\backslashbox{$\nu$\kern-1em}{\kern-1em$\mu$}}&t&np&pp \\
\hline
t&1&0&0\\
np&0&1&-1\\
nn&0&2&0
\end{array}
\end{equation}
and \begin{equation}\label{eq_a_fermi}
\mbox{ $a'^{ij}_{\mu\nu}$}=\begin{array}{c|ccc}
\mbox{\backslashbox{$\mu$\kern-1em}{\kern-1em$\nu$}}&t&np&pp \\ \hline
t&-1&3&3\\
np&1&1&1\\
nn&2&-2&-2
\end{array}~,
\end{equation}
$\mu, \nu$ denote the different channels of the three-nucleon wave function ($t, np, pp$ for $^3$He and $t, np,nn$ for $^3$H), where $\psi_{\mu}, \psi_{\nu}$ are the three-nucleon wave functions for the different channels, defined using the homogeneous solution of the three-nucleon scattering amplitude \cite{Big_paper} and $q_0=E_{^3\text{He}}-E_{^3\text{H}}$ is the energy transfer.
\begin{comment}
\begin{equation}\label{eq_psi}
\bra{ E, q}\psi\rangle=\int dq_0\mathcal{B} (E+q_0, q)S(-q_0, -{\bf q})
\mathcal{D} (E+q_0, {\bf q}), 
\end{equation}

where $\mathcal{B}(E,q)$ is the solution of the homogeneous three-nucleon integral equation and $ S (q_0, {\bf q})$ is the single-nucleon propagator: 
\begin{equation}
i S (q_0, {\bf q})=i\left[q_0-\frac{{\bf
q}^2}{2M}+i\varepsilon\right]^{-1}.
\end{equation}
and $\mathcal{D} (E+q_0, {\bf q})$, is the two-nucleon propagator \cite{Big_paper}:
\begin{equation}
\label{NLO_correction_triplet}
\begin{split}
i\mathcal{D} ^{\text{NLO}}_{t,s} (q_0,{\bf q})=
&i\frac{4\pi}{M y_{t,s}^2}
\left (\frac{1}{a_{t,s}}-\sqrt{-M q_0+\frac{{\bf q}^2}{4}}\right)^{-1}\\
&\cdot\left[1-\frac{\rho_{t,s}}{2}\left (\sqrt{{\bf
q}^2/4-Mq_0}+\frac{1}{a_{t,s}}\right)\right].
\end{split}
\end{equation}
\end{comment}
The general diagrammatic form of $^3$H $\beta$-decay, shown in Fig.~\ref{fig_weak_topo}, is similar to the general matrix element introduced in Chapter~\ref{general_matrix}. For both $\langle F\rangle$ and
$\langle GT\rangle$ transitions, the left hand side
bubbles of the diagrams are $^3$H, while the right hand side bubbles are $^3$He. 
\cblack

The one-body diagrams which contain a one-body weak
interaction are coupled to the weak axial LEC, $g_A$, and contribute to both $\langle F\rangle$ and $\langle GT\rangle$ transitions. Similarly to $A=3$ magnetic moments, this one-body diagrams are taken up to NLO, and therefore contain the NLO insertion for the one-body diagrams as discussed in Chapter \ref{NLO_corrections}. The two-body diagrams include a two-body term originating from the ERE term in the Lagrangian: $\frac{1}{2}\frac{\rho_t+\rho_s}{\sqrt{\rho_t\rho_s}}g_A$, and a two-body weak interaction which is coupled to the
unknown LEC
 $l_{1, A}$. The two-body diagrams contribute to the $\langle GT\rangle$ transition only.
% ** Fermi transition 
\section{Fermi operator}
%\cblack
In the absence of the Coulomb interaction, $^3$H is identical to $^3$He and the Fermi
transition is equal to the triton wave function normalization as defined in Chapter~\ref{general_matrix}:
\begin{multline}\label{Eq_F_triton}
\langle F\rangle^{0}=\frac{\langle\psi^{^3\text{H}}\|{\tau}^0\|\psi^{^3\text{H}}\rangle}{\sqrt{2}}={\Bigl\langle\frac{1}{2}\Bigl\|\tau^0\Bigr\|\frac{1}{2}\Bigr\rangle}\cdot\\\sum\limits_{\mu,\nu}
\bra{\psi^{^3\text{H}}_\mu(E_{^3\text{H}},p')}y_\mu y_\nu\left[d'^{ij}_{\mu\nu} \mathcal{I}^{q=0}(E_{^3\text{H}},q_0=0,p,p')+a'^{ij}_{\mu\nu}\mathcal{K}^{q=0}(p,p',E_{^3\text{H}},q_0=0)\right] \ket{\psi^{^3\text{H}}_\nu(E_{^3\text{H}},p)}=1~,
\end{multline}
where: 
\begin{align}
d'^{ij}_{\mu,\nu}&=\delta_{\mu,\nu}\label{eq_d_mu_nu} \\
a'^{ij}_{\mu\nu}&=a_{\mu\nu}
\end{align}

%Where $\psi$ is the defined using the homogeneous solution to three-nucleon scattering amplitude (\cref{eq_psi}). 

\begin{comment}
\begin{multline}
\langle
F\rangle_{\alpha=0}=
\sum_{A, B}\Bigg[\int\frac{d^3p_i}{ (2\pi)^3}\int\frac{d^3p_j}{ (2\pi)^3}{\psi_A (E_{{^3\text{H}}}, p_i, )}\cdot\\
\mathcal{O}^q_{j,i} (p_i, p_j)\psi_B (E_{{^3\text{H}}}, p_j)\Bigg]\delta^{S_j}_{S_i}. 
=1
\end{multline}

Where 
\end{comment}

From comparison between \cref{eq_a_mu_nu,eq_d_mu_nu} and \cref{eq_a_fermi,eq_d_fermi} we expect that $\langle F\rangle=1-\epsilon$ \cite{fermi_reference_1998}, where $\epsilon\ll1$, originates mostly from the isospin breaking due to the Coulomb interaction. We can therefore isolate the effect of isospin breaking and the one-photon exchange diagrams on the Fermi transition and compare them later to the Gamow-Teller calculations. In this section, we present our calculations of the Fermi transition. First, we calculate the Fermi transition in the absence of Coulomb interaction, but and under the assumption that $a_{nn, np}\neq a_{pp}$. Second, we calculate the Fermi transition with $\alpha \neq 0$, and, obviously, $a_{nn, np}\neq a_{pp}$, as a result for both Z- and ERE-parameterization. All these calculations result from the left hand side of the diagrams in Fig.~\ref{fig_weak_topo}.

\begin{figure}[h!]
	\centerline{
		\includegraphics[width=0.75\linewidth]{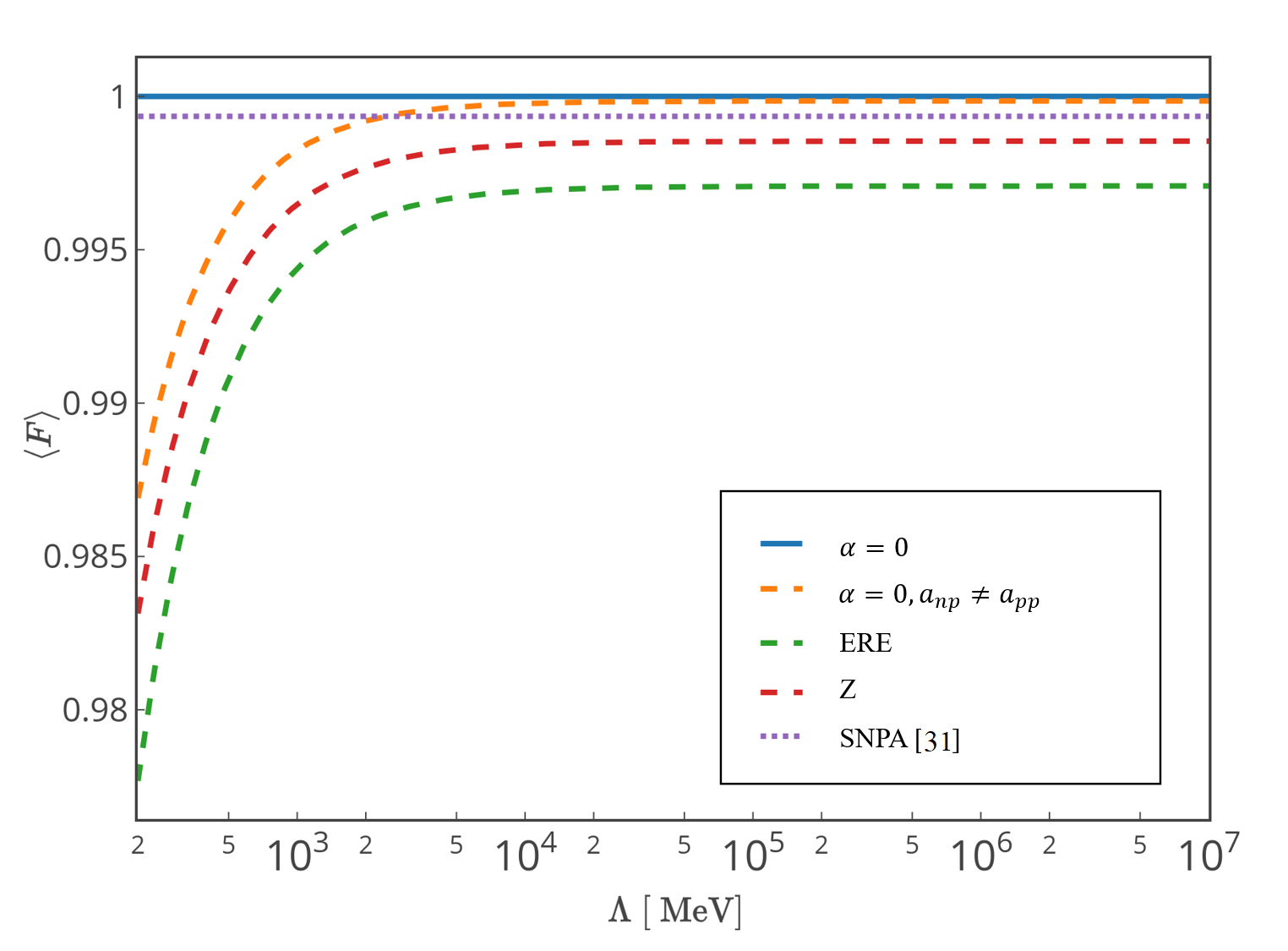}}
	\caption{\label{fig_tauminus} \footnotesize{Numerical results
			for the Fermi transition. The solid line is the LO result for
			$\langle F \rangle=1$ with $\alpha=0$. The dashed
			line is the numerical result for $\alpha=0$ with isospin
			breaking effects in the scattering lengths $a_{np}\neq a_{pp}$
			(LO). The dashed-dotted line shows the numerical result at NLO
			with $\alpha\neq0$. The dotted line gives the value of
			$\langle F \rangle=0.9993$ from Ref.~\cite{fermi_reference_1998}.}}
\end{figure}

We use the experimental data shown in Tab.~\ref{table_1} as input for our numerical calculations shown in Fig.~\ref{fig_tauminus} and Tab.~\ref{table_F}.

\begin{table}[H]
	\begin{center}
		\begin{tabular}{l |c}
			\centering
			&$\langle F\rangle$\\
			\hline
			One-body, LO $\alpha=0$& 1\\
			One-body, LO $\alpha=0, a_{np} \neq a_{pp}$& 0.9999\\
			LO, ERE &0.9971\\
			LO, Z &0.9985\\
			SNPA \cite{fermi_reference_1998}&0.9993
		\end{tabular}
		\caption{\footnotesize{Numerical results of $\langle F \rangle$. Note
				that the second row is without explicit Coulomb force
				($\alpha=0$), but with isospin breaking in the scattering
				lengths, {\it i.e.,} with the physical values for the scattering
				lengths $a_{np} \neq a_{pp}$.}}
		\label{table_F}
	\end{center}
\end{table}

\cblack
Our numerical result compares well to the $\langle F \rangle$ standard nuclear
physics approach (SNPA) calculation by Schiavilla {\it et al.}~
\cite{fermi_reference_1998}. The SNPA calculation involves nuclear wave functions
derived from high-precision phenomenological nuclear potentials, and
one-nucleon and two-nucleon electro-weak currents. 

% ** GT operator
\section{Gamow-Teller operator}

% *** GT calculation
In contrast to the Fermi transition, the Gamow-Teller transition also involves two-body operators at NLO. The diagrams which contain a one-body weak interaction are coupled to $g_A$ and contain one ERE insertion up to NLO. The two-body diagrams are coupled to $\frac{\rho_t+\rho_s}{2\sqrt{\rho_s\rho_t}}-l_{1,A}(\Lambda)$. By summing over all diagrams and comparing the resulting sum to the triton half-life, \cite{Chou}, $l_{1, A}(\Lambda)$ can be extracted, as will be discussed later in this chapter. We used the experimental input parameters shown in Tab. \ref{table_1} for all numerical calculations.

\subsection{Gamow-Teller Transition for $\alpha=0$}
As mentioned before, in the absence of Coulomb interaction, the Fermi transition matrix element at LO is 1 ({\it i.e.} $\alpha=0$). Similarly, the LO matrix element of the Gamow-Teller transition with $\alpha=0$, easily found to be: 
\begin{align}
\nonumber
&\langle GT\rangle^{\text{LO}}_{\alpha=0}=
\dfrac{\langle\frac{1}{2}\|\boldsymbol{\sigma}\|\frac{1}{2}\rangle}{\sqrt{2}}{\Bigl\langle\frac{1}{2}\Bigl\|\tau^0\Bigr\|\frac{1}{2}\Bigr\rangle}\cdot\\
\nonumber
&\sum\limits_{\mu,\nu}
\bra{\psi^{^3\text{H}}_\mu(E_{^3\text{H}},p')}y_\mu y_\nu\left[\delta_{\mu,\nu} \mathcal{I}^{q=0}(E_{^3\text{H}},q_0=0,p,p')+a_{\mu\nu}\mathcal{K}^{q=0}(p,p',E_{^3\text{H}},q_0=0)\right] \ket{\psi^{^3\text{H}}_\nu(E_{^3\text{H}},p)}=\\
&\qquad\qquad\dfrac{\sqrt{6}}{\sqrt{2}}=\sqrt{3},
\end{align}
where $a_{\mu\nu}$ is given in \cref{eq_a_mu_nu}. We have performed this
calculation in two ways: one with $\alpha=0$ for both the scattering
amplitude and the matrix element, and second with $\alpha= 0$ for
the matrix element, but for different scattering lengths, 
similarly to the Fermi case. From Tab.~\ref{table_GT}, one sees that the effect of different scattering lengths leads to $< 0.6\%$ difference for both Fermi and Gamow-Teller transitions. These results imply that for the Fermi and Gamow-Teller transitions, the Coulomb interaction has to be taken non-perturbatively, since its contribution for the matrix element is not neglectable. As for the scattering length difference, Tabs.~\ref{table_F} and \ref{table_GT}
indicate that up to LO, the scattering length difference has a small effect on both transitions, which can be estimated perturbatively. 

% *** numerical results
\subsection{Numerical results for GT strength of triton decay}
\begin{figure}[h!]
	\centerline{
		\includegraphics[width=.7\linewidth]{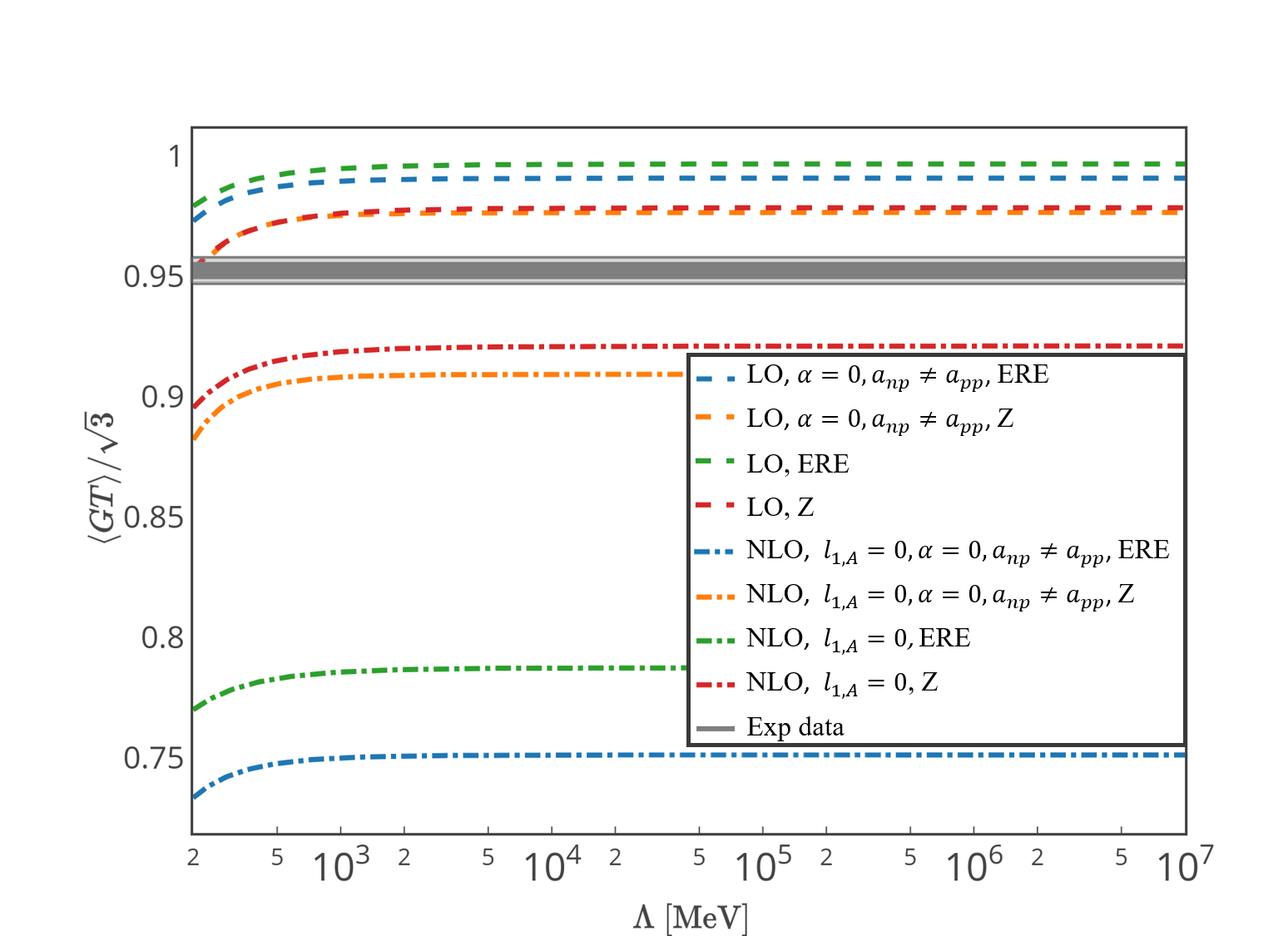}}
	\caption{\label{fig_full_GT} \footnotesize{Numerical results of the
			Gamow-Teller transition. %the solid line is LO calculating without
			%Coulomb interaction$ (\sqrt{3})$.
			The gray area is the full
			$\langle GT \rangle$ matrix element with $g_A=1.273\pm 0.003\pm0.005$ \cite{gA, 
				doi:10.1063/1.4983578}. The short dashed lines are the numerical results of
			$\langle GT\rangle^{\text{LO}} $, the dashed-dotted lines arew $\langle GT \rangle$ with $,l_{1, A}=0$.
			%, %while the long dashed line is the NLO calculating without
			%Coulomb interaction.
			$Z$ refers to numerical results
			with the $Z$-parameterization: $\rho'_t=\frac{Z_d-1}{\gamma_t}$. }}
\end{figure}

Our $\langle GT \rangle$ numerical results for both NLO arrangements are shown in Tab.~\ref{table_GT} and in
Fig.~\ref{fig_full_GT}. The full NLO result with
$l_{1, A}=0$ includes both, one-body and two-body terms which contribute
to $\langle GT \rangle$, without the diagrams that are coupled to
$l_{1, A}$.
%\begin{comment}

%\ebd}
%\begin{comment}
%\end{comment}
\cblack
\begin{table}[H]
	\begin{center}
		\begin{tabular}{l| c|c}
			\centering
			&$\langle GT \rangle$,ERE&$\langle GT \rangle$, $Z$\\
			\hline
			One-body, LO $\alpha=0$& $\sqrt{3}$&$\sqrt{3}$\\
			One-body, LO $\alpha=0$, $a_{np}\neq a_{pp}$&1.716&1.692 \\
			One-body, LO& 1.727& 1.695
			\\
			Full NLO, $l_{1, A}=0$, $\alpha=0$, $a_{np}\neq a_{pp}$&1.301&1.575\\
			Full NLO, $l_{1, A}=0$&1.383&1.596\\
		\end{tabular}
		\caption{\footnotesize{Numerical results for $\langle GT \rangle$. Note that the rows with the comment
				''$\alpha=0$, $a_{np}\neq a_{pp}$'' are without explicit Coulomb force
				($\alpha=0$), but with isospin breaking in the scattering
				lengths, {\it i.e.,} with physical values for the scattering
				lengths $a_{np} \neq a_{pp}$.}}
		\label{table_GT}
	\end{center}
\end{table}
\cblack
\section{Empirical extraction of Gamow-Teller strength and fixing $l_{1, A}$}
The GT matrix element can be extracted from a triton half-life calculation using \cref{eq_GT}. The axial
coupling constant $g_A$ has been remeasured recently, leading to results
whose range is much bigger than the current recommendation. To be on
the safe side, we take $g_A=1.273\pm 0.003\pm0.005$ \cite{doi:10.1063/1.4983578, 
	gA}. The first uncertainty in $g_A$ arises from the difference
between the measurements of Refs.~\cite{gA, doi:10.1063/1.4983578}, and the second
uncertainty is the statistical experimental uncertainty. In order to
extract the Gamow-Teller strength, we use our prediction for the Fermi transition: $\langle F\rangle=0.9993$ \cite{fermi_reference_1998}. At large cutoff values we
find that the empirical GT strength is
$\langle GT \rangle_\text{emp}= \sqrt{3}\frac{1.213\pm0.002}{g_A}$ \cite{fermi_reference_1998}.
The uncertainty here originates mainly from the uncertainty in the
triton half-life.

\begin{figure}[h!]
	\centerline{
		\includegraphics[width=0.7\linewidth]{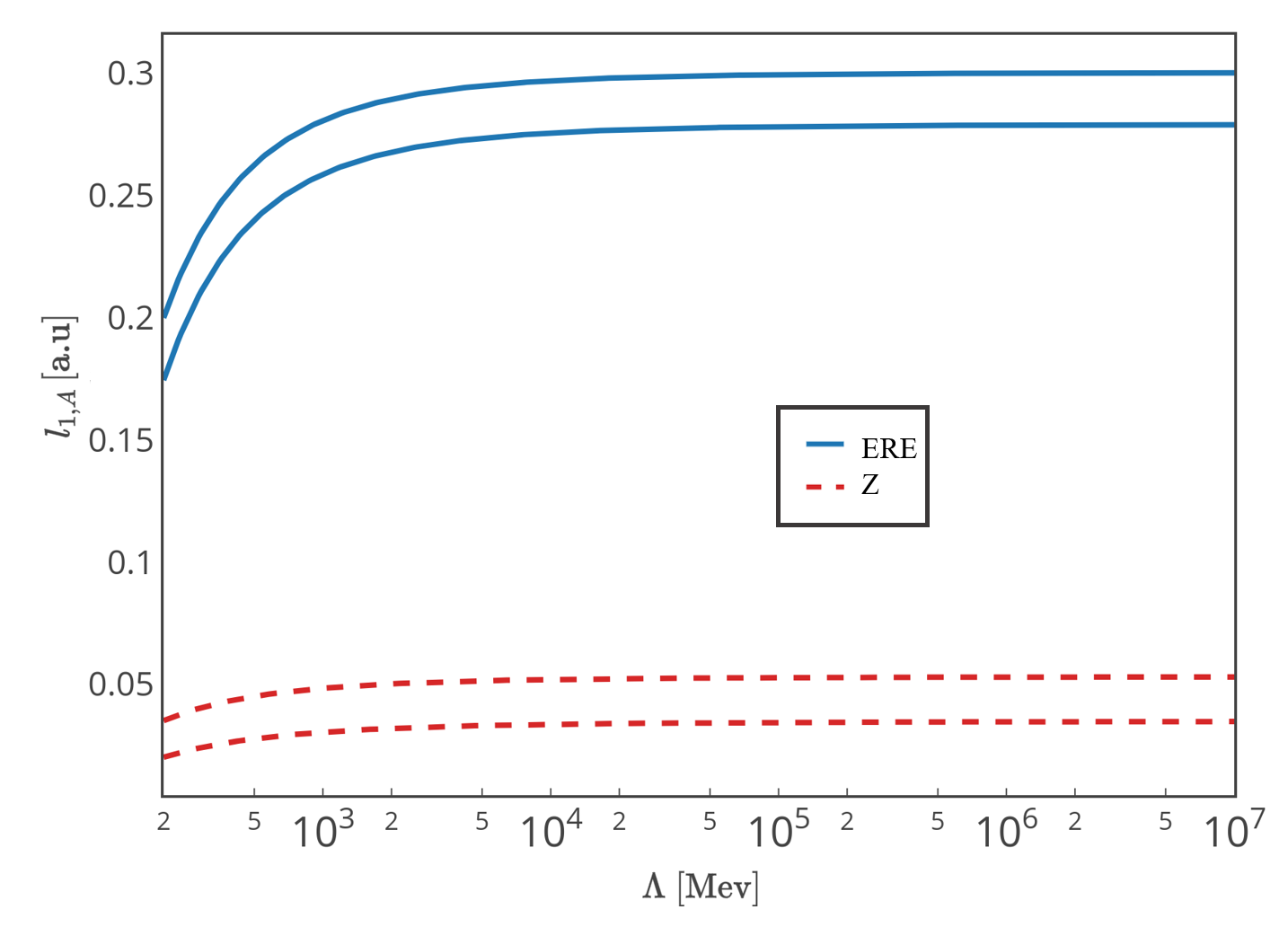}}
	\caption{\label{fig_L1A} \footnotesize{Numerical results of the RG invariant, 
			$l_{1, A}$, with $g_A=1.273(03)(05)$ \cite{gA, 
				doi:10.1063/1.4983578}. The solid lines are the upper and lower limits of the ERE calculations while the dashed lines are the upper and lower limits of the $Z$ calculations.}}
\end{figure}

The difference between the empirical GT strength and the numerical
result for the GT-transition at NLO is used to fix $l_{1, A}$ such that: 
\begin{equation}\label{eq_L1A_emp}
\l_{1,A}(\Lambda)= \frac{\langle GT \rangle_{\text{emp}}-\langle GT\rangle_{l_{1, A}=0}^{\text{NLO}}}{\langle GT\rangle_{l_{1, A}}^{\text{NLO}} }, 
\end{equation}
where $\langle GT\rangle_{l_{1, A}}^{\text{NLO}}$ are the two-body diagrams that contribute to the triton $\beta$-decay and coupled to $l_{1, A}$, while $\langle GT\rangle_{l_{1, A}=0}^{\text{NLO}}$ is the sum over all the diagrams that contribute to the triton $\beta$-decay \textbf{without} the diagrams coupled to $l_{1, A}$. 
%שונו בהתא %\cblack
The numerical results for $l_{1,A}$ for both ERE- and
$Z$-parameterizations are shown in Fig.~\ref{fig_L1A}.

Importantly, we find numerically, that for both parameterizations, $l_{1,A}$ is RG invariant, a fact that has been already
predicted by theory \cite{2000NuPhA.675..575B}, where for $\Lambda \rightarrow \infty$:
\begin{subequations}\label{eq_fineal_l1A}
	\begin{align}
	l_{1, A}^{\infty}
	&=0.055(0.317)\pm0.004\pm0.001\label{eq_L1A_ERE_1},\qquad g_A=1.2701~,\\ 
		l_{1, A}^{\infty}&=0.047(0.308)\pm0.004\pm0.001\label{eq_L1A_Zd_1},\qquad g_A=1.2767~,
	\end{align}
\end{subequations}
The nominal value is calculated using $Z$-parameterization, while the number in brackets is calculated using the ERE-parameterization. The first uncertainty comes from $g_A$ experimental uncertainty and the second uncertainty comes from the rest of the 
experimental uncertainties, such as the statistical uncertainties in the measured triton half-life.

\cblack
\chapter{The proton-proton fusion matrix element}
The energy generated in the Sun comes from an exothermic set of reactions, named the proton-proton ($pp$) chain, which fuses four Hydrogen ions into $^4$He: 
\begin{equation}\label{eq_Sun}
4p\rightarrow ^4\text{He}+2e^++2\nu_e.
\end{equation}
The leading reaction in this chain is the $pp$ fusion: 
\begin{equation}\label{eq_pp1}
p+p\rightarrow d+e^++\nu_e.
\end{equation}
This reaction is governed by the weak interaction, which makes it the slowest reaction (by far) in the whole chain, and therefore it determines the Sun's lifetime.
This fusion is the main contributor to the solar neutrinos \cite{solar1, solar2, solar3}.
Since a direct measurement of its cross-section is impossible, the fusion rate estimates depend entirely on theory \cite{pp_review}. %The above reaction (\cref{eq_pp1}) is characterized by very low-energy, and therefore can be described accurately using \pilesseft~\cite{Kong1, Kong2, Ando_proton, Proton_Proton_Fifth_Order}.% An addition of a weak interaction to nuclear system entails two universal coupling constants: 
%$g_{A}$, the axial coupling constant for a single nucleon, known from neutron $\beta$-decay, and analogously, $L_{1, A}$ for two-body interaction.
 Since the $pp$ fusion is a weak reaction, calibrating $L_{1, A}$ with sufficient accuracy is crucial for predicting its rate.

The $pp$ fusion rate, i.e. the number of reactions per unit of time per volume, is given by: 
\begin{equation}
R_{pp}=\frac{N_p}{2}\langle \sigma\nu\rangle,
\end{equation}
where
$N_p$ is the proton number density, $\sigma$ is the cross-section in the center of the mass and $\nu$ is the proton center-of-mass thermal average velocity, assuming a Boltzmann distribution with temperature of the solar core, {\it i.e.,} approximately 15 million degrees. This defines low-energy scale in the nuclear regime ($\approx 1.5 keV \ll E_{d}$).

The astrophysical $S$-Factor is defined in the terms of the cross-section \cite{solar2}: 
\begin{equation}
S^{11}(E)=\left(\sigma E\right)e^{2\pi\eta(E)}
\end{equation} 
where $\eta(E)=\frac{2\pi}{\hbar\nu}$ is the Coulomb parameter in terms of $E$, where $E$ is the kinetic energy of the center of mass of the interacting protons, dictated by the temperature. In solar conditions, the magnitude of $E$ is only a few $\kev$s and therefore $S^{11}(E)$ can be expanded in a power series in $E$:
\begin{equation}
S^{11}(E)=S^{11}(0)+{S^{11}}'(0)E+\frac{{S^{11}}''(0)E^2}{2}+...
\end{equation}
where (\cite{pp_review}): 
\begin{equation}
S^{11}(0)=4.01\left(\frac{g_A}{1.2695}\right)^2\frac{\Lambda_{pp}(0)^2}{7.035}10^{-23}\mev\cdot\fm^2, \end{equation}
where $\Lambda_{pp}(0)$ is the weak interaction matrix element:
\begin{equation}
\Lambda_{pp}(0)\propto\bra{d}\mathcal{L}_{weak}\ket{pp}, 
\end{equation}
 $\mathcal{L}_{weak}$ is the \pilesseft interaction Lagrangian, introduced in Chapter \ref{beta_decay}.

\section{{\it pp} fusion matrix element}
\begin{figure}[h!]
\begin{center}
\includegraphics[width=0.995\linewidth]{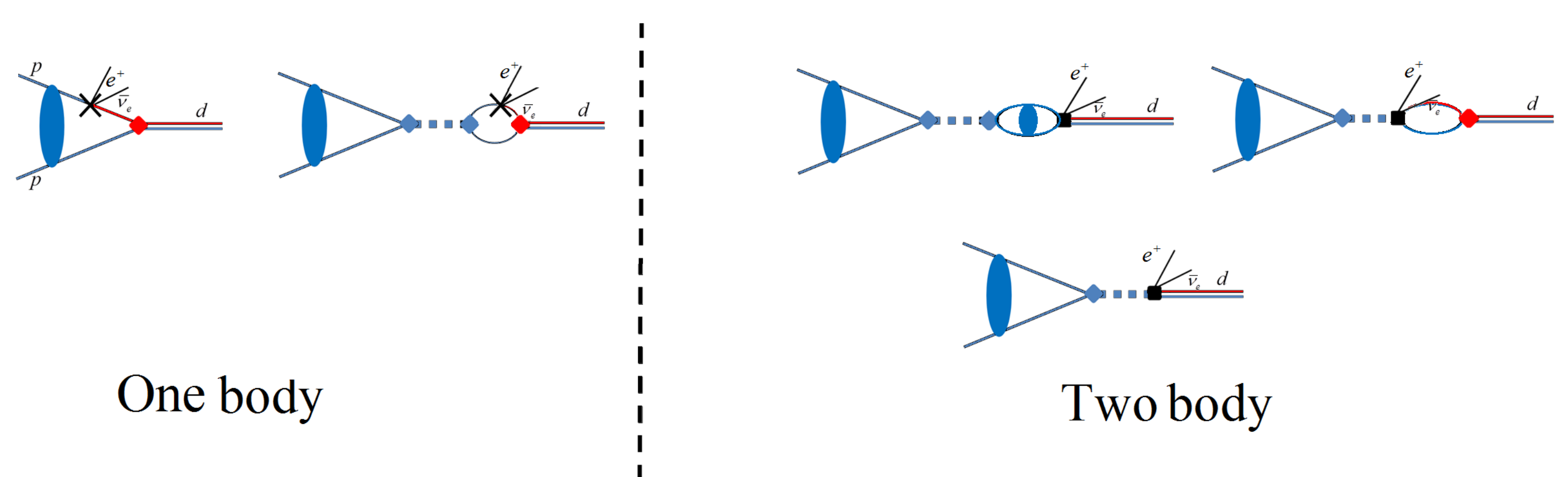}
\caption{\footnotesize{Diagrams for the $pp$ fusion process, \cref{eq_pp1}, up to NLO.
The double lines are the NLO propagators of the two dibaryon fields: $D_t$ (solid) and $D_{pp}$ (dotted). Two-nucleon propagator with
a blue bubble denotes the Green$'$s function including the Coulomb potential. A spin-singlet dibaryon-nucleon-nucleon (s-NN) vertex is proportional to $y_s$ (blue diamond) while a spin-triplet dibaryon-nucleon-nucleon (t-NN) vertex is proportional to $y_t$ (red diamond). The one-body diagrams are coupled to the one-body LEC, $g_A$ where the two-body diagrams are coupled to the two-body LEC, $L'_{1, A}$.}}
\label{pp_fusiton}
\end{center}
\end{figure}

 The diagrams contributing to the $pp$ fusion matrix element, $\Lambda_{pp}(0)$ (up to NLO) are shown in Fig \ref{pp_fusiton}. %(see \cite{konig3, konig4} for Coulomb-modified-effective range expansion)
The sum over all $pp$ fusion diagrams up to NLO in $\frac{Q}{\Lambda_{cut}}$ is given by (see Refs.~\cite{Proton_Proton_Fifth_Order, Ando_proton}): 
\begin{equation}
T_{fi}=\bra{d}\vec{\mathcal{A}}^-\ket{pp} =\sqrt{\frac{8\pi}{\gamma_t^3}}{C_\eta e^{i\sigma_0}}\biggl\{\sqrt{Z_d^{\text{NLO}}}\left[e^\chi -M\alpha a_pI(\chi)\right]+\sqrt{Z_d^{\text{LO}}}\frac{a_p\gamma_t^2}{2}\sqrt{\rho_t\rho_s}\left(\dfrac{\rho_t+\rho_s}{2\sqrt{\rho_t\rho_s}}-l_{1,A}\right)\biggr\},
\end{equation}
where $Z_d^{\text{LO}}, Z_d^{\text{NLO}}$ are the deuteron LO and NLO normalizations, respectively (\cref{eq_Zd_Q,eq_Zd_2}), $\chi=\frac{\alpha M}{\gamma_t}$ and $\sigma_0$, $C_\eta,a_{p},\rho_{t,s},\alpha,M$ and $\gamma_t$ were introduced in Chapter \ref{formalism}. The integral $I(\chi)$ is defined as: 
\begin{equation}
I(\chi)=\int dx\frac{\chi e^{\frac{x}{\pi} \arctan\left(\frac{\pi \chi}{x}\right)}}{\left(e^x-1\right) \left(\pi^2\chi^2+x^2\right)},
\end{equation}
and $\Lambda_{pp}(0)$ is given by \cite{Kong1, Kong2}:
%we get that(for $\mu=\mu$ ): 
\begin{equation}\label{eq__Lambda_fineal}
\begin{split}
\Lambda_{pp}(0, \mu) &=\gamma_t^2\sqrt{\frac{\rho_t}{8\pi C_\eta^2}}\left|T_{fi}(0,\mu)\right|=\\
&\sqrt{Z_d^{\text{NLO}}}\left[e^\chi -M\alpha a_pI(\chi)\right]+\sqrt{Z_d^{\text{LO}}}\frac{a_p\gamma_t^2}{2}\sqrt{\rho_t\rho_s}\left(\dfrac{\rho_t+\rho_s}{2\sqrt{\rho_t\rho_s}}-l_{1,A}\right).
\end{split}
\end{equation}
For the ERE-parameterization, \cref{eq__Lambda_fineal} becomes:
\begin{equation}
\begin{split}
\Lambda_{pp}(0, \mu)=
\left(1+\frac{\gamma_t\rho_t}{2}\right)\left[e^\chi -M\alpha a_pI(\chi)\right]+\frac{a_p\gamma_t^2}{2}\sqrt{\rho_t\rho_s}\left(\dfrac{\rho_t+\rho_s}{2\sqrt{\rho_t\rho_s}}-l_{1,A}\right)~,
\end{split}
\end{equation}
while for the $Z$-parameterization, we find that: 
\begin{equation}
\begin{split}
\Lambda_{pp}(0, \mu)=
\sqrt{Z_d}\left[e^\chi -M\alpha a_pI(\chi)\right]+\frac{a_p\gamma_t^2}{2}\sqrt{\rho_t\rho_s}\left(\dfrac{\rho_t+\rho_s}{2\sqrt{\rho_t\rho_s}}-l_{1,A}\right).
\end{split}
\end{equation}
Similarly to the electromagnetic observables we define $l'_{1, A}$ such that: 
\begin{eqnarray}
\Lambda^{\text{ERE}}_{pp}(0, \mu)&=&\left(1+\frac{\gamma_t\rho_t}{2}\right)\left[e^\chi -a_p\gamma_t\chi I(\chi)\right]+\dfrac{a_p\gamma_t^2}{4}(\rho_t+\rho_s)-{l'_{1, A}}(\mu)\\
\Lambda_{pp}^{Z}(0, \mu)&=&\sqrt{Z_d}\left[e^\chi -a_p\gamma_t\chi I(\chi)\right]+\dfrac{a_p\gamma_t^2}{4}(\rho_t+\rho_s)-{l'_{1, A}}(\mu),
\end{eqnarray} 
with
%\cblack
 \begin{equation}\label{l1A}
 l'_{1, A}(\mu)=\frac{a_p\gamma_t^2}{2}\sqrt{\rho_t\rho_s}l_{1,A}(\mu).
 \end{equation}
 \cblack

%The astrophysical factor $S^{11}(0)$ is given by: 
%\begin{equation}
%S^{11}(0)=4.01\left(\frac{g_A}{1.2695}\right)^2\frac{\Lambda^2}{7.035}
%\end{equation}

\section{Numerical results}
We calculate the matrix element, $\Lambda_{pp}(0)$, of \cref{eq__Lambda_fineal} using the parameters introduced in Tab.~\ref{table_1} and for $\chi=0.1494$, so $I(\chi)=4.97311$. 

For $l'_{1, A}=0$ we get: 
\begin{eqnarray}
\Lambda_{pp}(0)&=&2.67(2.55	), 
\end{eqnarray}
where the nominal value is the one produced using the $Z$-parameterization, and the ERE-parameterization value is given in brackets.
%\cblack
For $l'_{1, A}=-0.025
(-0.146
)\pm0.002\pm0.001 \quad g_A=1.2701$, $l'_{1, A}=-0.021(-0.141)\pm0.002\pm0.001 \quad g_A=1.2769$, calculated from \cref{eq_fineal_l1A}, \cblack we can predict the $pp$ fusion rate, accompanied by 
theoretical and empirical uncertainties. In the next chapter, we will analyze the \pilesseft uncertainties due to both sources using the electromagnetic observables, which will enable us to predict the $pp$ fusion.
\chapter{Analysis of the results and a prediction for $pp$ fusion}\label{discussion}
The purpose of this work is to establish a consistent method for calculating electro-weak (EW) matrix elements of $A<4$ nuclei, focusing on the prediction for the $pp$ fusion rate. The approach used in this work, the \pilesseft, like any other EFT, is useful for robust and reliable theoretical uncertainty estimates due to the neglected orders in the EFT expansion. The results presented in the previous chapters show that \pilesseft is both viable and attractive theory for describing electromagnetic observables of the $A=2, \, 3$ systems and for predicting RG invariant electro-weak LECs up to NLO with high accuracy. In this chapter we use these results to provide a theoretical uncertainty estimate for the electromagnetic observables \cite{magnetic_moments, doron_fake}. 
\cblack Also, this chapter presents an analysis of the similarity between the \pilesseft characteristics of electromagnetic and weak observables which enables us to estimate the theoretical uncertainty for the weak observables and to predict the $pp$ fusion rate with sufficient certainty. 
\cblack

\section{An order-by-order analysis of the results and theoretical uncertainty}

%\cblack
The \pilesseft presented here is purely perturbative in effective range expansion, \cblack {\it i.e.,} consistently organizes the expansion in a perturbative manner without including any higher order terms. Moreover, an order-by-order renormalization was obtained, as shown numerically, by the cutoff invariance. In chiral effective field theory ($\chi$EFT), a cutoff variation is frequently used to obtain an uncertainty estimate for any calculation. A main advantage of using
\pilesseft is the cutoff invariance, which is obtained at a natural scale of no more than a few times the physical breakdown scale ($\Lambda\sim \text{few } m_\pi$). This is a measure of the irrelevance of regularization effects and does not only remove questions regarding residual cutoff dependencies that might contribute to the total uncertainty \cite{Kirscher:2017fqc, 2016PhLB..755..253K}, but also allows giving a physical meaning to the size of the NLO contribution. \cblack Since, in a non-renormalizable theory, strength can be shifted from one order to the other via similarity transformations \cite{Bogner:2009bt}. %We note that, by the theoretical expectations, the strength of the pure two-body low-energy constants is RG invariant~\cite{Chen:1999tn}.

\cblack The order-by-order RG invariance indicates that analyzing the results can be accomplished by studying the order-by-order contributions. We divide the calculation to three contributions: LO, NLO range corrections, and short-range contributions, proportional to $l'^\infty_{1, 2}$ and $l_{1, A}$.
\cblack 
Table.~\ref{Tab_NLO_parts} presents these three contributions to the different electro-weak matrix elements, for two NLO arrangements: the ERE- and $Z$-parameterizations (see section \ref{Zd}). For the $pp$ fusion ($A=2,\Lambda_{pp}(0)$), the short-range corrections were calculated using $l_{1,A}$ fixed by the $A=3$ observables ($\langle GT \rangle,\langle F \rangle$) while the electromagnetic $A=2,3$ observables were calculated using ${l'}^\infty_{1,2}$ fixed by comparing our numerical results with experimental data of all combinations of two of the four $A=2,3$ observables (see Chapter~\ref{Magnetic}). The Gamow-Teller (GT) range correction is given by the difference between $\langle GT \rangle_{\text{emp}}$ and our NLO predictions as discussed in Chapter \ref{beta_decay}.

	\begin {table}[h!]
\centering
\begin{tabular}{c|c| c| c|c}
	&\multirow{1}{*}{EW}&\multirow{1}{*}{NLO/LO}&NLO/LO&NLO/LO, \\ 
	&matrix&\multirow{2}{*}{total}&range &LECs\\
	&element& & corrections&contributions\\
	\hhline{=|=|=|=|=}	\multirow{4}{*}{\rotatebox[origin=c]{90}{\parbox[c]{2cm}{\centering Electro-magnetic observable}}}
	&$\langle\hat{\mu}_{^3\text{H}}\rangle$ [nNM]&7\% (0.1\%)&3\% (11\%)&5\% (10\%)\\
	&$\langle\hat{\mu}_{^3\text{He}}\rangle$[nNM]&16\% (4\%)&4\% (25\%)&12\% (29\%)\\
	&$\langle\hat{\mu}_{d}\rangle$ [nNM]			&	1\% (1\%)&0\% (0\%)&1\% (1\%)\\
	&$Y'_{np}$									&6\% (9\%)&2\% (2\%)&4\% (12\%)\\
	\hline
	\multirow{4}{*}{\rotatebox[origin=c]{90}{\parbox[c]{2cm}{\centering Weak observable}}}&$A=3$ $\langle GT \rangle$, $g_A=1.2701$ &2\%(4\%)&6\%(21\%)&3\% (16\%)\\
	&$A=3$ $\langle GT \rangle$, $g_A=1.2769$ &3\%(5\%)	&6\%(21\%)&3\% (16\%)\\
	&$\Lambda_{pp}(0),$ $g_A=1.2701$&6\%	(6\%)&	4.8\%(0.3\%)&1.3\% (5.6\%)\\
	&$\Lambda_{pp}(0),$ $g_A=1.2769$&6\%	(6\%)&	4.8\%(0.3\%)&1.1\% (5.4\%)\\
\end{tabular}
\vspace{0.2 cm}
\caption{ \footnotesize{The order-by-order expansion of the electro-weak matrix elements. The order-by-order expansion of the electromagnetic matrix elements is based on their mean values given in Tab.~\ref{table_all_M1}. The nominal value is calculated using $Z$-parameterization, while the number in brackets is calculated using the ERE-parameterization. Mean denotes the mean value of the $M_1$ observable based on its three predictions while $\Delta$ denotes the standard divination of these predictions.}}
\label{Tab_NLO_parts}
\end{table}
From Tab~\ref{Tab_NLO_parts} we find that the NLO contributions for all electro-weak matrix elements are smaller than the LO part, and in particular for the case of the Z-parameterization, which leads to a better convergence up to NLO.

Thus, in the next subsections, we examine the \pilesseft NLO contributions and estimate its truncation error using the well-measured electromagnetic observables: $\langle\hat{\mu}_{^3\text{H}}\rangle$, $\langle\hat{\mu}_{^3\text{He}}\rangle$, $\langle\hat{\mu}_d\rangle$, $\sigma_{np}$ only for the Z-parameterization, which shows more natural convergence pattern, and therefore has a better predictive power. The large variations and fluctuations of the ERE-parameterization raise questions about its relevance at
NLO for predictions of electro-weak observables.
\cblack

\cblack
\subsection{$M_1$ observables as a case study}
The electromagnetic observables: the three magnetic monuments, $\langle\hat{\mu}_{^3\text{H}}\rangle$, $\langle\hat{\mu}_{^3\text{He}}\rangle$, $\langle\hat{\mu}_d\rangle$, and the radiative capture cross-section, $\sigma_{np}$, are all well-measured, which makes them a perfect tool for verifying and validating \pilesseft at NLO. In this section, we examine the two-body electromagnetic LECs calibrated from $A = 2,3$ observables 
(see Tab.~\ref{table_all_M1} in Chapter~\ref{Magnetic}) for better understanding the consistency of \pilesseft in making predictions, as well as in estimating theoretical uncertainties and truncation errors.

\subsection{Pionless EFT expansion and setting ${l'}^\infty_{2}$ to zero}

We found (in Chapter~\ref{Magnetic}) qualitative differences between the ${l'_{1}}^{\infty}$ and ${l'_{2}}^{\infty}$ standard deviation. While $\Delta{l'_{1}}^{\infty}/{l'_1}\approx3\%$, the standard deviation of ${l'_{2}}^{\infty}$ is of the same order of magnitude as ${l'_{2}}^{\infty}$, \textit{i.e.,} $ \Delta{l'_{2}}^{\infty}/{l'_{2}}^{\infty}\approx70\%$, so $l'^\infty_2$ is consistent with zero. In addition, the NLO contribution to $\langle \hat{\mu}_{d}\rangle$ (the $M_1$ observable that depends on ${l'}^\infty_{2}$ only) originates from LECs, is much smaller compared to other $M_1$ observables, as shown in Tab.~\ref{Tab_NLO_parts}. This means that $l'^\infty_2$ might be regarded as higher order than NLO. This is an amazing result, which will be discussed also in future publications.

Table~\ref{table_all_M1_l2_0} shows our predictions for ${l'}^\infty_{1}$and $M_1$ observables up to NLO for the case in which ${l'_{2}}^{\infty}=0$. Similarly to Tab.~\ref{table_all_M1}, for each row, the $'\star'$ notes the $M_1$ observable used for ${l'}^\infty_{1}$ calibration.
\begin {table}[H]
\centering
\begin{tabular}{c| c| c|c|c}
	& ${l'_{1}}^{\infty}\cdot 10^{-2}$&$\langle\hat{\mu}_{^3\text{H}}\rangle$%[nNM]
	& $\langle\hat{\mu}_{^3\text{He}}\rangle$%[nNM]
	&$Y'_{np}$\\ 
	\hhline{=|=|=|=|=}
	&4.36	&	$\star$		&-2.10 	&1.250\\
	&4.97	&2.996 	&$\star$		&1.256\\
	&4.66	&2.987 	&-2.11 	&$\star$\\
	\hline								
	\multicolumn{1}{c|}{Mean}	 	& 4.7	&2.99&-2.11	&1.253\\
	\hline								
	\multicolumn{1}{c|}{$\Delta$}	&0.6	&0.01 	&0.01 	&0.006 	\\
	\hline
	\multicolumn{1}{c|}{$\%$NLO/LO}	&				&7\% 	&13\% 	&	5\% \\
	\hline
	\multicolumn{1}{c|}{Exp data}&					&	2.979		&	-2.128		&	1.253
\end{tabular}
\vspace{0.2 cm}
\caption{ \footnotesize{Numerical results for our prediction for ${l'_{1}}^{\infty}$ and $A=2,3$ $M_1$ observables with ${l'_{2}}^{\infty}=0$. The $A=3$ magnetic moments are given in units of nuclear magnetons ([nNM]).}}
\label{table_all_M1_l2_0}
\end{table}
It is clear that setting ${l'}^\infty_{2}$ to zero does not reduce the precision and accuracy of ${l'}^\infty_{1}$ and $M_1$ predictions. 
This implies that the ${l'}^\infty_{2}$ contribution to the $M_1$ matrix element is suppressed compared to NLO contributions. This fundamental point would be studied better in
a study following this thesis. 
\cblack
\subsection{Estimating pionless EFT theoretical uncertainty}

As aforementioned, EFT is a systematic expansion in some small parameter. In order to judge whether the EFT describes well the observables, one needs to make sure that the expansion converges order-by-order. 
The theory presented here is purely perturbative, i.e., consistently organizes the expansion in a perturbative manner.

Since \pilesseft is RG invariance, the \pilesseft expansion for any $M_1$ observable can be written as: 
\begin{equation} \label{eq: expansion}
\langle{M_1}\rangle=\langle{M_1}\rangle_{\text {LO}} \cdot \left ( 1+a^{\text {NLO}}_{M_1}+ {\mathcal{O}} (\delta^2)\right).
\end{equation}
EFT suggests that $c^{\text {NLO}}_{M_1}=a^{\text {NLO}}_{M_1}/\delta$ are of natural size, and thus the truncation error is dictated by $\delta$. In \pilesseft, the na\"ive expansion parameter is $\delta\approx \frac{1}{3}$. This expansion parameter, however, can be estimated directly from the numerical results we have presented. Firstly, in Tab.~\ref{table_all_M1_l2_0} we calculate the ratio of NLO to LO contribution for each matrix element. These ratios are found to be in the range of $0.05-0.13$. Secondly, since $\langle\hat{\mu}_d\rangle$ has vanishing NLO contribution, its deviation from experiment can be regarded as N$^2$LO, and assuming a natural convergence we expect the ratio of this contribution to the LO contribution to be $({\text{N}^2\text{LO}}/{\text{LO}})\approx ({\text{NLO}}/{\text{LO}})^2$. This leads to $({\text{NLO}}/{\text{LO}})\approx 0.1$. Lastly, the variation of ${l'}^\infty_{1}$ (see Tab.~\ref{table_all_M1_l2_0}) leads to a variation in the predictions for the different ${l'}^\infty_{1}$ dependent observables. This variation represents the contribution of higher orders, thus the ratio of the variation to the NLO contribution should be of the order of the expansion parameter. Using Tab.~\ref{table_all_M1_l2_0} this leads to $({\text{N}^2\text{LO}}/{\text{NLO}})\approx 0.04-0.1$. 

If one assumes that the expansion parameter $\delta$ is common to all $M_1$ observables, then one can use the results to assess the value of $\delta$. In order to do this, let us take the log average of all the aforementioned samples: $\log\left| a^{\text{N}^k\text{LO}}_{M_1}/a^{\text{N}^{k-1}\text{LO}}_{M_1}\right|=\log\delta+\log R$. The numbers $R$ are positive natural numbers and are not biased, thus should be distributed about "1", and summing over their logarithms should vanish. The log average of many such samples should converge to $\log \delta$. The fact that this is a finite sized sample, means that there remains a measure of uncertainty, represented in a distribution, which we find that at a 95\% degree of belief the expansion parameter is in the range, $0.05<\delta<0.1$.% as shown in Fig.\ref{fig_delta_CDF}.

These suggest that the expansion converges faster, by about a factor of 3, than the na\"ive \pilesseft estimate. %We discuss this result further in Sec.~\ref{discussion}???. 

The truncation error of a given expansion, given a prior which represents the naturalness of the expansion, follows a posterior that was calculated in Refs.~\cite{Griesshammer:2015ahu, PhysRevC.92.024005, Cacciari2011}. In the current case, since the expansion parameter is unknown,\textit{ i.e.,} follows the aforementioned prior distribution, one should fold these two distributions to find the posterior distribution of these two distribution. The formalism is further explained in Appendix~\ref{ap_Bayesian approach}. 

\begin{figure}[h!]
	\centering
	\includegraphics[width=0.75\linewidth]{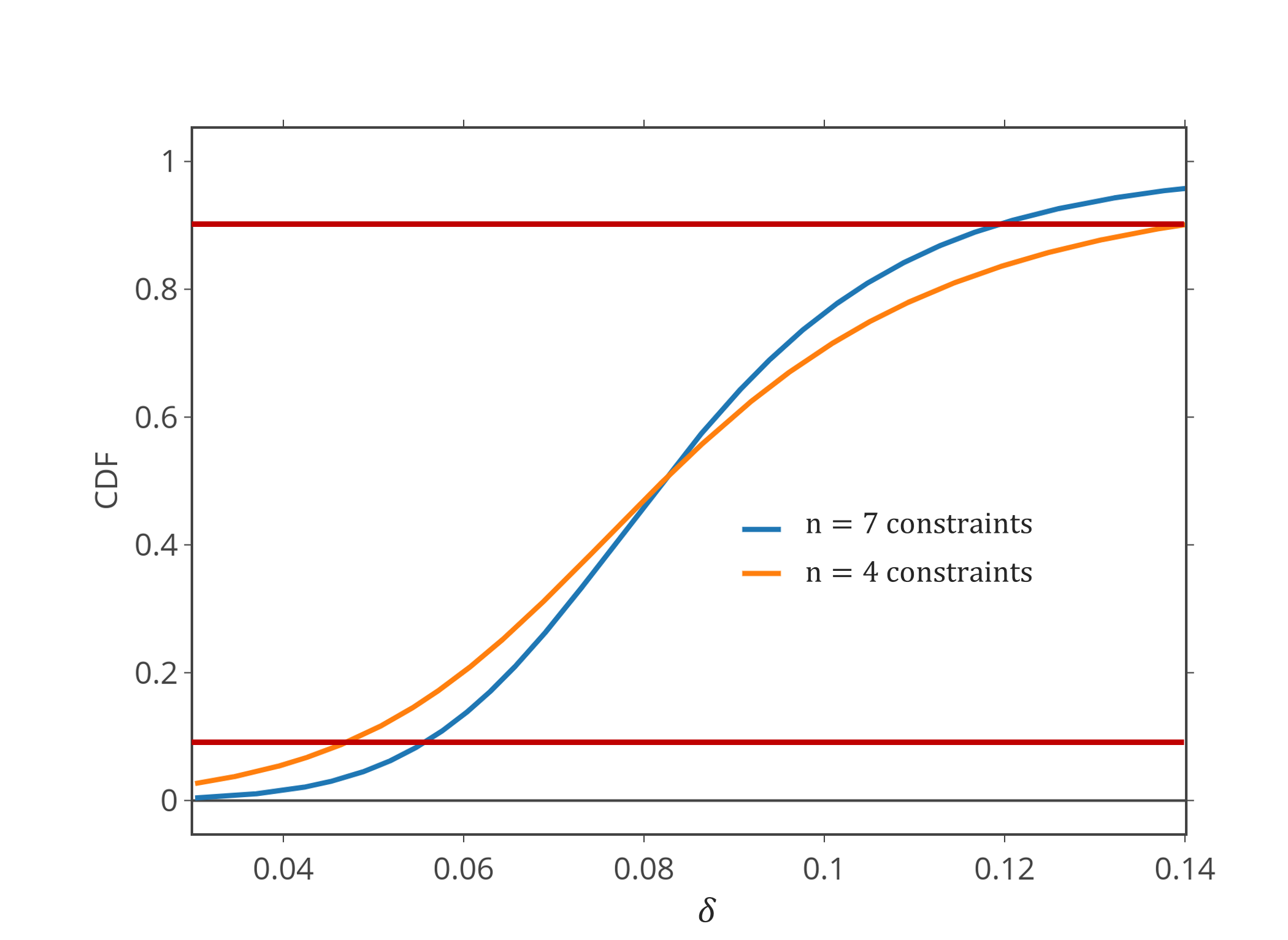}\\
	\caption{\footnotesize{
			Cumulative Density Functions (CDFs) of $\delta$, the expansion parameter. The blue curve represents a calculation which takes into account the constraints of the NLO contributions of $\langle\hat{\mu}_{^3\text{H}}\rangle$, $\langle\hat{\mu}_{^3\text{He}}\rangle$, $Y_{np}$, the N$^2$LO contribution of $\langle\hat{\mu}_d\rangle$, and the variation of $l’^\infty_1$. The orange curve takes into account only the first four former constraints. The red lines limit the $10-90\%$ probability range.}} 
	\label{fig_delta_CDF2}
\end{figure}

In order to check the sensitivity of the expansion parameter to the number of observables, we calculate the \cblack Cumulative Density Functions (CDF) of $\delta$, the expansion parameter, with all the $n=7$ constraints: the NLO contributions of $\langle\mu_{^3\text{H}}\rangle$, $\langle\hat{\mu}_{^3\text{He}}\rangle$, ${Y}_{np}$, the N$^2$LO contribution of $\langle\hat{\mu}_d\rangle$, and the variation of $l’^\infty_1$ stems from the three electromagnetic observables. Also, we calculate the CDF of $\delta$ only with the $n=4$ first constraints stemming from the order of the calculation and not from the LEC variation. As shown in Fig.~\ref{fig_delta_CDF2}, at 70\% degree of belief, the effect of the change is rather small (a change of about 20\% in the estimated truncation error). At higher degrees of belief, especially above 90\%, the truncation error depends significantly on the number of constraints, as can be expected.

\begin{figure}[h!]
	\centering
	\includegraphics[width=0.75\linewidth]{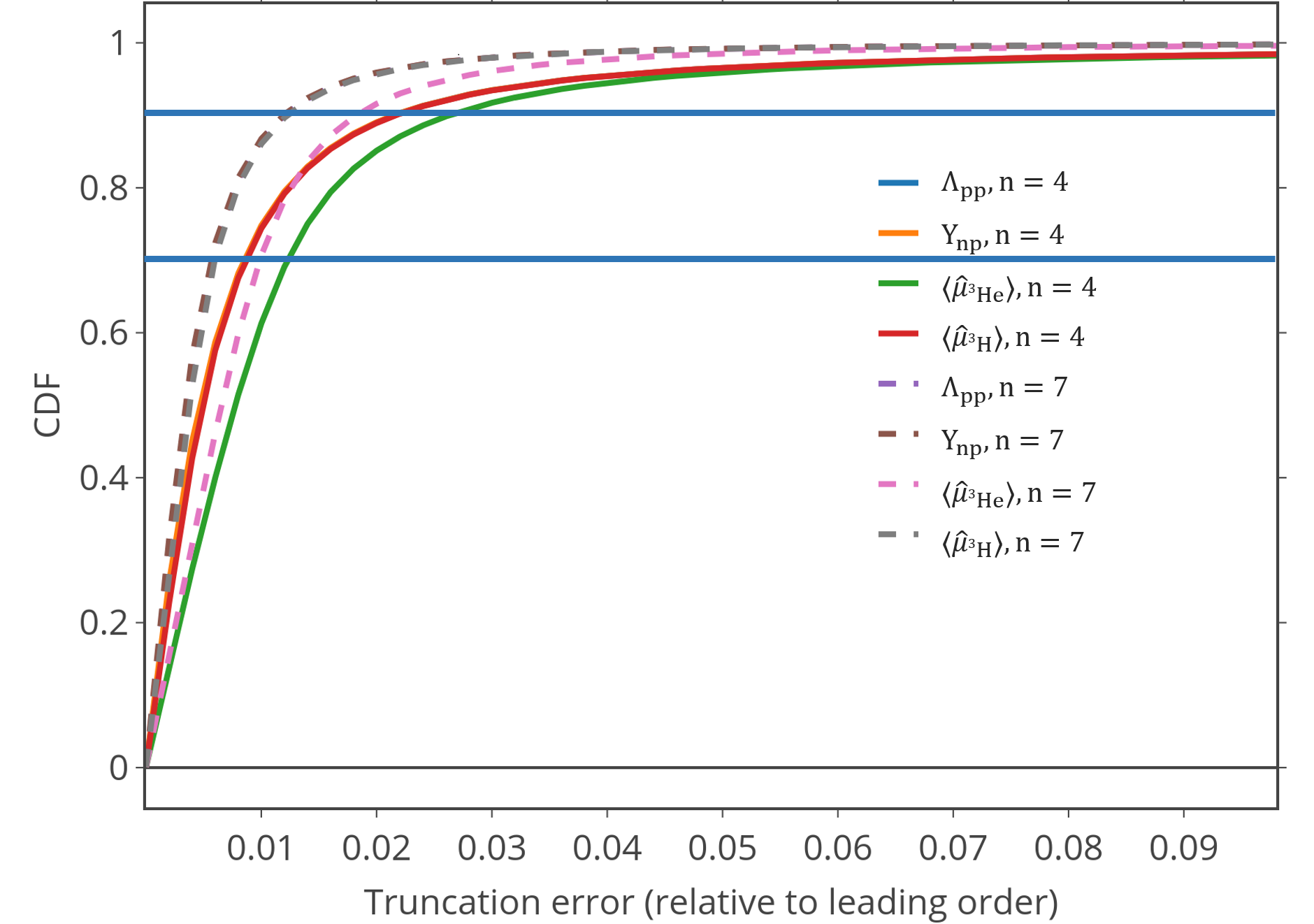}\\
	\caption{\footnotesize{
			Cumulative Density Functions (CDFs) for the different observables, as calculated using \cref{eq_delta}. Horizontal lines are the 70\% and 90\% degree of beliefs. We show CDFs relevant to expansion parameter priors with $n=4$ (solid lines) and $n=7$ (dashed lines) constraints, as explained in Fig.~\ref{fig_delta_CDF2}. }} 
	\label{fig_delta_CDF}
\end{figure}

Figure~\ref{fig_delta_CDF} shows that at about 70\% degree of belief, the $M_1$ obesrables' uncertainty does not exceed 1\%~\footnote{The uncertainty varies with the observable. 1\% is the maximal value, to be on the conservative side.}. Also shown a similar analysis for the proton-proton fusion, assuming that this reaction has the same expansion parameter as the magnetic observables. This will be
discussed further in the next section.

\begin{figure}[h!]
	\centering
	\includegraphics[width=0.75\linewidth]{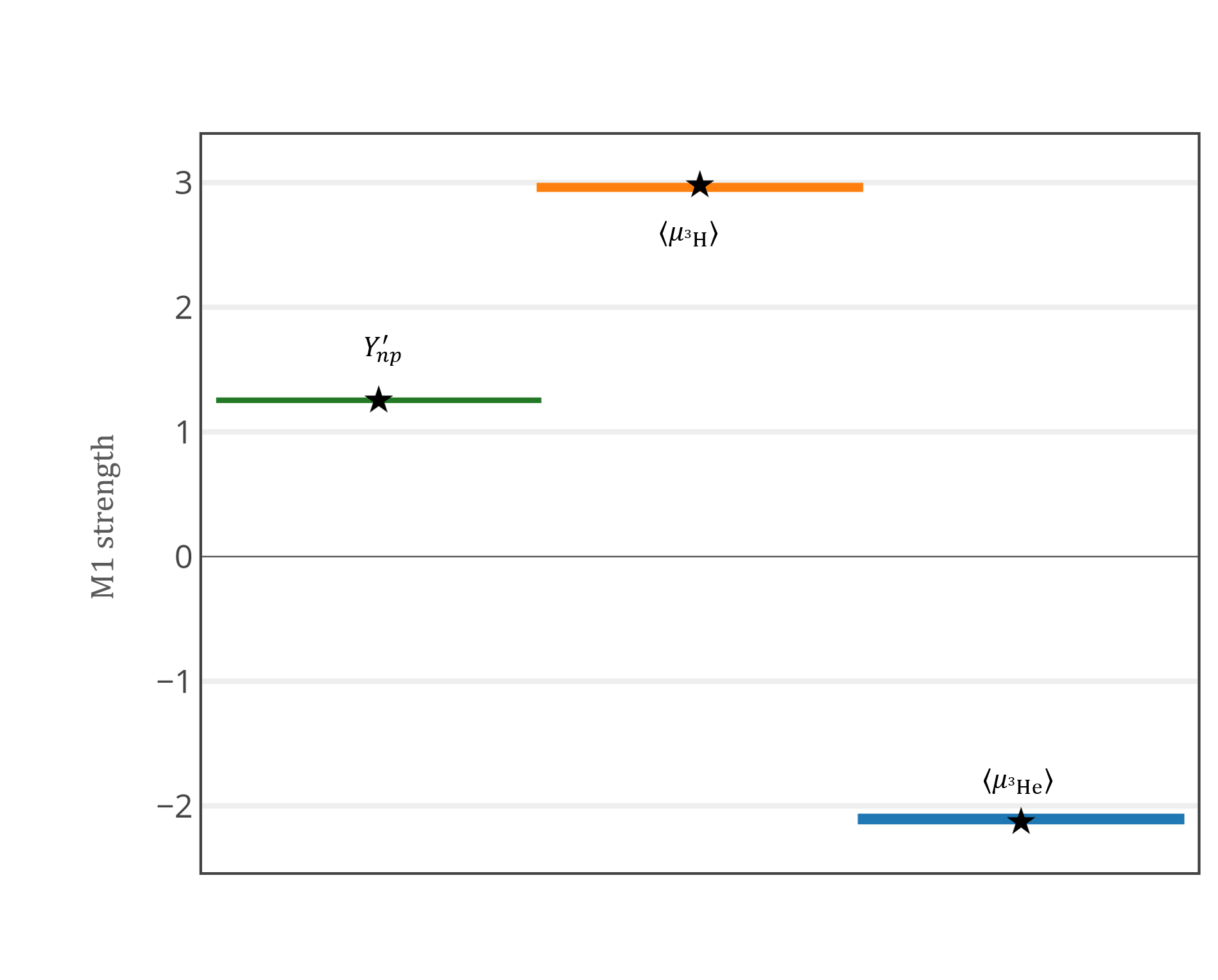}\\
	\caption{\footnotesize{The strength of $A=2, \, 3$ $M_1$ observables with 70\% degree of belief. The bands correspond to the theoretical Z-parameterization uncertainty as shown in from Fig.~\ref{fig_delta_CDF}. The stars are the experimental values.}} 
	\label{fig_A_2_3}
\end{figure}

Comparing our theoretical prediction (including the uncertainty calculated above) with experimental results, we find that the theory well reproduces the electromagnetic observables within the estimated uncertainties, as is graphically summarized in Fig.~\ref{fig_A_2_3}.

\cblack
\section{Analogy between $M_1$ observables and weak reactions}
A significant conclusion arises in this work, is the similar structure of the weak and electromagnetic observables in \pilesseft (see Tab.~\ref{table_EW} and Chapter \ref{general_matrix}). 
As shown in Tab.~\ref{table_EW}, the one-body operators of the weak and electromagnetic interactions have the same structure, \textit{i.e.,} one-body LECs coupled to Pauli matrices.

The electromagnetic LEC, $l_1$, which couples the spin-singlet to spin-triplet channel, has a weak analogue, $l_{1,A}$. We found that electromagnetic LEC, $l_2$, which couples two spin-triplet channels (and has no weak analogue) to be of a higher order than $l_{1}$ and $l_{1,A}$. This result emphasizes the qualitative similarity between the two interactions.

This similarity goes beyond this qualitative statement. 
In Chapter \ref{beta_decay} we have calculated the triton $\beta$-decay into $^3$He, which contains both the Fermi and the Gamow-Teller transitions. The Fermi transition, which is the weak analogues to the three-body normalization is found to be nearly one. Also, the Gamow-Teller transition, which is the weak analogue to the three-body magnetic moments is also found to be nearly $\sqrt{3}$ at LO, \textit{i.e.,} $\langle GT^{\text{LO}}\rangle=\langle\psi^{\text{LO}}(E_{^3\text{He}},p)|\boldsymbol{\sigma}\tau^+|\psi^{\text{LO}}(E_{^3\text{H}},p)\rangle=0.99(1)\sqrt{3}.$ This is consistent with the electromagnetic matrix moments at LO where $\langle \hat{\mu}_{^3\text{H}}^{\text{LO}}\rangle \approx \mu_p$ and $\langle \hat{\mu}_{^3\text{He}}^{\text{LO}}\rangle \approx \mu_n$. 

These similar characters of the electromagnetic and weak interactions along with the perturbative character of the Coulomb interaction, presented in Chapters \ref{general_matrix} and \ref{beta_decay} are reflecting the small effect of the Coulomb interaction on the $^3$He wave-function and implies that the expansion parameter, $\delta$, calculated from the electromagnetic observables is the same for the weak interactions and in particular, the $pp$-fusion.

\cblack
\section{Theoretical uncertainty of $pp$ fusion due to pionless EFT truncation}
As mentioned above, for the weak interaction only the triton $\beta$-decay is low-energy, well-measured reaction. Therefore, we cannot estimate the theoretical uncertainty of the weak \pilesseft calculation in the same method used for the electromagnetic case. Nevertheless, we can use the analogy between the electromagnetic to weak observables for estimating the theoretical uncertainty of the $pp$ fusion and alluding to their nominal NLO arrangement. In the previous sections we showed that both $pp$ fusion and $np$-radiative capture have similar small NLO contributions as shown in Tab.~\ref{Tab_NLO_parts}. This behavior, as well as the similar \pilesseft characteristics presented above, indicate that the electromagnetic theoretical uncertainty originating in \pilesseft truncation is valid also for the weak observables, which is 1\% for the $Z$-parameterization (see Fig.~\ref{fig_delta_CDF}). We note that this is not inconsistent. The uncertainty estimate relies upon the small expansion parameter, implied by the ratio of the NLO contribution, compared to the LO result. The uncertainty should then be of the order of the expansion parameter squared. The similarities between the electromagnetic and the weak calculations bring 
us to two separate conclusions: (i) the expansion parameter of both should be the same, and thus the uncertainty analysis should lead to the same results, (ii) the $Z$-parameterization is more appropriate for such 
long distance observables at NLO. Thus, in the following we consider only the $Z$-parameterization predictions for the $pp$ fusion, and assume 0.6\% of theoretical uncertainty in the calculation of the weak matrix element with 70\% degree of belief. %Moreover, we will show below, that choosing $Z$-parameterization for the $pp$ fusion rate is consistent with previous $\chi$EFT prediction.

\section{Final results and comparison to previous calculations of $l_{1, A}$ and the $pp$ fusion}
\cblack
As discussed above, the analogy between the electromagnetic and the weak observables enables us to use the Z-parameterization as the preferred parameterization for LECs calibration up to NLO. Therefore we find that: 

\begin{align}
l_{1, A}&=0.055\pm0.004\pm0.001, \quad g_A=1.2701\\%\cite{gA_1.2701}\\
l_{1, A}&=0.049\pm0.004\pm0.001, \quad g_A=1.2767%\cite{1674-1137-38-9-090001}
\end{align}
where the first uncertainty comes from $g_A$ experimental uncertainty and the second uncertainty comes from the rest of the 
experimental uncertainties, such as the statistical uncertainties in the measured triton half-life.
\cblack

The $pp$ fusion matrix element is given by: 
\cblack
\begin{equation}
\Lambda_{pp}(0)=2.66+0.6l_{1,A}.
\end{equation}

Since the astrophysical $S^{11}(0)$ is given by: 
\begin{equation}
S^{11}(0)=4.01\left(\frac{g_A}{1.2695}\right)^2\frac{\Lambda_{pp}^2(0)}{7.035}~, 
\end{equation}
one gets
\begin{equation}
S_{pp}(0, g_A=1.2701)=\left(4.14 \pm 0.05 \pm 0.012 \pm 0.002 \right) \cdot 10^{-23}\text{MeV}\cdot\text{fm}^2,
\end{equation}
and
\begin{equation}
S_{pp}(0, g_A=1.2766)=\left(4.17 \pm 0.05 \pm 0.012 \pm 0.002 \right) \cdot 10^{-23}\text{MeV}\cdot\text{fm}^2,
\end{equation}
The first uncertainty originates in the theoretical uncertainty estimated as 0.6\% for the weak matrix element with 70\% degree of belief (see Fig.~\ref{fig_delta_CDF}). The second uncertainty originates in the experimental uncertainty of $g_A$ (1$\sigma$), while the third uncertainty originates from the experimental value of $^3$H 
half-life (1$\sigma$). 
Since the theoretical uncertainty of $S_{pp}(0)$ is more dominant than the other two, we get at about 70\% degree of belief, combining them quadratically (assuming Gaussian pdfs for the experimental uncertainties):
\begin{subequations}
\begin{alignat}{2}\label{eq_fineal_pp1}
S_{pp}(0, g_A=1.2701)&=\left(4.14 \pm 0.05 \pm 0.012 \pm 0.002 \right) \cdot 10^{-23}\text{MeV}\cdot\text{fm}^2,\\
\label{eq_fineal_pp2}
S_{pp}(0, g_A=1.2766)&=\left(4.17 \pm 0.05 \pm 0.012 \pm 0.002 \right) \cdot 10^{-23}\text{MeV}\cdot\text{fm}^2,
\end{alignat} 
\end{subequations}
\cref{eq_fineal_pp1,eq_fineal_pp2} are the main results of this thesis.

\cblack

\subsection{Previous extractions of $L_{1, A}$}
Due to the importance of $L_{1, A}$, as the only counterterm that appears in the pionless description of $pp$ fusion, its evaluation has attracted much attention in the literature. In this subsection, we review the previous extraction of $l_{1, A}$ in the \pilesseft and the latest predictions of the $pp$ fusion rate. 

Two main approaches were taken in previous studies to match
$l_{1, A}$. In the first, an experimental value of a two-body weak
interaction process, usually at cutoff of $\mu=m_\pi$ has been used for
matching. Among these reactions are deuteron dissociation by
anti-neutrinos from reactors \cite{2002PhLB..549...26B}, and neutrino
reactions with the deuteron, as measured in SNO
 \cite{PhysRevC.67.025801}, both lead to similar RG invariant
combinations of the two-body axial
strength, $l_{1, A}\approx 0.13 \pm 0.26$. In both cases, the large uncertainties
originate from statistical errors in the experiments, due to the small cross-section for neutrino-deuteron reactions. The authors of
Ref.~\cite{L1A} proposed, therefore, a precision measurement of muon
capture on the deuteron, with the aim of reducing the uncertainties by a factor of 3, reflecting an estimated 2-3\% experimental uncertainty in
the (then proposed) ongoing MuSun experiment \cite{Andreev:2010wd}. It
is important to note that the $\mu^-d$ capture has a large energy
transfer, possibly too large for an application of \pilesseft, which might impact the convergence pattern of any calculation. In all
these, the uncertainties are mainly experimental, due to the
uncertainty in the observable, i.e. neglecting the truncation error.

A different approach was taken by Ando and collaborators in
Ref.~\cite{Ando_proton}. They used the hybrid calculation of the
$pp$ fusion rate from Park, Marcucci {\it et al.}~\cite{PhysRevC.67.055206}. As a
constraint, the ratio of the two-body strength over the one-body strength was taken from that calculation and matched $L_{1, A}$ to reproduce
this ratio in the \pilesseft regime. Their result was
$l_{1, A}= 0.038\pm0.002$. Their small uncertainties is due to the
accurate triton half-life measurement that is used to fix undetermined
counterterms in Ref.~\cite{PhysRevC.67.055206}. However, this work
highlights a few criticisms of Ref.~\cite{Ando_proton}. Firstly, a three-body constraint should be taken at a sufficiently large cutoff, to make sure no regularization effects affect the cutoff
invariance. Secondly, they take a phenomenological approach which leads to contributions from higher orders to the NLO calculation, as they take the non-perturbative deuteron normalization at all orders. These two reasons
entail that a robust theoretical uncertainty estimate is very hard within this approach, and indeed it is missing from the $L_{1, A}$ result in this case. Lastly, and most important, the constraint they take is not an observable, as one can use hermitian transformations to shift strengths from the two-body sector to the one-body, and vice versa.

In 2017, the Nuclear Physics with Lattice Quantum Chromo Dynamics (NPLQCD) collaboration has calibrated $L_{1, A}$ using the triton $\beta$-decay \cite{PhysRevLett.119.062002}, with 
\[l_{1,A}=0.068
(0.003
)
(0.017
)
(0.006)
(0.016
)
.\] The four uncertainties
are: the statistical uncertainty, the fitting and analysis systematic uncertainty, the mass extrapolation systematic uncertainty, and a power-counting estimate of higher order corrections in \pilesseft, respectively. Their value matches our $Z$-parameterization value, while considering all uncertainties. Similarly to the electromagnetic calculation of $l_1$, we find that their value for the two-body LEC is larger than our $Z$-parameterization prediction. We assume that this difference between these values comes from the different power-counting of the two works. As discussed in Chapter~\ref{Magnetic}, the power-counting used in that paper, is different from the one we use here, e.g., the deuteron normalization is taken at its nature value ($Z_d\approx 1.69$) at all orders, and not perturbatively, as in this work. For both electromagnetic weak interaction, the lattice QCD (LQCD) matrix elements have the form: $Y=\sqrt{Z_d}\left(Y^{\text{LO}}+Y^{\text{NLO}}\right)$, where $\sqrt{Z_d}=\frac{1}{\sqrt{1-\gamma_t\rho_t}}$. 
This form is inconsistent with our power-counting that assumes $\frac{\rho}{a}\ll1$. 
The disadvantage of the power-counting used in Ref.~\cite{PhysRevLett.119.062002}, arises from the fact that for this power-counting, the deuteron propagator is taken at all orders, (\cref{full}) and therefore it contains a pole at $q\sim150 \mev$, which is around the breakup scale $\mu= m_\pi$. This pole makes it very difficult to calculate the $A=3$ observables at cutoffs greater than the breakup scale $(\mu \sim \text{ few } m_\pi$), and therefore the two-body LEC, $l_{1, A}$, calibrated from the $^3$H $\beta$-decay was calculated at $\mu=m_\pi$. This calculation stands in contrast to the \pilesseft used here, which was found to be RG invariant for $\Lambda \sim \text{ few } m_\pi$ and according to what is known today, cannot be achieved without assuming that $\frac{\rho}{a}<<1$. 

\subsection{$pp$ fusion past predictions}
The validity of our \pilesseft prediction for the $pp$ fusion rate is obtained also by benchmarking with previous calculations, where the relevant past predictions were calculated by Marcucci {\it et al.} in 2013 \cite{PhysRevLett.110.192503}, Acharya {\it et al.} in 2016 \cite{Acharya:2016kfl} and LQCD in 2017 \cite{PhysRevLett.119.062002}.

The comparison between our predictions to that of Marcucci {\it et al.} \cite{PhysRevLett.110.192503} (pure Coulomb $\chi$EFT S-calculation) has to be made for the same $^3$H decay rate, $g_A$ and for the same $\langle F\rangle$ value. Our prediction, assuming their values for the needed parameters, is:
\begin{eqnarray}
S_{pp}^{\text{\pilesseft}}(0, g_A=1.2695)=(4.14\pm0.05)\cdot10^{-23}\text{MeV}\cdot \text{fm}^2, \label{eq_pp_1.2695}
%&& S_{pp}^{\text{\pilesseft}}(0), \ Z_d=4.06
%\cdot10^{-23}\text{MeV}\cdot \text{fm}^2 
\end{eqnarray}
while that of Marcucci {\it et al.} is: 
\begin{eqnarray}
S_{pp}^{\chi EFT}(^3S_1, \text{pure Coulomb})=\left (4.030\pm 0.006\right)\cdot10^{-23}\text{MeV}\cdot \text{fm}^2.
\end{eqnarray}

This interesting tension was found also by Acharya {\it et al.} \cite{Acharya:2016kfl} and was explained in the following way. According to Acharya {\it et al.} \cite{Acharya:2016kfl}, this inconsistency can arise from the infrared convergence of the matrix elements between
a bound-state and a scattering state wave function of Marcucci {\it et al.} calculation. In their work, Acharya {\it et al.} \cite{Acharya:2016kfl} have recalculated the S-factor of $pp$ fusion reaction using $\chi$EFT up to NNLO: 
\begin{equation}
S_{{pp}}^{\text{NNLO}}(0) = \left(4.081^
{+0.024}_
{-0.032}
\right) \cdot 10
^{-23}
\text{MeV}\cdot \text{fm}^2, 
\end{equation}
where the uncertainty originates from optimizing 42 different
NNLO interactions used in that work, as well as the statistical uncertainties of all LECs used for that calculation, including $g_A$. %Comparing to their calculations, and taking all the experimental uncertainties we find: 

Finally, for the LQCD calculating \cite{PhysRevLett.119.062002}, we find that: 
\begin{equation}
S_{pp}^
{\text{LQCD}}(0) = 4.052(0.006)(0.03)(0.012)(0.027)\cdot 10
^{-23}\mev\cdot \fm^2, 
\end{equation}

(where the four uncertainties are the statistical uncertainty, the fitting and analysis systematic uncertainty, the mass extrapolation systematic uncertainty, and a power-counting estimate of higher order corrections in \pilesseft, respectively \cite{PhysRevLett.119.062002}). As discussed above, the different power-counting of Ref.~\cite{PhysRevLett.119.062002} makes the comparison between the two predictions, problematic. Moreover, the NPLQCD method, based on extrapolating the quark mass up to its natural value is still partially controlled and has big uncertainties.

Among the four estimates (Marccuci, Acharya, LQCD and this work), only the predictions done by Acharya {\it et al.} can serve as a benchmark for our prediction, due to wrong calculation of wave functions in Ref.~\cite{PhysRevLett.110.192503} and the uncontrolled extrapolation and approximations in the lattice calculation of Ref.~\cite{PhysRevLett.119.062002}. The comparison between the two calculations shows a good agreement, taking the uncertainties into account (\cref{eq_fineal_pp1}). Moreover, since both calculations have used the triton $\beta$-decay for calibrating the two-body LEC ($l_{1,A}$), the overlapping of the two predictions is another confirmation to the validity of the theory. This consistency has a great significance due to the distinct differences of the recent predictions for the $pp$ fusion in the last decade.

In conclusion, by comparison to the electromagnetic predictions, we find that the $Z$-parameterization is the more appropriate parameterization for calculating $A=2, 3$ observables in \pilesseft at NLO, for these long distance observables. %We choose not to use the ERE-parameterization due to the significant uncertainties arise from this parameterization. Also, 
We note that for very low-energy interactions, the deuteron normalization must include the deuteron tail of the wave function, which is not possible up to NLO in the ERE-parameterization.

Our final predictions for the $pp$ fusion rate derived in this work, are:
\cblack
\begin{alignat}{2}
S_{pp}(0, g_A&=&1.2701)&=(4.14\pm 0.05)\cdot 10^{-23}\text{ MeV}\cdot\text{ fm}^2~,\\
S_{pp}(0, g_A&=&1.2766)&=(4.17\pm 0.05)\cdot 10^{-23}\text{ MeV}\cdot\text{ fm}^2~.
\end{alignat} 
\cblack which are consistent with chiral EFT predictions, and especially with Acharya {\it et al.} \cite{Acharya:2016kfl}. However, the current calculation, apart from including an estimate for the theoretical uncertainty, is RG invariant, fully perturbative, and includes about a factor of 3 less parameters. In addition, each parameter is related to a measured specific observable, such as the scattering lengths, the effective ranges, and the axial constant.

\chapter{Summary and outlook}

In this thesis, we presented a detailed study of low-energy electro-weak $A=2, \ 3$ observables using \pilesseft to next-to-leading order (NLO). The main purpose of the thesis was to calculate and estimate the theoretical and empirical uncertainties of the $pp$ fusion rate.

This was done by establishing a perturbative and consistent framework to calculate transitions between three-body bound-states, up to NLO. This method is based on a diagrammatic expansion of both the nuclear amplitudes (wave functions) and the matrix elements which are proportional to the electro-weak observables.
For these electro-weak matrix elements, the NLO contributions amount less than 10\% correction, which is smaller than the na\"ive expansion parameter of \pilesseft. The similar NLO contributions of both the electromagnetic and weak observables have strengthened the premise that the two sectors (weak and electromagnetic) share similar properties. Therefore we have used our predictions for the well-measured electromagnetic observables to estimate the theoretical and empirical \pilesseft uncertainties of $A=2, 3$ electro-weak interactions.

We showed that by using the so-called $Z$-parameterization, that correctly describes the deuteron wave function at large distances, the values of the short-range strengths are consistent in the $A=2$ and $A=3$ systems, up to 1\%, as estimated by the robust order-by-order convergence. We found that \pilesseft predictions for the electromagnetic observables have unprecedented precision and accuracy, which, as summarized in Fig.~\ref{fig_A_2_3}, verify and validate the way we applied \pilesseft at NLO.
Using the $Z$-parameterization, our final predictions for the electromagnetic interactions are, to 70\% degree of belief:
\begin {table}[H]
\begin{center}
	\begin{tabular}{c c c c c}
		\centering
		&&This work [{nNM}]& Experiment $[{\rm nNM}]$&\\ 
		$Y'_{np}$ &=& $1.253\pm0.006$ & $1.2532 \pm 0.0019$&\cite{np_data} \\
		$\langle\hat{\mu}_{^3\text{H}}\rangle$ &=& $2.992\pm0.015$ & $2.97896...$ &\cite{3He_3H_data} \\
		$\langle\hat{\mu}_{^3\text{He}}\rangle$ &=& $ -2.11\pm0.02 $ & $-2.12750...$& \cite{3He_3H_data}
	\end{tabular}
\end{center}
\end{table} 
Our predictions for the electromagnetic observables, which perfectly match the experimental data, point out that by using the $Z$-parameterization, \pilesseft is consistent for a simultaneous description of $A=2$ and $A=3$ systems up to NLO. This consistency enabled us to calibrate the unknown weak low-energy constant $l_{1, A}$, needed for the calculation of the $pp$ fusion rate, from the well-measured $^3$H $\beta$-decay.
\begin{align}
S_{pp}(0)=(4.15\pm 0.06)\cdot 10^{-23}\text{ MeV}\cdot\text{ fm}^2,\label{eq_pp_1}
\end{align}
%where $g_A = 1.2701$ is a result of the neutron half-life measurements, taken from Ref.~\cite{gA_1.2701} and used by Ref.~\cite{Acharya:2016kfl}, which their prediction for $S_{pp}$ is consistent with \cref{eq_pp_1}, while $g_A=1.2766$ is based on the latest $g_A$ prediction as presented in Ref.~\cite{doi:10.1063/1.4983578}.
 Our result is 1.5\%-4.5\% higher than recent predictions of the $pp$ fusion rate \cite{pp_review, PhysRevLett.110.192503}, which have been used for theoretical computations of stellar models \cite{Vinyoles:2016djt, TOGNELLI2015189}. Our new predictions will affect the stellar models and will be checked in future work. 
 \section{Outlook}
 A natural continuation of this Ph.D. research outlined above, is a study of other aspects of \pilesseft. The NLO contributions of the electro-weak observables are smaller than our na\"{i}ve estimations 
 and in our opinion, this requires further examination which is beyond the scope of this thesis.
 
 An interesting result, which was found at the ending stages of this thesis, is that the isoscalar two-body strength $l_2$, should enter at higher order than $l_1$, \textit{i.e.,} than its na\"{i}ve dimensional analysis suggested order. This might be related to the fact that in chiral EFT the isovector two-body electromagnetic current is an NLO term, while the isoscalar two-body contribution is further suppressed. In addition, the small NLO contributions to the $M_1$ $A=2,3$ observables lead to the famous approximate shell model structure of the magnetic moments. These unexpected signatures of pion physics and shell model naturally stem from pionless EFT. This seems to be a first example for a connection between these three approaches to nuclear structure. This fundamental point would be studied better in a study following this thesis.
%\end{document}
\appendix

\renewcommand{\theequation}{A-\arabic{equation}}
\renewcommand{\thesection}{\Roman{section}}
\renewcommand{\thesubsection}{\alph{subsection}}
% redefine the command that creates the equation no.
\setcounter{section}{0}
\setcounter{subsection}{0}
\setcounter{equation}{0}
\raggedright
\chapter{The Faddeev equations three-body scattering}
\label{whateverB}
\input{try}

\renewcommand{\theequation}{B-\arabic{equation}}
\renewcommand{\thesection}{\Roman{section}}
\renewcommand{\thesubsection}{\alph{subsection}}
% redefine the command that creates the equation no.
\setcounter{section}{0}
\setcounter{subsection}{0}
\setcounter{equation}{0}
\raggedright

\chapter{The analytic form of the NLO correction to the three-body force}
\label{whateverE}
\input{appendixH_NLO}
\renewcommand{\theequation}{C-\arabic{equation}}
\renewcommand{\thesection}{\Roman{section}}
\renewcommand{\thesubsection}{\alph{subsection}}
% redefine the command that creates the equation no.
\setcounter{section}{0}
\setcounter{subsection}{0}
\setcounter{equation}{0}
\chapter{The Hubbard-Stratonovich transformation with two-body electro-weak interaction}
\label{whatever}
\input{ap_H_S_EW}

\setcounter{subsection}{0} 
\setcounter{equation}{0} 
\setcounter{section}{0} 
\setcounter{figure}{0}
\renewcommand{\thesection}{D}
\renewcommand{\theequation}{D-\arabic{equation}}
\renewcommand{\thesubsection}{D.\Roman{subsection}}
\renewcommand{\thefigure}{D.\arabic{figure}}
\chapter{Appendix D - A Bayesian approach to estimate the convergence rate} %

\label{ap_Bayesian approach}
\input{Bayesian_approach2}

\bibliography{references1,bibliography1,bibliography}
\bibliographystyle{unsrt}
\end{document}
\begin{comment}
\clearpage
\phantomsection\addcontentsline{toc}{chapter}{Hebrew Abstract}
\pagenumbering{Alph}
\setcounter{page}{1}

\includepdf[pages=2]{hebrew_abstract_thesis.pdf}
\includepdf[pages=1]{hebrew_abstract_thesis.pdf}
\includepdf[pages=2]{hebrew_front_page.pdf}
\includepdf[pages=1]{hebrew_front_page.pdf}
\end{comment}

%% file: try.tex
 \section{The Faddeev equations for $n-d$ scattering } % use *-form to suppress numbering
 
 \subsection{The Quartet channel for $n-d$ scattering}
 
 Using the Feynman rules that follow from the Lagrangian, \cref{H_S_lag2}, and inserting appropriate
 symmetry factors, we get:
 \begin{equation}\label{quartet1}
 \begin{split}
 i T^{ij}(E;{\bf p, }{\bf k})^{\alpha a}_{\beta b}&=(-2 i
 y_t)^2\frac{1}{\sqrt{8}}(\sigma^2\sigma^i)^\alpha_{\alpha'}(\tau^2)^a_{a'}
 i
 S(k_0-p_0, {\bf k - p})\delta^{\alpha'}_{\beta'}\delta^{a'}_{b'}\frac{1}{\sqrt{8}}(\sigma^j\sigma^2)^{\beta'}_{\beta}(\tau^2)^{b'}_{b}\\
 &+\int\frac{\hbox{d}^4 q}{(2\pi)^4}(-2 i
 y_t)^2\frac{1}{\sqrt{8}}(\sigma^2\sigma^i)^\alpha_{\alpha'}(\tau^2)^a_{a'}
 iS\left(q_0-p_0, {\bf q-p}\right)i\mathcal{D}_t\left(q_0, q\right)\delta^{a'}_{b'}\times\\
 & \qquad \qquad \qquad iT^{ik}(E;{\bf q, }{\bf k})^{\alpha' c}_{\beta' b}
 \frac{1}{\sqrt{8}}(\sigma^k\sigma^2)^{\beta'}_{\beta}(\tau^2)^{b'}_{b}= \\
 i T^{ij}(E;{\bf p, }{\bf k})^{\beta a}_{\alpha b}=&(- i
 y_t)^2\frac{1}{{2}}(\sigma^j\sigma^i)^\beta_{\alpha}
 iS(k_0-p_0, {\bf k - p}))\delta^{a}_{b}\\
 &+\int\frac{\hbox{d}^4 q}{(2\pi)^4}(- i
 y_t)^2\frac{1}{{2}}(\sigma^j\sigma^k)^{\beta}_{\alpha'}
 iS(q_0-p_0, {\bf q-p})i\mathcal{D}_t\left(q_0, q\right)\delta^{a}_{c}i T^{ik}(E;{\bf q, }{\bf k})^{\alpha' c}_{\alpha b}.
 \end{split}
 \end{equation}
 Integration over $q_0$ yields:
 \begin{equation}\label{quartet3}
 \begin{split}
 i T^{ij}(E;{\bf p, }{\bf k})^{\beta a}_{\alpha b}=&- i\frac{My_t^2}{{2}}(\sigma^j\sigma^i)^\beta_{\alpha}
 \delta^{a}_{b}\frac{1}{{\bf k}^2+{\bf p}^2+{\bf k}\cdot {\bf p}-ME}\\
 &+i\int\frac{\hbox{d}^3 q}{(2\pi)^3}\frac{My_t^2}{{2}}(\sigma^j\sigma^k)^\beta_{\alpha'}
 \delta^{a}_{c}\mathcal{D}_d\left(E-\frac{q^2}{2M}, q\right)\frac{i T^{ik}(E;{\bf q, }{\bf k})^{\alpha' c}_{\alpha b}}{{\bf q}^2+{\bf p}^2+{\bf q}\cdot {\bf p}-ME}.
 \end{split}
 \end{equation}
 \subsection{The Doublet channel for $n-d$ scattering}
 The Faddeev equations for $n-d$ scattering are given by:
 \begin{equation}\label{STM_doublet}
 \begin{split}
 (iT^{i, j})^{\beta b}_{\alpha a}(E;{\bf k}, {\bf
 	p})=&-\frac{i M
 	y_t^2}{2}\cdot(\sigma^j\sigma^i)^\beta_\alpha\delta^b_a\, 
 \frac{1}{{\bf k}^2+{\bf k}\cdot{\bf p}+{\bf p}^2-M E}\\
 &+\frac{M
 	y_t^2}{2}\int\frac{\hbox{d}^3q}{(2\pi)^3}
 {D}_t\left(E,q\right)
 \frac{(\sigma^j \sigma^k)^\beta_\gamma\delta^b_c}{{\bf q}^2+{\bf
 		q}\cdot{\bf p}+{\bf p}^2-M
 	E}(iT^{ik})_{\alpha a}^{\gamma c}\\
 &+\frac{My_ty_s}{2}\int\frac{\hbox{d}^3q}{(2\pi)^3}
 {D}_s^{np}(E, q)
 \frac{(\sigma^k)^\beta_\gamma (\tau^{3})^b_c}{{\bf q}^2+{\bf q}\cdot{\bf
 		p}+{\bf p}^2- M E}(iS_{nn}^i)_{\alpha
 	a}^{\gamma c}\\
 &+\frac{M y_t y_s}{2}\int\frac{\hbox{d}^3q}{(2\pi)^3}
 \mathcal{D}_t^{nn}(E-\frac{{\bf q}^2}{2M}, {\bf q})
 \frac{(\sigma^k)^\beta_\gamma (\tau^{1+2i})^b_c}{{\bf q}^2+{\bf q}\cdot{\bf
 		p}+{\bf p}^2- M
 	E}(iS_{nn}^i)_{\alpha
 	a}^{\gamma c}\\
 (iS_{np})^{\beta b}_{\alpha a}({\bf k}, {\bf
 	p};E)=&-\frac{i M
 	y_s^2}{2}\cdot(\sigma^i)^\beta_\alpha(\tau^{3})^b_a\cdot
 \frac{1}{{\bf k}^2+{\bf k}\cdot{\bf p}+{\bf p}^2-M M
 	E}\\
 &+\frac{M
 	y_ty_s}{2}\int\frac{\hbox{d}^3q}{\left(2\pi\right)^3}
 \mathcal{D}_d(
 E-\frac{{\bf q}^2}{2M}, {\bf q})
 \frac{(\sigma^k)^\gamma_\alpha(\tau^{3})^b_c}{{\bf q}^2+{\bf
 		q}\cdot{\bf p}+{\bf p}^2-M
 	E}(iT^{ik})_{\alpha a}^{\gamma c}\\
 &+\frac{M
 	y_s^2}{2}\int\frac{\hbox{d}^3q}{\left(2\pi\right)^3}
 {D}_s^{np}(E, q)
 \frac{(\tau^{3})^b_c(\tau^{3})^c_d}{{\bf q}^2+{\bf
 		q}\cdot{\bf p}+{\bf p}^2-M
 	E}(iS_{np}^i)_{\alpha a}^{\beta d}\\
 &+\frac{M
 	y_s^2}{2}\int\frac{\hbox{d}^3q}{\left(2\pi\right)^3}
 \mathcal{D}_t^{nn}(E-\frac{{\bf q}^2}{2M}, {\bf q})
 \frac{(\tau^{C=i+i\cdot 2})^b_c(\tau^{3})^c_d}{{\bf q}^2+{\bf
 		q}\cdot{\bf p}+{\bf p}^2-M
 	E-}(iS_{nn}^i)_{\alpha a}^{\beta d}\\
 (iS_{nn}^i)^{\beta b}_{\alpha a}({\bf k}, {\bf
 	p};E)=&-\frac{i M
 	y_ty_s}{2}\cdot(\sigma^i)^\beta_\alpha(\tau^{B=1+i\cdot 2})^b_a\cdot
 \frac{1}{{\bf k}^2+{\bf k}\cdot{\bf p}+{\bf p}^2-M E-}\\
 &+\frac{M
 	y_ty_s}{2}\int\frac{\hbox{d}^3q}{\left(2\pi\right)^3}
 {D}_t\left(E,q\right)
 \frac{(\sigma^k)^\gamma_\alpha(\tau^{B=1+i\cdot 2})^b_c}{{\bf q}^2+{\bf
 		q}\cdot{\bf p}+{\bf p}^2-M
 	E}(iT^{ik})_{\alpha a}^{\gamma c}\\
 &+\frac{M
 	y_s^2}{2}\int\frac{\hbox{d}^3q}{\left(2\pi\right)^3}
 {D}_s^{np}(E, q)
 \frac{(\tau^{3})^b_c(\tau^{B=1+i\cdot 2})^c_d}{{\bf q}^2+{\bf
 		q}\cdot{\bf p}+{\bf p}^2-M
 	E}(iS_{nn}^i)_{\alpha a}^{\beta d},\\
 \end{split}
 \end{equation}
 
 where the superscripts $i, j$ denote the Cartesian indexes of the incoming
 and outgoing dibaryon fields and the amplitudes are not projected on
 definite $J$ and $T$. The projector is given by $\frac{1}{\sqrt{3}}\sigma_i$
 for deuteron channel and  by $\frac{1}{\sqrt{3}}\tau_i$ for the triplet channel \cite{Parity-violating}:
 \begin{equation}\label{int_field_t}
 iT(E;{\bf p}, {\bf k})^{\alpha a}_{\beta
 	b}=\frac{1}{3}(\sigma^i)^\alpha_{\alpha'}iT(E;{\bf p}, {\bf k})^{ij})^{\alpha' a}_{\beta'
 	b}(\sigma^j)^{\beta'}_\beta, 
 \end{equation}
 \begin{equation}\label{int_field_s}
 iS(E;{\bf p}, {\bf k})^{\alpha a}_{\beta
 	b}=\frac{1}{3}(\sigma^i)^\alpha_{\alpha'}i(S(E;{\bf p}, {\bf k})^{iC})^{\alpha' a}_{\beta
 	a'}(\tau^C)^{a'}_b, 
 \end{equation}
 
 where for $n-d$ scattering $a=b=1$, $\alpha=\beta$ and $a_{nn}=a_{np}=a_{s}$ and $S_{np}=S_{nn}$.
 \section{The Faddeev equations for $p-d$ scattering} % use *-form to suppress numbering
\subsection{The Quartet channel for $p-d$ scattering}

Using the Feynman rules (Tab.~\ref{tbl: feynman_ruls}) which follow from the Lagrangian \cref{H_S_lag2} and inserting appropriate
symmetry factors, we get:
\begin{equation}\label{quartet_p_d}
\begin{split}
& i T^{ij}(E;{\bf p, }{\bf {\bf k}})^{\alpha a}_{\beta b}=(-2 i
 y_t)^2\frac{1}{\sqrt{8}}(\sigma^2\sigma^i)^\alpha_{\alpha'}(\tau^2)^a_{a'}
 i
 S(k_0-p_0, {\bf {\bf k} - p})\delta^{\alpha'}_{\beta'}\delta^{a'}_{b'}\frac{1}{\sqrt{8}}(\sigma^j\sigma^2)^{\beta'}_{\beta}(\tau^2)^{b'}_{b}+\\
 &(-2 iy_t)^2\frac{1}{\sqrt{8}}(\sigma^2\sigma^i)^\alpha_{\alpha'}(\tau^2)^a_{a'}
 i\mathcal{I}_{B}(E,{\bf {\bf k}}, {\bf p})i\mathcal{D}_{photon}({\bf {\bf k} - {\bf p}}) \delta^{ij}\delta^{\alpha'}_{\beta'}\delta^{a'}_{b'}\frac{1}{\sqrt{8}}(\sigma^j\sigma^2)^{\beta'}_{\beta}(\tau^2)^{b'}_{b}+\\
 &(-2 iy_t)^2\frac{1}{\sqrt{8}}(\sigma^2\sigma^i)^\alpha_{\alpha'}(\tau^2)^a_{a'}
 i\mathcal{I}_b({\bf k}, {\bf p})\delta^{\alpha'}_{\beta'}\delta^{a'}_{b'}\frac{1}{\sqrt{8}}(\sigma^j\sigma^2)^{\beta'}_{\beta}(\tau^2)^{b'}_{b}+\\
&+\int\frac{\hbox{d}^4 q}{(2\pi)^4}(-2 i
 y_t)^2\frac{1}{\sqrt{8}}(\sigma^2\sigma^i)^\alpha_{\alpha'}(\tau^2)^a_{a'}
 iS\left(q_0-p_0, {\bf q-{\bf p}}\right)i\mathcal{D}_t\left(q_0, q\right)\delta^{a'}_{b'}\times\\
 & \qquad \qquad \qquad iT^{ik}(E;{\bf q, }{\bf {\bf k}})^{\alpha' c}_{\beta' b}
 \frac{1}{\sqrt{8}}(\sigma^{\bf k}\sigma^2)^{\beta'}_{\beta}(\tau^2)^{b'}_{b}+\\
 &+\int\frac{\hbox{d}^4 q}{(2\pi)^4}(-2 i y_t)^2\frac{1}{\sqrt{8}}(\sigma^2\sigma^i)^\alpha_{\alpha'}(\tau^2)^a_{a'}
 iS\left(q_0-p_0, {\bf q-{\bf p}}\right)i\mathcal{D}_t\left(q_0, q\right)\delta^{a'}_{b'}\times\\
 & \qquad \qquad \qquad iT^{ik}(E;{\bf q, }{\bf {\bf k}})^{\alpha' c}_{\beta b}i\mathcal{I}_{B}(E,{\bf {\bf q}}, {\bf p})i\mathcal{D}_{photon}\left({\bf q} \right)
 \frac{1}{\sqrt{8}}(\sigma^{\bf k}\sigma^2)^{\beta'}_{\beta}(\tau^2)^{b'}_{b}\\
 &+\int\frac{\hbox{d}^4 q                                                                                                                                                                                                                                                                                                                                                                                                                                                                                                                                                                                                                                                                                                                                                                                                                                                                                                                                                                                                                                                                                                                                                                                          }{(2\pi)^4}(-2 i
 y_t)^2\frac{1}{\sqrt{8}}(\sigma^2\sigma^i)^\alpha_{\alpha'}(\tau^2)^a_{a'}
 iS\left(q_0-p_0, {\bf q-p}\right)i\mathcal{D}_t\left(q_0, q\right)\delta^{a'}_{b'}\times\\
 & \qquad \qquad \qquad iT^{ik}(E;{\bf q, }{\bf {\bf k}})^{\alpha' c}_{\beta' b}i\mathcal{I}_b(E,)
 \frac{1}{\sqrt{8}}(\sigma^{\bf k}\sigma^2)^{\beta'}_{\beta}(\tau^2)^{b'}_{b},
\end{split}
\end{equation}
where
\begin{equation}\label{eq_d_photon}
i\mathcal{D}_{photon}({\bf k})=\frac{i}{{\bf k}^2+\lambda}
\end{equation}
where $\lambda$ is a small photon added to regulate the singularity of the propagator at zero momentum transfer.

The bubble integral (Figure \ref{Coulomb_correction}(a)) is given by:
\begin{equation}\label{eq_first}
\begin{split}
 i\mathcal{I}_{B}({\bf k}, p)&=\int\frac{d^4q}{(2\pi)^4}iS(q_0, -{\bf q})\cdot iS(E-k_0+q_0, {\bf k}+q_0)\cdot iS(p_0-q_0, p-{\bf q})=\\
 &\int\frac{d^3q}{(2\pi)^3}\frac{M^2}{\left(2 ME-p^2+2 Mp_0-2 p {\bf q}-2 {\bf q}^2\right) \left({\bf k}^2-2 ME-2 k_0 M+2 {\bf k} {\bf q}+2 {\bf q}^2\right)},
\end{split}
\end{equation}
where \cref{eq_first} can be written as:
\begin{equation}\label{1}
i\mathcal{I}^{ij}_{B}(E,{\bf k}, {\bf p})=\int\frac{d^3q}{(2\pi)^3}\frac{4M^2}{A({\bf q})\cdot B({\bf q})},
\end{equation}
and $A(q),B(q)$ are:
\begin{equation}\label{A}
 A({\bf q})=\left(2 ME-{\bf p}^2+2 Mp_0-2 {\bf p} \cdot{\bf q}-2 {\bf q}^2\right)
\end{equation}
\begin{equation}\label{B}
B({\bf q})=-\left(\left({\bf k}^2-2 ME-2 k_0 M+2 {\bf k}\cdot {\bf q}+2 {\bf q}^2\right)\right).
\end{equation}
Using Feynman integrals, one find that:
\begin{equation}\label{eq_IB2}
 i\mathcal{I}_{B}({\bf k}, {\bf p})=\int\frac{d^3q}{(2\pi)^3}\int_0^1dx\frac{4M^2}{\left(xA({\bf q})+ B({\bf q})\cdot(1-x)\right)^2},
\end{equation}
by performing a d-dimensional integrals in Minkowski space, the final expression for \cref{eq_first}, is:
\begin{equation}\label{IB3}
\begin{split}
 i\mathcal{I}_{B}({\bf k}, {\bf p})&=-\int_0^1dx\frac{iM^2}{
 2 \pi}\times \big(\left.-{\bf k}^2 +4ME + 4 k_0 M
 - 4 k_0 Mx +
 2 {\bf k} {\bf p} x - 2 {\bf p}^2 x + 4 M
 p_0 x \right. \\
 &\qquad\qquad\qquad\qquad\qquad \left. + {\bf k}^2 x^2 - 2 {\bf k} {\bf p} x^2 + {\bf p}^2 x^2\right. \big)^{-1/2}=\\
 &\frac{i M^2}{2 ({\bf k}-{\bf p}) \pi } \biggr\{\left.\log\left[2k_0 M+({\bf k}-{\bf p}) \left(\sqrt{4 (E+k_0) M}-{\bf k}^2-{\bf p}\right)-2 Mp_0\right]\right.\\
 &\qquad\left.-\log\left[2 M (k_0-p_0)-({\bf p}-{\bf k}) \left(-{\bf {\bf k}}+\sqrt{-{\bf p}^2+4 M (E+p_0)}\right)\right]\right.\biggr\}.
 \end{split}
\end{equation}

Using the residue theorem and performing the integral over $k_0=\frac{-{\bf k}^2}{2M}$ and $p_0=\frac{-{\bf p}^2}{2M}$, we get that:
\begin{equation}\label{IB5}
\begin{split}
i\mathcal{I}_{B}(E,{\bf {\bf k}}, {\bf p})=\frac{i M
^2}{2 ({\bf k}-{\bf p}) \pi } \biggr(&\left.\log\left[\left(\sqrt{4 ME-3 {\bf k}^2
}-{\bf k}-2 {\bf p}\right) ({\bf k}-{\bf p})\right]\right.\\
&\left.-\log\left[(-{\bf k}+{\bf p}) \left(2 {\bf k}+{\bf p}-\sqrt{4 ME
-3 {\bf p}^2}\right)\right]\right.\biggr).
\end{split}
\end{equation}

Finally we get:
\begin{equation}\label{IB6}
 i\mathcal{I}_{B}({\bf k, \bf {\bf p}})=\frac{M^2 \left(\arctan\left(\frac{2 {\bf k}+{\bf p}}{\sqrt{3 {\bf p}^2-4 ME}}\right)-\arctan\left(\frac{{\bf k}+2 {\bf p}}{\sqrt{3 {\bf k}^2-4 ME}}\right)\right)}{2 \pi ({\bf k}-{\bf p})}.
\end{equation}

The one-nucleon exchange coupled to one photon exchange, $K_b^{A, B}$ (Figure \ref{Coulomb_correction}(b)), is given by:

\begin{equation}\label{second}
\begin{split}
 K^{A,B}_{b}({\bf k}, {\bf p})&=\int\frac{d^4q}{(2\pi)^4}S(-k_0, -k) S(-p_0, -p) \mathcal{D}_{photon}(k+p+{\bf q}) \mathcal{D}_{A}(E+k_0, k)\times\\
 &\qquad\qquad\mathcal{D}_{B}(E+p_0, p) S(E-q_0, -{\bf q}) S(k_0+q_0, k+{\bf q}) S(p_0+q_0, p+{\bf q})=\\
 &-\int\frac{d^3q}{(2\pi)^3}\frac{i \alpha M^2 \mathcal{D}_{A}\left(E-\frac{k^2}{2 M}, k\right) \mathcal{D}_{B}\left(E-\frac{p^2}{2 M}, p\right)}{\left(-ME+k^2+k {\bf q}+{\bf q}^2\right) \left(-ME+p^2+p {\bf q}+{\bf q}^2\right) \left[[\lambda ^2+(k+p+{\bf q})^2\right]}
\end{split}
\end{equation}
\cref{second} can be written as:
\begin{equation}\label{2}
 K_{b}^{A, B}(E,{\bf k}, {\bf p})=\int\frac{d^3q}{(2\pi)^3}\frac{i \alpha M^2 {D}_A\left(E, k\right) {D}^{B}\left(E, p\right)}{A({\bf q})\cdot B({\bf q})\cdot C({\bf {\bf q}})}
\end{equation}
where
\begin{eqnarray}
 A({\bf q})&=&-ME+k^2+k {\bf q}+{\bf q}^2\\
B({\bf {\bf q}})&=&-ME+p^2+{\bf p} {\bf q}+{\bf q}^2\\
C({\bf q})&=&\lambda ^2+(k+{\bf p}+{\bf q})^2,
\end{eqnarray}

using Feynman integrals we get:
\begin{equation}\label{IB2}
K_{b}^{A, B}({\bf k}, {\bf p})=\int\frac{d^3q}{(2\pi)^3}\int_0^1dx\int_0^{1-x}dy\frac{i \alpha M^2 {D}_{A}\left(E, k\right) {D}_{B}\left(E, p\right)}{\left(xA({\bf q})+y B({\bf q})+ C({\bf q})\cdot(1-x-y)\right)^3}, 
\end{equation}
and by performing a d-dimensional integrals in
Minkowski space we get:
\begin{equation}\label{Ib2}
\begin{split}
&iK_{b}^{A, B}(E,{\bf k}, {\bf p})=\int_0^1dx\int_0^{1-x}dy{\alpha M^2 {D}_{A}\left(E, {\bf k}\right) {D}_{B}\left(E, {\bf p}\right)}\times\\
&\frac{1}{4 \pi} \left\{2 y \left[2 ME-2 \left({\bf k}^2+{\bf k} {\bf p}+{\bf p}^2\right)+x (2 {\bf k}+{\bf p}) ({\bf k}+2 {\bf p})\right]+x \left[4 ME-4 \left({\bf k}^2+{\bf k} {\bf p}+{\bf p}^2\right)\right.\right.\\
&\left.\left. x ({\bf k}+2 {\bf p})^2\right]+y^2 (2 {\bf k}+{\bf p})^2\right\}^{-3/2}={D}_{A}\left(E, {\bf k}\right)\mathcal{I}_b(E,{\bf k},{\bf p}){D}_{B}\left(E, {\bf p}\right),
\end{split}
\end{equation}
where
\begin{multline}
\mathcal{I}_b(E,{\bf k},{\bf p})=\frac{M^2\alpha}{4\left({\bf k}^2-ME+kp+{\bf p}^2\right)}
Q\left(\frac{{\bf k}^2+{{\bf p}}^2-ME}{kp}\right)\times\\
\left[\frac{ \arctan\left(\frac{{\bf k}+2 {\bf p}}{\sqrt{3 {\bf k}^2-4 ME}}\right)-\arctan\left(\frac{2 {\bf k}+{\bf p}}{\sqrt{3p-4 ME}}\right) }{{\bf k}-{\bf p}}\right].
\end{multline}
$K_{c, d}^{A, B}$ (Figure \ref{Coulomb_correction}(c, d)) is given by:
\begin{align}\label{_1}
K_{c, d}^{A, B}=&-\int\frac{d^4q}{(2\pi)^4}i \mathcal{D}_{photon}(-{\bf q}) \mathcal{D}^{A, B}(-p_0, -{\bf p}) S(-k_0, -{\bf k}) S(-q_0, -{\bf q})\times\\
\nonumber
 &\qquad\mathcal{D}^{A, B}(E+k_0, ) S(E+p_0, {\bf p}) S(k_0-p_0+q_0, {\bf k}-{\bf p}+{\bf q})=\\
 \nonumber
 &\int\frac{d^4q}{(2\pi)^4}\frac{i \alpha M \mathcal{D}^{A, B}\left(E-\frac{{\bf p}^2}{2 M}, -{\bf p}\right) \mathcal{D}^{A, B}\left(E-\frac{{\bf k}^2}{2 M}, {\bf q}\right)}{\left(\lambda ^2+{\bf q}^2\right) \left(-ME+{\bf k}^2+{\bf k} ({\bf q}-{\bf p})+{\bf p}^2-{\bf p} {\bf q}+{\bf q}^2\right)}=\\
 \nonumber
 &-\int_0^1dx\frac{\alpha M\mathcal{D}^{A, B}\left(E-\frac{{\bf p}^2}{2 M}, -{\bf p}\right) \mathcal{D}^{A, B}\left(E-\frac{{\bf k}^2}{2 M}, {\bf q}\right)}{4 \pi \sqrt{4 ME x+{\bf k}^2 (x-4) x-4 \lambda ^2-2 {\bf k} {\bf p} (x-2) x+{\bf p}^2 x^2-4 {\bf p}^2 x+4 \lambda ^2 x}}=\\
 &{D}_{A}\left(E, {\bf k}\right)\mathcal{I}_{c,d}(E,{\bf k},{\bf p}){D}_{B}\left(E, {\bf p}\right),
\end{align}where
\begin{equation}
\begin{split}
&\mathcal{I}_{c, d}(E,{\bf k},{\bf p})=\frac{\alpha M^2}{8 \pi ({\bf k}- {\bf p})kp}Q\left(\frac{{\bf k}^2+{{\bf p}}^2-ME}{kp}\right)\left[\log \left(2 ME-2 {\bf k}^2+2 {\bf k} {\bf p}-2 {\bf p}^2\right)\right.\\
&\left.-\log \left(-{\bf k} \sqrt{4 ME-3 {\bf k}^2+2 {\bf k} {\bf p}-3 {\bf p}^2}+{\bf p} \sqrt{4 ME-3 {\bf k}^2+2 {\bf k} {\bf p}-3 {\bf p}^2}+2 ME-{\bf k}^2-{\bf p}^2\right)\right].
\end{split}
\end{equation}
\subsection{The Doublet channel for the $p-d$ scattering}

\begin{equation}\label{p_d_STM_doublet}
\begin{split}
 &(iT^{i, j})^{\beta b}_{\alpha a}(E;{\bf {\bf k}}, {\bf
 p})=-i\frac{My_t^2}{2}(\sigma^j\sigma^i)^\beta_{\alpha}\delta^{a}_{b}
 \frac{1}{{\bf {\bf k}}^2+{\bf {\bf k}}^2+{\bf {\bf k}}\cdot {\bf p}-ME}\\
 &-i\frac{My_t^2}{2}(\sigma^j\sigma^i)^\beta_{\alpha}\delta^a_b
 \left(i\mathcal{I}_b-\mathcal{I}_B(E,{\bf q},{\bf p})\mathcal{D}_{photon}({\bf q}-{\bf p})\right)\\
 &+\frac{My_t^2}{2}\int\frac{\hbox{d}^3 {\bf q}}{(2\pi)^3}
 (\sigma^j\sigma^{\bf k})^\beta_{\gamma}
 \frac{ iT^{ik}(E;{\bf q, }{\bf {\bf {\bf k}}})^{\gamma c}_{\beta b}}{{\bf q}^2+{\bf {\bf p}}^2+{\bf q}\cdot {\bf p}-ME}{D}_t\left(E,{\bf q}\right)\delta^{a}_{c} \\
 &+\frac{My_t^2}{2}\int\frac{\hbox{d}^3 {\bf q}}{(2\pi)^3}(\sigma^j\sigma^{\bf k})^\beta_{\gamma}iT^{ik}(E;{\bf q, }{\bf {\bf {\bf k}}})^{\gamma c}_{\alpha b}
 {D}_t\left(E,{\bf q}\right)\delta^{a}_{c}i\mathcal{I}_b(E,{\bf q},{\bf p})\\
 &-\frac{My_t^2}{2}\int\frac{\hbox{d}^3 {\bf q}}{(2\pi)^3}T^{jk}(E;{\bf p, }{\bf {\bf {\bf k}}})^{\gamma c}_{\alpha b}
 {D}_t\left(E,{\bf q}\right)\delta^{ik}\delta^\beta_\gamma\left(\frac{1+\tau^3}{2}\right)^{a}_c\mathcal{I}_B(E,{\bf q},p)\mathcal{D}_{photon}({\bf q}-{\bf p})\\
&+\frac{My_ty_s}{2}\int\frac{\hbox{d}^3q}{(2\pi)^3}
{D}_s^{np}(E, {\bf q})
(\sigma^{\bf k})^\beta_\gamma (\tau^{3})^b_c(iS_{np}^i)_{\alpha
 a}^{\gamma c}\times\\
 &\qquad\qquad\qquad\left(\frac{1}{{\bf q}^2+{\bf q}\cdot{\bf
 p}+{\bf p}^2- M E}+i\mathcal{I}_b(E,{\bf q},{\bf p})\right)\\
 & \frac{My_ty_s}{2}\int\frac{\hbox{d}^3q}{(2\pi)^3}
{D}_{pp}\left(E,{\bf q}\right)
 (\sigma^{\bf k})^\beta_\gamma \left(\tau^{l=1-i\cdot 2}\right)^b_c\left(iP^{bl}\right)_{\alpha a}^{\gamma c}\\
 &\qquad\qquad\qquad\left(\frac{1}{{\bf q}^2+{\bf q}\cdot{\bf p}+{\bf p}^2- M E}+i\mathcal{I}_c(E,{\bf q},{\bf p})\right)\\
 \end{split}
\end{equation}

\begin{equation}
\begin{split}
&(iS_{np})^{\beta b}_{\alpha a}(E;{\bf k}, {\bf p})=-\frac{i M
 y_ty_s}{2}(\sigma^i)^\beta_\alpha(\tau^{3})^b_a\cdot
\frac{1}{{\bf k}^2+{\bf k}\cdot{\bf p}+{\bf p}^2-M
 E}\\
 &-iy_s^2\frac{1}{2}(\sigma^i)^\beta_\alpha(\tau^{3})^b_a
 \left(i\mathcal{I}_b(E,{\bf k},{\bf p})-\mathcal{I}_B(E,{\bf k},{\bf p})\mathcal{D}_{photon}({\bf k}-{\bf p})\right)\\
&+\frac{M
 y_ty_s}{2}\int\frac{\hbox{d}^3q}{\left(2\pi\right)^3}
{D}_t\left(E,{\bf q}\right)
{(\sigma^k)^\gamma_\alpha(\tau^{3})^b_c}(iT^{ik})_{\alpha a}^{\gamma c}\times\\
 &\qquad\qquad\qquad\left(\frac{1}{{\bf q}^2+{\bf
 q}\cdot{\bf p}+{\bf p}^2-M
 E}+i\mathcal{I}_b(E,{\bf q},{\bf p})\right)\\
&+\frac{M
 y_s^2}{2}\int\frac{\hbox{d}^3q}{\left(2\pi\right)^3}
{D}_s^{np}(E, {\bf q})
\frac{(\tau^{3})^b_c(\tau^{3})^c_d}{{\bf q}^2+{\bf
 q}\cdot{\bf p}+{\bf p}^2-M
 E}(iS_{np}^i)_{\alpha a}^{\beta d}\\
 &+\frac{My_s^2}{2}\int\frac{\hbox{d}^3 q}{(2\pi)^3}(\tau^{3})^b_c(\tau^{3})^c_d\left(iS_{np}\right)_{\alpha a}^{\beta d}
 {D}_t\left(E,{\bf q}\right)i\mathcal{I}_b(E,{\bf q},{\bf p})\\
 &-\frac{My_s^2}{2}\int\frac{\hbox{d}^3 {\bf q}}{(2\pi)^3}S_{np}^i(E;{\bf {\bf p}, }{\bf {\bf k}})^{\gamma c}_{\alpha b}
 {D}_t\left(E,{\bf q}\right)\delta^\beta_\gamma\left(\frac{1+\tau^3}{2}\right)^{a}_c\mathcal{I}_B(E,{\bf q},{\bf p})\mathcal{D}_{photon}({\bf q}-{\bf p})\\
&+\frac{My_s^2}{2}\int\frac{\hbox{d}^3q}{\left(2\pi\right)^3}
{D}_{pp}\left(E,{\bf q}\right)
(\tau^{1-i\cdot 2})^b_c(\tau^{3})^c_d(iP^i)_{\alpha a}^{\beta d}\times\\
 &\qquad\qquad\qquad\qquad\left(\frac{1}{{\bf q}^2+{\bf
 q}\cdot{\bf p}+{\bf p}^2-M
 E}+i\mathcal{I}_c(E,{\bf q},{\bf p})\right),
 \end{split}
 \end{equation}
 \begin{equation}
 \begin{split}
&(iP^i)^{\beta b}_{\alpha a}(E;{\bf k}, {\bf
 p})=-\frac{i M
 y_ty_s}{2}\cdot(\sigma^i)^\beta_\alpha(\tau^{B=1-i \cdot 2})^b_a\cdot
\frac{1}{{\bf k}^2+{\bf k}\cdot{\bf p}+{\bf p}^2-M E}\\
&+\frac{M
 y_ty_s}{2}\int\frac{\hbox{d}^3q}{\left(2\pi\right)^3}
{D}_t\left(E,{\bf q}\right)
{(\sigma^k)^\gamma_\alpha(\tau^{B=1+i\cdot 2})^b_c}(iT^{ik})_{\alpha a}^{\gamma c}\times\\
 &\qquad\qquad\qquad\left(\frac{1}{{\bf q}^2+{\bf
 q}\cdot{\bf p}+{\bf p}^2-M
 E}+i\mathcal{I}_d(E,{\bf q},{\bf p})\right)\\
&+\frac{M
 y_s^2}{2}\int\frac{\hbox{d}^3q}{\left(2\pi\right)^3}
{D}_s^{np}(E, {\bf q})
{(\tau^{3})^b_c(\tau^{B=1-i\cdot 2})^c_d}(iS_{np}^i)_{\alpha a}^{\beta d}\times\\
 &\qquad\qquad\qquad\left(\frac{1}{{\bf q}^2+{\bf
 q}\cdot{\bf p}+{\bf p}^2-M
 E}+i\mathcal{I}_d(E,{\bf q},{\bf p})\right)\\
&+\frac{M
 y_s^2}{2}\int\frac{\hbox{d}^3q}{\left(2\pi\right)^3}
\mathcal{D}_{pp}(E-\frac{{\bf q}^2}{2M}, {\bf q})
{(\tau^{1-i\cdot 2})^b_c(\tau^{B=1-i\cdot 2})^c_d}(iP^i)_{\alpha a}^{\beta d}\times\\
 &\qquad\qquad\qquad\left(\frac{1}{{\bf q}^2+{\bf
 q}\cdot{\bf p}+{\bf p}^2-M
 E}+i\mathcal{I}_b(E,{\bf q},{\bf p})\right)
\end{split}
\end{equation}

%% file: AppendixH_NLO.tex
Up to NLO, the three-body form has the form:
\begin{equation}
H(\Lambda)=H^{0}(\Lambda)+H^{(1)}(\Lambda)~
\end{equation}
where the NLO correction, $H^{(1)}(\Lambda)$, is giving by\cite{H_NLO,konig2}:
\begin{multline}
H^{(1)}(\Lambda)=\Lambda\left[-{3\pi}\frac{1+s_0^2}{128}\left(\rho_t+\rho_s\right)
\frac{1-\left(1/\sqrt{1+4s_0^2}\right)\sin(2s_0\log(\Lambda/\Lambda^*)+\arctan(1/2s_0))}{\sin^2\left(s_0\log(\Lambda/\Lambda^*)-\arctan(s_0)\right)}\right]-\\
-\frac{3\pi\left(1+s_0^2\right)}{64}\Biggl\{\frac{1}{\sqrt{3}}\left(\rho_s+\rho_t\right)\left(a_s^{-1}+\gamma_t\right)|B_{-1}|\mathcal{G}_1(B_{-1})-\frac{1}{2\sqrt{3}}\left(-\rho_s+\rho_t\right)\left(-a_s^{-1}+\gamma_t\right)|\tilde{B}_{-1}|\mathcal{G}_1(\tilde{B}_{-1})+\\
\frac{1}{2\sqrt{3}}\left(a_s^{-1}\rho_s+\rho_t\gamma_t\right)\mathcal{G}_1(0)-0.1252
\Biggr\}\sin^{-2}\left[s_0\log(\Lambda/\Lambda^*)-\arctan(s_0)\right]~,
\end{multline} 
where:
\begin{equation}
B_{-1}=\frac{I(is_0-1)}{1-I(is_0-1)}~,
\end{equation}
\begin{equation}
\tilde{B}_{-1}=\frac{I(is_0-1)}{1+\frac{1}{2}I(is_0-1)}~,
\end{equation}
\begin{equation}
	I(s) = \frac{8}{\sqrt{3}s} \frac{\sin(\frac{\pi s}{6})}{\cos(\frac{\pi s}{2})}~,
	\end{equation}
	and 
	\begin{equation}
	\mathcal{G}(x)=\cos\left[\arg(x)\right]\log(\Lambda)-\frac{1}{2s_0}\sin\left[2s_0\log(\frac{\Lambda}{\Lambda^*})+\arg(x)\right]
	\end{equation}

%% file: ap_H_S_EW.tex
\vspace{-0.2 cm}
In this appendix, we present the Hubbard-Stratonovich (H-S) transformation for a \pilesseft Lagrangian with an electro-weak interaction.

The two-body Lagrangian with electro-weak interaction has the form:
\begin{equation}
\mathcal{L}=\mathcal{L}_\text{strong}+\mathcal{L}_{\text{electroweak}},
\end{equation}
where $\mathcal{L}$ is two-body Lagrangian \cite{Chen_N_N}: 
\begin{equation}\label{lag2}
\mathcal{L}= N^\dagger\left(i\partial_0+\frac{\nabla^2}{2M}\right)N-\sum_{\mu}C_{0}^\mu\phi_\mu^\dagger\phi_\mu-
\frac{MC_{2}^\mu}{2}\left[\phi_\mu^\dagger\mathcal{O}_D\phi_\mu+h.c\right]~,
\end{equation}
where:
\begin{equation}\label{not}
\left(N^TP_{t, s}N\right)=\phi_{t, s}~,
\end{equation}
\begin{equation}
\mathcal{O}_D=\left(i\partial_0+\frac{\nabla^2}{4M}\right)~,
\end{equation}
and (see for example \cite{KSW1998_a}):
\begin{eqnarray}\label{eq_C0}
C_0^\mu&=&\dfrac{4\pi}{M}\dfrac{1}{\left(-\mu+\frac{1}{a_{\mu}}\right)}\\
C_2^\mu&=&C_2^{\mu}=\dfrac{4 \pi}{M \left(-\mu +\frac{1}{a_{\mu}}\right)^2}\dfrac{\rho_{\mu}}{2}~.
\end{eqnarray}

$\mathcal{L}_{\text{electroweak}}$ is the electro-weak part of the \pilesseft Lagrangian:
\begin{equation}\label{eq_lweak}
\mathcal{L}^\mu_{\text{electroweak}}\propto\mathcal{A}_{\mu }=\mathcal{A}_{\mu }^{\text{1B}}+\mathcal{A}_{\mu }^{\text{2B}}.
\end{equation}
the two-body part of the electro-weak current, $\mathcal{A}_{\mu }^{\text{2B}}$, has the form:
\begin{equation}\label{weak2}
\boldsymbol{ \mathcal{A}_{\mu }}=\sum_{\mu,\nu}L_{\mu\nu}\phi_\mu^\dagger\phi_\nu~,
\end{equation}
where $L_{\mu\nu}$ is the LEC that couples the two two-nucleon fields, e.g., for the weak interaction $L_{\mu\nu}=L_{ts}=L_{1,A}$. 
\begin{comment}
In order to find the right H-S transformation for $\mathcal{L}^1$ we will define:
\begin{eqnarray}
y_t&=&\frac{t}{\sqrt{c1}}\\
y_s&=&\frac{s}{\sqrt{c_2}}\\
\beta_1&=&\chi_1-{c_1} \varepsilon\\ 
\beta 2&=&\chi_2-{c_2} \varepsilon 
\end{eqnarray}
\end{comment}
In order to find the right H-S transformation for $\mathcal{L}$, we assume that after applying the H-S transformation, $\mathcal{L}$ is of the form:
\begin{equation}
\mathcal{L}^{H-S}_{\text{electroweak}}=\sum_{\mu,\nu=t, s}-\underbrace{\alpha_\mu\left(\mu^\dagger\phi_\mu+h.c\right)-\mu^\dagger\beta_\mu \mu}_{\text{strong part}}\underbrace{-\gamma_{\mu\nu}\phi_\nu^\dagger \mu-\gamma'_{\mu\nu}[\mu^\dagger \nu+h.c]}_{\text{electro-weak part}}~,
\end{equation}
where the H-S transformation is defined such that:
\begin{equation}
\int dt\int ds \exp\left(\mathcal{-L}^{H-S}_{\text{electroweak}}\right)=\exp\left[-\left(\sum_{\mu=t, s}A_\mu\phi_\mu^\dagger\phi_\mu+B_\mu\mathcal{O}_D\phi_\mu^\dagger\phi_\mu+L_{\mu\nu}\psi_\mu^\dagger\psi_\nu\right)\right]~.
\end{equation}

By setting:
\begin{eqnarray}
A_{\mu} &=& -C_0^{\mu}\\
B_{\mu} &=& -\frac{C_2^{\mu}M}{2}
\end{eqnarray}
and
\begin{eqnarray}
\alpha_{\mu} &=& y_{\mu}\\
\beta_{\mu} &=& \mathcal{O}_D-\sigma_{\mu}~,
\end{eqnarray}
we get that:
\begin{eqnarray}
\gamma_{\mu\nu}&=&\frac{L_{\mu\nu}}{\sqrt{C_2^\mu M}}+c\frac{C_2^\mu}{C_0^\mu\sqrt{ M C^\mu_2}}\\
\gamma'_{\mu\nu}&=&\frac{L_{\mu\nu}}{\sqrt{C_2^\mu C_2^\nu M^2}}-c\left(\frac{C_0^\mu C_2^\nu}{C_0^\nu \sqrt{C_2^\mu C_2^\nu}}
+\frac{C_0^\nu C_2^\mu}{C_0^\mu \sqrt{C_2^\mu C_2^\nu}}\right)~, 
\end{eqnarray}
where $c$ is an arbitrary constant that has to be determined by the original Lagrangian.

%% file: Bayesian_approach2.tex
In the following, we suggest an approach to study the truncation uncertainty and the size of the expansion parameter, of a set of observables whose expansion are $\langle{M_1}\rangle=\langle{M_1}\rangle_{\text LO} \cdot \left ( 1+c^{\text NLO}_{M_1} \delta + c^{\text N2LO}_{M_1} \delta^2+{\mathcal{O}} (\delta^3)\right)$.

\subsection{The Bayesian probability distribution of the expansion parameter}
We use information theory arguments to understand the expected behaviour of the expansion convergence rate. Let us look at the ratio of different expansion terms, 
\begin{equation}
r_{k-l}^{M_1}\delta^{k-l}=\left|\frac{c^{{\text {N}}^k\text{LO}}_{M_1}}{c^{{\text {N}}^l\text{LO}}_{M_1}}\right|.
\end{equation}
\pilesseft formalism suggests that $r_{k-l}^{M_1}$  should be a natural number. We interpret this as a statement regarding the nature of the distribution of these numbers. In laymen words, one would be surprised if these numbers deviate much from 1. In other words, these coefficients have some natural range of change $\frac{1}{\alpha} < r_{k-l}^{M_1} < \alpha$, where $\alpha$ is a measure of naturalness. One can expect that $\alpha$ is from a order  of 2-3, while bigger variations are acceptable as long as they are rare. From a Bayesian point of view, $r_{k-l}^{M_1}$ are independent and identically distributed random variable (i.i.d) with an average of about $1$ and their logarithm has (unknown) standard deviation $\log \alpha$. 

Information theory now states that the probability density function (pdf) $f(r)$ should maximize the entropy $S[f]=-\int dr f(r)\log f(r)$ subject to the constraints $\overline{\log r} =0$ and $\overline{ (\log r-\overline{\log r})^2}=\log \alpha$.
Thus, the log-average of $r_{k-l}^{M_1}\delta^{k-l}$ should be the expansion parameter $(k-l)\log{\delta}$. These lead to a pdf $f(r)$ which is a log-normal distribution.

One can now use a sample with size $n$ to estimate the expansion parameter, i.e., $\overline{\log{\delta}}=\frac{1}{n}\sum_{i=1}^n \frac{1}{k-l} \log\left( r_{k-l}^{M_1(i)}\delta^{k-l}\right) $. Then, by using Bayes theorem, the resulting distribution for the expansion parameter is Student's {\it t}-distribution with n-1 degrees of freedom $\frac{\log\delta-\overline{\log{\delta}}}{\overline{\sigma^2}/\sqrt{n}}\sim T (\overline{\sigma^2}, n-1)$.

\subsection{The Bayesian probability distribution of the truncation error of an expansion, given a prior for the expansion parameter}

A good estimate for the truncation error is the maximal coefficient in an expansion of order $k$, multiplied by $\delta^{k+1}$. In Refs.~\cite{Griesshammer:2015ahu, PhysRevC.92.024005, Cacciari2011} a Bayesian probability distribution is calculated for the truncation error under the assumption of a natural expansion, albeit in the case where the expansion parameter $\delta$ is known. In what follows, we combine their idea, with the probability distribution for the expansion parameter found above, to find the Bayesian probability distribution that the NLO result will deviate by $\Delta$ from the true value. Then, 
\begin{align}\label{eq_delta}
pr\left (\Delta \big | \left\{a^{\text NLO}_{M^k_1}\right\}_{k=1}^n\right)=
\int d\delta pr\left (\Delta \big | \left\{c^{\text NLO}_{M^k_1}\right\}_{k=1}^n, \delta\right)\cdot pr\left (\delta \big | \left\{a^{\text NLO}_{M^k_1}\right\}_{k=1}^n\right).
\end{align}
$pr (\Delta \big | \left\{c^{\text NLO}_{M^k_1}\right\}_{k=1}^n, \delta)$ is calculated in~Ref.~\cite{Griesshammer:2015ahu}, and at NLO, is roughly constant for $|\Delta|\le R_\xi (\delta)$, and decays as $1/|\Delta|^3$ for $|\Delta|\ge R_\xi (\delta)$, where $R_\xi (\delta)=max\left (1, \left\{c^{\text NLO}_{M^k_1}\right\}_{n=1}\right)\delta^2$. The exact functional form depends on the prior assumption for $\left\{c^{\text NLO}_{M^k_1}\right\}_{k=1}^n$, but a common behavior to all the checked priors is that at degree of belief of $\frac{k}{k+1}$ (translating to $\approx 67\%$ for NLO calculations) the resulting truncation error is less than $R_\xi (\delta) \delta^{k+1}$. 

As the pdf for $\delta$, we take the Student's {\it t}-distribution found in the previous subsection.